\newcommand{\lepton}{\ell}

\newcommand{\antinu}{\overline{\nu}}

\newcommand{\GeV}{\hbox{\rm GeV}}
\newcommand{\centi}{\hbox{\rm cm}}
\newcommand{\gt}{\rightarrow}

\def\cp-v{\mbox{${\hbox{$CP$\kern-0.6em\lower-.1ex\hbox{/}}}$}\, } 
\documentclass[aps,prstab,groupedaddress,floatfix,byrevtex,twoside,superscriptaddress,preprint]{revtex4}
\ifx\pdftexversion\undefined
  \usepackage[dvips]{graphicx}
\else
  \usepackage[pdftex]{graphicx}
\fi

\usepackage{color}
\usepackage{amsmath}
\usepackage[psamsfonts]{amssymb}
\usepackage{natbib}
\usepackage{hyperref}
\usepackage{pifont}
\graphicspath{{graphics-dir/}}
\setkeys{Gin}{width=\linewidth}

\begin{document}

\preprint{BNL-72369-2004}
\preprint{FNAL-TM-2259}
\preprint{LBNL-55478}

\title{Neutrino Factory and Beta Beam Experiments and Development}
\author{C.~Albright} \affiliation{Fermi National Accelerator Laboratory,
Batavia, IL 60510, USA}
\author{V.~Barger}\affiliation{Dept. of Physics, University of Wisconsin, Madison, WI 53706, USA}
\author{J.~Beacom} \affiliation{Fermi National Accelerator Laboratory, Batavia, IL 60510, USA} 
\author{J.S.~Berg} \affiliation{Brookhaven National Laboratory, Upton, NY 11973, USA}
\author{E.~Black} \affiliation{Illinois Institute of Technology, Physics Department, Chicago, IL 60616, USA} 
\author{A.~Blondel}\affiliation{DPNC, Section de Physique, Universit\'{e}
de Gen\`{e}ve, Switzerland} 
\author{S.~Bogacz}\affiliation{Jefferson Laboratory, 12000 Jefferson
Avenue, Newport News, VA 23606, USA}%
\author{S.~Brice} \affiliation{Fermi National Accelerator Laboratory,
Batavia, IL 60510, USA}
\author{S.~Caspi}\affiliation{Lawrence Berkeley National Laboratory, Berkeley, CA 94720, USA}
\author{W.~Chou} \affiliation{Fermi National Accelerator Laboratory, Batavia, IL 60510, USA} 
\author{M.~Cummings} \affiliation{Northern Illinois University, DeKalb, IL 60115, USA }
\author{R.~Fernow} \affiliation{Brookhaven National Laboratory, Upton, NY 11973, USA}
\author{D.~Finley} \affiliation{Fermi National Accelerator Laboratory, Batavia, IL 
60510, USA} 
\author{J.~Gallardo} \affiliation{Brookhaven National Laboratory, Upton, NY 11973, USA}
\author{S.~Geer} \affiliation{Fermi National Accelerator Laboratory,
Batavia, IL 60510, USA} 
\author{J.J.~Gomez-Cadenas}\affiliation{Departamento de F\'{i}sica Te\'{o}rica and IFIC, Universidad de Valencia, E-46100 Burjassot, Spain}
\author{M.~Goodman} \affiliation{Argonne National Laboratory, Argonne, IL 60439, USA}
\author{D.~Harris} \affiliation{Fermi National Accelerator Laboratory, Batavia, IL 
60510, USA} 
\author{P.~Huber}\affiliation{Technischen Universit\"{a}t M\"{u}nchen, Garching, Germany}
\author{A.~Jansson} \affiliation{Fermi National Accelerator Laboratory, Batavia, IL 60510, USA} 
\author{C.~Johnstone}\affiliation{Fermi National Accelerator Laboratory, Batavia, IL 60510, USA}
\author{S.~Kahn} \affiliation{Brookhaven National Laboratory, Upton, NY 11973, USA}
\author{D.~Kaplan} \affiliation{Illinois Institute of Technology, Physics Department, Chicago, IL 60616, USA}
\author{H.~Kirk}\affiliation{Brookhaven National Laboratory, Upton, NY 11973, USA}
\author{T.~Kobilarcik} \affiliation{Fermi National Accelerator Laboratory, Batavia, IL 
60510, USA} 
\author{M.~Lindner}\affiliation{Technischen Universit\"{a}t M\"{u}nchen, Garching, Germany}
\author{K.~McDonald} \affiliation{Princeton University, Joseph Henry Laboratories, Princeton, NJ 08544, USA} 
\author{O.~Mena}  \affiliation{Fermi National Accelerator Laboratory, Batavia, IL 
60510, USA}
\author{D.~Neuffer} \affiliation{Fermi National Accelerator Laboratory, Batavia, IL 
60510, USA} 
\author{V.~Palladino}\affiliation{INFN Napoli e Universit\`{a} Federico~II, Napoli, Italy} 
\author{R.~Palmer}  \affiliation{Brookhaven National Laboratory, Upton, NY 11973, USA}
\author{K.~Paul} \affiliation{University of Illinois at Urbana-Champaign, Urbana-Champaign, IL 60115, USA} 
\author{P.~Rapidis} \affiliation{Fermi National Accelerator Laboratory, Batavia, IL 
60510, USA} 
\author{N.~Solomey} \affiliation{Illinois Institute of Technology, Physics Department, Chicago, IL 60616, USA}
\author{P.~Spampinato}\affiliation{Oak Ridge National Laboratory, Oak Ridge, TN 37831, USA}
\author{D.~Summers}\affiliation{University of Mississippi, Oxford, MS
38677, USA}%
\author{Y.~Torun} \affiliation{Illinois Institute of Technology, Physics Department, Chicago, IL 60616, USA}
\author{K.~Whisnant} \affiliation{Iowa State University, Ames, IA  50011, USA} 
\author{W.~Winter}\affiliation{Technischen Universit\"{a}t M\"{u}nchen, Garching, Germany}
\author{M.~Zisman} \affiliation{Lawrence Berkeley National Laboratory,
Berkeley, CA 94720, USA}
\author{The Neutrino Factory and Muon Collider Collaboration}\noaffiliation%

\date{\today}

\begin{abstract}
The long-term prospects for fully exploring three-flavor mixing in the neutrino sector depend upon an ongoing and increased investment in the appropriate accelerator R\&D. Two new concepts have been proposed that would revolutionize neutrino experiments, namely the Neutrino Factory and the Beta Beam facility.
These new facilities would dramatically improve our ability to test the three-flavor mixing framework, measure \textsl{CP} violation in 
the lepton sector, and perhaps determine the neutrino mass hierarchy, and, 
if necessary, probe extremely small values of the mixing angle $\theta_{13}$.
The stunning sensitivity that could be achieved with a Neutrino Factory is 
described, together with our present understanding of the corresponding 
sensitivity that might be achieved with a Beta Beam facility. In the Beta Beam case,  
additional study is required to better understand the optimum Beta Beam 
energy, and the achievable sensitivity. Neither a Neutrino Factory nor 
a Beta Beam facility could be built without significant R\&D. An impressive Neutrino 
Factory R\&D effort has been ongoing in the U.S. and elsewhere over the 
last few years and significant progress has been made towards optimizing 
the design, developing and testing the required accelerator components, 
and significantly reducing the cost. The recent progress is described here. 
There has been no corresponding activity in the U.S. on Beta Beam facility design and, 
given the very limited resources, there is little prospect of starting 
a significant U.S. Beta Beam R\&D effort in the near future. However, the 
Beta Beam concept is interesting, and progress on its development in Europe should be followed. The Neutrino Factory R\&D program has reached a critical stage  in which support is required for two crucial international experiments 
and a third-generation international design study. If this support is 
forthcoming, a Neutrino Factory could be added to the Neutrino Community's 
road map in about a decade.

\end{abstract}
\maketitle


\begin{center}
\huge
\textbf{Preface}
\normalsize
\end{center}
In response to the remarkable recent discoveries in neutrino physics, the APS 
Divisions of Particles and Fields, and of Nuclear Physics, together with the 
APS Divisions of Astrophysics and the Physics of Beams, have organized a year
long \textit{Study on the Physics of Neutrinos}~\cite{aps-study} that began in the fall of 2003. 
Within the context of this study, the \textit{Neutrino Factory and Beta Beam Experiments and Development  Working Group} was charged with reviewing, and if possible advancing, our 
understanding of the physics capabilities and design issues for these two 
new types of future neutrino facilities. To fulfill this charge, the working 
group conducted a Workshop at ANL March 3--4, 2004. The presentations and 
discussion at this \textit{Neutrino Factory and Beta Beams Workshop}, 
together with the Neutrino Factory Design work 
of the \textit{Neutrino Factory and Muon Collider Collaboration}~\cite{MC}, form the basis 
for this report.   
Over the last few years, there have been a series of workshops that have 
explored the design and physics capabilities of Neutrino Factories. These 
meetings include the international NUFACT Workshop series~\cite{nufact99,nufact00,nufact01,nufact02,nufact03}, many smaller, more 
specialized, workshops focused on specific parts of Neutrino Factory design 
and technology, and two more detailed \textit{Feasibility Studies}~\cite{fs1,fs2}. In addition, 
a large body of literature documents the physics motivation for Neutrino 
Factories and the progress that has been made towards realizing this new 
type of neutrino facility. The Neutrino Factory related goals for the 
working group were therefore to (i) review and summarize the results of the 
extensive work already done, and (ii) to update the picture utilizing the 
design study resources of the \textit{Neutrino Factory and Muon Collider 
Collaboration}, and the latest results from those making detailed studies  
of the physics capabilities of Neutrino Factories.
The Beta Beam concept is several years younger than the Neutrino Factory
concept, and the community's understanding of both the physics capabilities 
and the required design parameters (particularly the beam energy) is still 
evolving. Beta Beam R\&D is being pursued in Europe, but there is no 
significant Beta Beam R\&D activity in the U.S. Hence, Beta Beam related 
goals of the working group were necessarily more modest than the equivalent 
Neutrino Factory related goals. We restricted our ambitions to reviewing 
the evolving understanding of the physics reach coming out of work from 
Beta Beam proponents in Europe, and the R\&D challenges that must 
be met before a Beta Beam facility could be built. Possibilities for 
Neutrino Factory and Beta Beam facilities seem to have caught the 
imagination of the community. We hope that 
this report goes some way towards documenting why, and what is required to 
make these new and very promising neutrino tools a reality.
\vspace{0.5in}
\large
\begin{flushright}
Steve Geer and Mike Zisman
\end{flushright}
\normalsize

\clearpage
\section{Introduction \label{sec1}}


Neutrino Factory~\cite{geer98,status_report,blondel} and Beta Beam~\cite{zucchelli} facilities  offer two exciting options for the 
long-term neutrino physics program. In the U.S. there has been a significant 
investment in developing the concepts and technologies required for a 
Neutrino Factory, but no equivalent investment in developing Beta Beams.
In the following we consider first the Neutrino Factory, and then the Beta Beam case. 

New accelerator technologies offer the possibility of building, in the not-too-distant future, an accelerator complex to produce and capture 
more than $10^{20}$ muons per year~\cite{status_report}.  
It has been proposed to build a Neutrino Factory by 
accelerating the muons from this intense source to 
energies of several tens of GeV, injecting them into a storage ring 
having long straight sections, and exploiting the 
intense neutrino beams that are produced by muons decaying 
in the straight sections. The decays
\begin{equation}
    \mu^{-}  \to  e^{-}\nu_{\mu}\bar{\nu}_{e}\; , \qquad 
    \mu^{+}  \to  e^{+}\bar{\nu}_{\mu}\nu_{e}
    \label{mumpdk}
\end{equation}
offer exciting possibilities to pursue the study of neutrino oscillations 
and neutrino interactions with exquisite precision. 

To create a sufficiently intense muon source, a 
Neutrino Factory requires an intense multi-GeV proton source capable of 
producing a primary proton beam with a beam power of 1~MW or more on target. 
This is just the proton source required in the medium term for Neutrino 
Superbeams. Hence, there is a natural evolution from Superbeam experiments 
in the medium term to Neutrino Factory experiments in the longer term. 

The physics case for a Neutrino Factory will depend upon results from the 
next round of planned neutrino oscillation experiments. If the unknown 
mixing angle $\theta_{13}$ is small, such that 
$\sin^{2}2\theta_{13} < O(10^{-2})$, 
or if there is a surprise and three-flavor mixing does not completely 
describe the observed phenomenology, then answers to some or all of the most 
important neutrino oscillation questions will require a Neutrino 
Factory. If $\sin^{2}2\theta_{13}$ is large, just below the present upper 
limit, and if there are no experimental surprises, the physics case for 
a Neutrino Factory will depend on the values of the oscillation parameters, 
the achievable sensitivity that will be demonstrated 
by the first generation of $\nu_e$ appearance experiments, 
and the nature of the second generation of basic physics questions that 
will emerge from the first round of results. 
In either case (large or small $\theta_{13}$), in about 
a decade the neutrino community may need to insert a Neutrino Factory into 
the global neutrino plan. The option to do this in the next 10--15~years 
will depend upon the accelerator R\&D that is done during the intervening period. 

In the U.S., the \textit{Neutrino Factory and Muon Collider Collaboration} (referred to
herein as the Muon Collaboration, or MC)~\cite{MC} is a 
collaboration of 130 scientists and engineers 
devoted to carrying out the accelerator R\&D that is needed before a Neutrino Factory 
could be inserted into the global plan. Much technical progress has been 
made over the last few years, and the required key accelerator experiments 
are now in the process of being proposed and approved. The 2001 HEPAP 
subpanel~\cite{HEPAP} recommended a level of support that is sufficient 
to perform the critical accelerator R\&D during the next 10--15 years. This 
support level 
significantly exceeds the present investment in Neutrino Factory R\&D. 
In addition to the U.S. effort, there are active Neutrino Factory R\&D 
groups in Europe~\cite{UK},~\cite{CERN} and Japan~\cite{JAPAN}, and much of the R\&D is performed and organized 
as an international endeavor. Thus, because a Neutrino Factory is 
potentially the key facility for the long-term neutrino program, Neutrino 
Factory R\&D is an important part of the \textit{present} global neutrino program.
Indeed, the key R\&D experiments are seeking funding now, and will need 
to be supported if Neutrino Factories are to be an option for the future.

Consider next Beta Beam facilities~\cite{zucchelli}, \cite{autin}. It has been proposed to modify the Neutrino Factory 
concept by injecting beta-unstable radioactive ions, rather than muons, into a storage ring with long straight sections. This would produce a pure 
$\nu_e$ or $\bar{\nu_e}$ beam, depending on the stored ion species. 
The very low \textsl{Q} value for the decay means that the resulting neutrino beam will have a very small divergence, but it also means that the parent ions 
must be accelerated to high energies to produce neutrinos with even modest 
energies. The baseline Beta Beam concept involves accelerating 
the radioactive ions in the CERN SPS, which yields neutrino beams with 
energies of a few hundred MeV. The sensitivity of these Beta Beams to 
small values of $\theta_{13}$ appears to be comparable with the ultimate 
sensitivity of Superbeam experiments. Better performance might be 
achieved with higher energy Beta Beams, requiring the ions to be accelerated 
to at least TeV energies. This requires further study. 
 This R\&D is currently being 
pursued in Europe, where the proponents hope that a Beta Beam facility together with 
a Superbeam at CERN and a very massive water Cerenkov detector in the 
Fr\'{e}jus tunnel, would yield a very exciting neutrino program.

In this report, we summarize the expected sensitivities of Neutrino Factory 
and Beta Beam neutrino oscillation experiments, and the status of the 
R\&D required before these exciting facilities could become a part of 
the neutrino community's global plan. Exploiting the enthusiastic involvement 
of the Muon Collaboration in the study, we also describe an updated 
Neutrino Factory design that demonstrates significant progress toward 
cost reduction for this ambitious facility. The report is organized as 
follows. 
Section~\ref{sec2} describes in some detail the Neutrino Factory and Beta Beam 
design concepts.
 In Section~\ref{sec3}, Neutrino Factory and Beta Beam properties are 
described and compared with conventional neutrino beams. 
The neutrino oscillation physics reach is presented in Section~\ref{sec4}.  
Progress on Neutrino Factory designs along with some comments on the possibility of a U.S.-based Beta Beam facility are  
discussed in Section~\ref{sec5}. The Neutrino Factory and Beta Beam 
R\&D programs are described in Section~\ref{sec6}. A summary is given in
Section~\ref{sec7} and some recommendations are presented in
Section~\ref{sec8}. Finally, in Appendix A a cost scaling with respect to
the Feasibility Study-II cost numbers is presented.

\section{Machine Concepts \label{sec2}}

In this Section we describe the basic concepts that are used to create a
Neutrino Factory or a Beta Beam facility. Though the details of the two
facilities are quite different, many of the required features have common
origins. Both facilities are ``secondary beam'' machines, that is, a
production beam is used to create the secondary beam that eventually provides
the neutrino flux for the detector.

For a Neutrino Factory, the production beam is a high intensity proton beam of
moderate energy (beams of 2--50 GeV have been considered by various groups)
that impinges on a target, typically a high-$Z$ material. The collisions
between the proton beam and the target nuclei produce a secondary pion beam
that quickly decays into a longer-lived (2.2 $\mu$s) muon beam. The remainder
of the Neutrino Factory is used to condition the muon beam (see Section~\ref{neufact}), accelerate it rapidly to the desired final energy of a few
tens of GeV, and store it in a decay ring having a long straight section
oriented such that decay neutrinos produced there will hit a detector located
thousands of kilometers from the source.

A Beta Beam facility is one in which a pure electron neutrino 
$\left(\text{from }\beta^{+}\right)$ or antineutrino $\left(\text{from
  }\beta^{-}\right)$ 
beam is produced from the decay of beta unstable
radioactive ions circulating in a storage ring. As was the case for the
Neutrino Factory, current Beta Beam facility concepts are based on using a
proton beam to hit a layered production target. In this case, nuclear
reactions are used to produce secondary particles of a beta-unstable nuclide.
The proposed approach uses either spallation neutrons from a high-$Z$ target
material or the incident protons themselves to generate the required reactions
in a low-$Z$ material. The nuclide of interest is then collected, ionized,
accumulated, and accelerated to its final energy. The process is relatively
slow, but this is acceptable as the lifetimes of the required nuclides, of
order 1~s, are sufficiently long.
\subsection{Neutrino Factory}
\label{neufact}
The various components of a Neutrino Factory, based in part on the most recent
Feasibility Study (Study-II, referred to herein as FS2)~\cite{fs2}
that was carried out jointly by BNL and the U.S. \textit{Neutrino Factory
  and Muon Collider Collaboration}, are
described briefly below. Details of the design discussed here are based on the
specific scenario of sending a neutrino beam from BNL to a detector in
Carlsbad, New Mexico. More generally, however, the design exemplifies a
Neutrino Factory for which two Feasibility Studies~\cite{fs1,fs2} have demonstrated technical
feasibility (provided the challenging component specifications are met),
established a cost baseline, and established the expected range of physics
performance. It is worth noting that the Neutrino Factory design we envision
could fit comfortably on the site of an existing laboratory, such as BNL or
FNAL. As part of the current Study, we have developed improved methods for
accomplishing some of the needed beam manipulations. These improvements are
included in the description below.

The main ingredients of a Neutrino Factory include:
\begin{itemize}
\item{\textbf{Proton Driver:}} Provides 1--4~MW of protons on target from an
upgraded AGS; a new booster at Fermilab would perform equivalently.
\item{\textbf{Target and Capture:}} A high-power target immersed in a 20~T
superconducting solenoidal field to capture pions produced in proton-nucleus
interactions. The high magnetic field at the target is smoothly tapered down
to a much lower value, 1.75~T, which is then maintained through the
bunching and phase rotation sections of the Neutrino Factory.
\item{\textbf{Bunching and Phase Rotation:}} We first accomplish the bunching
with rf cavities of modest gradient, whose frequencies change as we proceed
down the beam line. After bunching the beam, another set of rf cavities, with
higher gradients and again having decreasing frequencies as we proceed down
the beam line, is used to rotate the beam in longitudinal phase space to
reduce its energy spread.
\item{\textbf{Cooling:}} A solenoidal focusing channel, with high-gradient
201.25~MHz rf cavities and LiH absorbers, cools the transverse normalized rms
emittance from 17~mm$\cdot$rad to about 7~mm$\cdot$rad. This takes place at a
central muon momentum of 220~MeV/c.
\item{\textbf{Acceleration:}} A superconducting linac with solenoidal focusing
is used to raise the muon beam energy to 1.5~GeV, followed by a Recirculating
Linear Accelerator (RLA), arranged in a ``dogbone'' geometry, to provide a 5~GeV muon beam. Thereafter, a pair of cascaded Fixed-Field, Alternating
Gradient (FFAG) rings, having quadrupole triplet focusing, is used to reach 20~GeV. Additional FFAG stages could be added to reach a higher beam energy, if
the physics requires this.
\item{\textbf{Storage Ring:}} We employ a compact racetrack-shaped
superconducting storage ring in which $\approx35$\% of the stored muons decay
toward a detector located some 3000~km from the ring. Muons survive for
roughly 500 turns.
\end{itemize}
\subsubsection{Proton Driver}
\begin{figure}[tbh]
\includegraphics[width=5.5in]{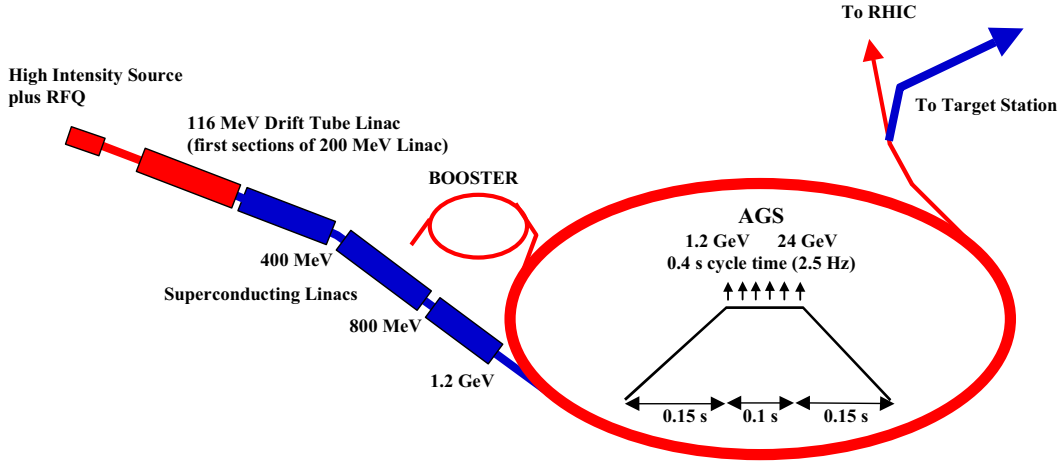}
\caption{(Color) AGS proton driver layout.}%
\label{Proton:bnl}%
\end{figure}

The proton driver considered in FS2, and taken here as well, is an upgrade of
the BNL Alternating Gradient Synchrotron (AGS) and uses most of the existing
components and facilities; parameters are listed in Table~\ref{Proton:tb1}. To
serve as the proton driver for a Neutrino Factory, the existing booster would
be replaced by a 1.2~GeV superconducting proton linac. The modified layout is
shown in Fig.~\ref{Proton:bnl}. 
The AGS repetition rate would be increased from 0.5~Hz to 2.5~Hz
by adding power supplies to permit ramping the ring more quickly. No new
technology is required for this---the existing supplies would be replicated
and the magnet strings would be split into six sectors rather than the two
used presently. The total proton charge ($10^{14}$ ppp in six bunches) is only
40\% higher than the current performance of the AGS. However, the bunches
required for a Neutrino Factory are shorter than those used in the AGS at
present, so there is a large increase in peak current and concomitant need for
an improved vacuum chamber; this is included in the upgrade. The six bunches
are extracted separately, spaced by 20~ms, so that the target and rf systems
that follow need only deal with single bunches at an instantaneous repetition
rate of 50~Hz (average rate of 15~Hz). The average proton beam power is 1~MW.
A possible future upgrade to $2\times10^{14}$~ppp and 5~Hz could give an
average beam power of 4~MW. At this higher intensity, a superconducting bunch
compressor ring would be needed to maintain the rms bunch length at 3~ns.

If the facility were built at Fermilab, the proton driver would be newly
constructed. A number of technical options are presently being explored~\cite{fnal1},\cite{fnal2}.
\begin{table}[tbh]
\caption{Proton driver parameters for BNL design.}%
\label{Proton:tb1}
\begin{ruledtabular}
\begin{tabular}[c]{lc}
\multicolumn{2}{c}{AGS}\\%
\hline
Total beam power (MW) & 1\\
Beam energy (GeV) & 24\\
Average beam current ($\mu$A) & 42\\
Cycle time (ms) & 400\\
Number of protons per fill & $1\times10^{14}$\\
Average circulating current (A) & 6\\
No. of bunches per fill & 6\\
No. of protons per bunch & $1.7\times10^{13}$\\
Time between extracted bunches (ms) & 20\\
Bunch length at extraction, rms (ns) & 3\\
\end{tabular}
\end{ruledtabular}
\end{table}
\subsubsection{Target and Capture}
\begin{figure}[tbh]
\includegraphics*{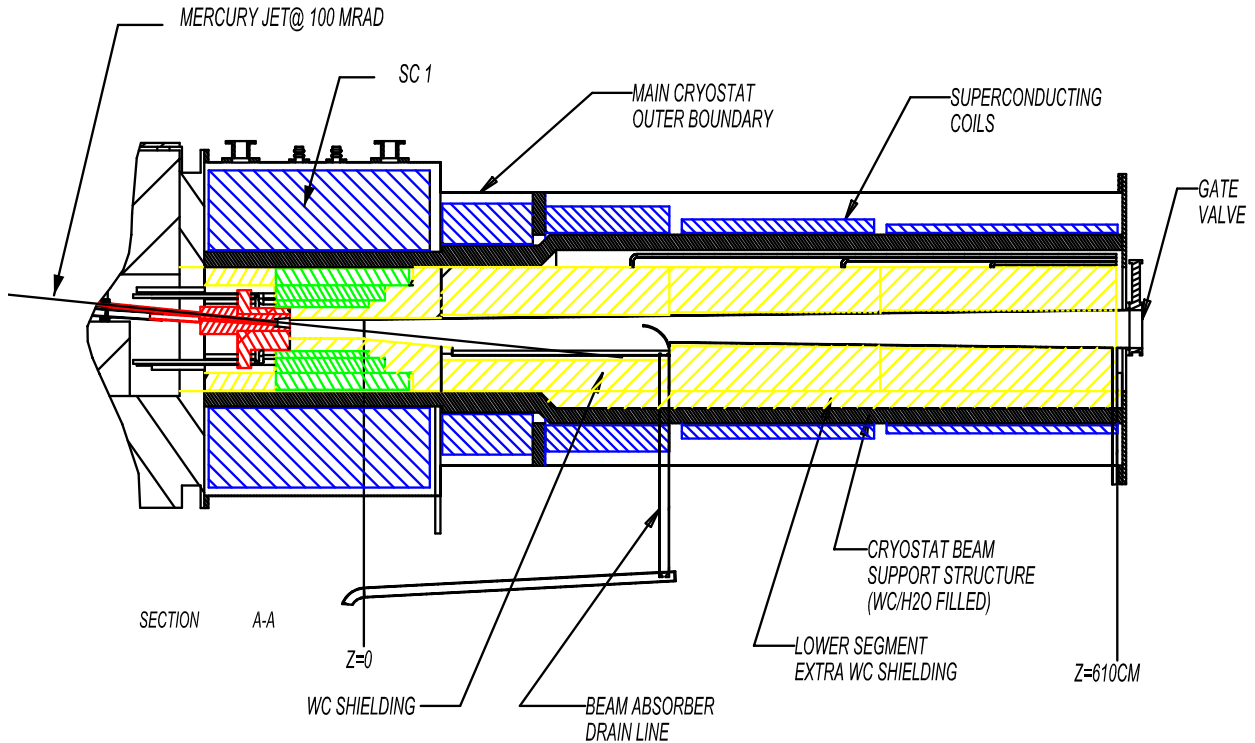}
\caption{(Color) Target, capture solenoids and mercury containment.}%
\label{tgtc}%
\end{figure}

A mercury-jet target is chosen to give a high yield of pions per MW of
incident proton power. The 1-cm-diameter jet is continuous, and is tilted with
respect to the beam axis. The target layout is shown in Fig.~\ref{tgtc}.
We assume that the thermal shock from the interacting proton bunch
fully disperses the mercury, so the jet must have a velocity of 20--30 m/s to
allow the target material to be renewed before the next proton bunch arrives.
Calculations of pion yields that reflect the detailed magnetic geometry of the
target area have been performed with the MARS code~\cite{mars1} and are reported in Section~\ref{sec5}. The FS2 design was updated for the present study to improve muon
throughput. To avoid mechanical fatigue problems, a mercury pool serves as the
beam dump. This pool is part of the overall target system---its mercury is
circulated through the mercury jet nozzle after passing through a heat exchanger.
Pions emerging from the target are captured and focused down the decay channel
by a solenoidal field that is 20~T at the target center, and tapers down, over
12~m, to 1.75~T. The 20~T solenoid, with a resistive magnet insert and
superconducting outer coil, is similar in character to higher field (up to 45~T), but smaller bore, magnets existing at several laboratories~\cite{ITERmag}.
The magnet insert is made with hollow copper conductor having ceramic
insulation to withstand radiation. MARS simulations~\cite{mars2} of
radiation levels show that, with the shielding provided, both the copper and
superconducting magnets will have reasonable lifetime.
\subsubsection{Buncher and Phase Rotation}
Pions, and the muons into which they decay, are generated in the target over a
very wide range of energies, but in a short time pulse ($\approx3$~ns rms).
 To prepare the muon beam for acceleration thus requires significant
``conditioning.'' First, the bunch is drifted to develop an energy
correlation, with higher energy particles at the head and lower energy
particles at the tail of the bunch. Next, the long bunch is separated into a
number of shorter bunches suitable for capture and acceleration in a 201-MHz
rf system. This is done with a series of rf cavities having frequencies that
decrease along the beam line, separated by suitably chosen drift spaces. The
resultant bunch train still has a substantial energy correlation, with the
higher energy bunches first and progressively lower energy bunches coming
behind. The large energy tilt is then ``phase rotated'', using additional rf
cavities and drifts, into a bunch train with a longer time duration and a
lower energy spread. The beam at the end of the buncher and phase rotation
section has an average momentum of about 220~MeV/c.
The proposed system is based on standard rf technology, and is expected to be
much more cost effective than the induction-linac-based system considered in
Ref.~\cite{fs2}. A fringe benefit of the rf-based system is the ability to transport both
signs of muon simultaneously.
\subsubsection{Cooling}
Transverse emittance cooling is achieved by lowering the beam energy in LiH
absorbers, interspersed with rf acceleration to keep the average energy
constant. Both transverse and longitudinal momenta are lowered in the
absorbers, but only the longitudinal momentum is restored by the rf system.
The emittance increase from Coulomb scattering is controlled by maintaining
the focusing strength such that the angular spread of the beam at the absorber
locations is reasonably large. In the present cooling lattice, the energy
absorbers are attached directly to the apertures of the rf cavity, thus
serving the dual purposes of closing the cavity apertures electromagnetically
(increasing the cavity shunt impedance) and providing energy loss. Compared
with the approach used in FS2, the absorbers are more distributed, and do not
lend themselves to being located at an optical focus. Therefore, the focusing
is kept essentially constant along the cooling channel, but at a beta function
somewhat higher than the minimum value achieved in FS2. A straightforward
Focus-Focus (FOFO) lattice is employed. The solenoidal fields in each
half-cell alternate in sign, giving rise to a sinusoidal field variation along
the channel.
Use of solid absorbers instead of the liquid-hydrogen absorbers assumed in FS2
will considerably simplify the cooling channel, and the new magnet
requirements are also more modest, since fewer and weaker components are
needed compared with FS2. Together, these features reduce the cost of the
cooling channel with respect to the FS2 design. Although the cooling
performance is reduced, the overall throughput is comparable to that in FS2
due to the increased acceptance built into the downstream acceleration system.
Here too, the ability to utilize both signs of muon is available.
\subsubsection{Acceleration}
\begin{figure}[tbhp!]
\includegraphics{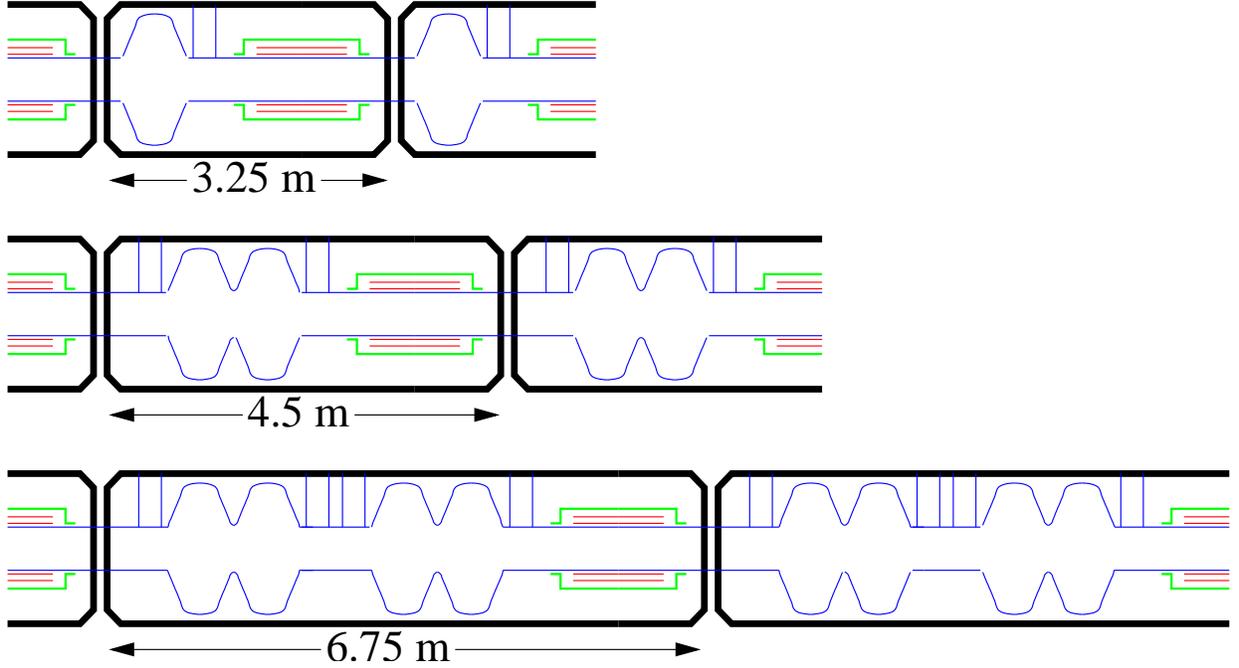}
\caption{(Color) Layouts of superconducting linac
pre-accelerator cryomodules. Blue lines are the SC walls of the cavities and
solenoid coils are indicated in red. The dimension of the cryomodules are
shown in Table~\ref{tab:acc:cryo} and Table~\ref{tab:acc:linac} summarizes parameters for the linac.}%
\label{fig:acc:cryomod}%
\end{figure}

Parameters of the acceleration system are listed in Table~\ref{tab:acc:parm}.
A matching section, using normal conducting rf systems, matches the cooling
channel optics to the requirements of a  superconducting rf linac with
solenoidal focusing which raises the energy to 1.5~GeV. The linac is in three parts (see Section~\ref{sec5-sub2}). The first part has only a
single-cell 201~MHz cavity per period. The second part, with longer period,
has a 2-cell rf cavity unit per period. The third part, as a still longer
period becomes possible, accommodates two 2-cell cavity units per
period. Figure~\ref{fig:acc:cryomod} shows the three cryomodule types that make
up the pre-accelerator linac. 

\begin{figure}[tbh!]
\includegraphics{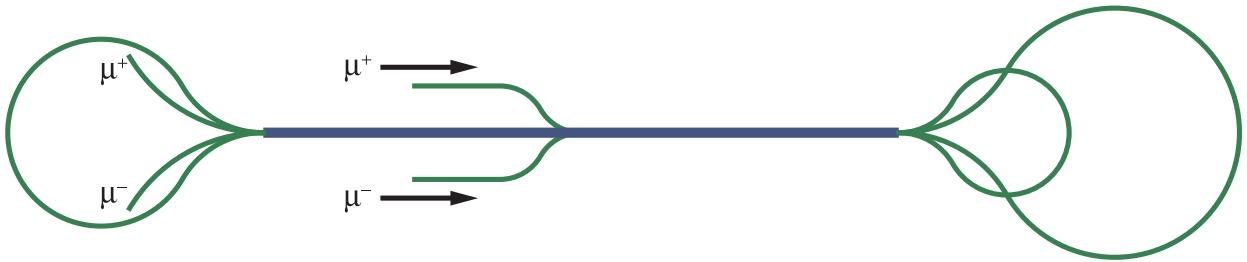}

\caption{(Color) Layout of the RLA.}%
\label{fig:acc:rlalinac}%
\end{figure}

This linac is followed by a 3.5-pass \textsl{dogbone} RLA (see Fig.~\ref{fig:acc:rlalinac}) that raises the energy from
1.5 to 5~GeV. The RLA uses four 2-cell superconducting
rf cavity structures per cell, and utilizes
quadrupole triplet (as opposed to solenoidal) focusing. 

Following the RLA are two cascaded FFAG rings that increase the beam energy
from 5--10~GeV, and 10--20~GeV, respectively. Each ring uses combined-function
magnets arranged in a triplet (F-D-F) focusing arrangement. The lower energy
FFAG ring has a circumference of about 400~m; the higher energy ring is about
500~m in circumference. As discussed in Section~\ref{sec5-sub2}, an effort was made to
achieve a reasonably cost-optimized design. Without detailed engineering, it
is not possible to fully optimize costs, but we have employed general formulae
that properly represent the cost trends and that were considered adequate to
make choices at the present stage of the design. As the acceleration system
was one of the dominant cost items in FS2, we are confident that the approach
adopted here will result in a less expensive Neutrino Factory facility with
essentially the same performance as calculated for the FS2 design.
Achieving a higher beam energy would require additional FFAG acceleration stages.
\begin{table}[tbh]
\caption{Main parameters of the muon accelerator driver.}%
\label{tab:acc:parm}
\begin{ruledtabular}
\begin{tabular}[c]{lc}
Injection momentum (MeV/c) & 273\\
Injection kinetic energy (MeV) & 187\\
Final total energy (GeV) & 20\\
Initial normalized acceptance (mm-rad) & 30\\
\quad rms normalized emittance (mm-rad) & 3.84\\
Initial longitudinal acceptance, $\Delta pL_{b}/m_{\mu}c$ (mm) & 150\\
\quad Total energy spread, $\Delta E$ (MeV)& $\pm45.8$\\
\quad Total time-of-flight (ns) & $\pm1.16$\\
\quad rms energy spread (MeV) &19.8\\
\quad rms time-of-flight (ns) &0.501 \\
Number of bunches per pulse & 89\\
Peak number of particles per bunch & $1.1\times10^{11}$\\
Number of particles per pulse (per charge) & $3\times10^{12}$\\
Bunch frequency\textbf{/}accelerating frequency (MHz) & 201.25\textbf{/}201.25\\
Average beam power (per charge) (kW) & 144\\
\end{tabular}
\end{ruledtabular}
\end{table}
\subsubsection{Storage Ring}
After acceleration in the final FFAG ring, the muons are injected into the
upward-going straight section of a racetrack-shaped storage ring with a
circumference of 358~m. Parameters of the ring are summarized in
Table~\ref{SRING:tb}. High-field superconducting arc magnets are used to
minimize the arc length and maximize the fraction (35\%) of muons that decay
 in the downward-going straight, generating neutrinos headed toward the
detector located some 3000~km away.

All muons are allowed to decay; the maximum heat load from their decay
electrons is 42~kW (126~W/m). This load is too high to be dissipated in the
superconducting coils. For FS2, a magnet design was chosen that allows the
majority of these electrons to exit between separate upper and lower
cryostats, and be dissipated in a dump at room temperature. To maintain the
vertical cryostat separation in focusing elements, skew quadrupoles are
employed in place of standard quadrupoles. In order to maximize the average
bending field, Nb$_{3}$Sn pancake coils are employed. One coil of the bending
magnet is extended and used as one half of the previous (or following) skew
quadrupole to minimize unused space. 
 For site-specific reasons, the ring is kept
above the local water table and is placed on a roughly 30-m-high berm. This
requirement places a premium on a compact storage ring. In the present study,
no attempt was made to revisit the design of the FS2 storage ring. For further
technical details on this component, see FS2, Ref.~\cite{fs2}.
\begin{table}[tb]
\caption{Muon storage ring parameters.}%
\label{SRING:tb}
\begin{ruledtabular}
\begin{tabular}[c]{ll}
Energy (GeV) & 20\\
Circumference (m) & 358.18\\
Normalized transverse acceptance (mm-rad) & 30\\
Energy acceptance (\%) & 2.2\\
\hline
\multicolumn{2}{c}{Arc}\\
\hline
Length (m) & 53.09\\
No. cells per arc & 10\\
Cell length (m) & 5.3\\
Phase advance ($\deg$) & 60\\
Dipole length (m) & 1.89\\
Dipole field (T) & 6.93\\
Skew quadrupole length (m) & 0.76\\
Skew quadrupole gradient (T/m) & 35\\
$\beta_{\text{max}}$ (m) & 8.6\\
\hline
\multicolumn{2}{c}{Production Straight}\\
\hline
Length (m) & 126\\
$\beta_{\text{max}}$ (m) & 200\\
\end{tabular}
\end{ruledtabular}
\end{table}

The footprint of a Neutrino Factory is reasonably small, and such a
machine would fit easily on the site of an existing laboratory.

\subsection{Beta Beam Facility}
The idea of a Beta Beam facility was first proposed by P. Zucchelli in 2002~\cite{zucchelli}. As the name suggests, it employs beams of beta-unstable
nuclides. By accelerating these ions to high energy and storing them in a
decay ring (analogous to that used for a muon-based Neutrino Factory) a very
pure beam of electron neutrinos (or antineutrinos) can be produced. As the
kinematics of the beta decay is well understood, the energy distribution of
the neutrinos can be predicted to a very high accuracy. Furthermore, as the
energy of the beta decay is low compared with that for muon decay, the
resulting neutrino beam has a small divergence.

For low-$Z$ beta-unstable nuclides, typical decay times are measured in
seconds. Thus, there is not so high a premium on rapid acceleration as is true
for a Neutrino Factory, and conventional (or even existing) accelerators could
be used for acceleration in a Beta Beam facility.
Two ion species, both having lifetimes on the order of 1 s, have been 
identified as optimal candidates: $^{6}$He for producing antineutrinos and
$^{18}$Ne for neutrinos.

Following the initial proposal, a study group was formed at CERN to
investigate the feasibility of the idea, and, in particular, to evaluate the
possibility of using existing CERN machines to accelerate the radioactive
ions. Their study took an energy of $\gamma=150$ for $^{6}$He, which
corresponds to the top energy of the SPS for this species and also matches the
distance to the proposed neutrino laboratory in the Fr\'{e}jus tunnel rather well.

In the spring of 2003, a European collaboration, the Beta Beam Study Group,
was formed. Eventually, they obtained funding from the European Union to
produce a conceptual design study. Here, we take our information from recent
presentations made by members of this group~\cite{BetaBeamWGpage}.

The EU Beta Beam Study Group has undertaken the study of a Beta Beam facility
with the goal of presenting a coherent and realistic scenario for such a
device. Their present ``boundary conditions'' are to re-use a maximum of
existing (CERN) infrastructure and to base the design on known technology---or
reasonable extrapolations thereof. In this sense, the approach taken is
similar to that of the Neutrino Factory feasibility studies.

For practical reasons, the Beta Beam study was included in the larger context
of the EURISOL study, due to the large synergies between the two at the low
energy end. (The EURISOL study aims to build a next-generation facility for
on-line production of radioactive isotopes, including those needed for the
Beta Beam facility.)

The basic ingredients of a Beta Beam facility are:
\begin{itemize}
\item{\textbf{Proton Driver:}} A 2.2\textbf{\ }GeV proton beam from the
proposed Super Proton Linac (SPL) at CERN would be used to initiate the
nuclear reactions that ultimately generate the required beta-unstable nuclides
($^{6}$He is used as the antineutrino source and $^{18}$Ne is used as the
neutrino source).
\item{\textbf{ISOL Target and Ion Source:}} The target system is patterned
after that of the EURISOL facility~\cite{EURISOLref}. For $^{6}$He production,
the target core would be a heavy metal (mercury~\cite{helge})
that converts the incoming proton beam into a neutron flux. Surrounding the
core is a cylinder of BeO~\cite{koster} that produces $^{6}$He via the $^{9}$Be(n,$\alpha$)
reaction. $^{18}$Ne would be produced via direct proton spallation on a MgO
target~\cite{ravn2}. The nuclides of interest will be extracted from the target as neutral
species, and so must be ionized to produce the beam to be accelerated. The
proposed ion source technology, shown in Fig.~\ref{ISOLtargetFig}, is based on
a pulsed ``ECR-duoplasmatron.''
\begin{figure}
\includegraphics[width=3.5in,angle=-90]{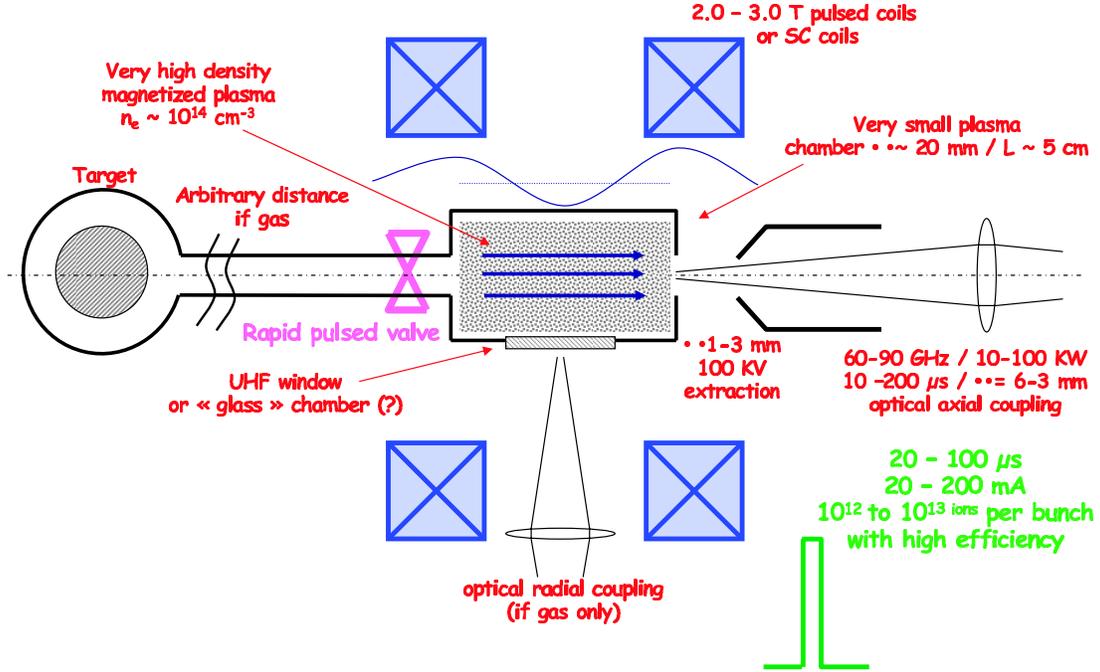}
\caption{(Color) Proposed ion source system for production of $^{6}$He beam.}
\label{ISOLtargetFig}
\end{figure}
\item{\textbf{Acceleration:}} Low energy acceleration would make use of a linac
to accelerate the nuclide of interest to 20--100~MeV/u, followed by a Rapid
Cycling Synchrotron (RCS) with multi-turn injection that would accelerate the
ion beam to 300~MeV/u. This system would feed the CERN PS with 16~bunches
(2.5 $\times$ 10$^{12}$ ions per bunch), which would be merged to 8 bunches
during the acceleration cycle to $\gamma=9$. Finally, the bunches would be 
transferred to the SPS and accelerated to $\gamma\thickapprox150$, which
corresponds to the maximum magnetic rigidity of that accelerator.
\item{\textbf{Decay Ring:}} The racetrack decay ring would have the same
circumference as the SPS (6880 m), with a long straight section, some 2500~m,
aimed at the detector. At the final energy, the lifetime of the beam becomes
minutes rather than seconds. Stacking is required to load the ring with enough
ions to get an acceptable neutrino flux. 
\end{itemize}
The parallels with the Neutrino Factory are obvious. The main difference
between the two types of facility is in the initial capture and beam
preparation. In the Neutrino Factory, the beam must be bunched, phase rotated,
and ionization cooled. In the Beta Beam facility, the beam must be collected,
ionized, and bunched.
\subsubsection{Proton Driver}
\begin{figure}[hptb!]
\includegraphics{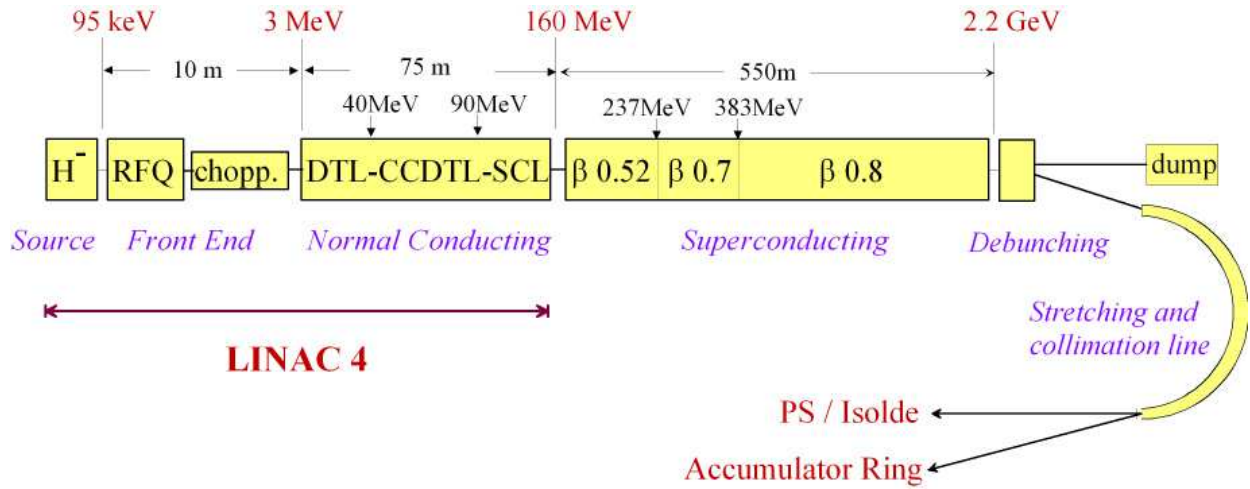}%
\caption{(Color) Baseline layout of the SPL facility at CERN.}%
\label{fig:SPLlayout}%
\end{figure}

The proposed proton driver for the Beta Beam facility is the SPL, a 2.2~GeV
Superconducting Proton Linac~\cite{SPLref}\ presently being designed at CERN
to serve both the LHC and the EURISOL facility. The machine will operate at 50
Hz and will be designed to provide up to 4~MW of proton beam power. The
present scenario is illustrated in Fig.~\ref{fig:SPLlayout}. It is anticipated
that the ISOL target will require only about 5\% of the proton beam power,
i.e., about 200~kW. %
\subsubsection{ISOL Target and Ion Source}
\begin{figure}[hptb!]
\includegraphics[width=3.6391in]{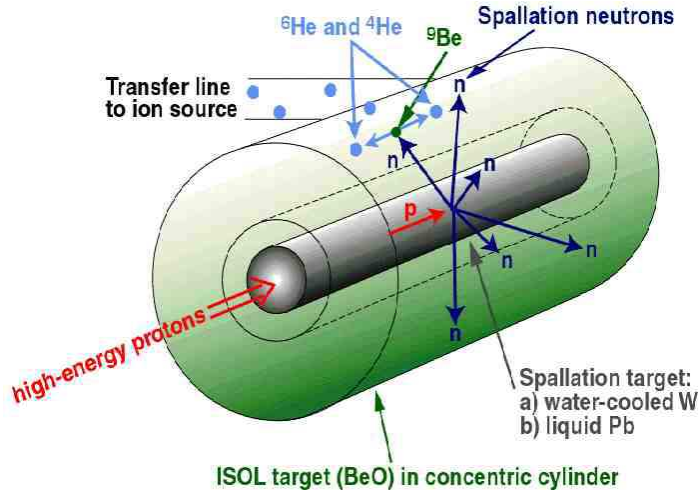}%
\caption{(Color) Proposed ISOL-type target for production of $^{6}$He beam.}%
\label{ISOLtgtPict}%
\end{figure}

As noted earlier, the target for $^{6}$He will use a water-cooled tungsten or
a liquid-lead core to serve as a proton-to-neutron converter. Surrounding this
core will be a cylinder of BeO, as shown in Fig.~\ref{ISOLtgtPict}. In the
case of $^{18}$Ne, a more straightforward approach will suffice. The proton
beam will impinge directly on a MgO target, producing the required nuclide via spallation.
An ion source capable of producing the required intense pulses is proposed;
development work on this device (see Fig.~\ref{ISOLtargetFig}) is under way at
Grenoble~\cite{sortais}. The device uses a very high density plasma
($n_{e}\thicksim10^{14}$ cm$^{-3}$) in a 2--3~T solenoidal field and operates
at 60--90~GHz. It is expected to provide pulses of 10$^{12}$--10$^{13}$ ions
per bunch.%

\subsubsection{Acceleration}
The proposed acceleration scheme is based on the existing CERN machines (PS
and SPS). Initial acceleration would be via a linac, followed by a rapid
cycling synchrotron that would be filled by multiturn injection. The RCS
would provide a single bunch, 150~ns long, at 300~MeV/u. The PS is a
relatively slow machine, and this results in substantial radiation levels due
to decays while the beam energy, and hence the lifetime, is low. A
rapid-cycling PS replacement would be of considerable benefit in this regard,
though it is not part of the baseline scenario. Another idea that merit
consideration is the use of a FFAG, which perhaps could be used to
accelerate muons at a latter time.

The SPS space-charge limit at
injection is another issue to deal with, and will likely require a transverse
emittance blowup to manage it. A new 40~MHz rf system will be added to the
existing 200~MHz system in the SPS to accelerate the beam to $\gamma=150.$
\subsubsection{Decay Ring}
The beam is transferred at full energy to a racetrack-shaped Decay Ring having
the same circumference as the SPS. The length of the decay straight section
(the one aimed at the detector) is chosen to permit about 35\% of the decays
 to occur there. At full energy, the lifetime is minutes rather than seconds.
This allows---and also \textit{requires}---the beam to be stacked in the Decay
Ring to provide the required decay intensity. The proposed stacking technique,
asymmetric bunch merging, is based on somewhat complicated rf gymnastics, but
has already been shown to work experimentally in tests~\cite{BunchMergeRef}.
An interesting possibility that has arisen only recently is the idea of
storing both $^{6}$He and $^{18}$Ne in the ring simultaneously. This requires
that the neon beam have the same rigidity as the helium beam, which
corresponds to $\gamma_{Ne}=250.$ For a detector at Fr\'{e}jus, the optimum
energies~\cite{mezzetto} are $\gamma_{He}=60$ and $\gamma_{Ne}=100.$

\section{Beam properties \label{sec3}}

The most important neutrino oscillation physics questions that 
we wish to address in the coming decades require the study of 
$\nu_e \leftrightarrow \nu_\mu$ transitions in long baseline 
experiments. Conventional neutrino 
beams are almost pure $\nu_\mu$ beams, which therefore permit the 
study of $\nu_\mu \to \nu_e$ oscillations. The experiments 
must look for $\nu_e$ CC interactions in a distant detector. Backgrounds 
that fake $\nu_e$ CC interactions, together with a small $\nu_e$ 
component in the initial beam, account for $O(1\%)$ of the total interaction 
rate. This makes it difficult for experiments using conventional neutrino 
beams to probe very small oscillation amplitudes, below the $0.01-0.001$ 
range. This limitation motivates new types of neutrino facilities that  
provide $\nu_e$ beams, permitting the search for $\nu_e \to \nu_\mu$ 
oscillations, and if the beam energy is above the $\nu_\tau$ CC interaction 
threshold, the search for $\nu_e \to \nu_\tau$ oscillations. 
Neutrino Factory and Beta Beam facilities both provide $\nu_e$ (and 
$\bar{\nu}_e$) beams, but with somewhat different beam properties.
We will begin by describing Neutrino Factory beams, and then describe 
Beta Beam facility beams.
\subsection{Neutrino Factory Beams}
\begin{figure}[hbtp!]
\includegraphics[width=4in]{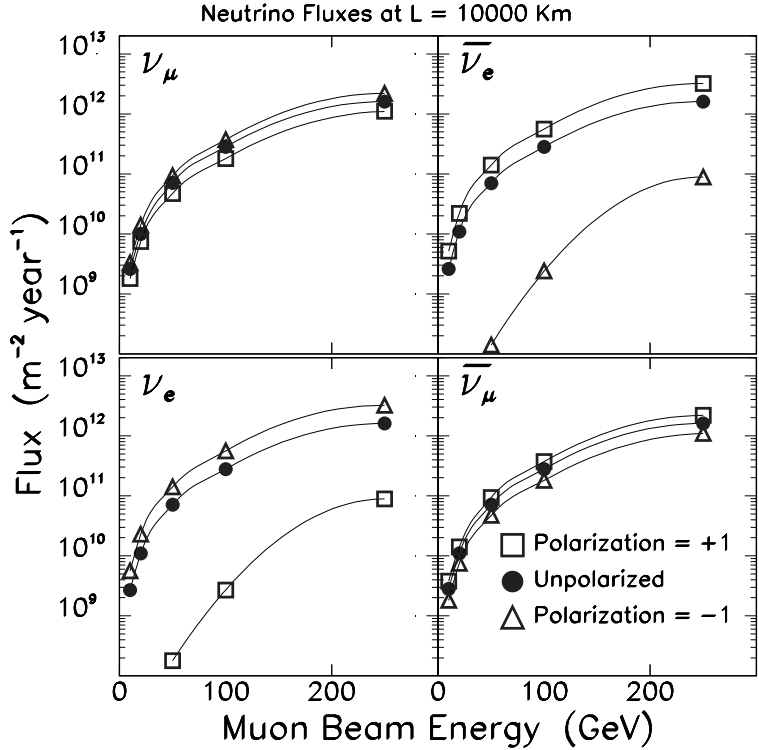}
\caption{Calculated $\nu$ and $\bar{\nu}$ fluxes in the
absence of oscillations at a far site located 
10000 km from a Neutrino Factory in which 
$2 \times 10^{20}$ muons have decayed in the storage ring straight section
pointing at the detector. 
The fluxes are shown as a function of the energy of the stored 
muons for negative muons (top two plots) 
and positive muons (bottom two plots), and for three muon polarizations
as indicated. The calculated fluxes are averaged over a circular area 
of radius 1~km at the far site. 
Calculation from Ref.~\cite{geer98}.}
\label{fluxes}
\end{figure}

\begin{figure}[hbtp!]
\includegraphics[width=4in]{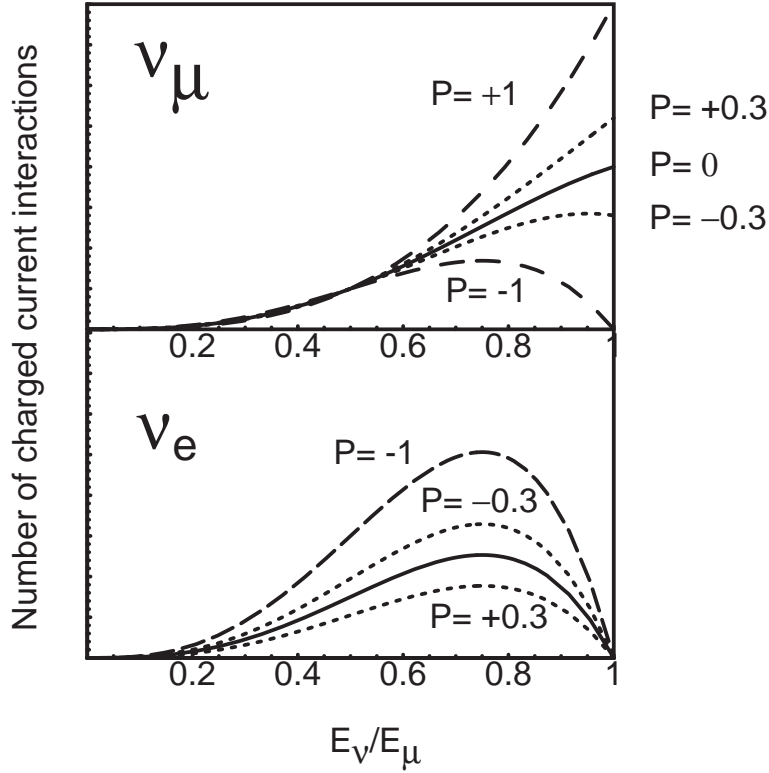}
\caption{Charged current event spectra at a far detector. 
The solid lines
indicate zero polarization, the dotted lines indicate polarization of $\pm 0.3$ and
the dashed lines indicate full polarization.  The $P=1$ case for electron neutrinos
results in no events and is hidden by the $x$ axis.}
\label{polarization}
\end{figure}

\begin{figure}[hbtp!]
\mbox{
\includegraphics*[width=3.in]{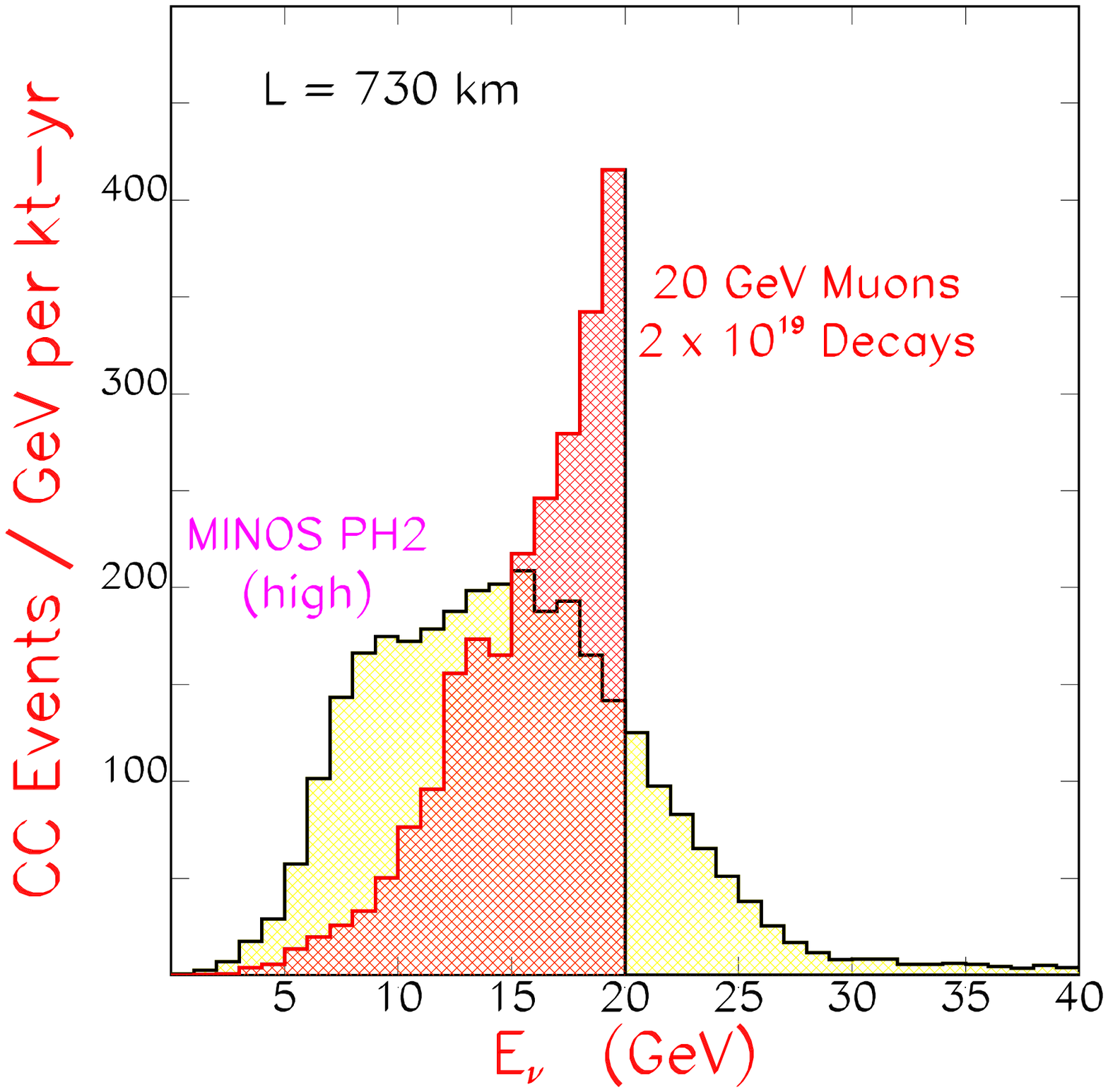}
}
\mbox{
\includegraphics*[width=3.in]{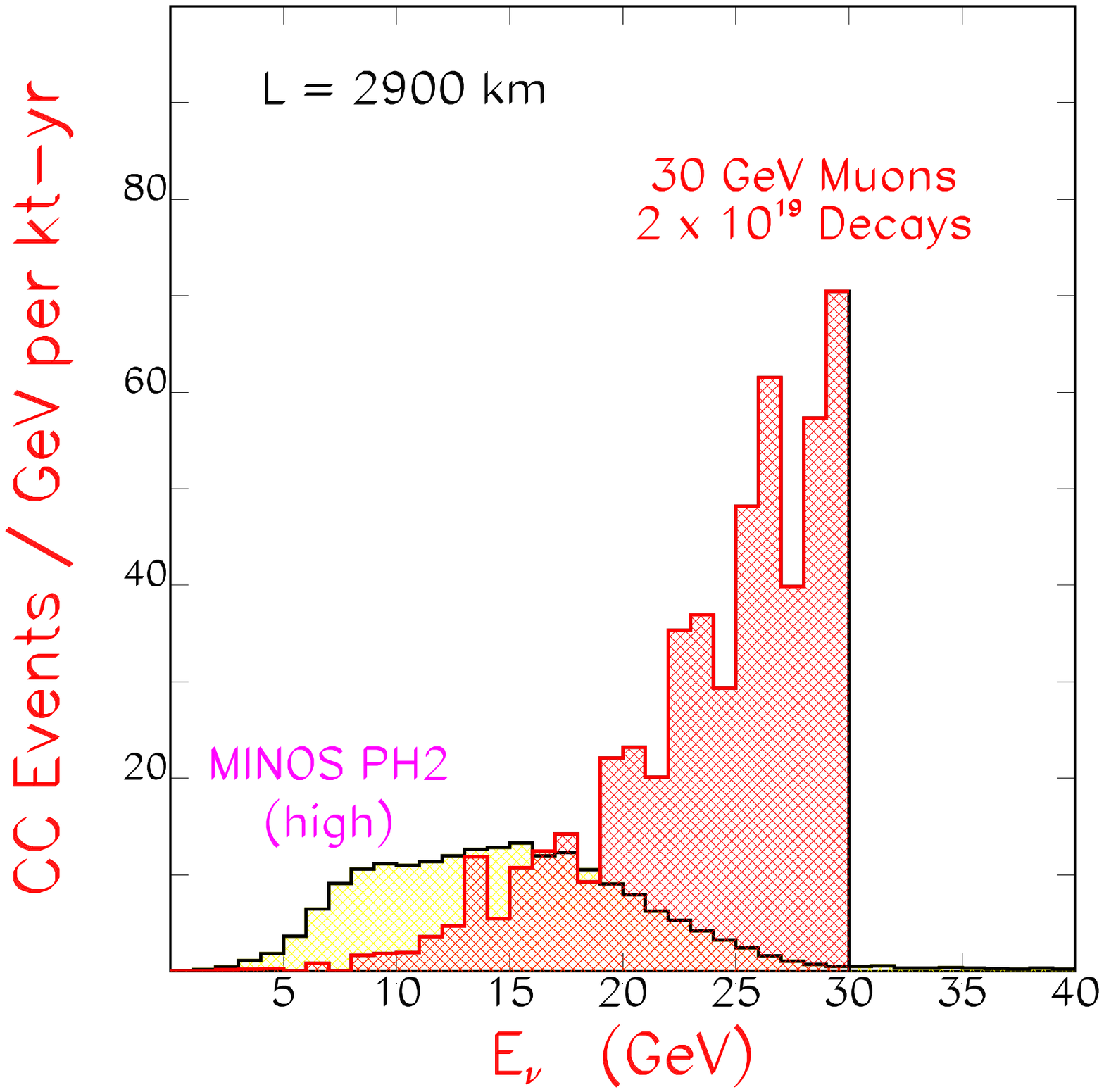}
}
\caption{(Color) Comparison of interacting $\nu_\mu$ energy distributions for
the NUMI high energy wide-band beam (Ref.~\cite{numi}) with
a 20~GeV Neutrino Factory beam (Ref.~\cite{geer98}) at $L = 730$~km and
a 30~GeV Neutrino Factory beam at $L = 2900$~km.
The Neutrino Factory distributions have been calculated based on
Eq.~(\ref{eq:n_numu}) (no approximations), and include realistic
muon beam divergences and energy spreads.
}
\label{minos_wbb}
\end{figure}

Neutrino Factory beams are produced from muons decaying in a storage 
ring with long straight sections. 
Consider an ensemble of polarized negatively-charged muons. 
When the muons decay they produce muon neutrinos with a distribution 
of energies and angles  in the muon rest--frame described 
by~\cite{gaisser}:
\begin{eqnarray}
\frac{d^2N_{\nu_\mu}}{dxd\Omega_{c.m.}} &\propto& \frac{2x^2}{4\pi}
 \left[ (3-2x) + (1-2x) P_\mu \cos\theta_{c.m.} \right] \, ,
\label{eq:n_numu}
\end{eqnarray}
where $x\equiv 2E_\nu/m_\mu$, $\theta_{c.m.}$ is the angle between the neutrino
momentum vector and the muon spin direction, and $P_\mu$ is the average muon
polarization along the beam direction. 
The electron antineutrino distribution is given by: 
\begin{eqnarray}
\frac{d^2N_{\bar\nu_e}}{dxd\Omega_{c.m.}} &\propto&  \frac{12x^2}{4\pi}
\left[ (1-x) + (1-x) P_\mu\cos\theta_{c.m.} \right] \, ,
\label{eq:n_nue}
\end{eqnarray}
and the corresponding distributions for
$\bar\nu_\mu$ and $\nu_e$ from $\mu^+$ decay are obtained by
the replacement $P_{\mu} \to -P_{\mu}$. 
Only neutrinos and antineutrinos emitted in the forward
direction ($\cos\theta_{lab}\simeq1$) are relevant to the neutrino flux for
long-baseline experiments; in this limit
$E_\nu = x E_{max}$ and at high energies the maximum $E_\nu$ in the 
laboratory frame is given by 
$E_{max} = \gamma   (1 + \beta \cos\theta_{c.m.})m_{\mu}/2 $, 
where $\beta$ and $\gamma$ are the usual relativistic factors. 
The $\nu_\mu$ and $\bar{\nu}_{e}$ distributions as a function
of the laboratory frame variables are then given by:
 \begin{eqnarray}
\frac{d^2N_{\nu_{\mu}}}{dxd\Omega_{lab}} &\propto&  
\frac{1}{\gamma^2 (1- \beta\cos\theta_{lab})^2}\frac{2x^2}{4\pi}
\left[ (3-2x) + (1-2x)P_{\mu}\cos\theta_{c.m.} \right] , 
\label{eq:numu}
\end{eqnarray}
and
\begin{eqnarray}
 \frac{d^2N_{\bar{\nu}_{e}}}{dxd\Omega_{lab}} &\propto&
\frac{1}{\gamma^2 (1- \beta\cos\theta_{lab})^2}\frac{12x^2}{4\pi}
\left[ (1-x) + (1-x)P_{\mu}\cos\theta_{c.m.} \right] \; .
\label{eq:nue}
\end{eqnarray}
Thus, for a high energy muon beam with no beam divergence, 
the neutrino and antineutrino energy and angular distributions depend upon the parent muon
energy, the decay angle, and the direction of the
muon spin vector.
With the muon beam intensities that could be provided by a 
muon--collider type muon source~\cite{status_report} 
the resulting neutrino fluxes 
at a distant site would be large. For example, Fig.~\ref{fluxes} shows 
as a function of muon energy and polarization, 
the computed fluxes per $2\times 10^{20}$ muon decays at a site on the 
other side of the Earth ($L = 10000$~km).
Note that the $\nu_e$ ($\bar{\nu}_e$) fluxes are suppressed 
when the muons have $P = +1\, (-1).$ This can be understood by examining 
Eq.~(\ref{eq:nue}) and noting that for $P = -1$ the two terms cancel 
in the forward direction for all $x$.
At low energies, the neutrino CC interaction cross section is dominated by
quasi-elastic scattering and resonance production. 
However, if $E_\nu$ is greater than $\sim10$~GeV, 
the total cross section is  dominated by deep inelastic scattering 
and is approximately~\cite{CCFRsigma}:
\begin{eqnarray}
\sigma(\nu +N \gt \lepton^- + X) &\approx& 0.67\times 10^{-38} \;
\times E_{\nu}(\GeV)\, \centi^2\, , \\
\sigma(\antinu +N \gt \lepton^+ + X) &\approx& 0.34\times10^{-38} \; 
\times E_{\antinu}(\GeV)\, \centi^2 \; .
\end{eqnarray}
The number of $\nu$ and $\bar{\nu}$ CC events per incident neutrino
 observed 
in an isoscalar target is given by:
\begin{eqnarray}
N(\nu +N \gt \lepton^- + X) 
&=& 4.0 \times 10^{-15}\times E_{\nu}(\GeV) \;\text{events per}\; \text{g/cm}^2, \\
N(\antinu +N \gt \lepton^+ + X) 
&=& 2.0 \times 10^{-15}\times E_{\antinu}(\GeV) \;\text{events per}\; \text{g/cm}^2.
\end{eqnarray}
Using this simple form for the energy dependence of the cross section, 
the predicted energy distributions for $\nu_e$ and $\nu_\mu$ 
interacting in a far detector $(\cos\theta = 1)$ at a Neutrino Factory are shown in Fig.~\ref{polarization}. The interacting $\nu_\mu$ 
energy distribution is compared in Fig.~\ref{minos_wbb} 
with the corresponding distribution arising from the high--energy 
NUMI~\cite{numi} wide-band beam. Note that neutrino beams from a 
Neutrino Factory have no high energy tail, and in that sense can be 
considered narrow-band beams. 
\begin{figure}[hbtp!]
\includegraphics*[width=3.5in]{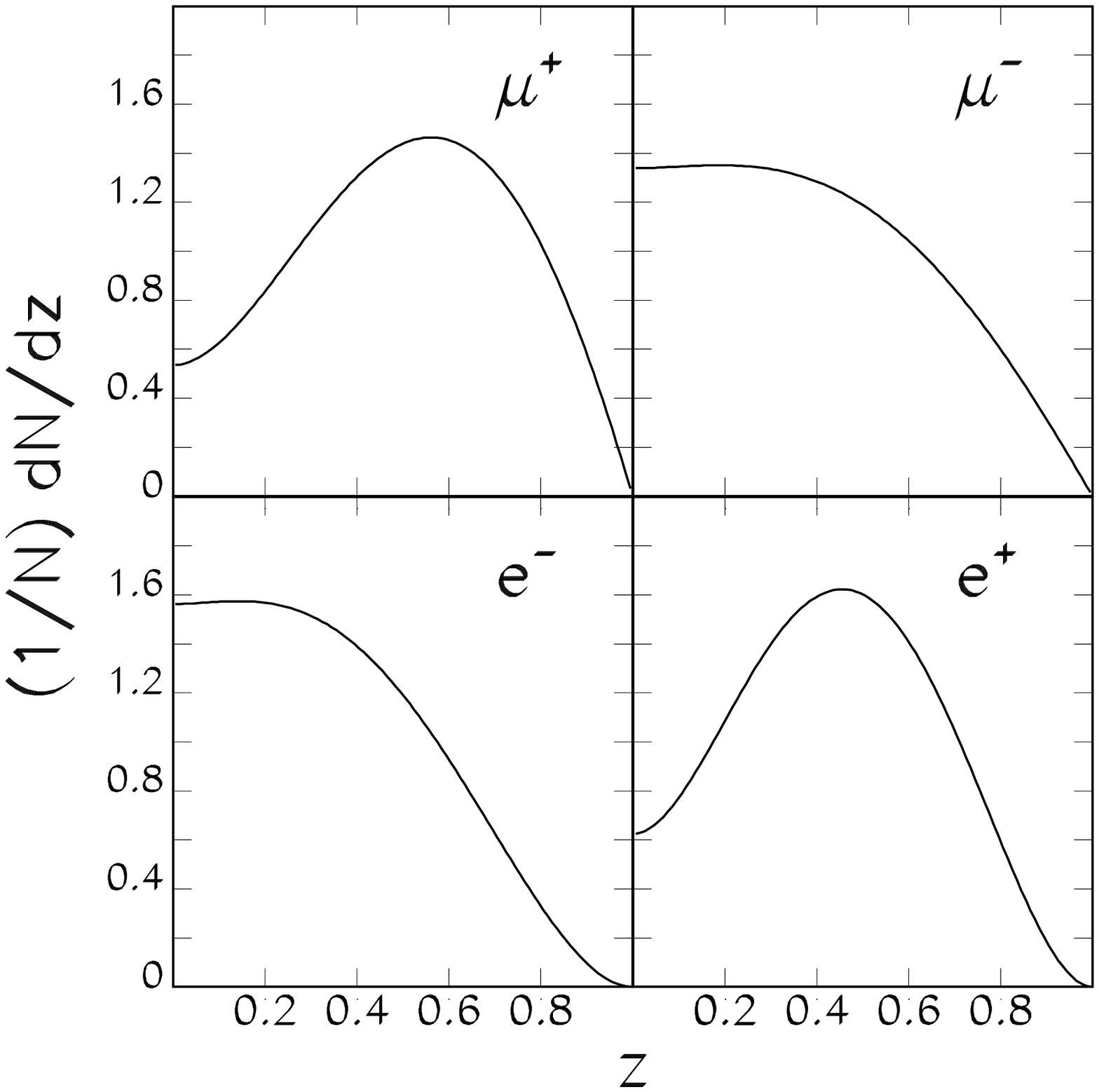}
\caption{Lepton energy spectra for CC $\bar{\nu}_\mu$ (top left),
$\nu_\mu$ (top right), $\nu_e$ (bottom left), and $\bar{\nu}_e$
(bottom right) interactions. Note that $z$ is the energy normalized
to the primary muon energy  $z = E_{\lepton}/E_\mu$. 
Calculation from Ref.~\cite{bgw99}.}
\label{fig:elept}
\end{figure}
In practice, CC interactions can only be cleanly 
identified when the final 
state lepton exceeds a threshold energy. The calculated final state lepton 
distributions are shown in Fig.~\ref{fig:elept}. 
Integrating over the energy distribution, the total $\nu$ and $\bar{\nu}$ 
interaction rates per muon decay are given by: 
\begin{eqnarray}
N_\nu &=& 1.2 \times 10^{-14} \; \biggr[\frac{E_{\mu}^3(\GeV)}{L^2(km)}\biggl] 
\times C(\nu) \;\; \hbox {events per kton}
\end{eqnarray}
and
\begin{eqnarray}
N_{\bar{\nu}}&=&0.6\times10^{-14} \;
\biggr[\frac{E_{\mu}^3(\GeV)}{L^2(km)}\biggl] 
\times C(\nu) \;\; \hbox{events per kton} \, ,
\end{eqnarray}
where
\begin{equation}
C(\nu_{\mu})= \frac{7}{10} + P_{\mu} \frac{3}{10}\quad , \quad 
C(\nu_{e})=\frac{6}{10} - P_{\mu} \frac{6}{10}.
\end{equation}
The calculated $\nu_e$ and $\nu_\mu$ CC interaction rates resulting from 
$10^{20}$ muon 
decays in the storage ring straight section of a Neutrino Factory 
are compared in Table~\ref{table:rates} with 
expectations for the corresponding rates at the next generation 
of accelerator--based neutrino experiments. Note that event rates 
at a Neutrino Factory increase as $E_\mu^3$, and are significantly 
larger than expected for the next generation of approved experiments 
if $E_\mu > 20$~GeV.
The radial dependence 
of the event rate is shown in Fig.~\ref{fig:radial} for a 20~GeV 
Neutrino Factory and three baselines. 
\begin{table}[hbtp!]
\caption{\label{table:rates}
Muon neutrino and electron antineutrino CC interaction rates in the
absence of oscillations, calculated for baseline length $L = 732$~km
(FNAL $\to$ Soudan), for MINOS using the wide-band beam and a
muon storage ring delivering $10^{20}$ decays 
with $E_\mu=10,20$, and $50$~GeV at three baselines.
The Neutrino Factory calculation includes a realistic muon beam divergence 
and energy spread.}
\begin{ruledtabular}
\begin{tabular}{c|cc|cc|cc}
Experiment& &Baseline & $\langle E_{\nu_\mu} \rangle$ & $\langle E_{\bar \nu_e} \rangle$& N($\nu_\mu$ CC) & N($\bar\nu_e$ CC) \\
 & &(km) & (GeV) & (GeV) & (per kton--yr) & (per kton--yr) \\
\hline
MINOS& Low energy    &732&  3 & -- &  458 & 1.3 \\
     & Medium energy &732&  6 & -- & 1439 & 0.9 \\
     & High energy   &732& 12 & -- & 3207 & 0.9 \\
\hline
Muon ring & $E_\mu$ (GeV) & & & & & \\
\hline
&  10 &732& 7.5 & 6.6 & 1400              & 620 \\
&  20 &732&  15 &  13 & 12000             & 5000\\
&  50 &732&  38 &  33 & 1.8$\times$10$^5$ & 7.7$\times$10$^4$ \\
\hline
Muon ring& $E_\mu$ (GeV)& & & & & \\
\hline
&  10 &2900& 7.6 & 6.5 & 91             &41\\
&  20 &2900&  15 &  13 & 740          & 330\\
&  50 &2900&  38 &  33 & 11000& 4900 \\
\hline 
Muon ring& $E_\mu$ (GeV)& & & & & \\
\hline
&  10 &7300& 7.5 & 6.4 & 14              & 6  \\
&  20 &7300&  15 &  13 & 110            & 51 \\
&  50 &7300&  38 &  33 & 1900 & 770 \\
\end{tabular}
\end{ruledtabular}
\end{table}
\begin{figure}[hbtp!]
\includegraphics[width=3in]{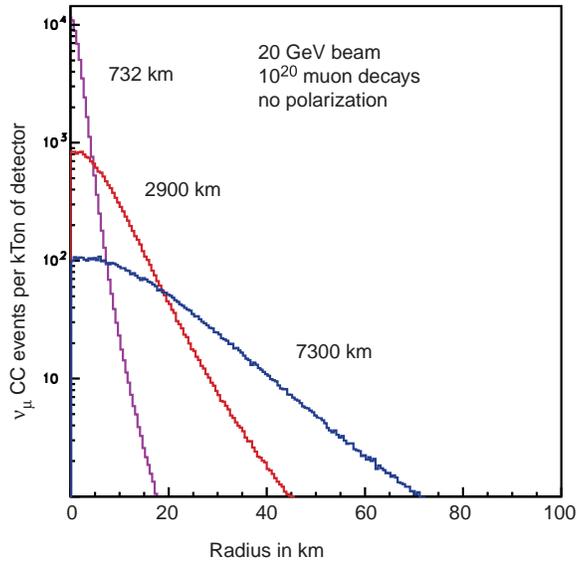}
\caption{(Color) Events per kton of detector as a function of distance from the beam center
for a 20 GeV muon beam.}
\label{fig:radial}
\end{figure}
\begin{table}[bhtp!]
\caption{Dependence of predicted charged current  event rates on muon 
beam properties at a Neutrino Factory. The last column 
lists the required precisions with which each beam property must 
be determined if the uncertainty on the neutrino flux at the 
far site is to be less than $\sim1$\%.  Here $\Delta$ denotes uncertainty
while $\sigma$ denotes the spread in a variable. Table from Ref.~\cite{cg00}.
\label{tab:flux}}
\begin{ruledtabular}
\begin{tabular}{c|c|cc} 
Muon Beam & Beam & Rate & Target\\
property & Type & Dependence & Precision \\
\hline
Energy ($E_\mu$) & $\nu$ (no osc.) 
  & $\Delta N / N = 3 \; \Delta E_\mu/E_\mu$ 
  & $\Delta(E_\mu)/E_\mu < 0.003$ \\
  & $\nu_{e} \to \nu_{\mu}$ 
  &$\Delta N / N = 2 \; \Delta E_\mu/E_\mu$ 
  & $\Delta(E_\mu)/E_\mu < 0.005$ \\
\hline
Direction ($\Delta\theta$) & $\nu$ (no osc.) 
  & $\Delta N/N \leq 0.01$ 
  & $\Delta\theta < 0.6 \; \sigma_\theta$ \\
  &  & (for $\Delta\theta < 0.6\; \sigma_\theta$) & \\
\hline
Divergence ($\sigma_\theta$)
  & $\nu$ (no osc.) 
  & $\Delta N / N \sim 0.03 \; \Delta\sigma_\theta / \sigma_\theta$
  & $\Delta\sigma_\theta / \sigma_\theta < 0.2$ \\
  & & (for $\sigma_\theta \sim 0.1/\gamma$) 
    & (for $\sigma_\theta \sim 0.1/\gamma$)\\
\hline
Momentum spread ($\sigma_p$) 
  & $\nu$ (no osc.) 
  & $\Delta N / N \sim 0.06 \; \Delta\sigma_p / \sigma_p$
  & $\Delta\sigma_p / \sigma_p < 0.17$ \\
\hline
Polarization ($P_\mu$) 
  & $\nu_e$ (no osc.) 
  & $\Delta N_{\nu_e} / N_{\nu_e} =  \Delta P_\mu$
  & $\Delta P_\mu < 0.01$ \\
  & $\nu_{\mu}$ (no osc.) 
  & $\Delta N_{\nu_\mu} / N_{\nu_\mu} = 0.4 \; \Delta P_\mu$
  & $\Delta P_\mu < 0.025$ \\ 
\end{tabular}
\end{ruledtabular}
\end{table}

\begin{figure}[hbtp!]
\includegraphics*[width=3.5in]{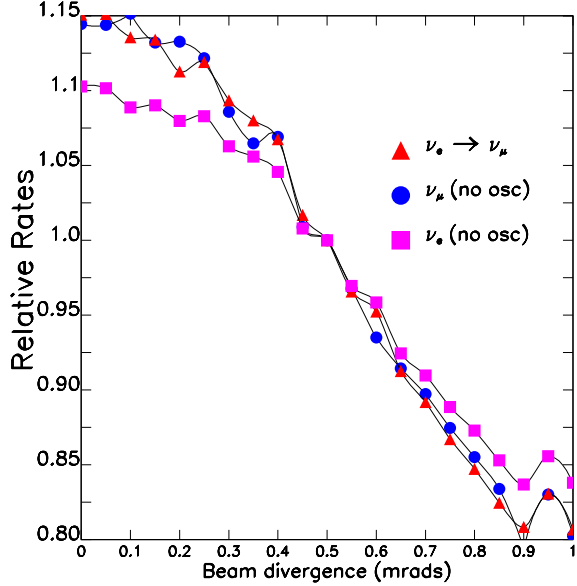}%
\vspace{-2.3cm}
\caption{(Color) Dependence of CC interaction rates on the muon beam divergence
for a detector located at
$L = 2800$~km from a muon storage ring containing 30~GeV unpolarized muons.
Rates are shown
for $\nu_e$ (boxes) and $\nu_\mu$ (circles) beams
in the absence of oscillations,
and for $\nu_e \to \nu_\mu$ oscillations (triangles) with the
three--flavor oscillation parameters, $\delta m_{12}^2=5\times10^{-5}\, \text{eV}^2/\text{c}^4,$
$\delta m_{32}^2=3.5\times10^{-3}\, \text{eV}^2/\text{c}^4,$ $s_{13}=0.10,$ $s_{23}=0.71,$
$s_{12}=0.53,$ $\delta=0.$ The calculation is from Ref.~\cite{cg00}.}
\label{fig:flux_xy}
\end{figure}

\begin{figure}[thbp!]
\vspace{1.5cm}
\includegraphics*[width=3.5in]{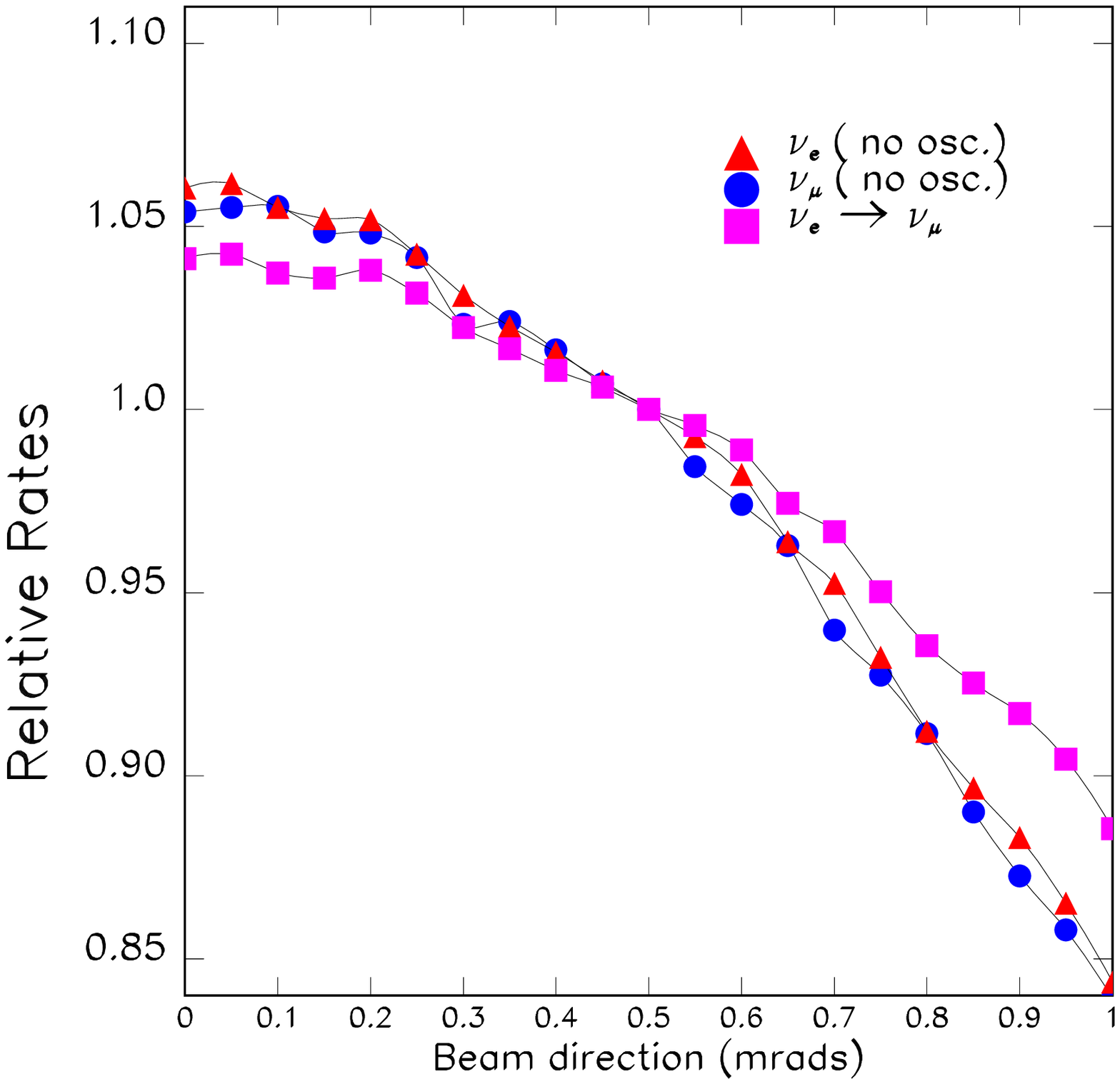}%
\vspace{-2.0cm}
\caption{(Color) Dependence of CC interaction rates on the neutrino beam direction. 
Relative rates are shown for 
a detector at a far site located downstream of 
a storage ring containing 30~GeV unpolarized muons, and 
a muon beam divergence of 0.33~mrad. Rates are shown
for $\nu_e$ (triangles) and $\nu_\mu$ (circles) beams
in the absence of oscillations,
and for $\nu_e \to \nu_\mu$ oscillations (boxes) with the
three--flavor oscillation parameters shown in Fig.~\ref{fig:flux_xy}. 
The calculation is from Ref.~\cite{cg00}.
}
\label{fig:flux_d}
\end{figure}

We next consider the systematic uncertainties on the neutrino 
flux. Since muon decay kinematics is very well understood, and the 
beam properties of the muons in the storage ring can be well determined, 
we expect the systematic uncertainties on the neutrino beam intensity 
and spectrum to be small compared to the corresponding uncertainties 
on the properties of conventional neutrino beams.  
In the muon decay straight section of a Neutrino Factory,  
the muon beam is designed to have an average divergence given by 
$\sigma_\theta =~O(\frac{0.1}{\gamma}).$ The neutrino beam divergence will 
therefore be dominated by muon decay kinematics, and uncertainties 
on the beam direction and divergence will yield only small 
uncertainties in the neutrino flux at a far site. However, if 
precise knowledge of the flux is required, the uncertainties 
on $\theta$ and $\sigma_\theta$ must be taken into account, along 
with uncertainties on the flux arising from uncertainties on 
the muon energy distribution and polarization. 
The relationships between the uncertainties on the muon beam 
properties and the resulting uncertainties on the neutrino flux 
are summarized in Table~\ref{tab:flux}. If, for example, we wish  
to know the $\nu_e$ and $\nu_{\mu}$ fluxes at a far site with a 
precision of 1\%, we must determine the beam divergence, $\sigma_\theta$,  
to 20\% (see, Fig.~\ref{fig:flux_xy}), and ensure that the beam direction 
is within $0.6\times \sigma_\theta$ of the nominal direction~\cite{cg00} (see,
Fig.~\ref{fig:flux_d}). We point out that it should be possible to do much better than this, 
and consequently, to know the fluxes at the far site with a precision much 
better than 1\%.

\begin{figure}[hbtp!]
\includegraphics[width=3.5in]{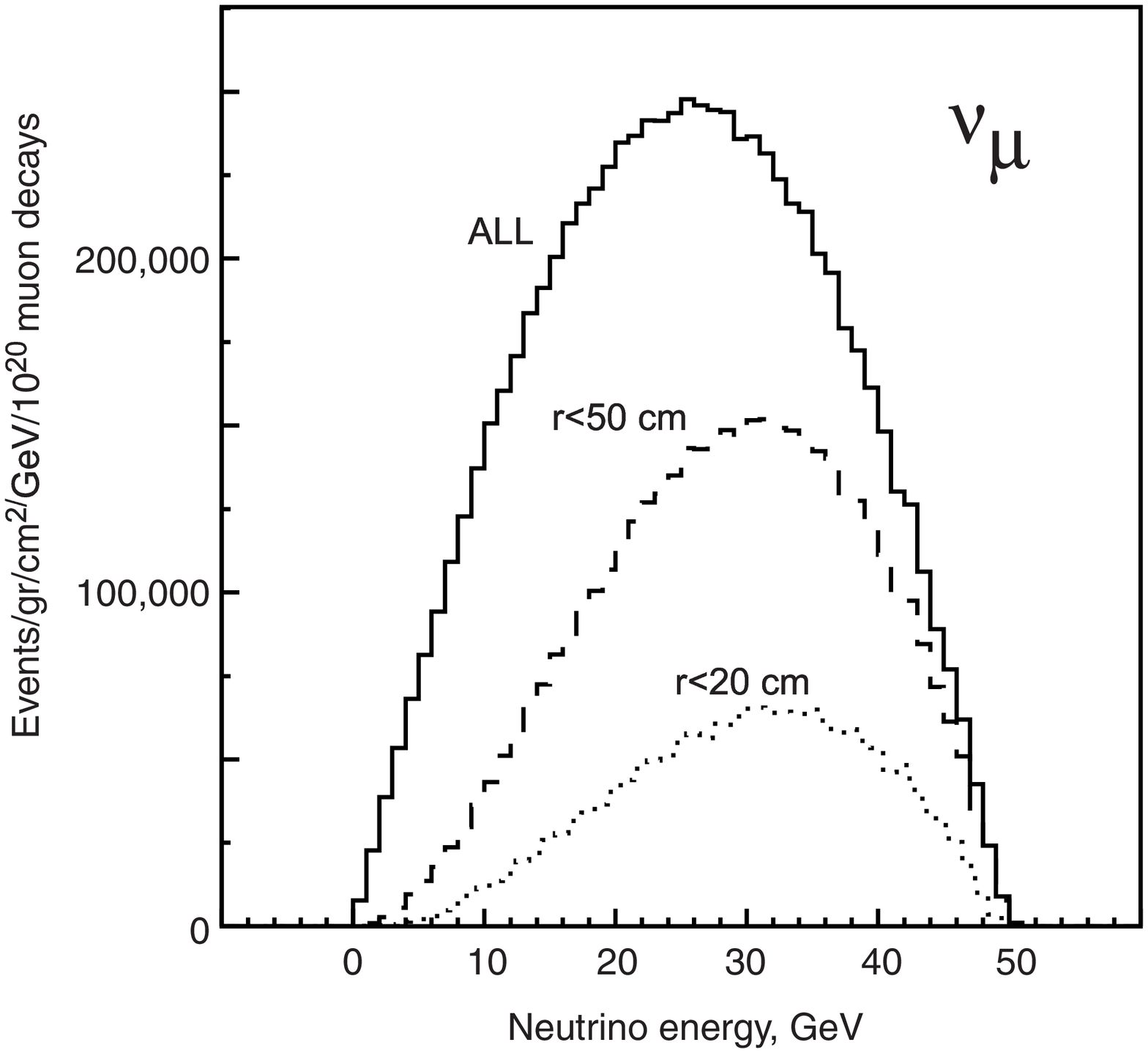}
\caption{Events per g/cm$^2$ per GeV for a detector 40~m from a
muon storage ring with a 600~m straight section.  The three curves show
all events and those falling within 50 and 20~cm of the beam center.}
\label{nearspectra}
\end{figure}

We now consider the event distributions in a detector at a near site, 
close to the Neutrino Factory, which 
will be quite different from the corresponding distributions 
at a far site. There are two main reasons for this difference. 
First, the near detector accepts neutrinos over a large range of 
muon decay angles $\theta$, not just those neutrinos traveling in the 
extreme forward direction. This results in a broader neutrino energy 
distribution that is sensitive to the radial size of the 
detector (Fig.~\ref{nearspectra}).
 
\begin{figure}[hbtp!]
\includegraphics[width=3in]{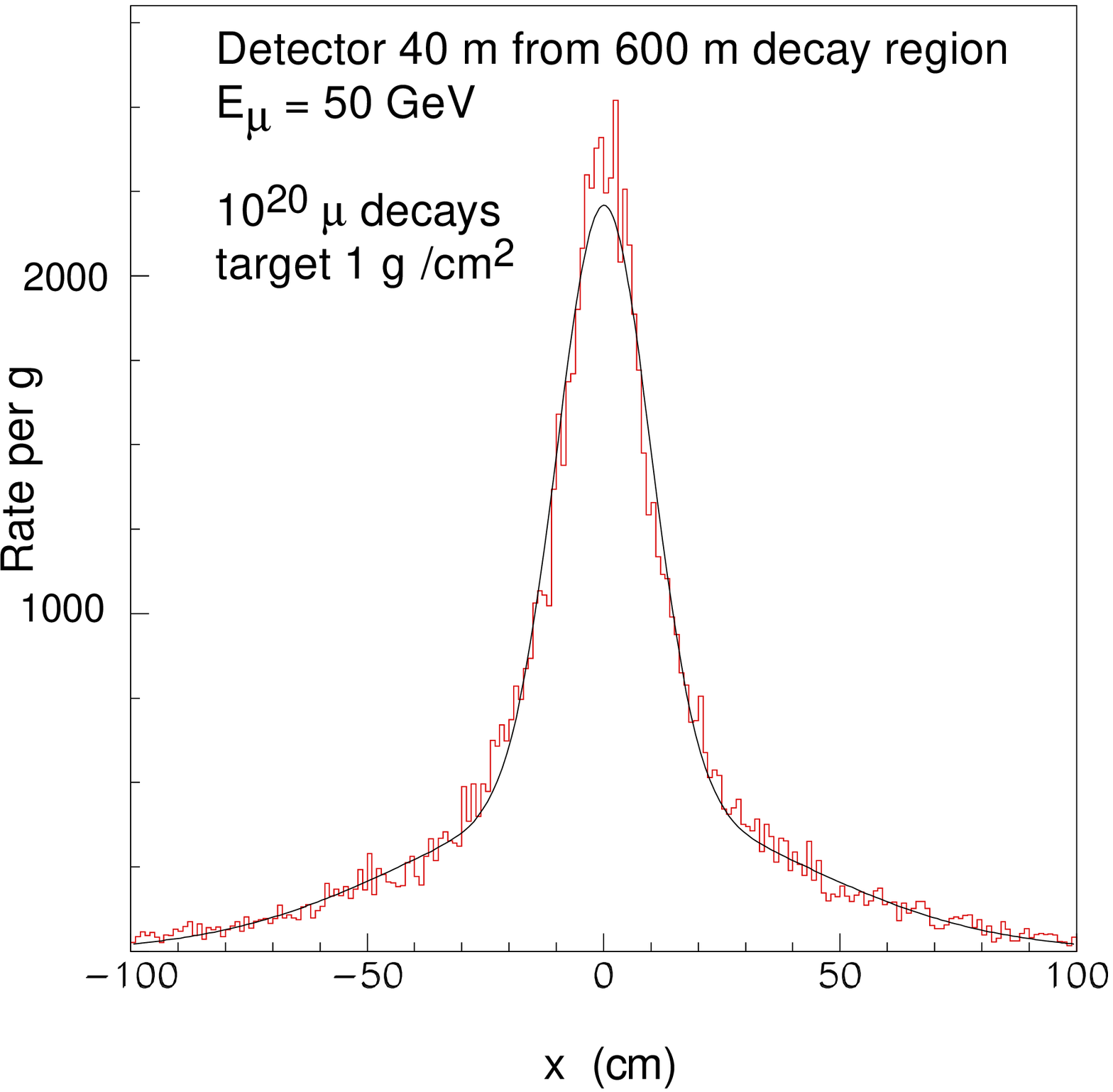}
\caption{(Color) Events per g/cm$^2$  as a function of the 
transverse coordinate, $x,$ 50~m downstream of a 50~GeV neutrino 
factory providing $10^{20}$ muon decays. 
The central peak is mainly due to decays 
in the last hundred meters of the decay pipe
while the large tails are due to upstream decays.}
\label{xplot}
\end{figure}

\begin{figure}[hbtp!]
\includegraphics[width=3in]{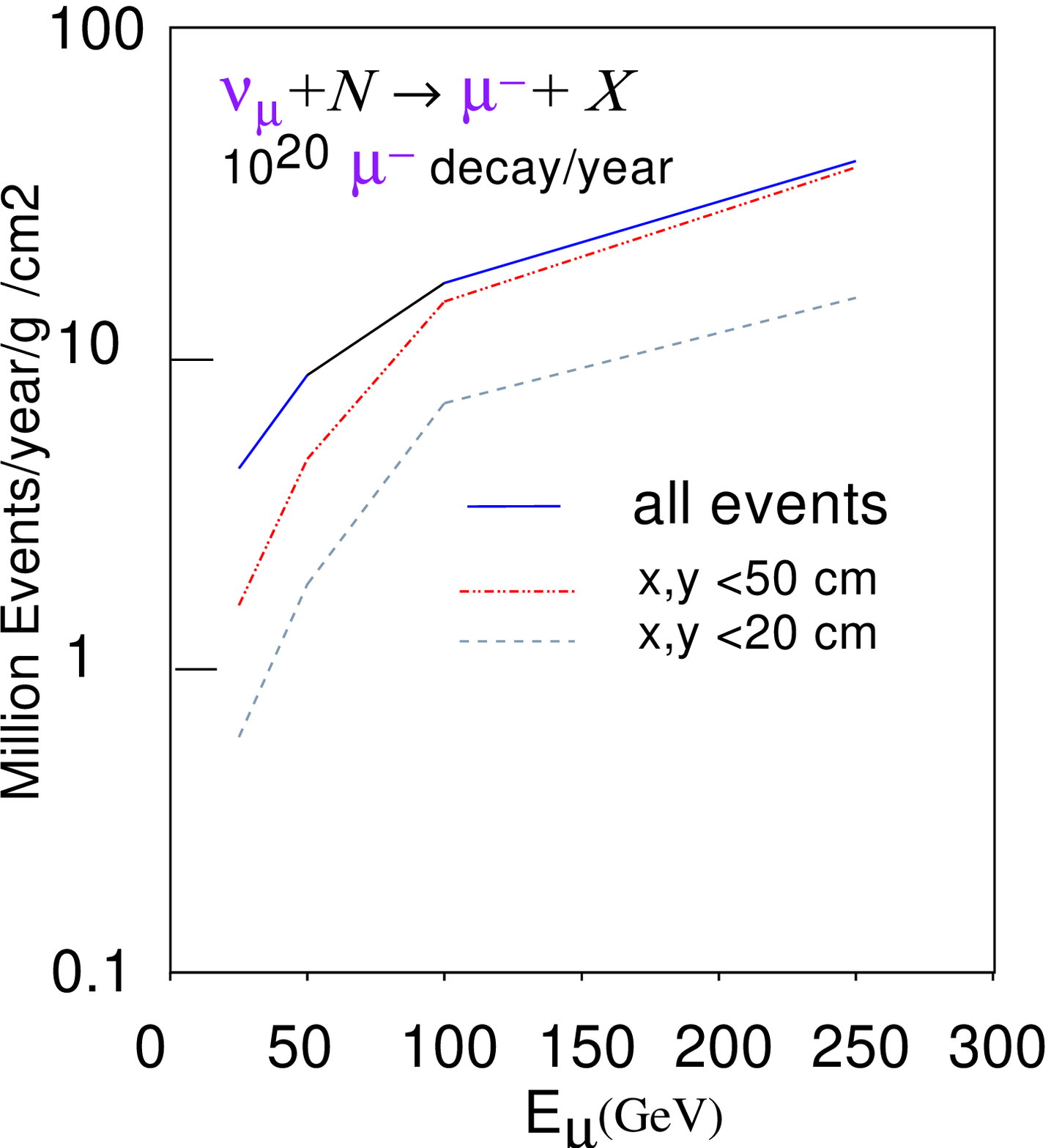}
\caption{(Color) Events per year and per g/cm$^2$ at a near detector as a function of
  muon beam energy in GeV.  The solid curves
indicate all events, the dashed and dotted curves show the effects of
radial position cuts.}
\label{eventrates}
\end{figure}

Second, if the distance of the near detector from the end of the decay  
straight section is of the order of the straight section length, 
then the $\theta$ acceptance of the detector varies with the position of the
muon decay along the straight section. This results in a more complicated 
radial flux distribution than expected for a far detector. However,
since the dominant effects are decay length and muon decay kinematics,
 it should be modeled quite accurately (Fig.~\ref{xplot}). 
 Note that, even in a limited angular range, the event rates in a 
near detector are very high.  Figure~\ref{eventrates} illustrates the 
event rates per g/cm$^2$ as a function of energy.  Because
most of the neutrinos produced forward in the center of mass traverse the 
detector fiducial volume, 
the factor of $\gamma^2$
present in the  flux for $\theta \sim0$ is canceled and the event rate 
increases linearly with $E_{\mu}$.  For a 50~GeV
muon storage ring, the interaction rate per 10$^{20}$ muon decays is 
$7\times10^6 \text{ events per g/cm}^2.$ 
Finally, in the absence of special magnetized shielding, 
the high neutrino event rates in any material
upstream of the detector will cause substantial backgrounds.  
The event rate in the last three interaction lengths ($300~\text{g/cm}^2$) 
of the shielding between
the detector and the storage ring would be 30 interactions per beam 
spill at a 15~Hz
machine delivering $2\times 10^{20}$ muon decays per year.  
These high background rates will require clever magnetized shielding 
designs and fast detector
readout to avoid overly high accidental rates in low mass experiments.
\subsection{Beta Beams}
\begin{figure}[bhtp!]
\mbox{
\includegraphics*[viewport=0 0 650 560,width=3.5in]{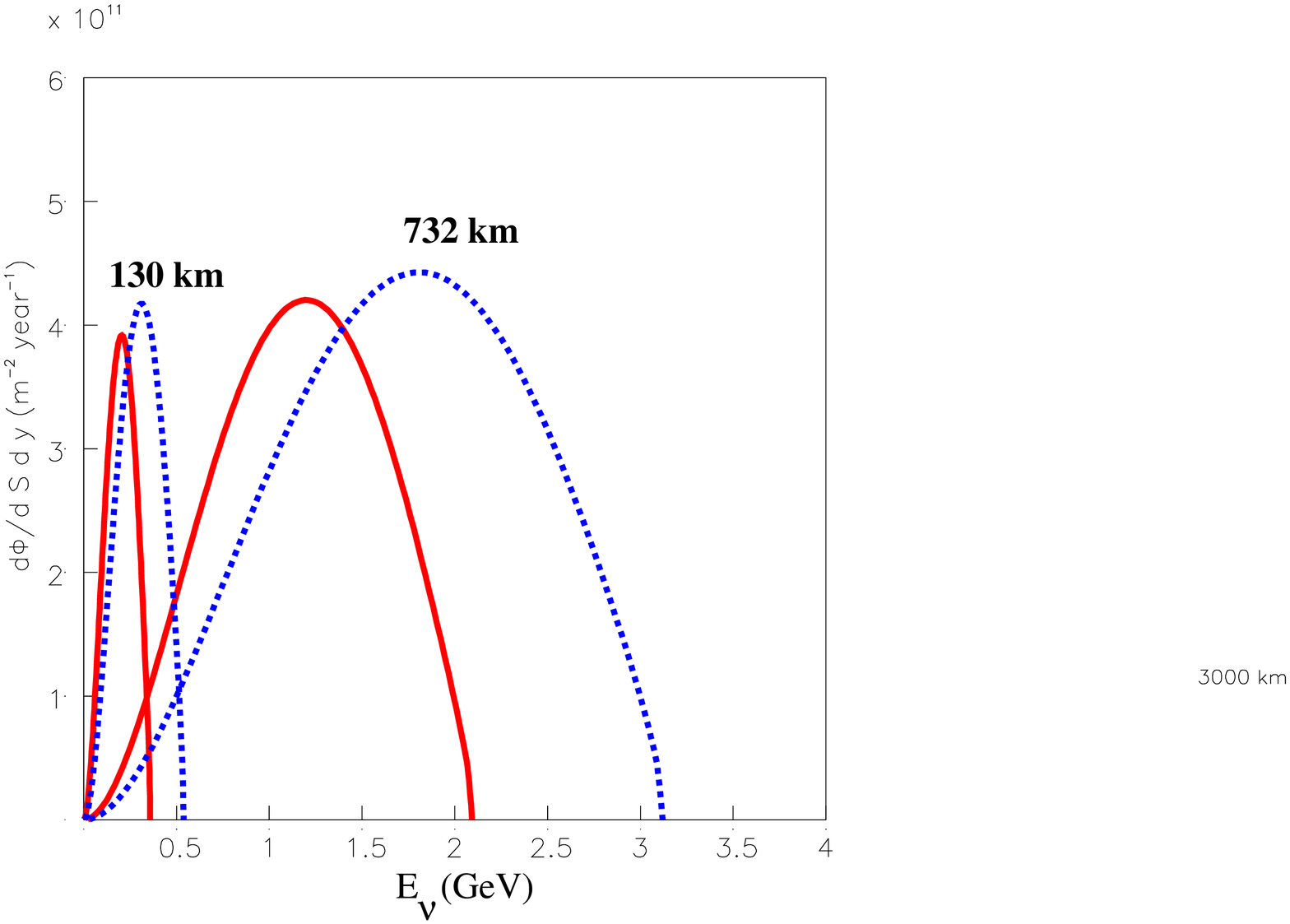}
}
\mbox{
\includegraphics*[viewport=0 0 540 557,width=3.0in]{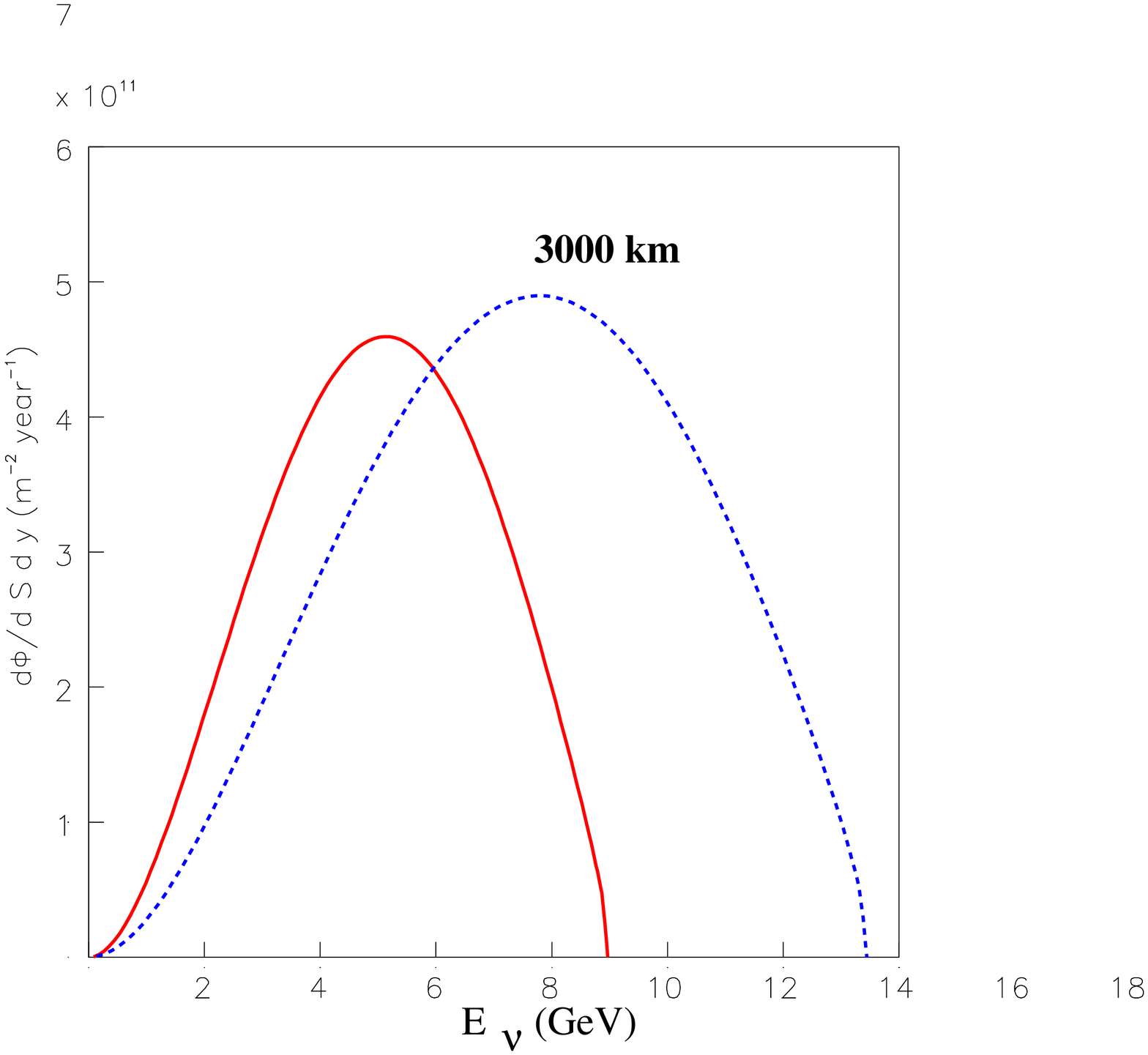}
}
\caption{(Color) Comparison of Beta Beam neutrino fluxes for the three setups 
described in the text, shown as a function of the neutrino energy 
for $\bar{\nu_e}$ (solid) and $\nu_e$ (dashed). Figures from Ref.~\cite{jj-beta}.
}
\label{fig:betabeam}
\end{figure}

We now consider the beam properties at a Beta Beam facility. In a 
Beta Beam facility the neutrinos are generated by the decay of radioactive
nuclei rather than muons.  The two ions deemed optimal are $^{18}$Ne for 
$\nu_e$ and
$^6$He for $\bar{\nu}_e$ production.  The resulting initial neutrino beam 
consists of a single flavor. In addition, since the decay kinematics is 
well known, the uncertainties on the neutrino energy spectrum are expected 
to be small. 
The electron energy spectrum produced by a nuclear $\beta$-decay at rest is
\begin{equation}
\frac{dN^{\rm rest}}{dE_e} \sim E^2_e (E_e-E_0)^2
\end{equation}
where $E_0$ is the electron end-point energy, which is 3.5~MeV for $^6$He
and 3.4~MeV for $^{18}$Ne.
In the rest frame of the ion, the spectrum of the neutrinos~\cite{jj-beta} is
\begin{equation}
\frac{dN^{\rm rest}}{d \cos\theta d E_{\nu}} \sim
E^2_{\nu} (E_0-E_{\nu})\sqrt{(E_0-E_{\nu})^2-m^2_e}.
\end{equation}
After performing a boost and normalizing to the total decays (in the 
straight section)
$N_\beta$, the neutrino flux per solid angle in a detector located at a 
distance $L$
and aligned with the straight section can be calculated as
\begin{equation}
\frac{d\Phi^{\rm lab}}{dS dy}\Bigg|_{\theta\simeq 0}
\approx
\frac{N_\beta}{\pi L^2}
\frac{\gamma^2}{g(y_e)} y^2 (1-y) \sqrt{(1-y)^2-y_e^2}
\end{equation}
where $0\leq y=\frac{E_\nu}{2 \gamma E_0} \leq 1-y_e$, $y_e=m_e/E_0$ and
\begin{equation}
g(y_e)=\frac{1}{60}
\left\{ \sqrt{1-y_e^2}(2-9y_e^2-8y_e^4)+15 y_e^4 \log
\left[\frac{y_e}{1-\sqrt{1-y_e^2}}\right] \right\}.
\end{equation}
The neutrino flux and energy distribution depend upon the boost $\gamma$, 
and hence upon the energies of the stored radioactive ions. The original 
Beta Beam proposal~\cite{autin} was to use the CERN SPS to accelerate the ions. 
The desire to simultaneously store ions of both species 
in the storage ring, and build a large detector in the Fr\'{e}jus tunnel in France 
(which fixes the baseline), has led to proposed Beta Beam 
energies corresponding to  
$\gamma \sim 60$ and 100 for the two ion species, yielding mean neutrino 
energies of 0.2~GeV and 0.3~GeV. Recently it has been suggested that 
these energies are too low for optimal sensitivity to the interesting 
physics, and hence higher energy scenarios are being considered, using 
the Fermilab Tevatron, or the CERN LHC to accelerate the ions. 
Figure \ref{fig:betabeam} shows the expected fluxes for the three  
scenarios,
``low" energy (e.g., SPS), ``medium" energy (e.g., Tevatron), and ``high" energy (e.g., LHC).
Although the integrated fluxes are similar, the cross section grows with 
energy, 
yielding more events for the higher energies. Table~\ref{tab:betabeam} shows 
the expected
charged current event rates for the three setups.
\begin{table}[thbp!]
\caption{Number of charged current events without oscillations per 
kton-year for
the three reference setups described in the text. Also is shown the average
neutrino energy. Table from Ref.~\cite{jj-beta}.
\label{tab:betabeam}}
\begin{ruledtabular}
\begin{tabular}{ccccc}
$\gamma$ & L (km) & $\bar{\nu}_e$ CC & $\nu_e$ CC & $<E_\nu>$ (GeV) \\
\hline
60/100 & 130 & 1.9 & 25.7 & 0.2/0.3 \\
350/580 & 730 & 48.6 & 194.2 & 1.17/1.87 \\
1500/2500 & 3000 & 244.5 & 800.2 & 5.01/7.55 \\
\end{tabular}
\end{ruledtabular}
\end{table}
 However, it should be noted that the higher energy options require both a 
TeV (or multi-TeV) accelerator and storage ring, which are expensive and 
introduce additional technical challenges. Finally, further study is needed
 to fully explore the systematic uncertainties 
on the beam properties of a Beta Beam facility. Note however, that 
the neutrino beam divergence is controlled by the \textit{Q} value of the beta decay,
and the beam divergence in the straight section. 
In the CERN (low energy) case, the typical decay angle is 7~mrad. 
By contrast, the parent
beam divergence would be O(100)~$\mu$rad, assuming a 200~m beta function 
in the
decay section. For higher energies, both inherent neutrino divergence, 
and the
parent beam divergence scale like 1/$\gamma$. Hence the decay kinematics 
is expected to dominate the beam divergence for all the Beta Beam scenarios.
A more detailed understanding of the systematics must await a detailed design 
for the storage ring and an understanding of the beam halo, etc. 
Background conditions for near detectors also deserve study.   

\section{Neutrino Oscillation Physics Reach of a Neutrino Factory and Beta Beam \label{sec4}}

Ultimately, to fully test the three-flavor mixing framework, 
determine all of the relevant neutrino oscillation parameters, 
and answer the most important neutrino-oscillation related 
physics questions, 
we would like to measure the oscillation probabilities 
$P(\nu_\alpha \to \nu_\beta)$ as a function of the 
baseline $L$ and neutrino energy $E$ (and hence $L/E$) 
for all possible initial and final flavors $\alpha$ and $\beta$.
This requires a beam with a well known initial flavor content, 
and a detector that can identify the flavor of the interacting 
neutrino. The neutrinos interact in the detector via charged 
current (CC) and neutral current (NC) interactions to produce 
a lepton accompanied by a hadronic shower arising 
from the remnants of the struck nucleon. 
In CC interactions, the final-state lepton 
tags the flavor ($\beta$) of the interacting neutrino. 
To accomplish our ultimate goal, we will need $\nu_e$ in addition to 
$\nu_\mu$ beams, and detectors that can distinguish between NC, 
$\nu_e$ CC, $\nu_\mu$ CC, and $\nu_\tau$ CC interactions. 
Conventional neutrino beams are $\nu_\mu$ beams, Beta Beams 
provide $\nu_e$ beams, and Neutrino Factories provide $\nu_e$ and 
$\nu_\mu$ beams. The sensitivities of experiments at the different 
facilities will depend on their statistical precision, the background 
rates, the ability of the experiments to discriminate between true and false 
solutions within the three-flavor mixing parameter space, and the ability of 
the experimental setups to detect as many of the oscillation modes as 
possible. In the following, we will first consider the experimental signatures 
and sensitivities at a Neutrino Factory, and then the corresponding signatures 
and sensitivities at a Beta Beam facility. 
\subsection{Neutrino Factory Sensitivity}
\subsubsection{Wrong-Sign Muons}
At a Neutrino Factory in which, for example, positive 
muons are stored, the initial beam consists of 50\% $\nu_e$ and 
50\% $\bar{\nu}_\mu$. 
In the absence of oscillations, the $\nu_e$ CC interactions 
produce electrons and the $\bar{\nu}_\mu$ CC interactions 
produce positive muons. 
Note that the charge of the final state lepton tags the flavor 
of the initial neutrino or antineutrino. 
In the presence of 
$\nu_e \to \nu_\mu$ oscillations, the $\nu_\mu$ CC interactions 
produce negative muons (i.e., wrong--sign muons).
This is a very clean experimental signature since, with a segmented magnetized 
iron-scintillator sampling calorimeter for example, 
it is straightforward to suppress 
backgrounds to 1 part in $10^4$ of the total CC interaction rate, or better.  
This means that at a Neutrino Factory backgrounds to the 
$\nu_e \to \nu_\mu$ oscillation signal are extremely small. The full 
statistical sensitivity can therefore be exploited down to values of 
$\sin^2 2\theta_{13}$ approaching $10^{-4}$ before backgrounds must be subtracted 
and further advances in sensitivity scale like $\sqrt{N}$ rather than $N$. 
This enables Neutrino Factories to go beyond the sensitivities achievable by 
conventional neutrino Superbeams, by about two orders of magnitude. 
A more complete discussion of backgrounds at a Neutrino Factory can be 
found in Refs.~\cite{fn-692,DeRujula:1998hd,Cervera:2000kp,Apollonio:2002en}.
\begin{figure}[!btph]
\includegraphics*[width=4in]{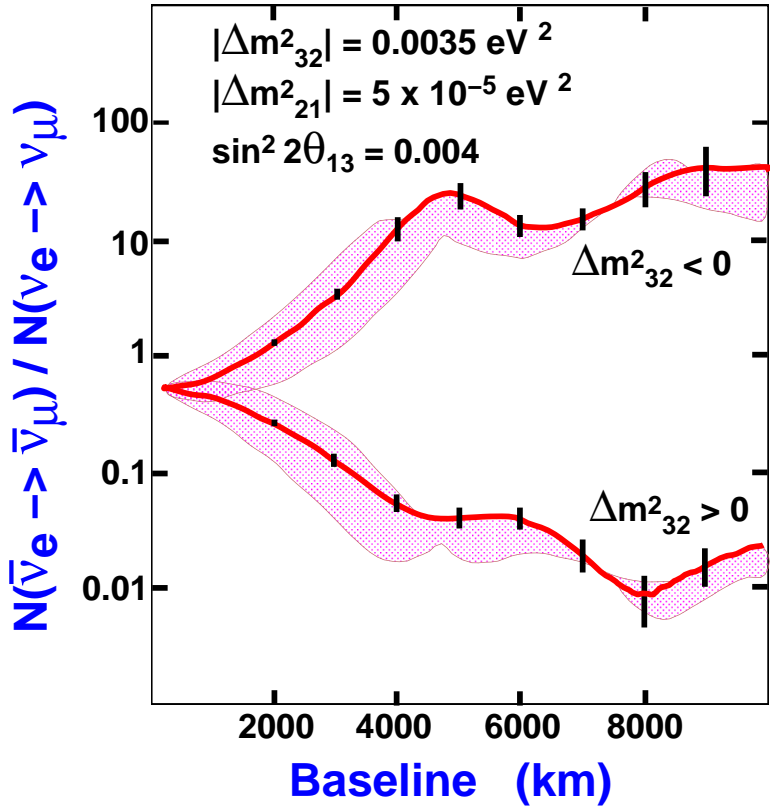}
\caption{(Color) Predicted ratios of wrong--sign muon event rates when positive and negative 
muons are stored in a 20~GeV Neutrino Factory, shown as a function of 
baseline. A muon measurement threshold of 4~GeV is assumed. 
The lower and upper bands correspond, respectively,  
to the two possible neutrino mass eigenstate orderings, as labeled. The 
widths of the bands show how the predictions vary as the \textsl{CP} violating 
phase $\delta$ is varied from $-\frac{\pi}{2}$ to $+\frac{\pi}{2}$, with the thick lines 
showing the predictions for $\delta = 0$. The statistical error bars 
correspond to a Neutrino Factory yielding a 
data sample of $10^{21}$ decays with a 50~kton detector. 
Figure from Ref.~\cite{comments}.}
\label{sec4:fig2}
\end{figure}

We now consider how wrong-sign muon measurements at a Neutrino Factory 
are used to answer the most important neutrino oscillation physics questions. 
Suppose  
we store positive muons in the Neutrino Factory, and measure the number of 
events tagged by a negative muon in a distant detector, and then store 
negative muons and measure the rate of events tagged by a positive muon. 
To illustrate the dependence of the expected measured rates on the chosen
baseline, 
the neutrino mass hierarchy, and the complex phase $\delta$, 
we will fix the other oscillation parameters and 
consider an experiment downstream of a 20~GeV Neutrino Factory. 
Let half of the 
data taking be with $\mu^+$ stored, and the other half with $\mu^-$ 
stored. In Fig.~\ref{sec4:fig2}, the predicted ratio of wrong-sign muon events 
$R \equiv N(\bar{\nu}_e \to \bar{\nu}_\mu) / 
N(\nu_e \to \nu_\mu)$ is shown as a function of baseline for
$\Delta m^2_{32} = +0.0035$~eV$^2$ and $- 0.0035$~eV$^2$, with 
$\sin^2 2\theta_{13}$ set to the small value 0.004. (Although these 
$\Delta m^2$ values 
are now a little different from those emerging from global analyses of the 
atmospheric and solar neutrino data, they are the ones used for the figure, 
which comes from Ref.~\cite{comments}, and are still useful to 
illustrate how the 
measurements can be used to determine the oscillation parameters.)    
Figure~\ref{sec4:fig2} shows two bands. The 
upper (lower) band corresponds to $\Delta m^2_{32} < 0\, (> 0).$ 
Within the bands, the \textsl{CP} phase $\delta$ is varying. 
At short baselines the bands converge, and the ratio $R = 0.5$ since 
the antineutrino CC cross section is half of the neutrino CC cross section. 
At large distances, matter effects enhance $R$ if $\Delta m^2_{32} < 0$ and 
reduce $R$ if $\Delta m^2_{32} > 0,$ and the bands diverge. Matter effects 
become significant for baselines exceeding about 2000~km. 
The error bars indicate 
the expected statistical uncertainty on the measured $R$ 
with a data sample of $5\times 10^{22}$~kton-decays. 
With these statistics, 
the sign of $\Delta m^2_{32}$ is determined with very high statistical significance.
With an order of magnitude smaller data sample 
(entry level scenario~\cite{entry-level}) or with 
an order of magnitude smaller $\sin^2 2\theta_{13}$ the statistical uncertainties 
would be $\sqrt{10}$ larger, but the 
sign of $\Delta m^2_{32}$ could still be determined with convincing precision.
\begin{figure}[tbph!]
\includegraphics*[width=3.5in]{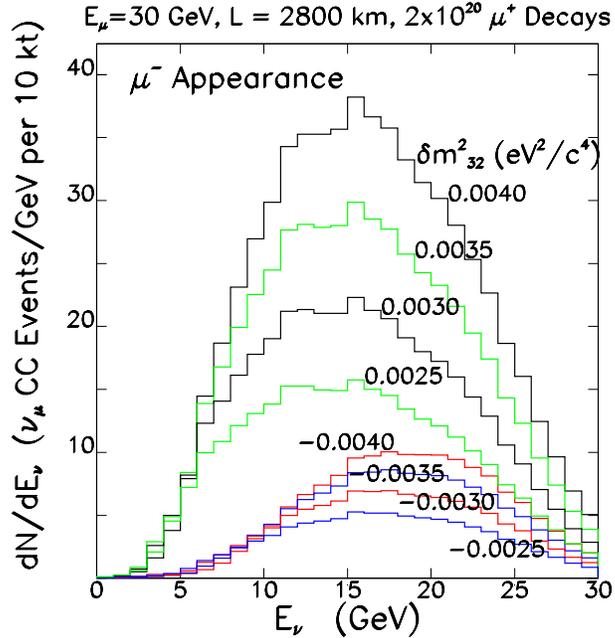}
\caption{(Color) Predicted measured energy distributions for CC events tagged by a wrong-sign (negative) muon from $\nu_e \to\nu_\mu$ oscillations (no cuts or backgrounds), shown for various $\delta m^2_{32}$, as labeled. The predictions correspond to $2 \times 10^{20}$ decays, $E_\mu = 30$~GeV, $L = 2800$~km, and a representative set of values for $\delta m^2_{12}$, $\sin^22\theta_{13}$, $\sin^22\theta_{23}$, $\sin^22\theta_{12}$, and $\delta.$ Results are from Ref.~\cite{bgrw99}.}
\label{fig:v3}
\end{figure}
\begin{figure}[btph!]
\includegraphics*[width=3.5in]{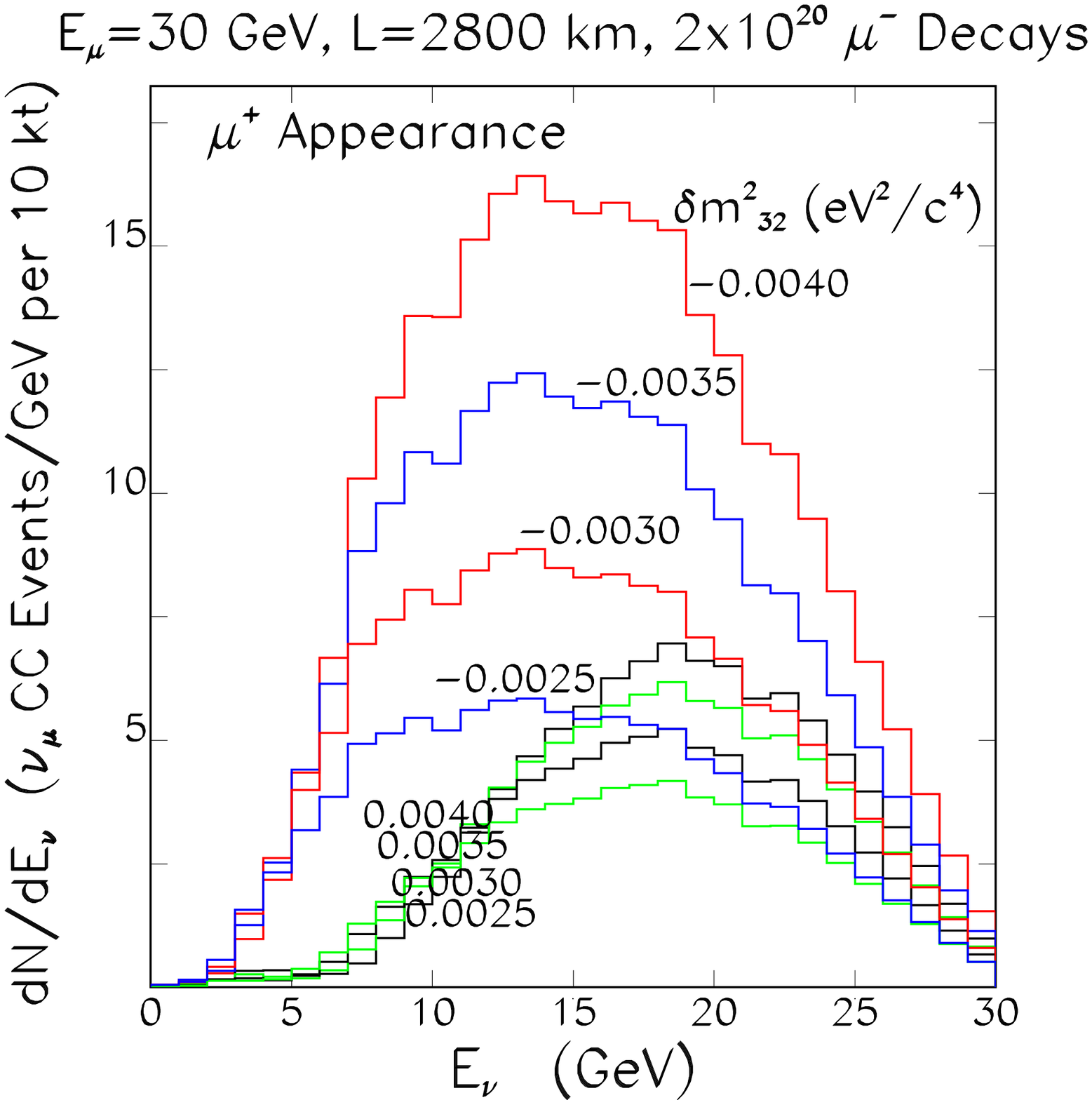}
\caption{(Color) Same as in Fig.~\ref{fig:v3}, for CC events
tagged by a wrong-sign (positive) muon from $\bar{\nu}_e \to
\bar{\nu}_\mu$ oscillations.
}
\label{fig:v4}
\end{figure}
In addition to the ratio of wrong--sign muon signal rates $R$, 
the two energy-dependent wrong-sign muon event energy distributions 
can be separately measured. To show how this additional information can help,  
the predicted measured energy distributions 2800~km downstream of 
a 30~GeV Neutrino 
Factory are shown in Figs.~\ref{fig:v3} and \ref{fig:v4} for, respectively,  
$\nu_e \to \nu_\mu$ and 
$\bar{\nu}_e \to \bar{\nu}_\mu$ wrong--sign muon 
events. 
The distributions are shown for 
a range of positive and negative values of $\delta m^2_{32}$. 
Note that, after allowing for the factor of two difference 
between the neutrino and antineutrino cross sections, 
for a given $|\delta m^2_{32}|$, if $\delta m^2_{32} > 0$ 
we would expect to observe a lower wrong--sign muon event rate and 
a harder associated spectrum when positive muons are stored in 
the Neutrino Factory than when negative muons are stored. 
On the other hand, if $\delta m^2_{32} < 0$
we would expect to observe a higher wrong--sign muon event rate and
a softer associated spectrum when positive muons are stored in
the Neutrino Factory than when negative muons are stored. Hence, 
measuring the differential spectra when positive and negative muons are 
alternately stored in the Neutrino Factory can both enable the sign of 
$\delta m^2_{32}$ to be unambiguously determined~\cite{bgrw99}, 
and also provide a measurement of $\delta m^2_{32}$ and a consistency check 
between the behavior of the rates and energy distributions.
\begin{figure}[!bhtp]
\includegraphics*[width=5in]{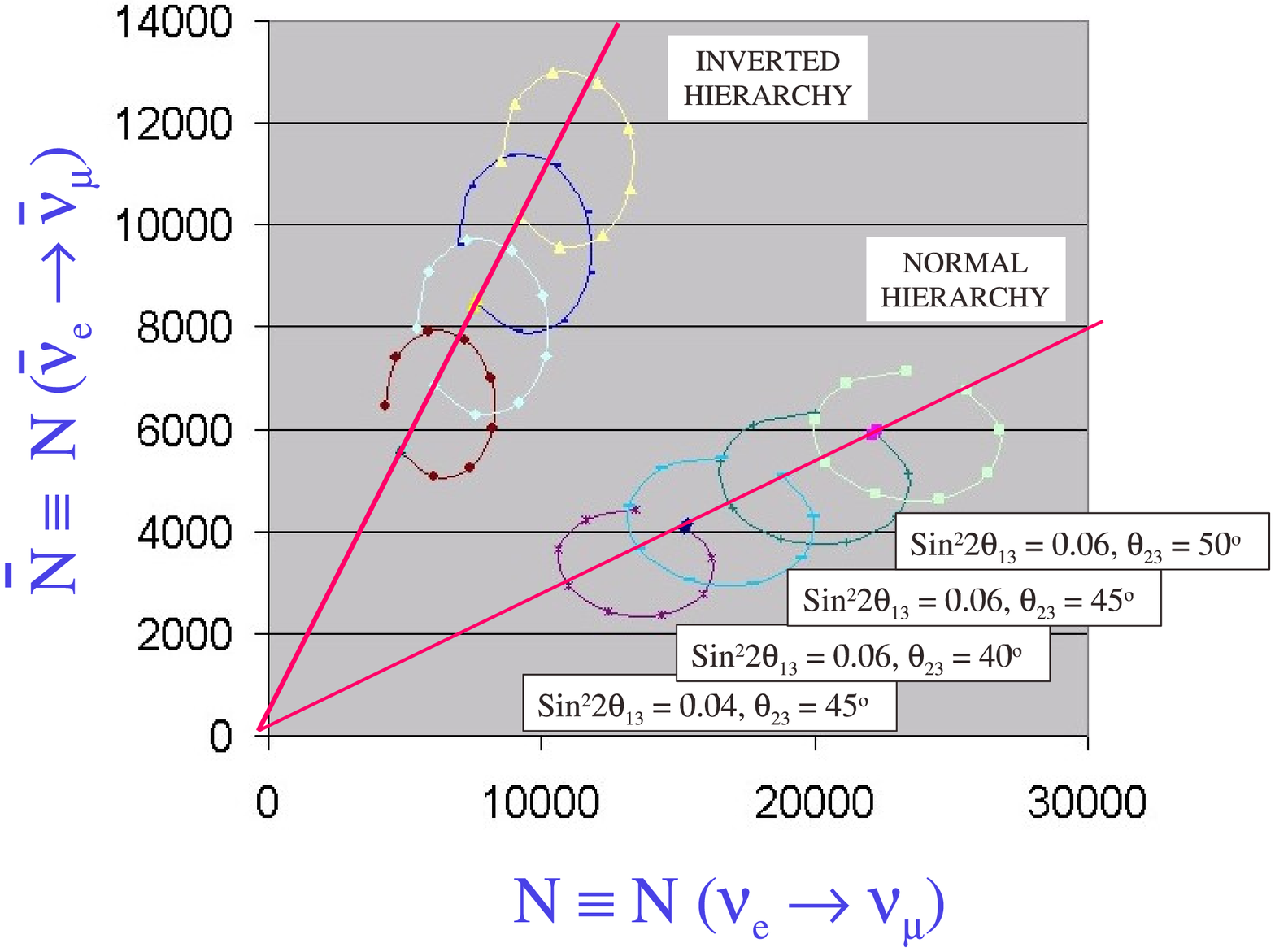}
\caption{(Color) The predicted number of wrong--sign muon events 
when negative muons 
are stored in the Neutrino Factory, versus the corresponding rate when 
positive muons are stored, shown as a function of 
$\theta_{13}, \theta_{23}, \delta$ and the assumed mass hierarchy, 
as labeled. The calculation corresponds to 
a 16~GeV Neutrino Factory 
with a baseline of 2000~km, and 10~years of data taking with a 
100~kton detector and $2 \times 10^{20} \; \mu^+$ and 
$2 \times 10^{20} \; \mu^-$ decays in the beam-forming straight section 
per year. The ellipses show how the predicted rates vary as the \textsl{CP} phase 
$\delta$ varies.}
\label{fig:ellipses}
\end{figure}
\subsubsection{Other Channels}
\begin{figure}[thbp!]
\includegraphics*[width=4.5in]{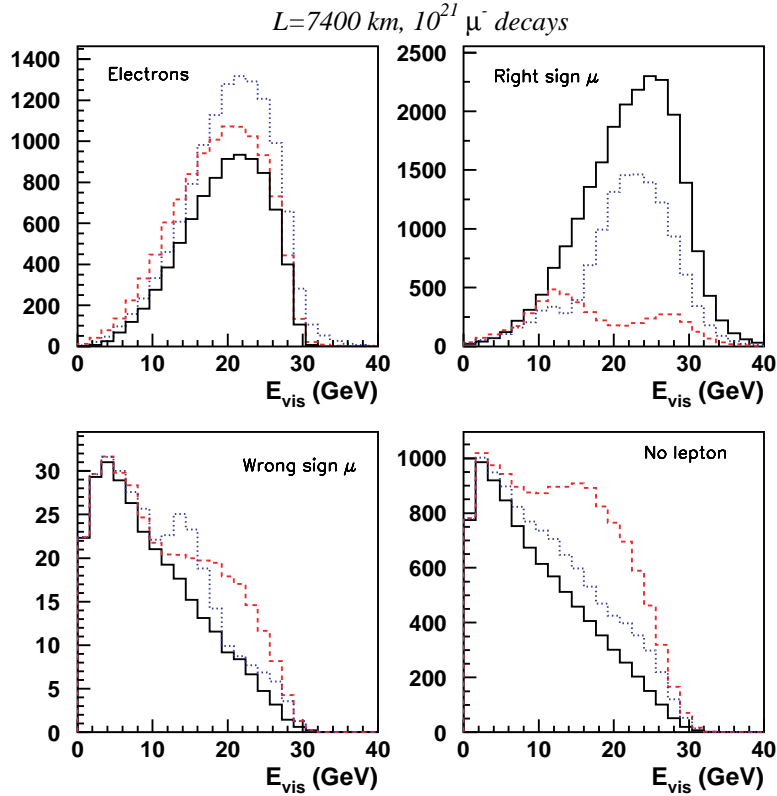}
\caption{(Color) Visible energy spectra for four event classes when 
$10^{21}\,\mu^-$ 
decay in a 30~GeV Neutrino Factory at $L = 7400$~km.
Black solid histogram: no oscillations. 
Blue dotted histogram: $\delta m^2_{32}=3.5\times 10^{-3}$~eV$^2$/c$^4$, 
$\sin^2\theta_{23}=1$. 
Red dashed histogram: $\delta m^2_{32}=7\times 10^{-3}$~eV$^2$/c$^4$, 
$\sin^2\theta_{23}=1$. 
The distributions in this figure and the following figure 
are for an ICANOE-type detector, and are 
from Ref.~\cite{camp00}.}
\label{fig:m1}
\end{figure}

\begin{figure}[bthp!]
\includegraphics*[width=4.5in]{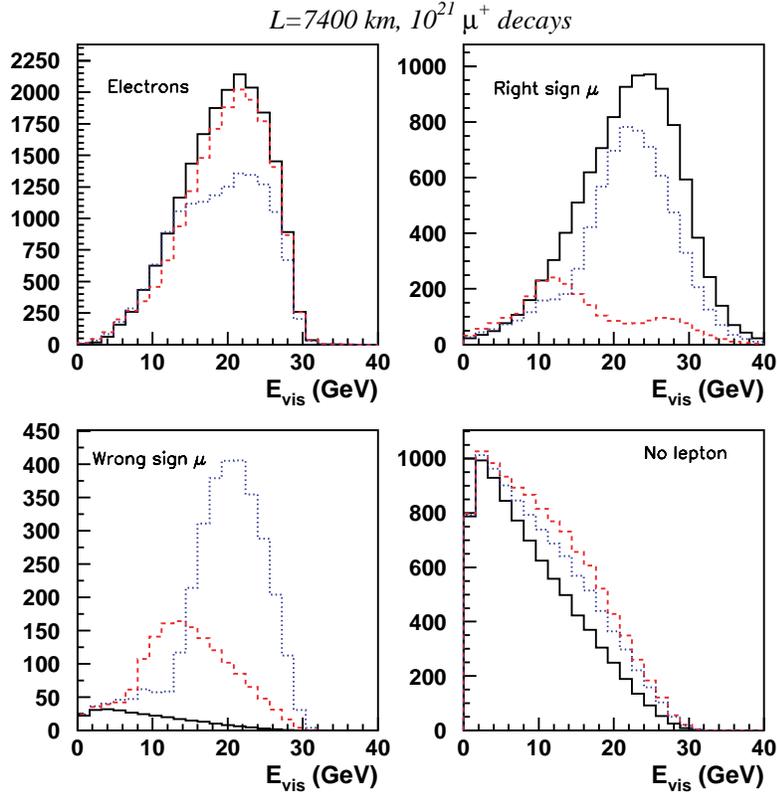}
\caption{(Color) Same as in Fig.~\ref{fig:m1}, but with positive muons circulating in the
storage ring. The difference between the two figures is due to the different
cross section for neutrinos and antineutrinos, and to matter effects.}
\label{fig:m2}
\end{figure}

\begin{figure}[thbp!]
\includegraphics*[width=3.5in]{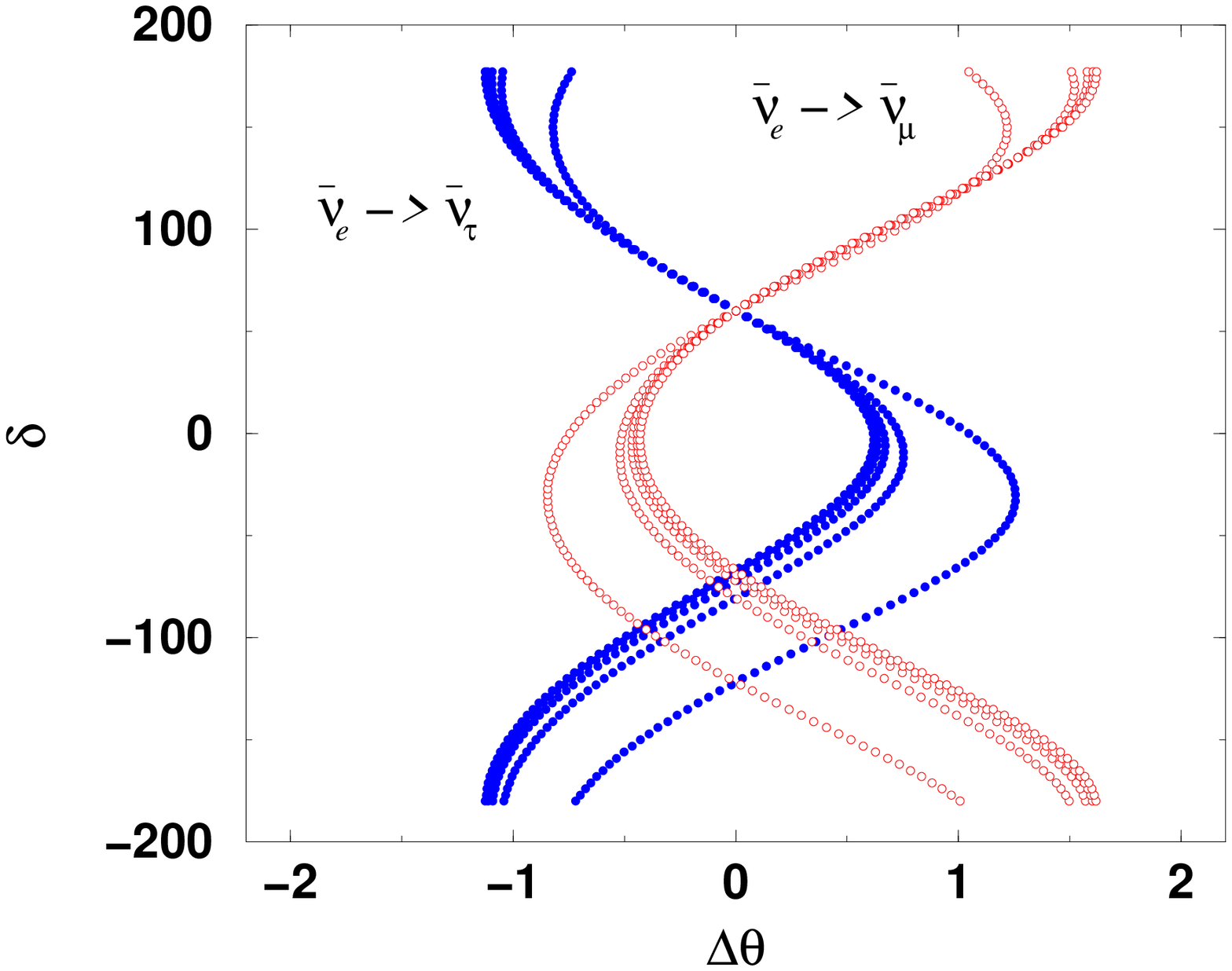}
\caption{(Color) Equiprobability curves in the ($\Delta \theta, \delta$)
plane, for $\bar \theta_{13} = 5^\circ,$ $ \bar \delta = 60^\circ$,
$E_\nu \in [5, 50] $ GeV and $L = 732$ km for the $\nu_e \to \nu_\mu$
and $\nu_e \to \nu_\tau$ oscillation (neutrinos on the left,
antineutrinos on the right). $\Delta \theta$ is defined as the
difference between the reconstructed parameter $ \theta_{13}$ and the
input parameter $\bar{\theta}_{13}$, i.e., $\Delta \theta = 
\theta_{13}- \bar{\theta}_{13}$. From Ref.~\cite{donini}.}
\label{fig:olga}
\end{figure}

In practice, to measure $\theta_{13}$, determine the mass hierarchy, and 
search for \textsl{CP} violation, 
the analysis of the wrong-sign muon rates must be performed allowing all of 
the oscillation parameters to simultaneously vary within their 
uncertainties. Since the relationship between the measured quantities 
and the underlying mixing parameters is complicated, with a minimal set 
of measurements it may not be 
possible to identify a unique region of parameter space consistent 
with the data. For Superbeams a detailed discussion of this problem 
can be found in Refs.~\cite{Minakata,Fogli:1996pv,Winter,Burguet-Castell:2001ez,Barger:2001yr,Burguet-Castell:2002qx}. 
To understand the nature of the challenge, Fig.~\ref{fig:ellipses} 
shows, as a function of $\theta_{13}, \theta_{23}, \delta$ and the 
assumed mass hierarchy, the predicted number of wrong--sign muon events 
when negative muons 
are stored in the Neutrino Factory, versus the corresponding rate when 
positive muons are stored. The example is for a 16~GeV Neutrino Factory 
with a baseline of 2000~km, and 10~years of data taking with a 
100~kton detector and $2 \times 10^{20} \; \mu^+$ and 
$2 \times 10^{20} \; \mu^-$ decays in the beam-forming straight section 
per year. The ellipses show how the predicted rates vary as the \textsl{CP} phase 
$\delta$ varies. All of the \textsl{CP} conserving points ($\delta = 0$ and $\pi$) 
lie on the diagonal lines. Varying the mixing angles moves the ellipses 
up and down the lines. Varying the mass hierarchy moves the family of 
ellipses from one diagonal line to the other. Note that the statistics 
are large, and the statistical errors would be barely visible if plotted 
on this figure. Given these statistical errors, for the parameter region 
illustrated by the figure, determining the mass hierarchy (which diagonal 
line is the measured point closest to) will be straightforward. Determining 
whether there is \textsl{CP} violation in the lepton sector will amount to 
determining whether the measured point is consistent with being on the 
\textsl{CP} conserving line. Determining the exact values for the mixing angles and 
$\delta$ is more complicated, since various combinations can result in the 
same predicted values for the two measured rates. This is the origin of 
possible false solutions in the three--flavor mixing parameter space.  
To eliminate those false solutions, event samples other 
than $\nu_e \to \nu_\mu$ transitions tagged by wrong-sign muons 
will be important. 
We have seen that, in the presence of 
$\nu_e \to \nu_\mu$ oscillations, the $\nu_\mu$ CC interactions 
produce negative muons (i.e., wrong--sign muons). Similarly, 
$\bar{\nu}_\mu \to \bar{\nu}_e$ oscillations 
produce wrong--sign electrons, 
$\bar{\nu}_\mu \to \bar{\nu}_\tau$ oscillations 
produce events tagged by a $\tau^+,$ and 
$\nu_e \to \nu_\tau$ oscillations 
produce events tagged by a $\tau^-$. 
Hence, there is a variety of information that can be used 
to measure or constrain neutrino oscillations at a Neutrino Factory, 
namely the rates and energy distributions of events tagged by
\begin{description} 
\item{(a)} right--sign muons 
\item{(b)} wrong--sign muons 
\item{(c)} electrons or positrons (their charge is difficult to determine in a massive detector) 
\item{(d)} positive $\tau$--leptons 
\item{(e)} negative $\tau$--leptons 
\item{(f)} no charged lepton. 
\end{description}
If these 
measurements are made when there are alternately positive and negative 
muons decaying in the storage ring, there are a total of 12~spectra 
that can be used to extract information about the oscillations. 
Some examples of the predicted measured spectra are shown as a function of the 
oscillation parameters in Figs.~\ref{fig:m1} and 
\ref{fig:m2} for a 10~kton detector sited 7400~km 
downstream of a 30~GeV Neutrino Factory. 
These distributions are sensitive to the oscillation parameters, and 
can be fit simultaneously to extract the maximum information. 
Clearly, the high intensity $\nu_e$, $\bar{\nu}_e$, $\nu_\mu$, and 
$\bar{\nu}_\mu$ beams at a Neutrino Factory would provide a wealth of 
precision oscillation data. 
The full value of this wealth of information has not been fully explored, 
but some specific things to be noted are:
\begin{enumerate}
\item It has been shown~\cite{donini,synergy1,huber,lindner} 
that the various measurements at a 
Neutrino Factory 
provide sufficient information to eliminate false solutions within the 
three--flavor parameter space. Indeed the wealth of information in the 
Neutrino Factory 
data is essential for this purpose.
\item If $\sin^2 2\theta_{13}$ exceeds $\sim 0.001$ 
the $\nu_e \to \nu_\tau$ channel is particularly 
important, 
both as a means to suppress the false solutions~\cite{donini,synergy1,donini2}, 
and also as the only direct 
experimental probe of $\nu_e \leftrightarrow \nu_\tau$ transitions. 
The ability of the $\nu_e \to \nu_\tau$ measurements to 
eliminate false solutions is illustrated in Fig.~\ref{fig:olga}, which, 
for a representative set of oscillation parameters, shows as 
a function of the \textsl{CP} phase $\delta$ the location of the false solution 
with respect to the correct solution in $\theta_{13}$--space 
(or more precisely, the distance between the two solutions $\Delta\theta$).
Note that, when compared to the 
$\bar{\nu_e} \to \bar{\nu_\mu}$ case, 
$\Delta\theta$ has the opposite sign for 
$\bar{\nu_e} \to \bar{\nu_\tau}$. In practice,  
this means that 
together the two measurements enable the false solution to be effectively eliminated.
\item Within the three--flavor framework, the relationship between 
the measured oscillation probabilities and the associated oscillation 
parameters is complicated. Experimental redundancy, permitting 
the over-determination of the oscillation parameters, 
is likely to prove essential, 
both to weed out misleading measurements 
and to ensure that the three-flavor framework is correct.
\end{enumerate}
\subsubsection{Neutrino Factory Calculations}
To understand how sensitive Neutrino Factory measurements will be 
in determining  $\theta_{13}$ and the neutrino mass hierarchy, and 
the sensitivity to \textsl{CP} violation in the lepton sector, we must consider 
the impact of statistical and systematic uncertainties, correlations 
between the parameters that vary within fits to the measured distributions, 
and the presence or absence of false solutions in the three-flavor 
mixing parameter space. To take account of these effects, and to see which 
different neutrino oscillation experiments best complement one another, a 
global fitting program has been created~\cite{lindner,globes} that  
uses simulated right-sign muon and wrong-sign muon data sets, and 
includes:
\begin{enumerate}
\item Beam spectral and normalization uncertainties.
\item Matter density variations of 5\% about the average value.
\item Constraint of solar neutrino oscillation parameters 
within the post-KamLAND LMA region.
\item Simulation of $\nu_\mu$ CC QE, $\nu_\mu$ and $\nu_e$ CC 
inelastic, and NC events for all flavors. Note that 
the NC events are included 
in the analysis as a source of background. The NC signal is not 
yet exploited as an additional constraint.
\item A check of the influence of cross section uncertainties 
(this mostly affects energies lower than those of interest for Neutrino Factories).
\item Energy-dependent detection efficiencies, enabling energy threshold
effects to be taken into account. 
\item Gaussian energy resolutions.
\item Flavor, charge, and event misidentification.
\item Overall energy-scale and normalization errors.
\item An analysis of statistical and systematic precisions, and the 
ability to eliminate false solutions.
\end{enumerate}
\begin{table}[thbp!]
\caption{Signal and background rates for a CERN SPS Beta beam, 
a high performance Superbeam (a 4~MW JHF beam with a 1~Mton water Cerenkov 
detector), and a Neutrino Factory. The numbers correspond to 
$\sin^2 2\theta_{13} = 0.1$ and $\delta = 0$. The rates have been calculated 
by the authors of Ref.~\cite{lindner}.
\label{tab:SoverB}}
\begin{ruledtabular}
\begin{tabular}{lccc}
 & $\beta$-Beam & JHF-HK& Nu-Factory\\
\hline
\multicolumn{4}{c}{$\nu$}\\
\hline
Signal & 4967 & 13171 & 69985 \\
Background & 397 & 2140 & 95.2 \\
Signal/Background & 12.5 & 6.2 & 735 \\
\hline
\multicolumn{4}{c}{$\bar{\nu}$}\\
\hline
Signal & 477 & 9377 & 15342 \\
Background & 1 & 3326 & 180 \\
Signal/Background & 477.5 & 2.8 & 85.2 \\
\end{tabular}
\end{ruledtabular}
\end{table}
\begin{figure}[thbp!]
\includegraphics*[width=3.5in]{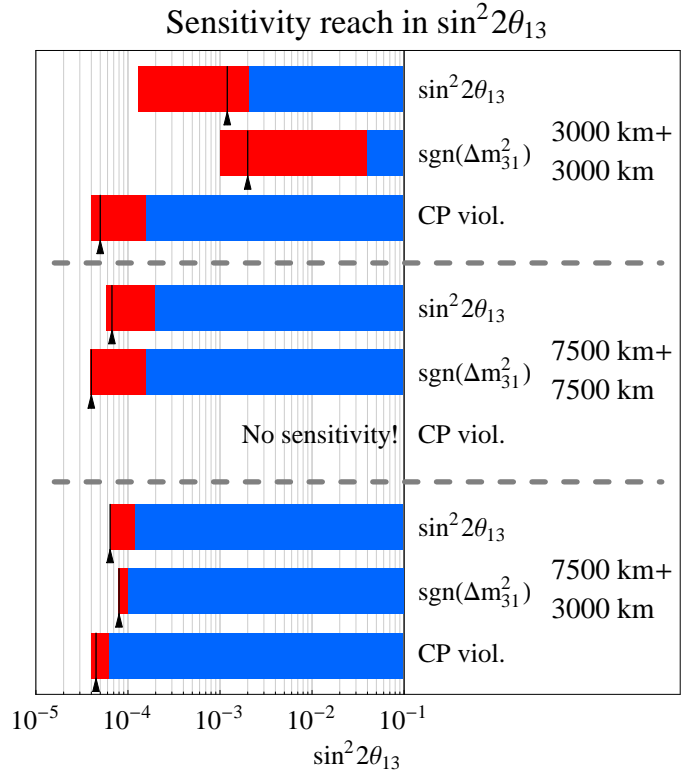}
\caption{(Color) The sensitivity reaches as functions of
$\sin^2 2 \theta_{13}$ for $\sin^2 2 \theta_{13}$ itself, the sign of
$\Delta m_{31}^2>0$, and (maximal) \textsl{CP} violation
$\delta_{\mathrm{CP}}=\pi/2$ for each of the indicated
baseline combinations. The bars show the ranges in $\sin^2 2
\theta_{13}$ where sensitivity to the corresponding quantity can be
achieved at the $3\sigma$
confidence level. The dark bars mark the variations in the sensitivity
limits by allowing the true value of $\Delta m_{21}^2$ to vary in the $3\sigma$
LMA-allowed range given in Ref.~\cite{Maltoni:2002aw} and others
$(\Delta m_{21}^2 \sim 4\times 10^{-5} \, \text{eV}^2 - 3\times 10^{-4} \, \text{eV}^2).$ The arrows/lines correspond to the LMA best-fit value. Figure from Ref.~\cite{huber}.}
\label{fig:lindner}
\end{figure}
The calculated signal and background rates are 
listed in Table~\ref{tab:SoverB}. The roughly two orders of magnitude 
improvement in the signal/background ratio at a Neutrino Factory, 
compared with the corresponding ratio at a high performance Superbeam, 
is evident. The 
results from the full calculations are shown in Fig.~\ref{fig:lindner}. 
The calculation is more fully described in Ref.~\cite{lindner}. 
The figure shows 
the minimum value of $\sin^2 2\theta_{13}$ for which three 
experimental goals could be achieved (with $3\sigma$ significance). 
First, the observation of a finite value of $\theta_{13}$. Second, 
the determination of the neutrino mass hierarchy. Third, the observation 
of non-zero \textsl{CP} violation in the lepton sector if the underlying 
$\delta$ corresponds to maximal \textsl{CP} violation. The three groups of 
bars correspond to three different experimental scenarios, with 
different baselines. The favored scenario is the one illustrated 
by the bottom group of three bars, for which there are two detectors, 
one at $L = 7500$~km and the other at $L = 3000$~km. Note that:

\textbf {At a Neutrino Factory ${\sin^2 2\theta_{13}}$ can be measured, the 
neutrino mass hierarchy determined, and a search for \textsl{CP} violation in the 
lepton sector made for all values of $\sin^2 2\theta_{13}$ down to 
\textit{O}$(10^{-4}$), or even a little less.}
%
\begin{figure}[thbp!]
\includegraphics*[width=4in]{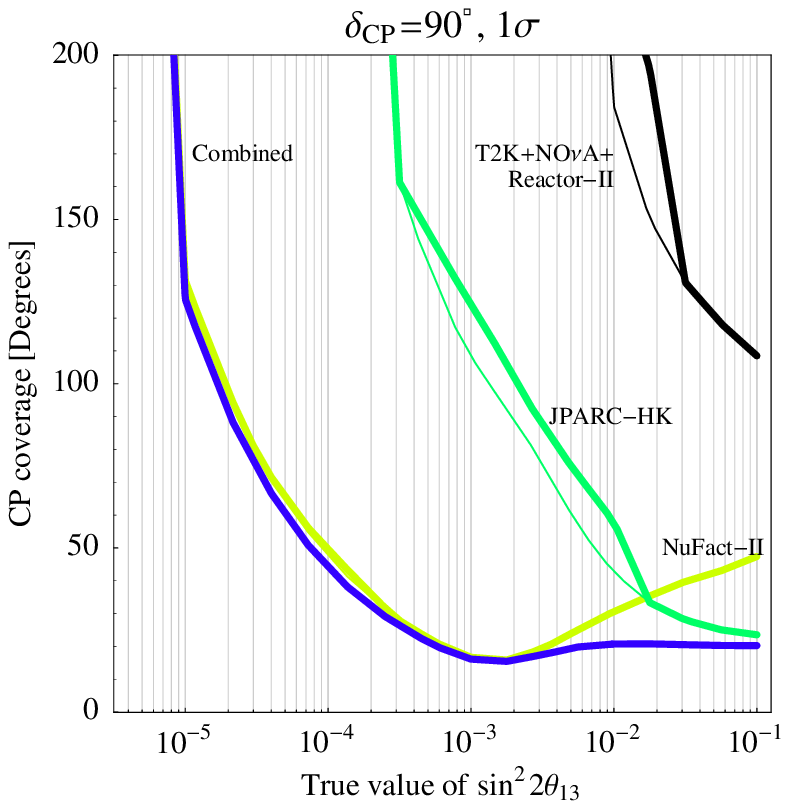}
\caption{(Color) The $1\sigma$
precision on the determination of the phase $\delta$ at a Neutrino Factory, 
and at a representative high-performance Superbeam, together with the 
combined Neutrino Factory plus Superbeam sensitivity. The sensitivities are 
shown as a function of the underlying value of $\sin^2 2\theta_{13}.$ The thin
curves correspond to cases where the the \textit{sign-degeneracy} is not taken into account. Calculation from the authors of Ref.~\cite{lindner}.}
\label{fig:delta}
\end{figure}

If $\sin^2 2\theta_{13}$ is fairly large, Superbeam experiments may also 
establish its value, and perhaps determine the mass hierarchy and begin 
the search for \textsl{CP} violation. Figure~\ref{fig:delta} 
illustrates the role of a Neutrino 
Factory over a broad range of $\sin^2 2\theta_{13}$ values. The figure shows,
as a function of the underlying value of $\sin^2 2\theta_{13}$, the $1\sigma$
precision on the determination of the phase $\delta$ at a Neutrino Factory, 
and at a representative high-performance Superbeam, together with the 
combined Neutrino Factory plus Superbeam sensitivity. Below values of 
$\sin^2 2\theta_{13} \sim 10^{-2}$ the Neutrino Factory sensitivity 
is significantly better than the sensitivity that can be achieved with 
Superbeams, and indeed provides the only sensitivity to the \textsl{CP} phase 
if $\sin^2 2\theta_{13}$ is significantly smaller than $10^{-2}$. Above 
$\sin^2 2\theta_{13} \sim 10^{-2}$ the Neutrino Factory measurements 
still enable a modest improvement to the \textsl{CP} violation measurement 
sensitivity, but the exact impact that a Neutrino Factory might have in 
this case is less clear. The uncertainty on the matter density, which 
is believed to be \textit{O}$(5\%),$ is likely to be a limiting uncertainty for 
\textsl{CP} violation measurements~\cite{Ohlsson}. Improved knowledge of the 
matter density along the neutrino flight-path would improve the expected 
Neutrino Factory sensitivity. In addition, Bueno \textit{et al.}~\cite{camp00}
have shown that the energy dependencies of matter and \textsl{CP} violating
 effects are different, and can be exploited to further separate the two 
effects. For $\sin^2 2\theta_{13} > 0.01,$ the case for 
a Neutrino Factory will depend upon just how well Superbeam experiments 
will ultimately be able to do, whether any new discoveries are made along 
the way that complicate the analysis, whether any theoretical progress is 
made along the way that leads to an emphasis on the type of measurements 
that a Neutrino Factory excels at, how important further tests of the 
oscillation formalism is in general, and the importance of observing 
and measuring $\nu_e \to \nu_\tau$ oscillations in particular. 

\textbf{We conclude there is a strong physics case for a Neutrino Factory if 
$\sin^2 2\theta_{13}$ is less than $\sim 0.01$. There may also be a strong 
case if $\sin^2 2\theta_{13}$ is larger than this, but it is too early to 
tell.}
\subsubsection{Special Case: $\theta_{13} = 0$}
The case $\theta_{13} = 0$ is very special. 
The number of mixing angles needed to describe the $3 \times 3$
unitary neutrino mixing matrix would be reduced from three to two, 
suggesting the existence of a new conservation
law resulting in an additional constraint on the elements of the 
mixing matrix. The discovery of a new
conservation law happens rarely in physics, and almost always leads 
to revolutionary insights in our
understanding of how the physical universe works. 
Hence, if it were possible to establish that  $\theta_{13} = 0$, 
it would be a major discovery. Note that in the limit $\theta_{13} \to 0,$ 
the oscillation
probability for $\nu_e \leftrightarrow \nu_\mu$ transitions is finite, 
and is given by:
\begin{eqnarray}
P\left( \nu_e \to \nu_\mu \right) & = &  
 \frac{\Delta m_{21}^2}{\Delta m_{31}^2} \sin^2 2\theta_{12} \cos^2 \theta_{23}
 \frac{\sin^2 A \Delta}{A^2} \, ,
\end{eqnarray}
where the matter parameter $A = 1$ if the neutrino energy corresponds 
to the matter resonance, which for a
long-baseline terrestrial experiment means neutrino energies 
$E \sim 12$~GeV. In addition, if the baseline $L$ is
chosen such that $L/E$ corresponds to the oscillation maximum, 
then $\sin^2\Delta$ = 1, and we have that
\begin{equation}
P(\nu_e \to \nu_\mu)\sim\sin^2 2\theta_{12} \cos^2 \theta_{23}  \frac{\Delta m_{21}^2}{\Delta m_{31}^2}.
\end{equation}
Substituting into this expression values for the oscillation parameters 
that are consistent with the present solar
and atmospheric neutrino data, we are led to conclude that even if 
$\theta_{13} = 0$, provided the neutrino
energy and baseline are chosen appropriately, 
$\nu_e \leftrightarrow \nu_\mu$ transitions are still directly
observable in an appearance experiment if oscillation probabilities of 
\textit{O}$(10^{-4})$ are observable. Hence, if
$\theta_{13}$ is very small, the ideal neutrino oscillation experiment 
will be a long baseline experiment that
uses neutrinos with energies close to 12~GeV, i.e., uses a baseline such 
that $L/E$ corresponds to the oscillation maximum, and
is sensitive to values of 
$P(\nu_e \leftrightarrow \nu_\mu)\sim 10^{-4}$ or smaller. 
Neutrino Factories provide
the only way we know to satisfy these experimental requirements.

\textbf{If $\theta_{13} = 0$ a Neutrino Factory experiment would enable 
(i) the first observation of $\nu_e \leftrightarrow \nu_\mu$ 
transitions in an appearance experiment, and (ii) an upper limit on 
$\sin^2 2\theta_{13}$ of \textit{O}($10^{-4}$) or smaller.}

These are major experimental results that would simultaneously 
provide a final confirmation the three-flavor
mixing framework (by establishing $\nu_e \leftrightarrow \nu_\mu$ 
transitions in an appearance experiment) while 
strongly suggesting the existence of a new conservation law.
In considering the case $\theta_{13} = 0,$ it should be noted that 
within the framework of GUT theories, radiative
corrections will change the value of $\sin^2 2\theta_{13}$ measured 
in the laboratory from the underlying value
of $\sin^2 2\theta_{13}$ at the GUT scale. Recent 
calculations~\cite{radiative} have 
suggested that these radiative corrections to
$\sin^2 2\theta_{13}$ will be \textit{O}$(10^{-4}).$
If this is the case, the ultimate Neutrino Factory experiment would not
only provide the first direct observation of 
$\nu_e \to \nu_\mu$ transitions, but would also
\begin{itemize} 
\item establish a finite value for $\theta_{13}$ at
laboratory scales consistent with being zero at the GUT scale, 
\item determine the sign of $\Delta m_{31}^2$, and
hence determine whether the neutrino mass hierarchy is normal or inverted, and 
\item detect maximal \textsl{CP} violation in the lepton sector.
\end{itemize} 
These would be tremendously important results.
\subsection{Beta Beam Calculations and Results}
The Beta Beam concept is more recent than the Neutrino Factory idea, and the performance of Beta Beam experiments is less well established. 
Recent calculations of the $\sin^2 2\theta_{13}$ sensitivity for
low energy Beta Beam scenarios~\cite{lindner-beta,Donini:2004hu} 
have included the effects of systematic
uncertainties, correlations, and false solutions in parameter
space. Expected signal and background rates are summarized in
Table~\ref{tab:SoverB}. The expected signal rates are relatively
modest. The neutrino Beta Beam signal would be a factor of 2--3 less than
expected at a high--performance Superbeam, and a factor of 14 less than at
a Neutrino Factory. The rates are even lower for an antineutrino Beta Beam;
a factor of 20 less than the rates at a high--performance Superbeam, and a
factor of 32 less than at a Neutrino Factory. In addition, it has been 
pointed out~\cite{jj-beta} that the neutrino
energies are comparable to the target nucleon kinetic energies due to
Fermi motion, and therefore there is no useful spectral information in the low
energy Beta Beam measurements. Hence, the useful information is restricted to 
the measured muon neutrino (and antineutrino) appearance
rates. Nevertheless, the signal/background ratios are good: 
12.5 for the
neutrino Beta Beam (compared with 6.2 for the Superbeam and 735 for the
Neutrino Factory), and an impressive 478 for the antineutrino Beta Beam
(compared with 2.8 for the Superbeam and 85 for the Neutrino Factory). Hence
the interest in Beta Beams.
\begin{figure}[hbtp!]
\includegraphics*[width=3in,angle=-90]{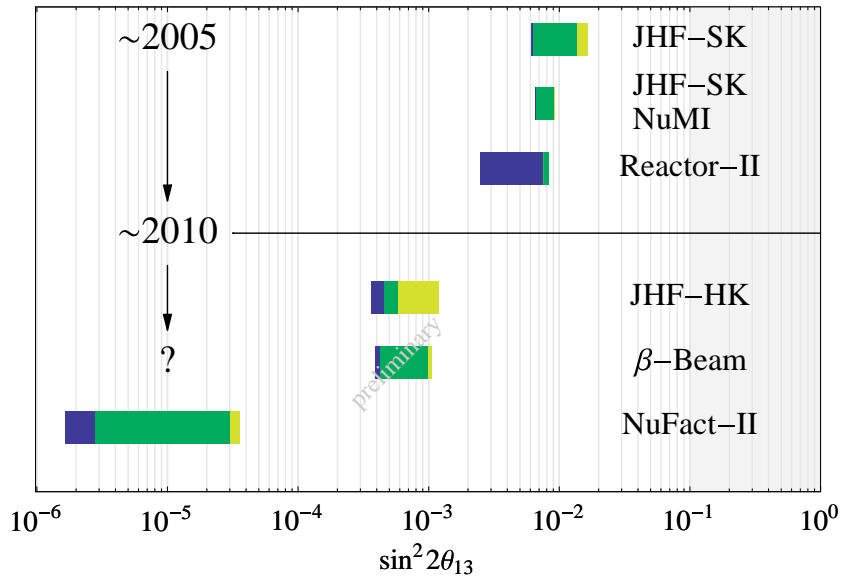}
\caption{(Color) CERN SPS Beta Beam sensitivity. 
}
\label{betabeam_comp}
\end{figure}
The ability of a low-energy Beta Beam to
discover a finite value for $\sin^2 2\theta_{13}$ is compared in
Fig.~\ref{betabeam_comp} 
with the corresponding $3\sigma$ sensitivities at a Neutrino Factory and high
performance Superbeam. The leftmost limits of each of the bars in  
Fig.~\ref{betabeam_comp} show the statistical sensitivities, and the shaded regions within the 
bars show the degradation of the sensitivities due to irreducible
experimental systematics, the effects of correlations, and the effects of
false solutions in the three-flavor mixing 
parameter space. The rightmost limit of
the bars therefore gives the expected sensitivities for each experiment. The
sensitivity of the low--energy Beta Beam experiment is expected to be
comparable to the corresponding Superbeam sensitivity.  A Neutrino Factory
would improve on the Beta Beam sensitivity by about a factor of
40. Combining low-energy Beta Beam results (the two measured rates) with
Superbeam results would enable the impact of correlations and ambiguities
to be reduced, which would potentially enable an improvement in the $\sin^2
2\theta_{13}$ sensitivity by a factor of 2--3 over the standalone
results. Hence, low energy Beta Beams offer only a
modest improvement in the $\sin^2 2\theta_{13}$ sensitivity beyond
that achievable with a high--performance Superbeam, and this realization 
has led to the
consideration of higher energy Beta Beams~\cite{jj-beta,Terranova:2004hu}. 
In particular, it has been
proposed that the energies be increased by at least a factor of a few so
that the neutrino and antineutrino energies are well above the Fermi motion
region, which would enable useful spectral information to be extracted from
the Beta Beam measurements. In addition, this would increase the signal 
rates (Table~\ref{tab:betabeam}), 
and if the energy were sufficiently high to result in significant
matter effects, then it would be possible (if $\theta_{13}$ is sufficiently
large) to use Beta Beams to determine the neutrino mass hierarchy.  The
particular scenarios that have been considered~\cite{jj-beta} are:
\begin{description}
\item{\textsl{Low Energy Beta Beam}:} This is the standard
  CERN scenario using the SPS for acceleration, and a 1~megaton water
  Cerenkov detector in the Fr\'{e}jus tunnel  ($\gamma = 60$, L =
  130~km).
\item{\textsl{Medium Energy Beta Beam}:} This would require the Fermilab
  Tevatron (or equivalent) for acceleration, and a 1~megaton water Cerenkov
  detector in the Soudan mine  ($\gamma = 350$, L = 730~km).
\item{\textsl{High Energy Beta Beam}:} This would require the LHC for acceleration, with
  $\gamma = 1500$, L = 3000~km.
\end{description}
In all three cases, the
running time is assumed to be 10~years. The improvement in statistical
precision enabled by the higher energy Beta Beam scenarios is illustrated
in Table~\ref{tab:betabeam} and Fig.~\ref{betabeam_fig}. 
\begin{figure}[thbp!]
\includegraphics*[width=3.5in]{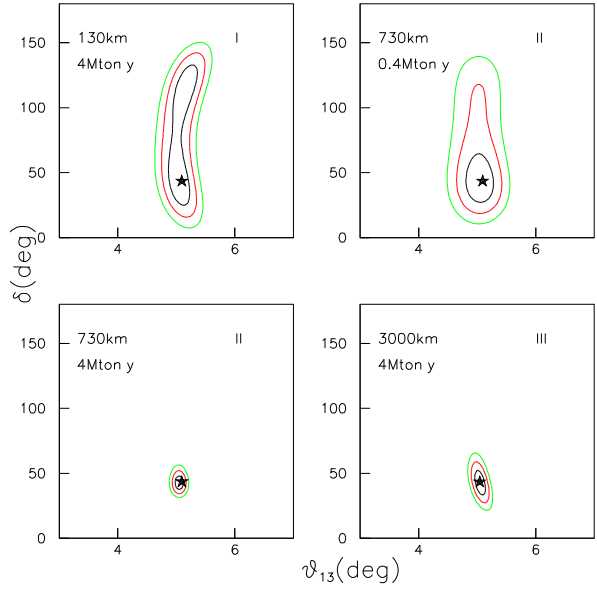}
\caption{(Color) Low-, Medium-, and High-Energy Beta Beam
  sensitivities. The estimated $\,1\sigma, 2\,\sigma$ and $3\,\sigma$ contours are shown for the setups described in the text. See Ref.~\cite{jj-beta}.}
\label{betabeam_fig}
\end{figure}
The figure shows,
for the three scenarios, the 1$\sigma$, 2$\sigma$, and 3$\sigma$ contours in
the ($\theta_{13},\delta$)--plane. Note that the expected sensitivity for
the medium energy case with a ``small'' water Cerenkov detector is
comparable to the low energy case with the megaton water Cerenkov
detector. However, the medium energy sensitivity is dramatically improved
with the much bigger detector. The further improvement obtained by going to
LHC energies seems to be marginal. Given the likelihood that the LHC would
not be available as a Beta Beam accelerator for a very long time, perhaps
the most interesting scenario is the medium energy one. To understand the
ability of medium energy Beta Beams to establish a finite value for
$\theta_{13}$, determine the neutrino mass hierarchy, and search for \textsl{CP}
 violation in the lepton sector, the full analysis must be performed, taking
care of all known systematic effects, and the impact of correlations and
degeneracies. Although this full analysis has not yet been done, a step
towards it has been made, and the results are
encouraging. 
\begin{figure}[bthp!]
\includegraphics*[width=3in]{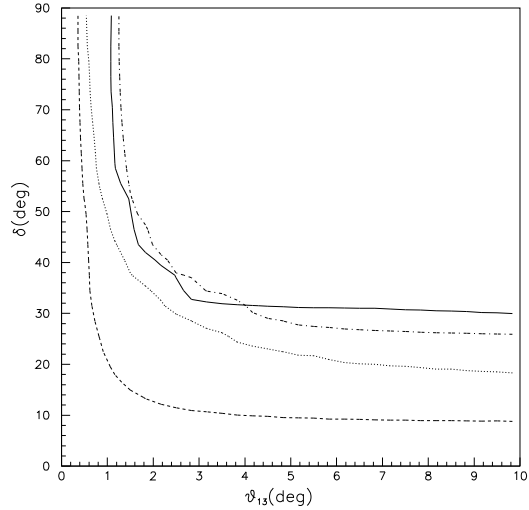}
\caption{Region where $\delta$ can be distinguished from $\delta=0^{\circ}$ or
  $\delta=180^{\circ}$ with a $99\%$ \textsl{CL} for the low energy Beta Beam(solid), 
medium energy Beta Beam with an
  UNO-type detector of 400~kton(dashed) and 
  with the same detector with a factor 10 smaller mass (dashed-dotted), and
  finally for the high energy Beta Beam (dotted) with a 40~kton tracking calorimeter. Figure from Ref.~\cite{jj-beta}.}
\label{betabeam_cpv_fig}
\end{figure}
Figure~\ref{betabeam_cpv_fig} shows the region of the
($\theta_{13},\delta$)--plane within which $\sin\delta = 1$ (maximal \textsl{CP}
 violation) can be separated from $\sin\delta = 0$ (no \textsl{CP} violation) at the
99\% C.L. The medium energy setup is sensitive to maximal \textsl{CP} violation for
values of $\theta_{13}$ exceeding $\sim 0.5$~degrees ($\sin^2 2\theta_{13}
\sim 3 \times 10^{-4}$). This is within a factor of a few of the expected
sensitivity that can be achieved at a Neutrino Factory. It will be
interesting to see if this calculated medium energy Beta Beam sensitivity
is significantly degraded when the uncertainties on all the oscillation
parameters and the systematic uncertainties on the neutrino cross sections,
etc., are included in the calculation. 
\begin{figure}[bhtp!]
\includegraphics*[width=3in]{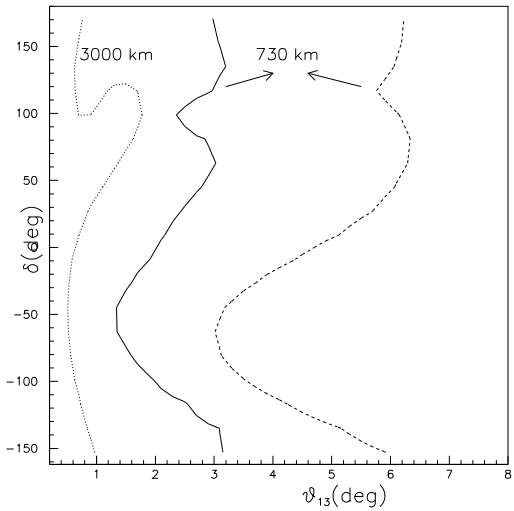}
\caption[a]{Regions where the true sign$(\Delta m^2_{23})=+1$ can be
  measured at $99\%$ C.L. (i.e., no solution at this level of confidence exists for the opposite sign). The lines correspond to the medium energy Beta Beam 
with a 400~kton water Cerenkov (solid), a 40~kton detector (dashed), 
and to the high energy Beta Beam(dotted). Figure from Ref.~\cite{jj-beta}.}
\label{betabeam_hierarchy_fig}
\end{figure}%
Finally, Fig.~\ref{betabeam_hierarchy_fig} shows, for the medium energy Beta Beam
scenario, the region of the ($\theta_{13},\delta$)--plane within which  the
neutrino mass hierarchy can be determined. The smallest value of
$\theta_{13}$ for which this can be accomplished is seen to be
$2-3$~degrees ($\sin^2 2\theta_{13} = 0.005 - 0.01$), which is perhaps a
little better than with a Superbeam, but is not competitive with a Neutrino
Factory.%

\section{Progress on Neutrino Factory and Beta Beam Facility Design\label{sec5}}
In this Section we describe the technical work accomplished as part of the
present Study. For the Neutrino Factory (Sections~\ref{sec5-sub1} and
\ref{sec5-sub2}) our focus was to update the FS2 design with some of the more
cost-effective approaches we have studied. In particular, a more optimized
capture section was designed, a shorter and less expensive bunching and phase
rotation scheme was developed, and a more optimized acceleration scheme based
on a combination of RLA and FFAG rings was worked out. Based on the improved
designs presented here, we worked out approximately what the savings with
respect to FS2 costs were. This is described in the Appendix.

For the Beta Beam facility, we have taken a brief look in Section~\ref{sec5-sub2-1} at the implications of using existing U.S. accelerator
facilities at BNL and Fermilab to provide the required beams. There is a real
motivation to explore this idea, because it appears that there are significant
scientific benefits associated with producing the neutrino beams from a Beta
Beam facility at higher energy than would be possible at CERN in the
foreseeable future.
\subsection{Neutrino Factory Front End\label{sec5-sub1}}
The front end of the neutrino factory (the part of the facility between the
target and the first linear accelerator) represented a large fraction of the
total facility costs in FS2~\cite{fs2}. However, several recent developments
have given hope that a new design for the front end may be possible that is
significantly less expensive: 
\begin{itemize}
\item A new approach to bunching and phase
rotation using the concept of adiabatic rf
bunching~\cite{adiab1,adiab2,adiab3,adiab4,adiab5} eliminates the very
expensive induction linacs used in FS2.
\item For a moderate cost, the
transverse acceptance of the accelerator chain could be doubled from its FS2
value. 
\item This diminished the demands on the transverse ionization cooling
section and allowed the design of a simplified cooler with fewer components
and reduced magnetic field strength.
\end{itemize}
 We denote as ``Study 2a'' the 
simulations that have been made of the performance of this new front end,
together with the new scheme for acceleration. The Monte Carlo simulations
were performed with the code ICOOL~\cite{icool}.
\begin{figure}[ptbh!]
\includegraphics*[viewport=20 275 570 750]{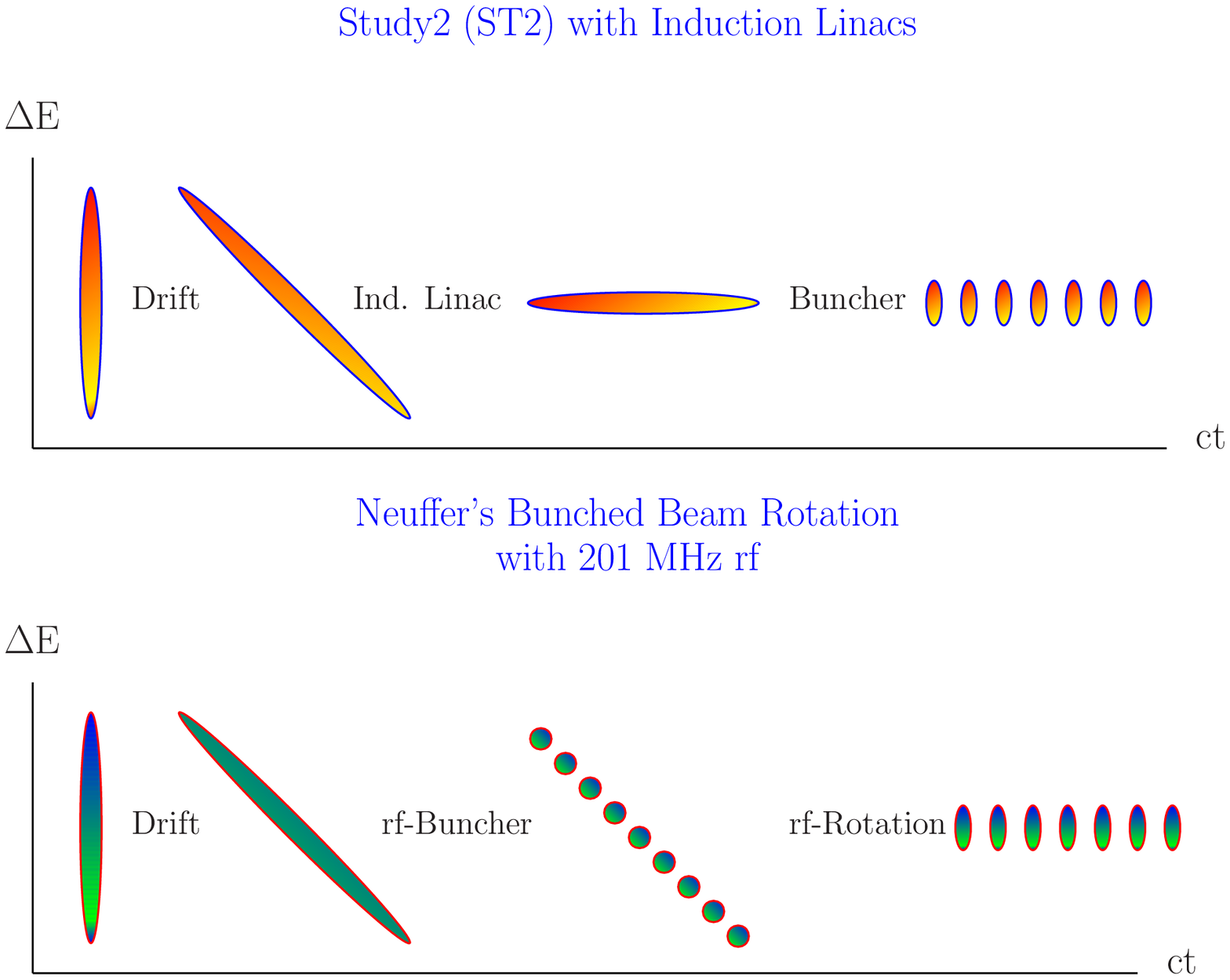}%
\caption{(Color) Comparison of the buncher concept used here with the bunching system
used in FS2.}%
\label{fig100}%
\end{figure}
The concept of the adiabatic buncher is compared with the system used in FS2
in Fig.~\ref{fig100}. The longitudinal phase space after the target is the
same in both cases. Initially, there is a small spread in time, but a very large spread
in energy. The target is followed by a drift space in both cases, where a strong
correlation develops between time and energy. In FS2, the energy spread in the
correlated beam was first flattened using a series of induction linacs. The
induction linacs did an excellent job, reducing the final rms energy spread to
4.4\%. The beam was then sent through a series of rf cavities for bunching, 
which increased the energy spread to $\approx8\%.$ In the new scheme, the
correlated beam is first adiabatically bunched using a series of rf cavities
with decreasing frequencies and increasing gradients. The beam is then phase
rotated with a second string of rf cavities with decreasing frequencies and
constant gradient. The final rms energy spread in the new design is 10.5\%.
This spread is adequate for the new cooling channel.
\begin{figure}[ptbh]
\includegraphics[width=4.5in,clip]{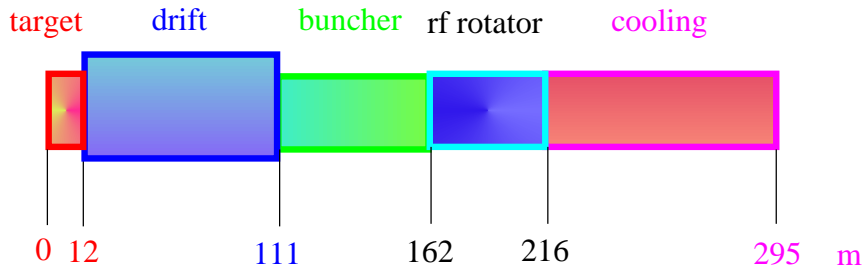}
\caption{(Color) Overall layout of the front-end.}%
\label{fig101}%
\end{figure}
The overall layout of the new front--end design is shown in Fig.~\ref{fig101}.
The first $\approx$12~m is used to capture pions produced in the target. The
field here drops adiabatically from 20~T over the target down to 1.75~T. At
the same time, the radial aperture of the beam pipe increases from 7.5~cm at
the target up to 25~cm. Next comes $\approx$100~m for the pions to decay into
muons and for the energy-time correlation to develop. The adiabatic bunching
occupies the next $\approx$50~m and the phase rotation $\approx$50~m
following that. Lastly, the channel has $\approx$80~m of ionization cooling.
The total length of the new front end is $295$~m. \begin{figure}[ptbh]
\includegraphics[angle=90,width=5in]{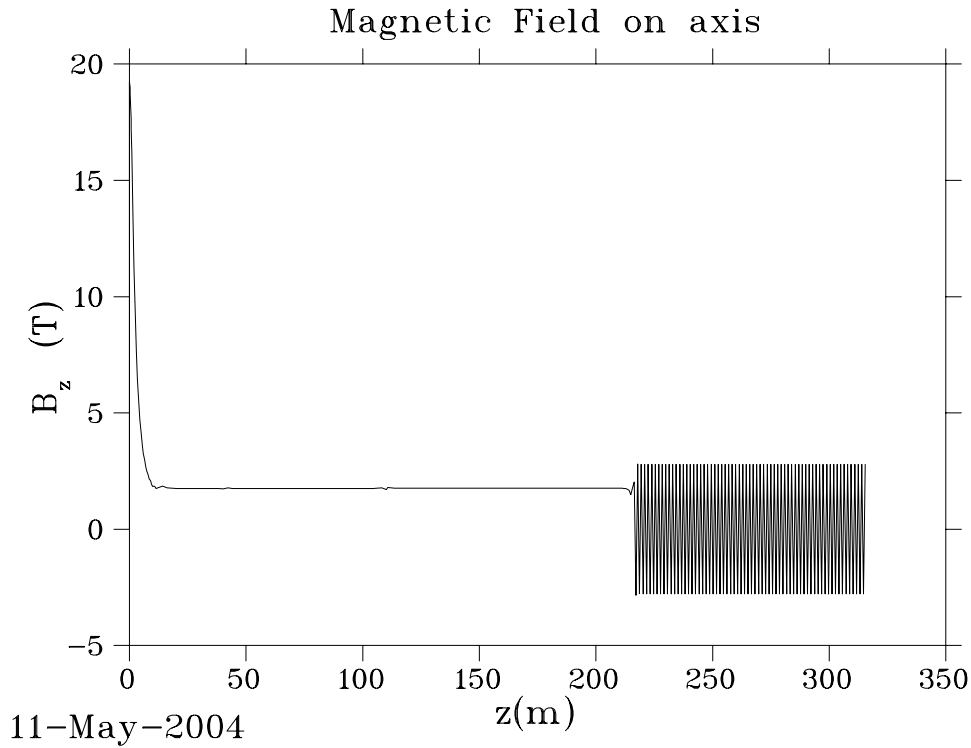}%
\caption{Longitudinal field component $B_{z}$ on-axis along the Study2a front-end.}%
\label{fig102}%
\end{figure}
The longitudinal field component on-axis is shown for the full front-end in
Fig.~\ref{fig102}. The field falls very rapidly in the collection region to a
value of 1.75~T. It keeps this value with very little ripple over the decay,
buncher and rotator regions. After a short matching section, the 1.75~T field
is changed adiabatically to the alternating field used in the cooler.
The beam distributions used in the simulations were generated using MARS~
\cite{mars1}. The distribution is calculated for a 24~GeV proton beam
interacting with a Hg jet~\cite{target}. The jet is incident at an angle of
100~mrad to the solenoid axis, while the beam is incident at an angle of
67~mrad to the solenoid axis. An independent study showed that the resulting
33~mrad crossing angle gives near-peak acceptance for the produced pions. An
examination of particles that were propagated to the end of the front-end
channel shows that they have a peak initial longitudinal momentum of $\approx$300~MeV/c with a long high-energy tail, and a peak initial transverse momentum
$\approx$180~MeV/c. 
\begin{figure}[ptbh]
\includegraphics{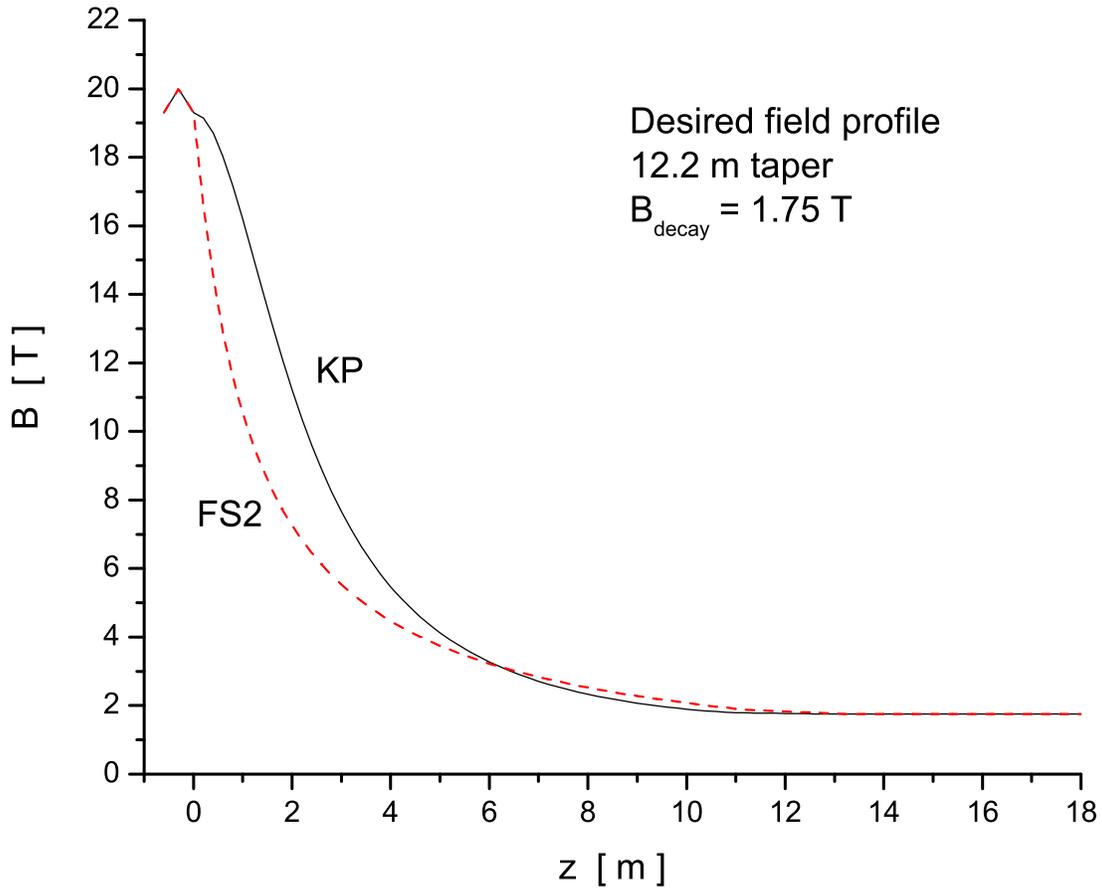}
\caption{(Color) Comparison of the
 capture region magnetic field used in the present simulation with that used
 in FS2.}%
\label{fig103}%
\end{figure}
We used an improved axial field profile in the capture region that
increased the final number of muons per proton in the accelerator acceptance
by $\approx$10\%. The new axial field profile (marked KP) is compared in
Fig.~\ref{fig103} with the profile used in FS2. Figure~\ref{fig104} shows the
actual coil configuration in the collection region. The end of the 60~cm long
target region is defined as $z = 0.$ The three small radius coils near $z=0$
are Cu coils, while the others are superconducting. The left axis shows the
error field on-axis compared with the desired field profile. We see that the
peak error field is $\approx0.07$~T. 
\begin{figure}[ptbh]
\includegraphics{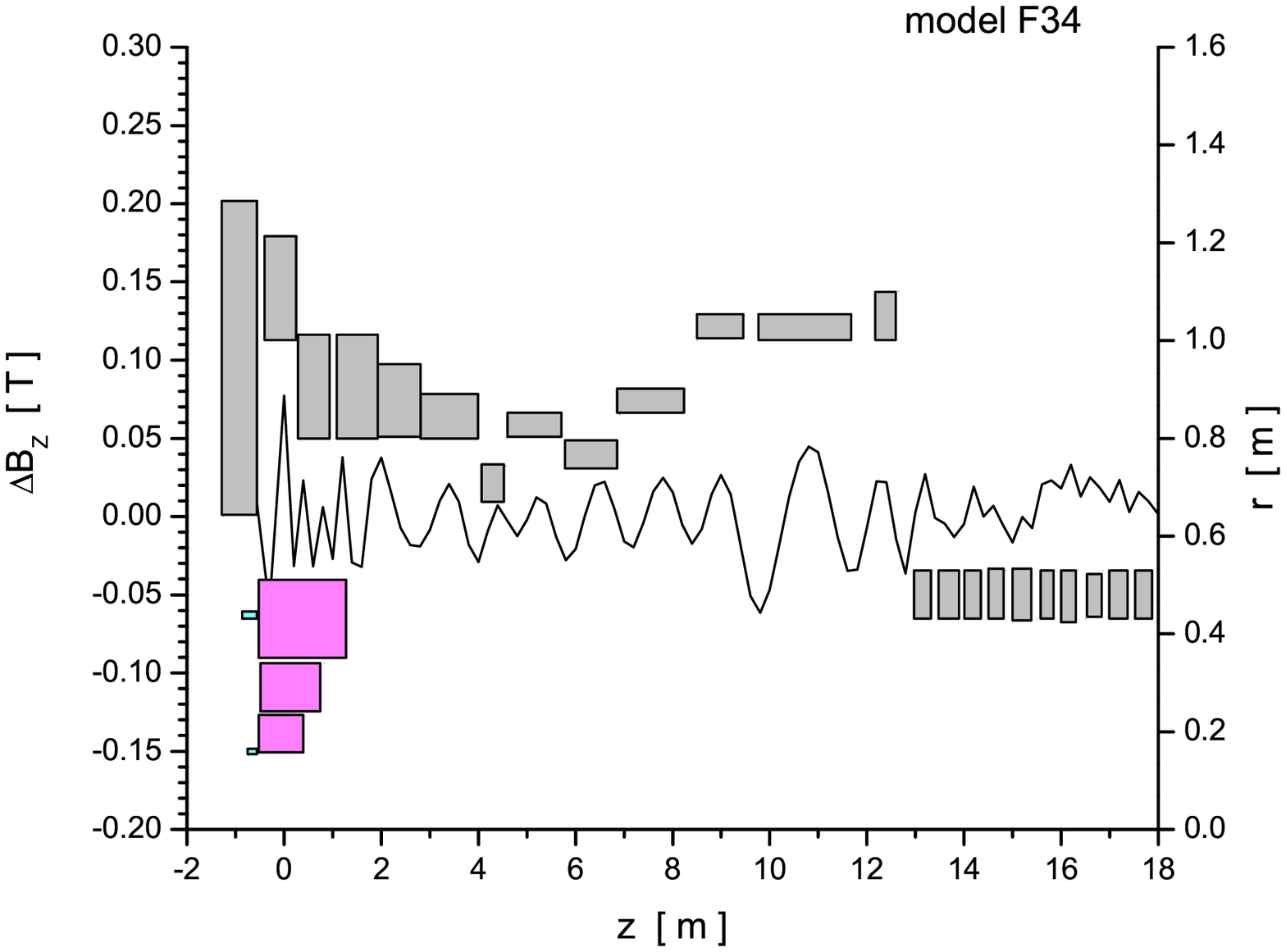}\caption{(Color) Actual coil configuration
in the collection region.The left axis shows the error field on-axis compared
with the optimal capture field profile, denoted KP in Fig.~\ref{fig103}. }%
\label{fig104}%
\end{figure}
Figure~\ref{fig105} shows a MARS calculation of the absorbed radiation dose in
the collection region. The peak deposition in the superconducting coils is
$\approx1$~Mgy/yr for a 1~MW beam running for 1~Snowmass year of
$10^{7}$~s. Assuming a lifetime dose for the insulation of 100~Mgy, there
should be no problem with radiation damage in the coils. 
\begin{figure}[ptbh]
\includegraphics[scale=1.75]{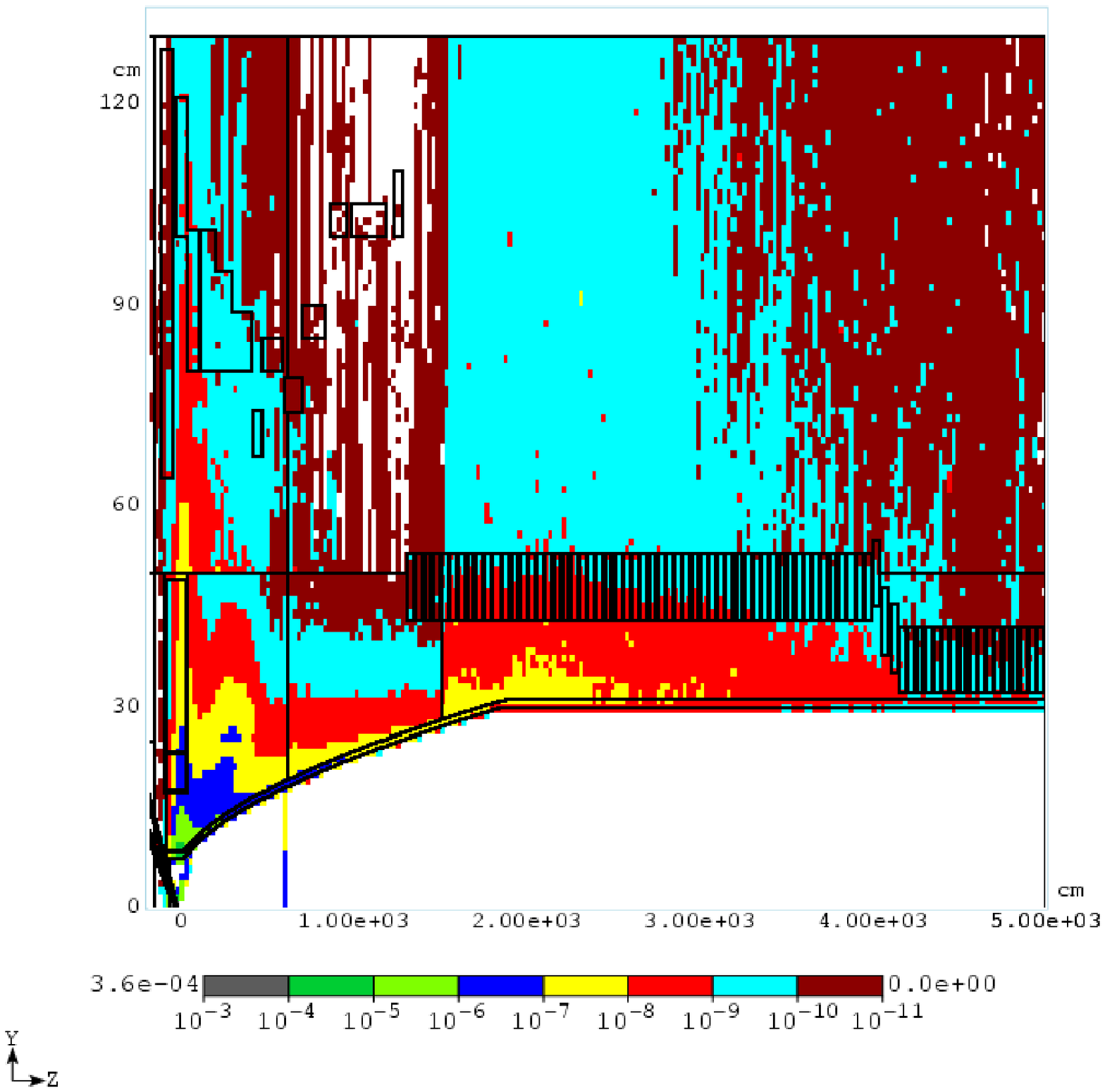}
\caption{(Color) MARS calculation
of the absorbed radiation dose in the collection region.}%
\label{fig105}%
\end{figure}

\begin{figure}[ptbh]
\includegraphics[width=4in]{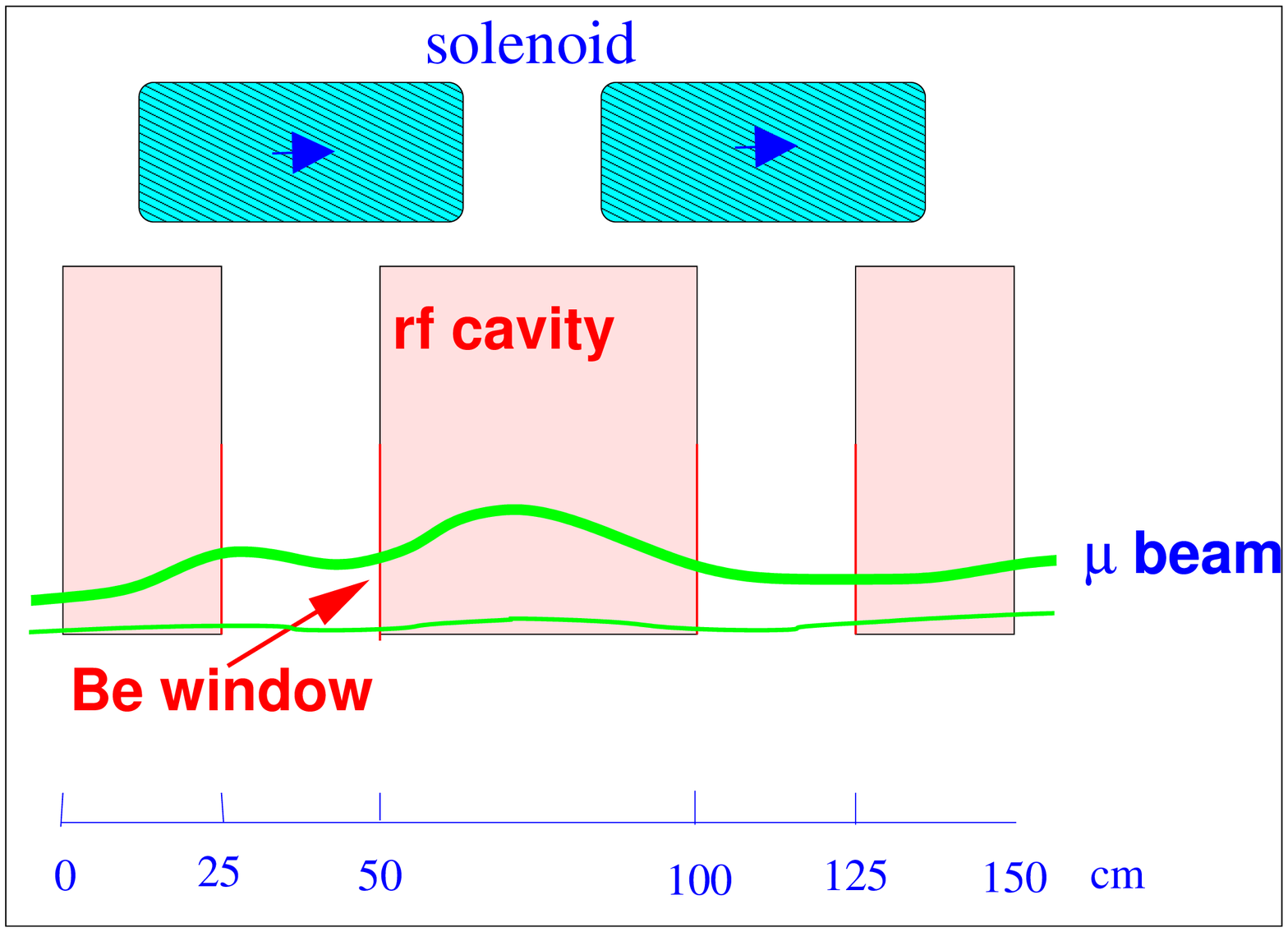}
\caption{(Color) Schematic of two cells of the buncher or phase rotator section.}%
\label{fig106}%
\end{figure}

Two cells of the buncher lattice are shown schematically in Fig.~\ref{fig106}.
Most of the 75~cm cell length is occupied by the 50-cm-long rf cavity. The
cavity iris is covered with a Be window. The limiting radial aperture in the
cell is determined by the 25~cm radius of the window. The 50-cm-long solenoid
was placed outside the rf cavity in order to decrease the magnetic field
ripple on the axis and minimize beam losses from momentum stop bands. The
buncher section contains 27~cavities with 13~discrete frequencies and
gradients varying from 5--10~MV/m.

\begin{figure}[ptbh]
\includegraphics[width=3in,angle=90]{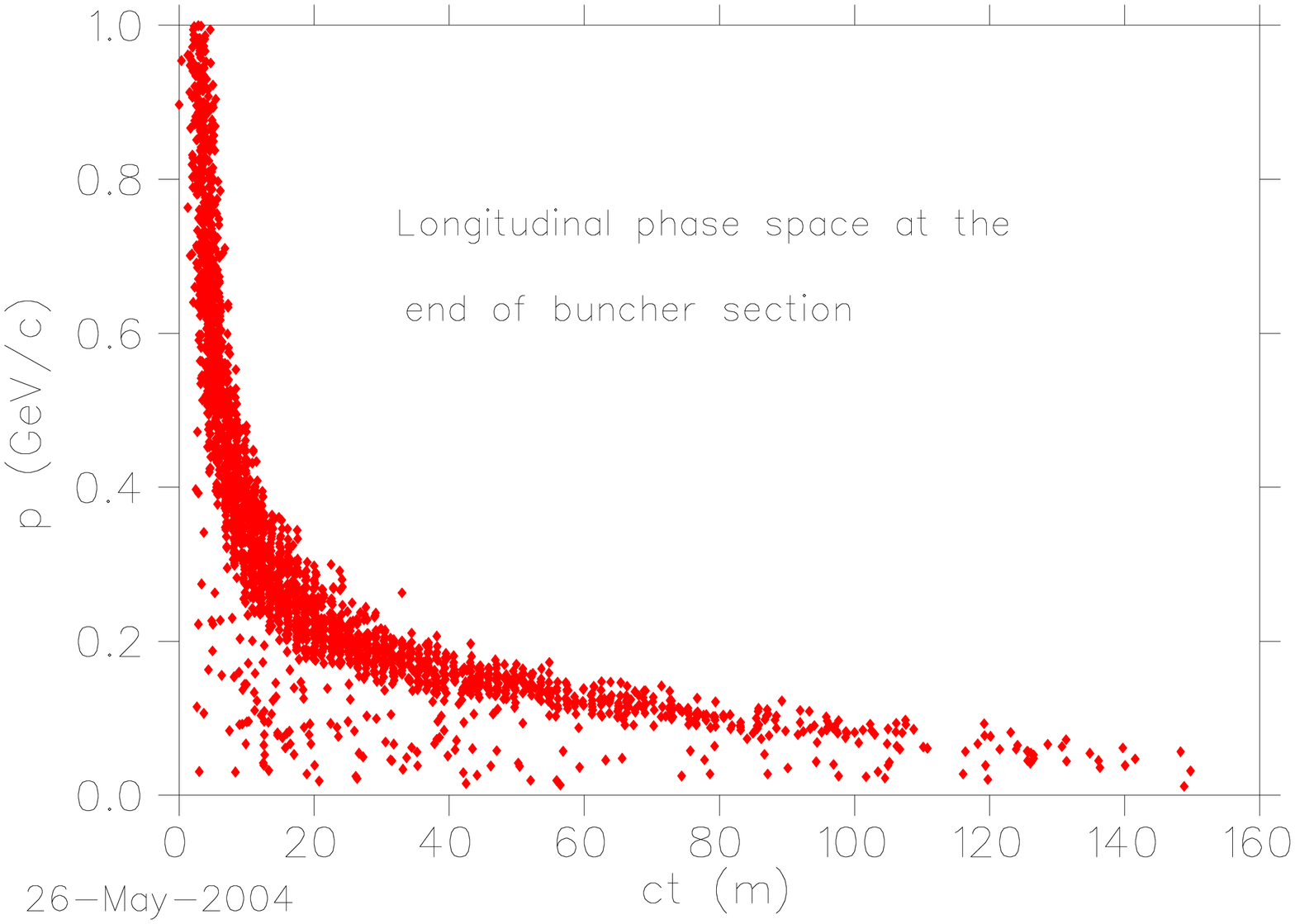}%
\caption{(Color) Longitudinal phase space after the buncher section.}%
\label{fig108}%
\end{figure}

The frequencies decrease from 333 to 234~MHz in the buncher region. The
cavities are not equally spaced. Fewer cavities are used at the beginning
where the required gradients are small. Figure~\ref{fig108} shows the
correlated longitudinal phase space and the bunching produced by the buncher.

\begin{figure}[ptbh]
\includegraphics[width=4in]{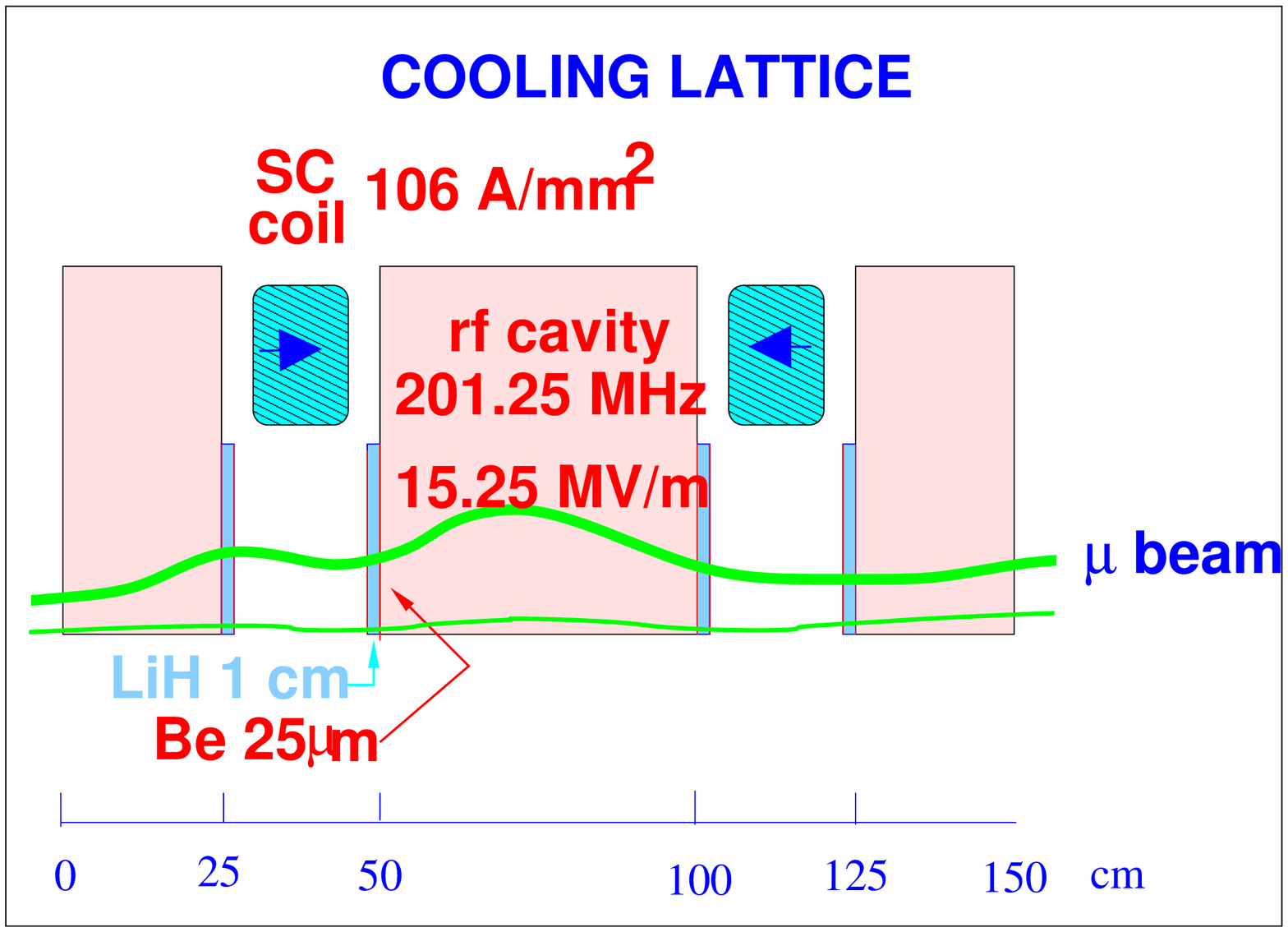}
\caption{(Color) Schematic of
one cell of the cooling section.}%
\label{fig111}%
\end{figure}

\begin{figure}[ptbhptbh]
\mbox{
\includegraphics[width=0.35\linewidth,angle=90]{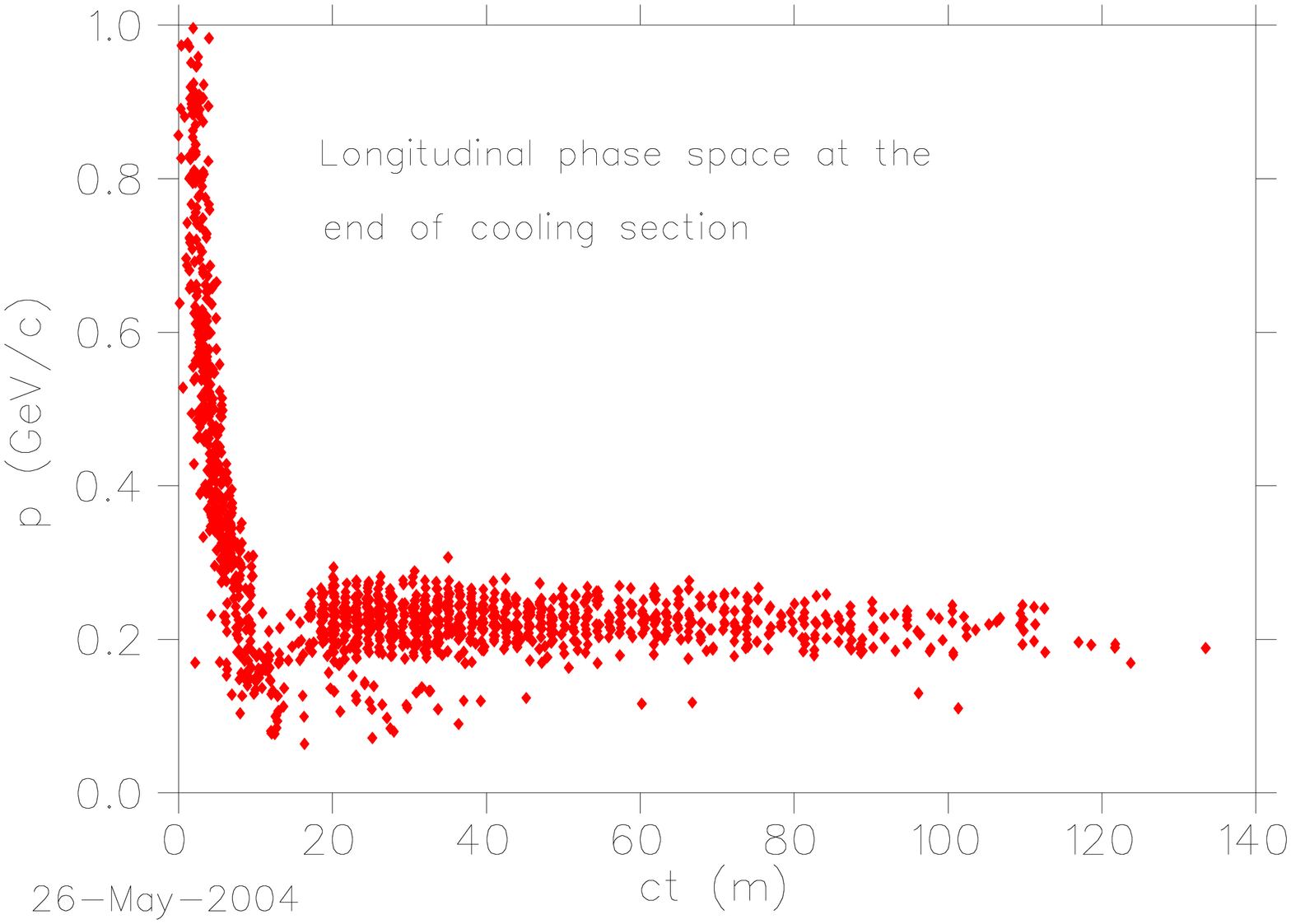}%
\includegraphics[width=0.35\linewidth,angle=90]{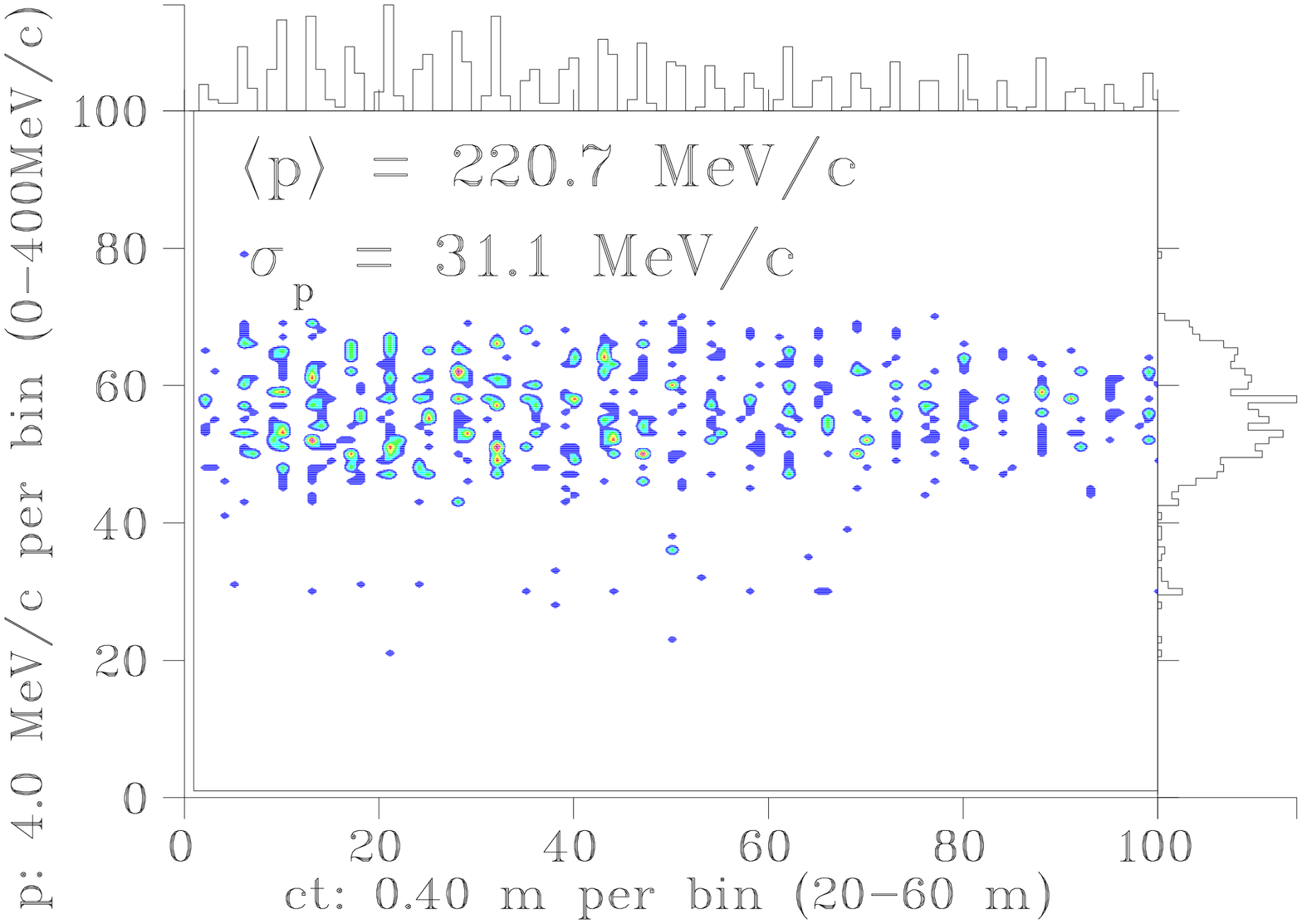}%
}\caption{(Color) Longitudinal phase space at the end of the channel.}%
\label{fig111a}%
\end{figure}

\begin{figure}[ptbh]
\mbox{
\includegraphics[angle=90,width=3in]{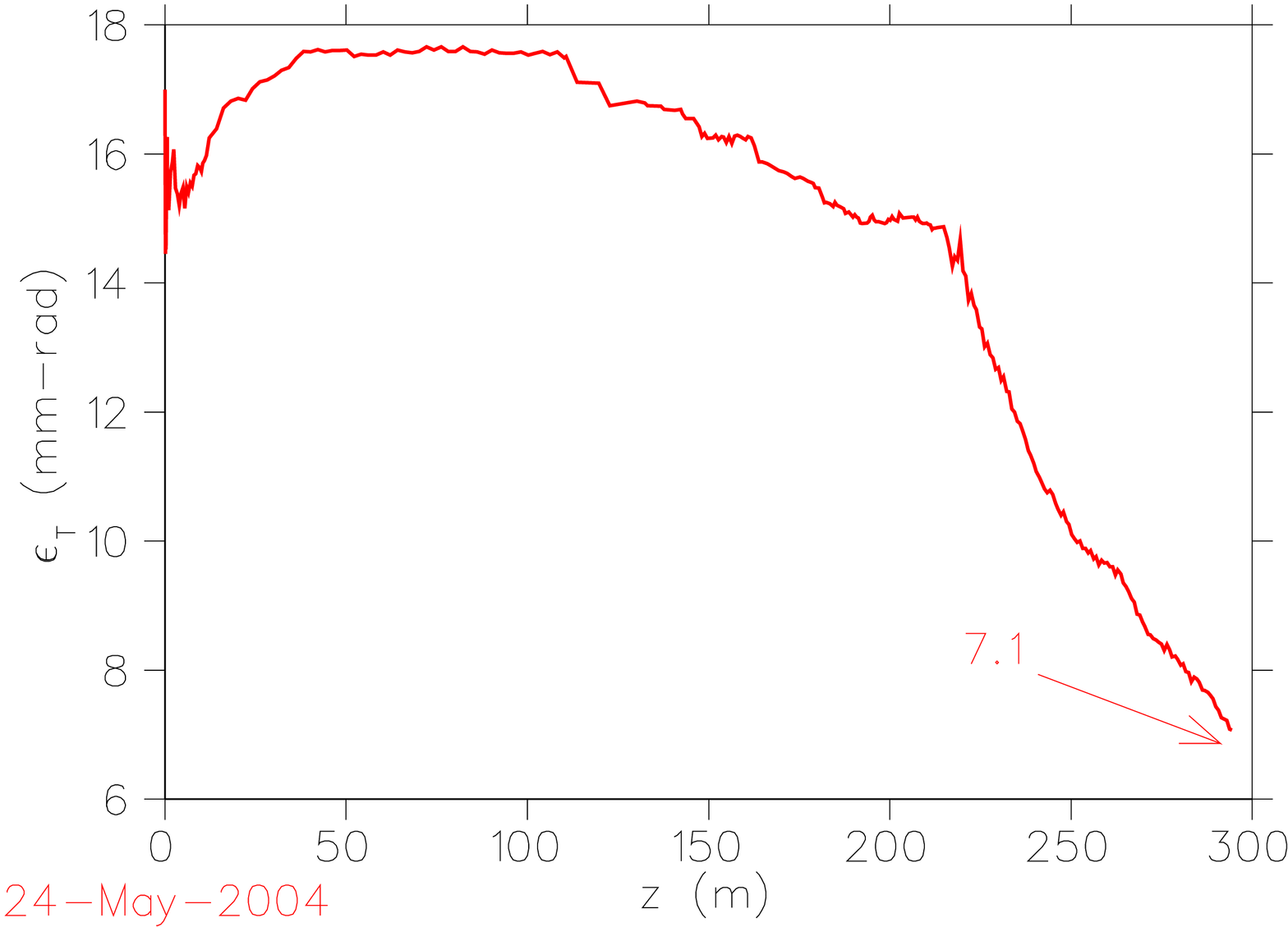}
\includegraphics[angle=90,width=3in]{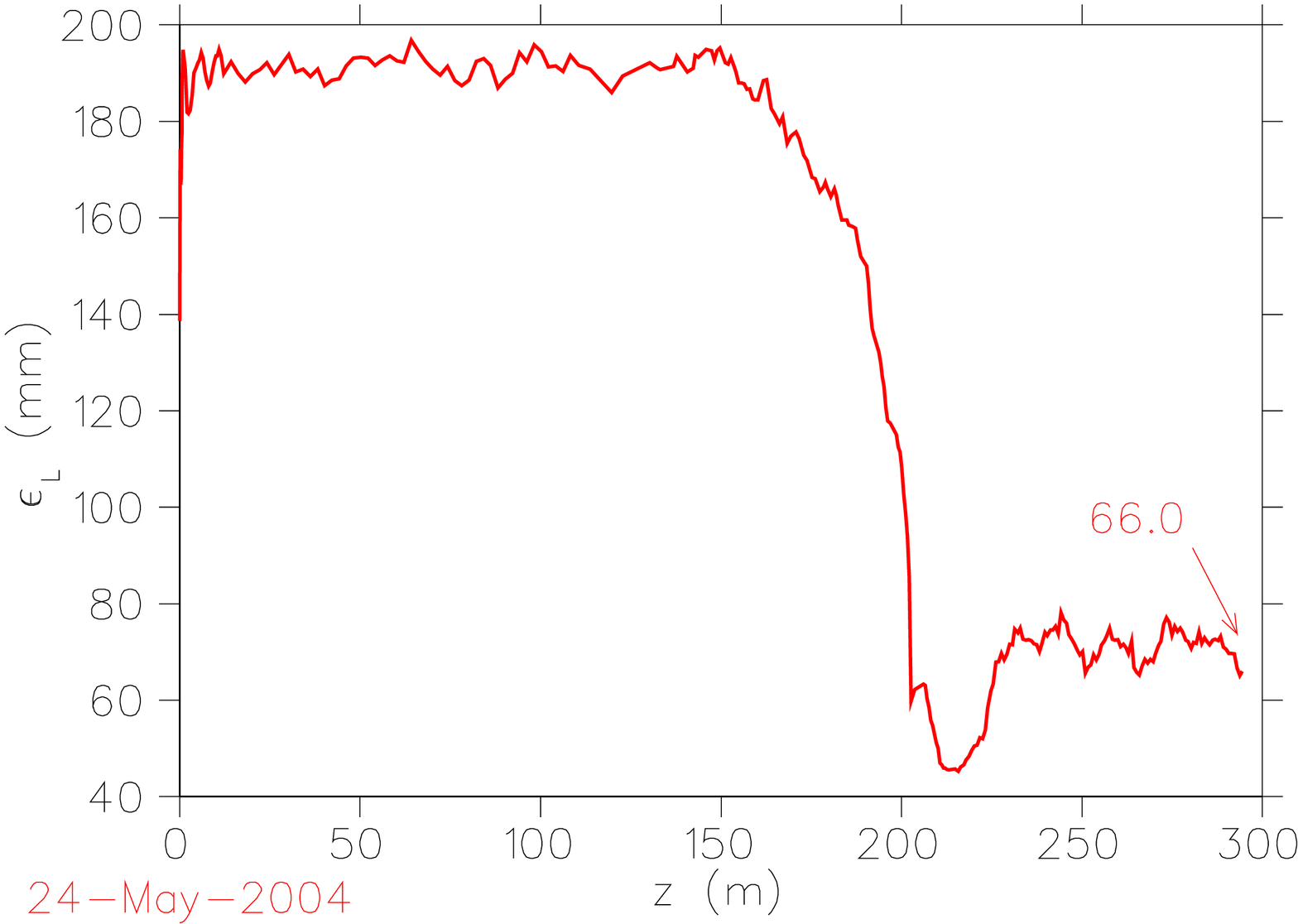}}
\caption{(Color) Normalized transverse emittance (left) and longitudinal emittance (right) along the front-end for a momentum cut $0.1 \leq p
\leq0.3$~GeV/c.}%
\label{fig112}%
\end{figure}

\begin{figure}[ptbh!]
\includegraphics[angle=90,width=4in]{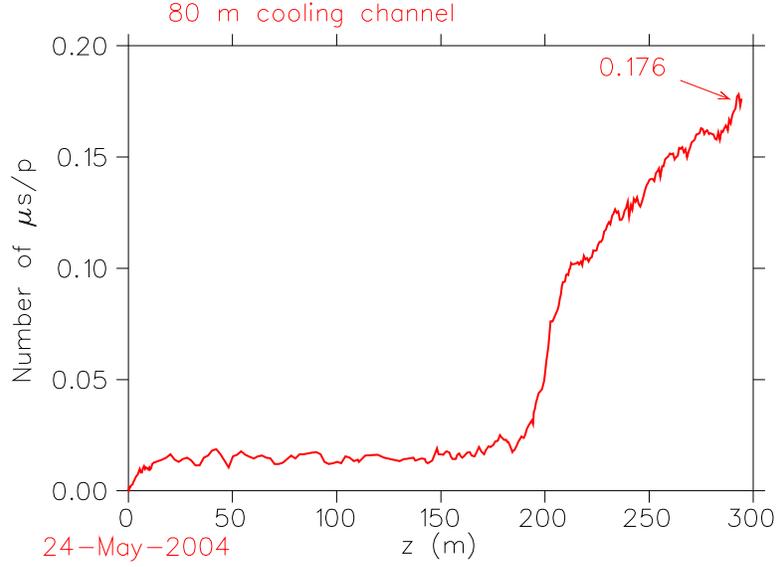}
\caption{(Color)
The muons per proton into the accelerator transverse normalized acceptance of
$A_{T}=30$~mm rad and normalized longitudinal acceptance of $A_{L}=150$~mm for
a momentum cut $0.1 \leq p \leq0.3$~GeV/c.}%
\label{fig113}%
\end{figure}

\begin{figure}[tpbh!]
\includegraphics[width=4in]{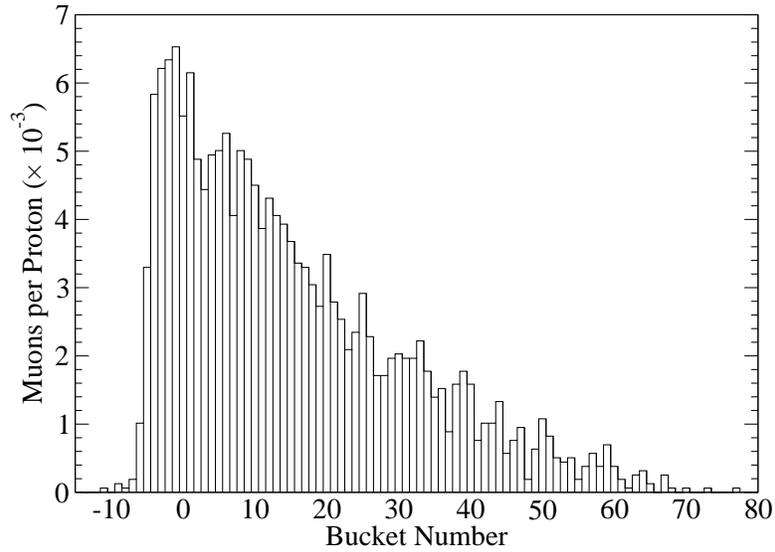}
\caption{Bunch structure
of the beam delivered to the accelerator transverse normalized acceptance of
$A_{T}=30$~mm rad and normalized longitudinal acceptance of $A_{L}=150$~mm
for a momentum cut $0.1 \leq p \leq0.3$~GeV/c.}%
\label{fig114}%
\end{figure}

\begin{figure}[ptbh!]
\includegraphics[width=4.5in]{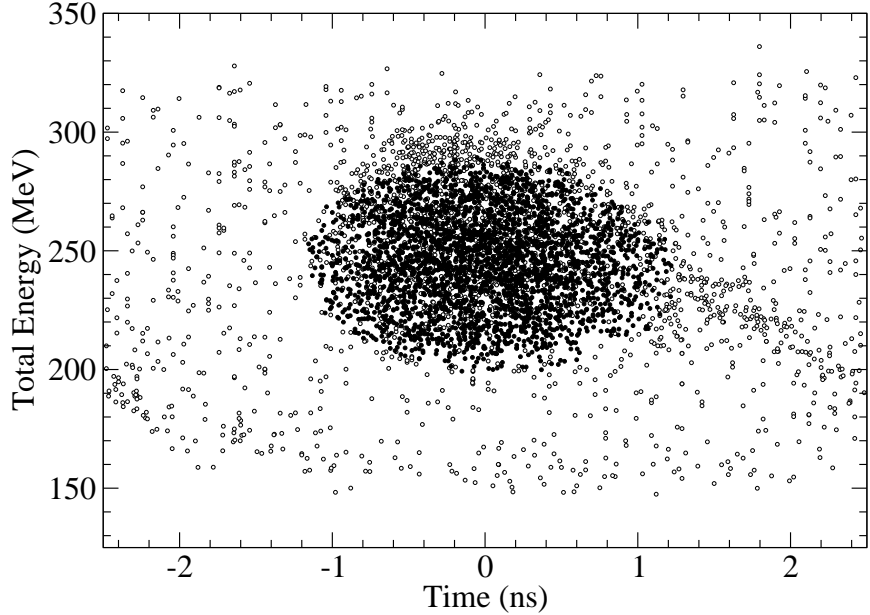}
\caption{Longitudinal phase
space of one bunch in the train at the end of the cooling section. The open
circles are all the particles that reach the end of the channel and the filled
circles are particles within the accelerator transverse normalized
acceptance of $A_{T}=30$~mm rad and normalized longitudinal acceptance of
$A_{L}=150$~mm for a momentum cut $0.1 \leq p \leq0.3$~GeV/c.}%
\label{fig115}%
\end{figure}

\begin{figure}[bpth!]
\includegraphics[width=4.5in]{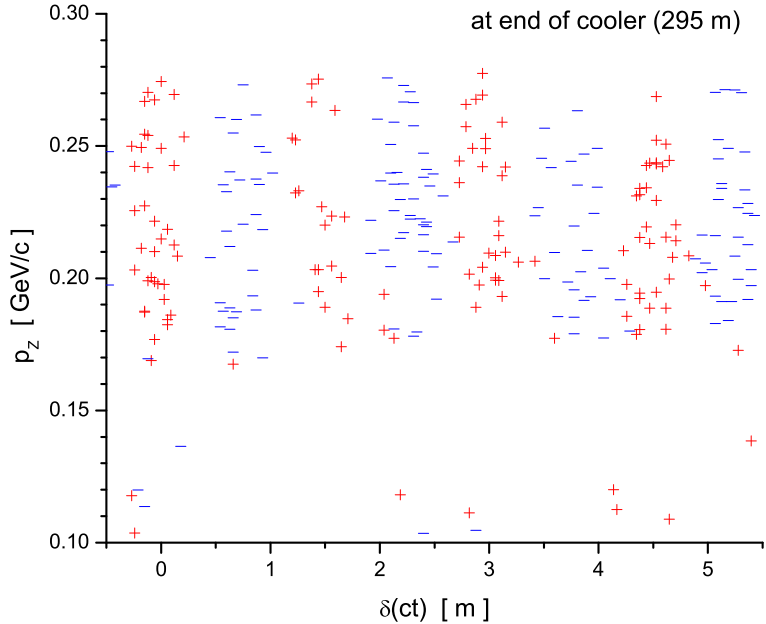}
\caption{(Color) A sample of the train of interleaved
  $\mu^{+}$~(red) and $\mu^{-}$~(blue) bunches exiting the
cooling section.}%
\label{fig116}%
\end{figure}

The rotator cell is very similar to the buncher cell. The major difference is
the use of tapered Be windows on the cavities because of the higher rf
gradient. There are 72~cavities in the rotator region, with 15~different
frequencies.
%
The frequencies decrease from 232 to 201~MHz in this part of the front end.
All cavities have a gradient of 12.5~MV/m.
%
The energy spread in the beam is significantly reduced. 
%
The cooling channel was designed to have a relatively flat transverse beta
function with a magnitude of about 80~cm. One cell of the channel is shown in
Fig.~\ref{fig111}. 
Most of the 150~cm cell length is taken up by the 50-cm-long rf
cavities. The cavities have a frequency of 201.25~MHz and a gradient of
15.25~MV/m. A novel aspect of this design comes from using the windows on the
rf cavity as the cooling absorbers. This is possible because the near constant
$\beta$ function does not significantly increase the emittance heating at the
window location. The window consists of a 1~cm thickness of LiH with $25~\mu$m
thick Be coatings (The Be will, in turn, have a thin coating of TiN to prevent
multipactoring~\cite{multipac}.) The alternating 2.8~T solenoidal field is
produced with one solenoid per half cell, located between the rf
cavities. Figure~\ref{fig111a} shows the longitudinal phase space at the end
of the cooling section. The reduction in normalized transverse emittance along
the cooling channel is shown in the left plot of Fig.~\ref{fig112} and the
right plot shows the normalized longitudinal emittance.
The channel produces a final value of $\epsilon_{T} =7.1$~mm rad, that is, more than a
 factor of two reduction from the initial value. The equilibrium value for a LiH
absorber with an 80~cm $\beta$ function is about $\epsilon_{T}^{\text{equil.}%
}\approx5.5$~mm rad. Figure~\ref{fig113} shows the muons per proton that fit
into the accelerator transverse normalized acceptance of $A_{T}=30$~mm rad
and normalized longitudinal acceptance of $A_{L}=150$~mm. The 80-m-long
cooling channel raises this quantity by about a factor of 1.7. The current
best value is $0.170\pm0.006.$ This is the same value obtained in FS2. Thus, we
have achieved the identical performance at the entrance to the accelerator as
FS2, but with a significantly simpler, shorter, and presumably less expensive
channel design. In addition, unlike FS2, this channel transmits both signs of
muons produced at the target. With appropriate modifications to the transport
line going into the storage ring, this design could deliver both (time tagged)
neutrinos and antineutrinos to the detector. 
 The beam at the end of the cooling section consists of a train of bunches (one
sign) with a varying population of muons in each one; this is shown in
Fig.~\ref{fig114}.
 Figure~\ref{fig115} depicts the longitudinal phase space of one of the
 bunches and Fig.~\ref{fig116} shows a few interleaved  $\mu^{+}$ and $\mu^{-}$ bunches exiting the cooling section.
\printfigures

\subsubsection{Other Variants}
The present scenario is not completely optimized in either performance or 
cost. In this section we discuss some of the options that have already been
studied briefly or that might be developed in future studies. 
Some variations we have considered:
\begin{itemize}
\item  Be absorbers in place of LiH absorbers
\item  Shorter buncher rotator
\item  Shorter bunch train
\item  Different rf frequency
\item  Gas-filled cavities
\item  Quadrupole-based cooling channel
\end{itemize}

The configuration of LiH absorbers with Be windows could be simplified by substituting thicker Be windows as the end plates of the cavities, with a
 thickness chosen to make the total energy loss the same as that in the
 baseline foil + LiH absorber case. This would eliminate the need for thin Be
 windows. Cooling would be a bit less effective because of the greater multiple
 scattering in Be absorbers. An initial evaluation~\cite{NeufferRef} of a
 Be-only scenario showed somewhat less capture into the acceleration channel
 acceptance ($\thickapprox$15\%). A scenario in which Be absorbers are
 initially installed and then upgraded later to more efficient LiH absorbers
 is, of course, also possible.

The baseline scenario requires a roughly 110~m drift, a 51~m buncher and a 52~m high-voltage phase rotation section. These parameters have not been
 optimized. For comparison, we considered an example having only a 26~m phase
 rotation section~\cite{NeufferRef}. The shorter rotation section would be
 significantly less expensive since it is not only shorter but provides about 200~MV
 less of high-gradient rf voltage. Initial evaluations indicate only small
 decreases in captured muons ($\thickapprox$10\%). 

The baseline case generates $\mu^{+}$ and $\mu^{-}$ bunch trains that are
 about 100~m long. These bunch trains are matched to the FS2 scenario
 requirements; in particular, they fit within the circumferences of the FS2, and
 the presently envisioned, accelerators and storage ring. However, other
 scenarios might make use of smaller circumference ring coolers, accelerators,
 and storage rings, and thus require shorter bunch trains. For example, a
 scenario with a 20~m drift, 20~m buncher and 20~m phase rotator has been
 explored~\cite{NeufferRef}. This produces a roughly 20~m long bunch train.

Although this shorter system would be much less expensive than the present
 roughly 200~m long system, an initial evaluation showed that the total number
 of captured muons was substantially reduced (by about 50\%). (On the other
 hand, a longer system, capturing longer bunch trains, might produce more
 muons, at a small incremental cost.)

Both FS2 and our present scenario use 201.25~MHz rf as the baseline final
 operating frequency, because of the availability of rf components at that
 frequency and because it is a plausible optimum for large-aperture and
 high-gradient operation. Other baseline frequencies could be considered, e.g.,
 scenarios at 50, 100, 300 or 400~MHz. Lower frequencies (larger bunches) may
 be desirable if the accelerator longitudinal motion requires larger phase-space buckets.

Muons Inc. has an STTR grant to explore the use of hydrogen gas-filled rf cavities for
 muon cooling~\cite{MuonsIncRef}. This approach simplifies the cooling channel
 design by integrating the energy-loss material into the rf system. Moreover,
 it may be more effective in permitting high-gradient operation of the
 cavities. Such cavities could also be used in the cooling and phase-rotation 
 (and possibly buncher) sections; an exploration with cost-performance
 optimization is planned. 

The transport and cooling system in the front-end scenario considered here
 uses high-field solenoids for focusing. A cooling system with similar
 performance parameters using large-aperture quadrupoles has also been examined~\cite{JohnstoneRef}, though a cost-performance comparison has not yet been
 made.

\subsection{Neutrino Factory Acceleration\label{sec5-sub2}}
\begin{table}[tbp]
\caption{Acceleration system requirements.}
\label{tab:acc:pars}%
\begin{ruledtabular}
    \begin{tabular}{lr}
      Initial kinetic energy (MeV)&187\\
      Final total energy (GeV)&20 \\
      Normalized transverse acceptance (mm)&30\\
      Normalized longitudinal acceptance (mm)&150\\
      Bunching frequency (MHz)&201.25\\
      Maximum muons per bunch&$1.1\times10^{11}$\\
      Muons per bunch train per sign&$3.0\times10^{12}$\\
      Bunches in train&89\\
      Average repetition rate (Hz)&15\\
      Minimum time between pulses (ms)&20
    \end{tabular}
  \end{ruledtabular}
\end{table}
The acceleration system takes the beam from the end of the cooling
channel and accelerates it to the energy required for the decay ring. Table~%
\ref{tab:acc:pars} gives the design  parameters of the acceleration system. 
Acceptance is defined such that if $A_\bot$ is the
transverse acceptance and $\beta_\bot$ is the beta function, then the
maximum particle displacement (of the particles we transmit) from the
reference orbit is $\sqrt{\beta_\bot A_\bot mc/p}$, where $p$ is the
particle's total momentum, $m$ is the particle's rest mass, and $c$ is the speed of light. 
 The acceleration system is able to accelerate bunch trains of both
signs simultaneously.

To reduce costs, the RLA acceleration systems from FS2 will be replaced, as
much as possible, by Fixed-Field Alternating Gradient (FFAG) accelerators.
\subsubsection{Initial Parameter Sets}
\begin{table}[tbp]
\caption{Parameters for FFAG lattices. See Fig.~\ref{fig:acc:ffaggeom} to
understand the signs of the parameters.}
\label{tab:acc:ffag}%
\begin{ruledtabular}
    \begin{tabular}{lrrrr}
      Maximum energy gain per cavity (MeV)&\multicolumn{4}{c}{7.5}\\
      Stored energy per cavity (J)&\multicolumn{4}{c}{368}\\
      Cells without cavities&\multicolumn{4}{c}{8}\\
      RF drift length (m)&\multicolumn{4}{c}{2}\\
      Drift length between quadrupoles (m)&\multicolumn{4}{c}{0.5}\\
      Initial total energy (GeV)&\multicolumn{2}{c}{5}&\multicolumn{2}{c}{10}\\
      Final total energy (GeV)&\multicolumn{2}{c}{10}&\multicolumn{2}{c}{20}\\
      Number of cells&\multicolumn{2}{c}{90}&\multicolumn{2}{c}{105}\\
      Magnet type&\multicolumn{1}{c}{Defocusing}&\multicolumn{1}{c}{Focusing}&
      \multicolumn{1}{c}{Defocusing}&\multicolumn{1}{c}{Focusing}\\
      Magnet length (m)&1.612338&1.065600&1.762347&1.275747\\
      Reference orbit radius of curvature (m)&15.2740&-59.6174&18.4002&-70.9958\\
      Magnet center offset from reference orbit (mm)&-1.573&7.667&1.148&8.745\\
      Magnet aperture radius (cm)&14.0916&15.2628&10.3756&12.6256\\
      Field on reference orbit (T)&1.63774&-0.41959&2.71917&-0.70474\\
      Field gradient (T/m)&-9.1883&8.1768&-15.4948&12.5874\\
    \end{tabular}
  \end{ruledtabular}
\end{table}
\begin{figure}[htbp]
\includegraphics[width=0.65\textwidth]{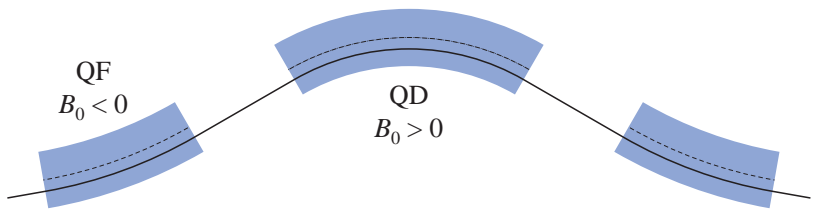}
\caption{(Color) Geometry of the Triplet Lattice. The ``magnet center offset from
reference orbit" listed in Table~\ref{tab:acc:ffag} is positive for both
magnets in this diagram.}
\label{fig:acc:ffaggeom}
\end{figure}
Based on an earlier version of the cost optimization process described
below, it was decided that two factor-of-two FFAG stages would be used: one
from 5 to 10~GeV, the other from 10 to 20~GeV total energy. Triplet lattices
were chosen for their good longitudinal performance and the extensive study
that had been done on them. The parameters that we adopted 
are given in Table~\ref{tab:acc:ffag}.

The 201.25~MHz cavities were taken to be single-cell superconducting
cavities. The energy gain per rf cavity was chosen based on the gradients
already achieved in the cavity studied at Cornell, namely 10~MV/m at 4.2~K
(they have achieved 11~MV/m at 2.5~K)~\cite{pac03:1309}. This is a very conservative number, as there is every reason to believe that improved
sputtering techniques will allow the cavity to achieve a gradient of 
 15~MV/m or higher. Energy gain and stored energy are computed by scaling
from the values for the 300~mm aperture FS2 cavities~\cite{fs2},\cite{pac03:1309}.

With the beam intensity given in Table~\ref{tab:acc:pars}, and both signs of muons,
about 16\% of the stored energy will be extracted from the cavities in the
5--10~GeV FFAG, and about 27\% will be extracted in the 10--20~GeV. While
this may seem substantial, it is easily handled. To keep the average voltage to 7.5~MV per cavity,
one need only increase the initial voltage to 7.8~MV for the 5--10~GeV FFAG and
8.1~MV for the 10--20~GeV FFAG. The most important effect is a differential
acceleration between the head and tail of the bunch train, which is about
1\% for both cases. This may be at least partially correctable by a phase offset  between the cavity and the bunch train.

\subsubsection{Low Energy Acceleration}
Based on cost considerations (see Section~\ref{sec5-sub2}) we have chosen
not to use FFAGs below 5~GeV total energy. Therefore, we
must provide alternative acceleration up to that point. As in FS2,
 we use a linac from the lowest energies to 1.5~GeV, followed by a
recirculating linear accelerator (RLA).

The linac turns out to be strongly constrained by the transverse acceptance.  In FS2, there were three types of cryomodules,
containing one, two, and four cavities, respectively.  With our larger
acceptance, the cryomodules from FS2 would require the beam to have a
momentum of at least 420~MeV/c, 672~MeV/c, and 1783~MeV/c, 
respectively. Note that the lowest momentum is much higher than
the average momentum in the cooling channel, which is about 220~MeV/c.
Thus, we need to make adjustments to the FS2 design to be able to accelerate this larger beam.

\begin{table}[tbp]
  \caption{Linac cryomodule structure.
    Numbers are lengths in m.\label{tab:acc:cryo}}
  \begin{ruledtabular}
    \begin{tabular}{rlrlrl}
\multicolumn{2}{c}{Cryostat I}&\multicolumn{2}{c}{Cryostat
  II}&\multicolumn{2}{c}{Cryostat III}\\
\hline
      End to solenoid&0.25&To solenoid&0.25&To solenoid&0.25\\
      Solenoid&1.50&Solenoid&1.50&Solenoid&1.50\\
      Input coupler&0.50&Input coupler&0.50&Input coupler&0.50\\
      Cavity&0.75&Cavity&1.50&Cavity&1.50\\
      To end&0.25&Input coupler&0.50&Between cavities&0.75\\
      \cline{1-2}
      Total&3.25&To end&0.25&Cavity&1.50\\
      \cline{3-4}
      &&Total&4.50&Input coupler&0.50\\
      &&&&To end&0.25\\
      \cline{5-6}
      &&&&Total&6.75
    \end{tabular}
  \end{ruledtabular}
\end{table}
In particular, to increase the acceptance, we must reduce the lengths of the
cryomodules.  We first construct a very short cryomodule by using
a single one-cell cavity instead of the two-cell cavities in the
FS2 cryomodules.  Not only does this shorten the cavity itself,
 it also eliminates one of the input couplers.  Secondly,
we  remove 50~cm between the solenoid and the input coupler.  We
intend to run the cavities with up to 0.1~T on them \cite{Ono99};
this is acceptable provided the cavities are cooled down before the
magnets are powered.  The field profile of the solenoids shown in
 FS2 indicates that the iron shield on the solenoids is sufficient
to bring the field down to that level even immediately adjacent
to the solenoid shield.  Finally, the FS2 cryomodules left 75~cm
for the end of the cryostat; we have reduced this to 50~cm.  Together,
these changes permit a total length for the first module type of 3.25~m.
Table~\ref{tab:acc:cryo} shows the dimensions of this cryostat.  The two
shortest cryostats from FS2 have been adjusted to meet these
specifications and, in addition, for the ``intermediate'' cryostat the 
spacing between the cavities was reduced to 75~cm, under the
assumptions that the cavities will be allowed to couple weakly, and that the
entire module will be tuned appropriately to take this into account.

\begin{table}[tbp]
  \caption{Linac cryomodule parameters.\label{tab:acc:linac}}
  \begin{ruledtabular}
    \begin{tabular}{lrrr}
&{Cryo I}&{Cryo II}&{Cryo III}\\
\hline
      Length (m)&3.25&4.50&6.75\\
      Minimum allowed momentum (MeV/c)&273&378&567\\
      Number of modules&18&12&23\\
      Cells per cavity&1&2&2\\
      Cavities per module&1&1&2\\
      Maximum energy gain per cavity (MeV)&11.25&22.5&22.5\\
      RF Frequency (MHz)&201.25&201.25&201.25\\
      Solenoid length (m)&1&1&1\\
      Solenoid field (T)&2.1&2.1&2.1
    \end{tabular}
  \end{ruledtabular}
\end{table}
Table~\ref{tab:acc:linac} summarizes parameters for the linac.  The
phase of the cavities in the linac will be varied linearly with length
from about 65$^\circ$ at the beginning of the linac to 0$^\circ$ at
the end.  As indicated in Table~\ref{tab:acc:linac}, we
must inject into the linac at a momentum of 273~MeV/c, which is
still higher than the average momentum in the cooling channel.  We deal
with this by designing a matching section from the cooling channel to the
linac in which sufficient acceleration will occur to reach the required 
momentum for the linac.  That matching section will consist of cavities similar
to those in the cooling channel, but with thinner windows.

\begin{figure}[tbp]
  \includegraphics[width=\textwidth]{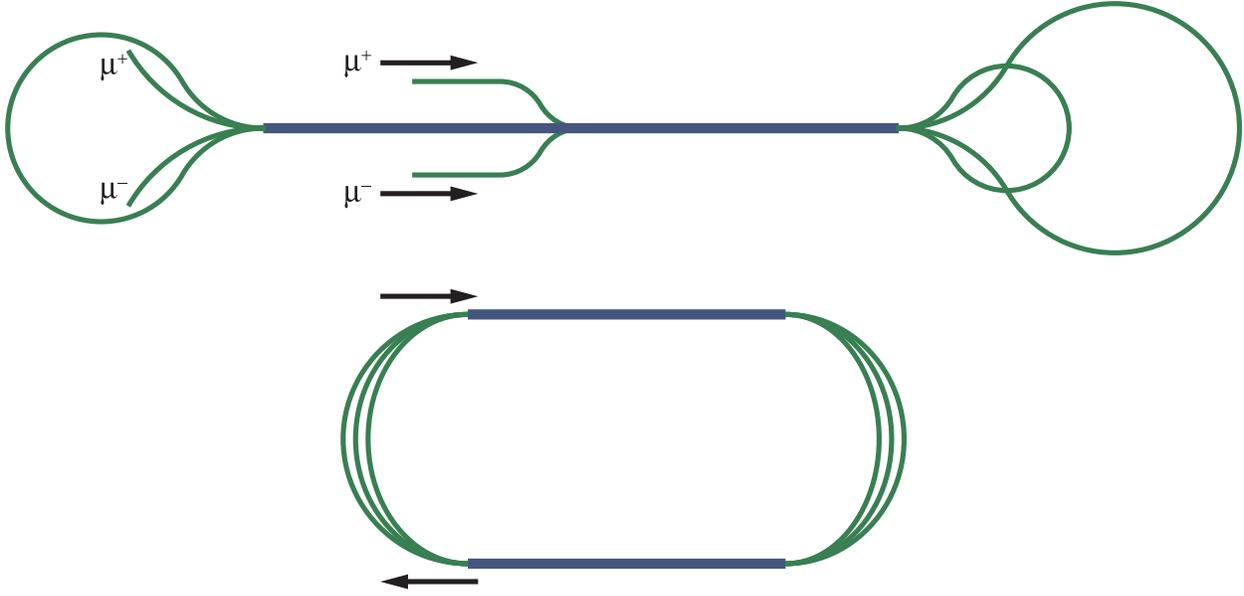}
  \caption{(Color) Dogbone (top) and racetrack (bottom) layout for the RLA.}
  \label{fig:acc:layout}
\end{figure}
Compared to FS2, we are injecting into the RLA at a lower energy
and are accelerating over a much smaller energy range.  This will make
it more difficult to have a large number of turns in the RLA.  To mitigate this, we choose a ``dogbone'' layout for the RLA
\cite{pac01:3323}.  For a given amount of installed rf, the dogbone layout
has twice the energy separation of the racetrack layout at the
spreaders and recombiners, making the switchyard much easier and
allowing more passes through the linac.

One disadvantage of the dogbone layout is that, because of the longer
linac and the very low injection energy, there is a significant
phase shift of the reference particle with respect to the cavity
phases along the length of the linac in the first pass (or the last
pass, depending on which energy the cavities are phased for).  To
reduce this effect, we inject into the center of the linac (as shown
in Fig.~\ref{fig:acc:layout}).

In the dogbone RLA, we have just over 1~GeV of linac, and we make
three and a half passes through that linac to accelerate from a total
energy of 1.5~GeV to 5~GeV.  The linac will use the
same cryomodules that were used in the RLA in FS2.

\begin{figure}[tbp]
  \includegraphics[width=\textwidth]{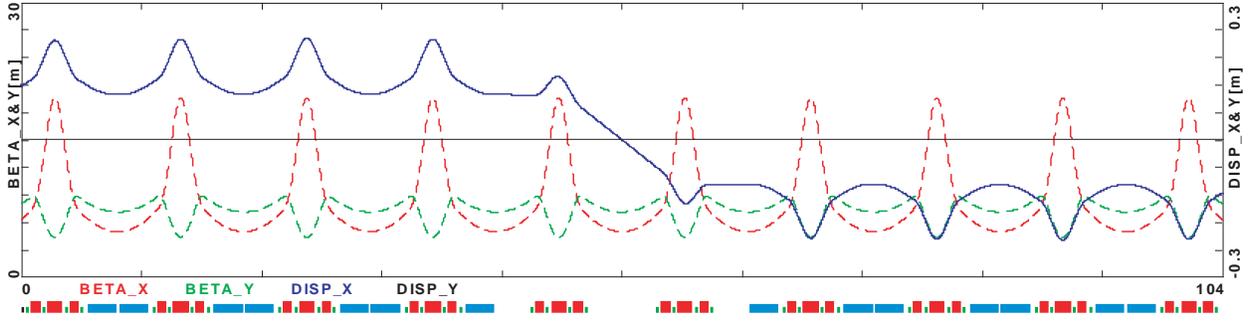}
  \caption{(Color) A section of the dogbone arc where the bend changes direction,
    showing the dispersion (solid) and beta functions (dashed).}
  \label{fig:acc:disp}
\end{figure}
Since the dogbone arc changes its direction of bend twice in each arc,
dispersion matching must be handled carefully.  This is done by having
a 90$^\circ$ phase advance per cell, and removing the dipoles from two
consecutive cells.  This will cause the dispersion to switch to the
other sign as desired, as shown in Figure~\ref{fig:acc:disp}.

\begin{figure}[tbp]
  \includegraphics[width=\textwidth]{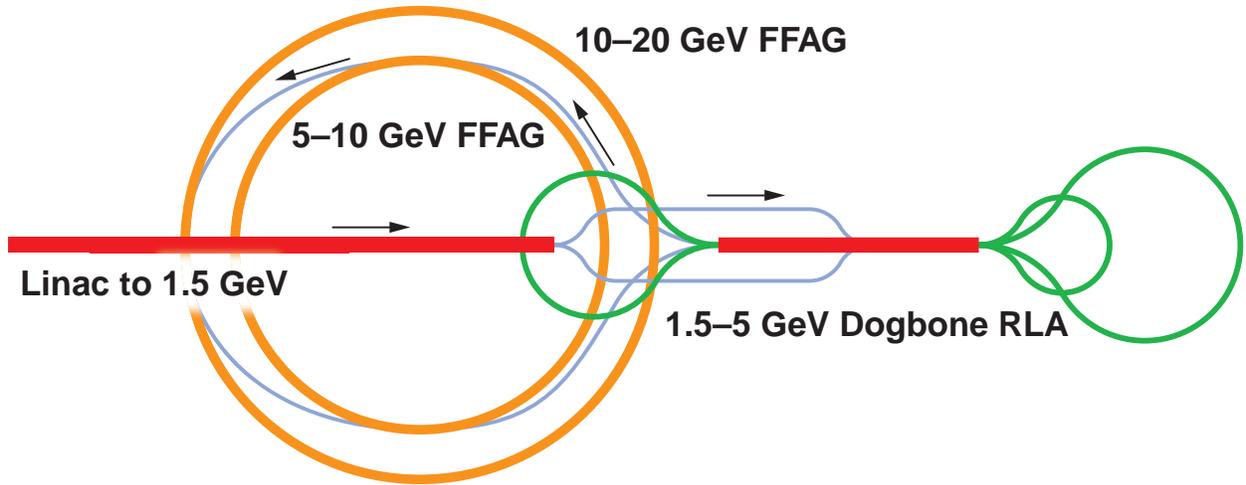}
  \caption{(Color) Potential layout for the acceleration systems.}
  \label{fig:acc:alllayout}
\end{figure}
Figure~\ref{fig:acc:alllayout} shows a compact potential layout for
all the acceleration systems described here.

\subsubsection{FFAG Tracking Results}
\begin{figure}[tbp]
\includegraphics[width=0.65\textwidth]{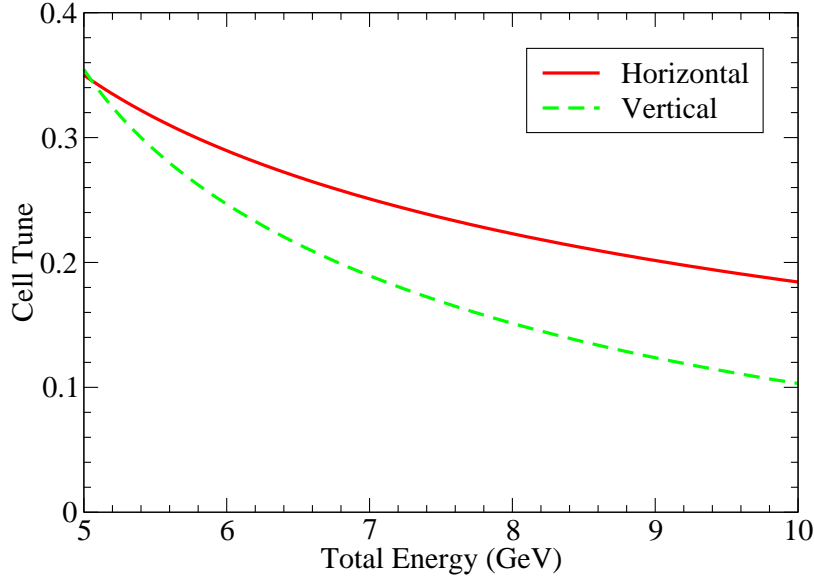}
\caption{(Color) Tunes as a function of energy in the 5--10~GeV FFAG reference
design.}
\label{fig:acc:nu}
\end{figure}
Initial experience with FFAG lattices having  a linear midplane field profile
has shown them to have a good dynamic aperture at fixed energies. We are
careful to avoid single-cell linear resonances to prevent beam loss.
However, since the tune is not constant (see Fig.~\ref{fig:acc:nu}), the
single-cell tune will pass through many nonlinear resonances. Nonlinearities
in the magnetic field due to end effects are capable of driving those
nonlinear resonances, and we must be sure that there is no beam loss and
minimal emittance growth because of this. Furthermore, there is the
potential to weakly drive multi-cell linear resonances because the changing
energy makes subsequent cells appear slightly different from each other.
These effects can be studied through tracking.

ICOOL~\cite{icool} is used for tracking for several reasons. It will
allow for a fairly arbitrary end-field description, it will attempt to make
that description consistent with Maxwell's equations, and it will track
accurately when the lattice acceptances, beam sizes, and energy spread are all large.

\begin{figure}[tbp]
\includegraphics[width=0.5\textwidth]{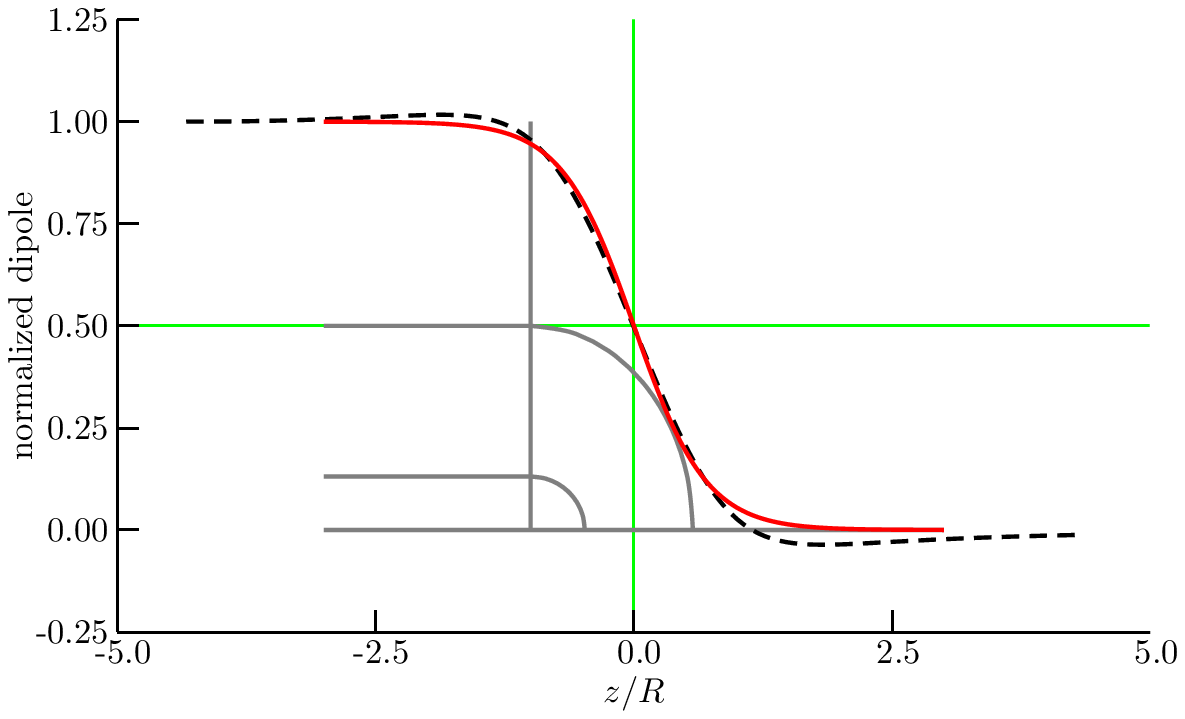}%
\includegraphics[width=0.5\textwidth]{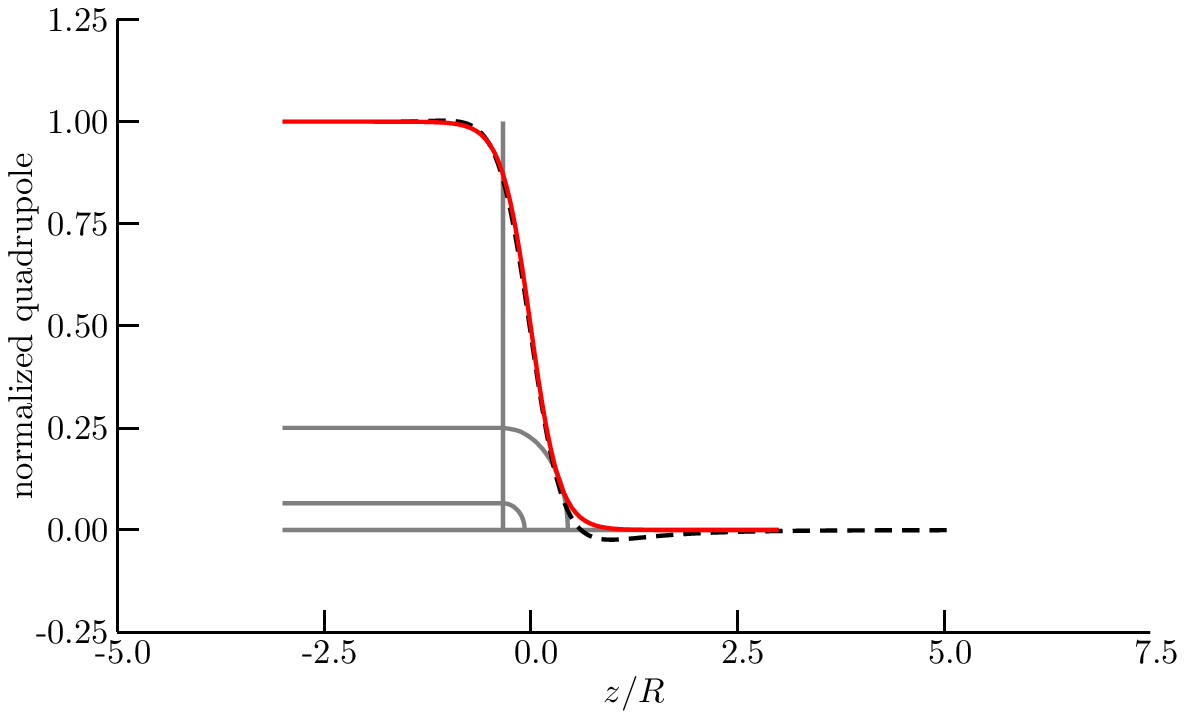}
\caption{(Color) Relative dipole field (left) and quadrupole field (right) near the
magnet end. The dashed line is the field from TOSCA, while the solid line is
our model.}
\label{fig:acc:end0}
\end{figure}
\begin{figure}[tbp]
\centering
\includegraphics[width=0.65\textwidth]{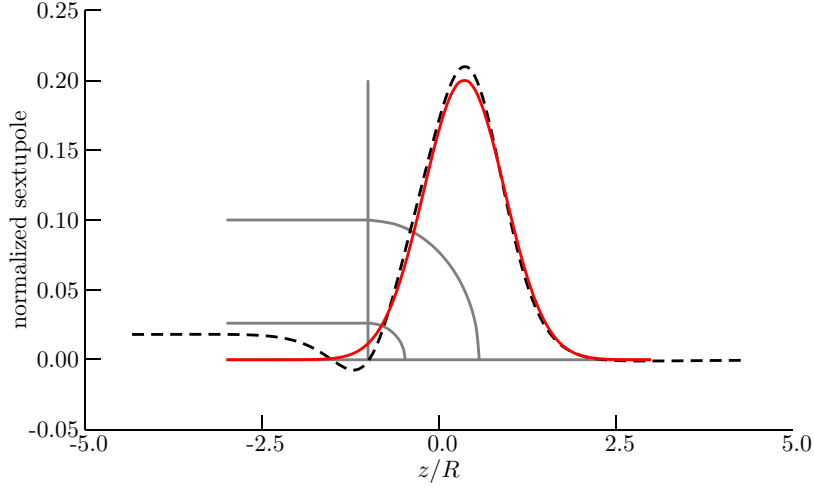}
\caption{(Color) Peak magnitude of the sextupole end field at radius $R$ (the magnet
aperture), divided by the dipole field. The dashed line is the field from
TOSCA, while the solid line is our model.}
\label{fig:acc:end2}
\end{figure}
We begin by constructing a simple model of both a quadrupole and dipole $%
\cos\theta$-type magnet, without iron, using TOSCA \cite{opera3d}. At the end
of the magnet, the field does not immediately drop to zero, but falls
gradually, as shown in Fig.~\ref{fig:acc:end0}. The end-field falloff  in the dipole and
 quadrupole generates nonlinear fields, which ICOOL calculates. In
addition, there are higher-order multipoles generated by breaking the
magnet symmetry at the ends where the coils form closed loops. We use
TOSCA to compute the sextupole components that arise from this effect, as shown in Fig.~\ref{fig:acc:end2}, and include them in our computation.

The TOSCA computation is done without iron, which leads to the overshoot in
the field values in Figs.~\ref{fig:acc:end0}--\ref{fig:acc:end2}. Iron in
the magnet will likely eliminate that overshoot. Thus, we approximate the
fields from TOSCA using functions without the overshoot. Fitting roughly to
the TOSCA results, the fields are approximated by 
\begin{equation}
\begin{gathered} B_0(z) = B_{00}\dfrac{1+\tanh{\dfrac{z}{0.7R}}}{2},\qquad
B_1(z) = B_{10}\dfrac{1+\tanh{\dfrac{z}{0.35R}}}{2}\\ B_2(z) =
-0.2B_{00}\exp\left[-\dfrac{1}{2} \left(\dfrac{z-0.36 R}{0.57
R}\right)^2\right], \end{gathered}
\end{equation}
where $R$ is the magnet aperture radius, $B_0(z)$ is the dipole field, $%
B_{00}$ is the dipole field in the center of the magnet, $B_1$ is the
quadrupole field, $B_{10}$ is the quadrupole field in the center of the magnet,
and $B_2$ is the maximum magnitude of the sextupole field at the radius $R$.
These fitted functions are shown in their corresponding plots in Figs.~\ref
{fig:acc:end0}--\ref{fig:acc:end2}.

\begin{figure}[tbp]
\centering
\includegraphics[width=0.65\textwidth]{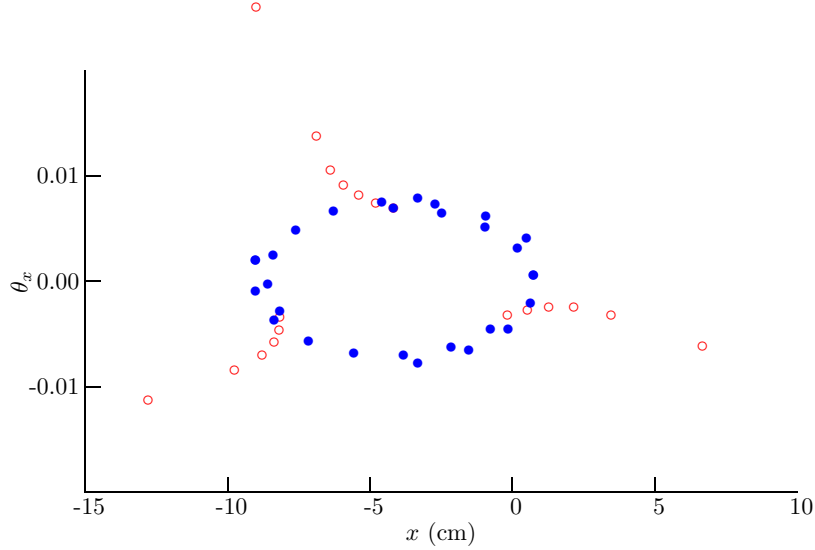}
\caption{(Color) Horizontal phase space of tracking at 5.1~GeV/c at the outer edge
of the acceptance. Open circles are without the body sextupole fields and
show a third-order resonance; filled circles are with the body sextupole
fields.}
\label{fig:acc:res3}
\end{figure}
\begin{figure}[tbp]
\centering
\includegraphics[width=0.65\textwidth]{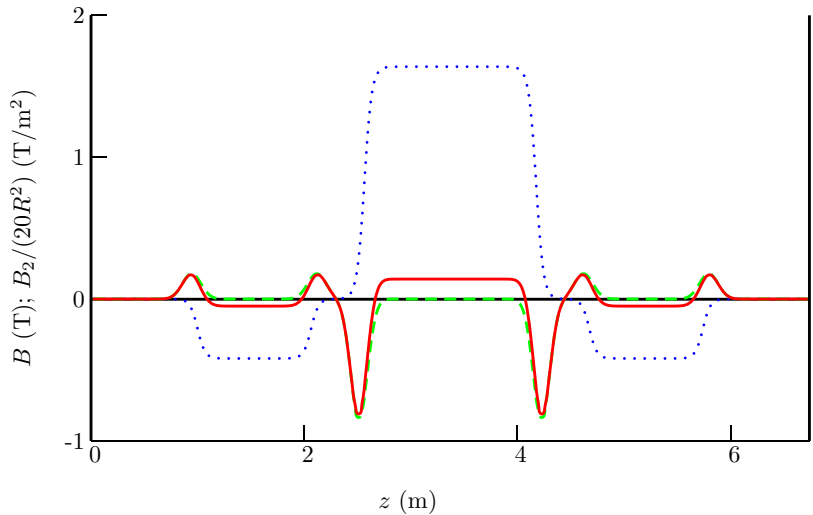}
\caption{(Color) Sextupole field components in the 5--10~GeV FFAG reference lattice.
The dotted line is the dipole field, the dashed line is $B_2/(20R^2)$ with
zero body sextupole field, and the solid line is with sufficient body
sextupole field to eliminate the third-order resonance.}
\label{fig:acc:sex}
\end{figure}
\begin{figure}[tbp]
\includegraphics[width=0.5\textwidth]{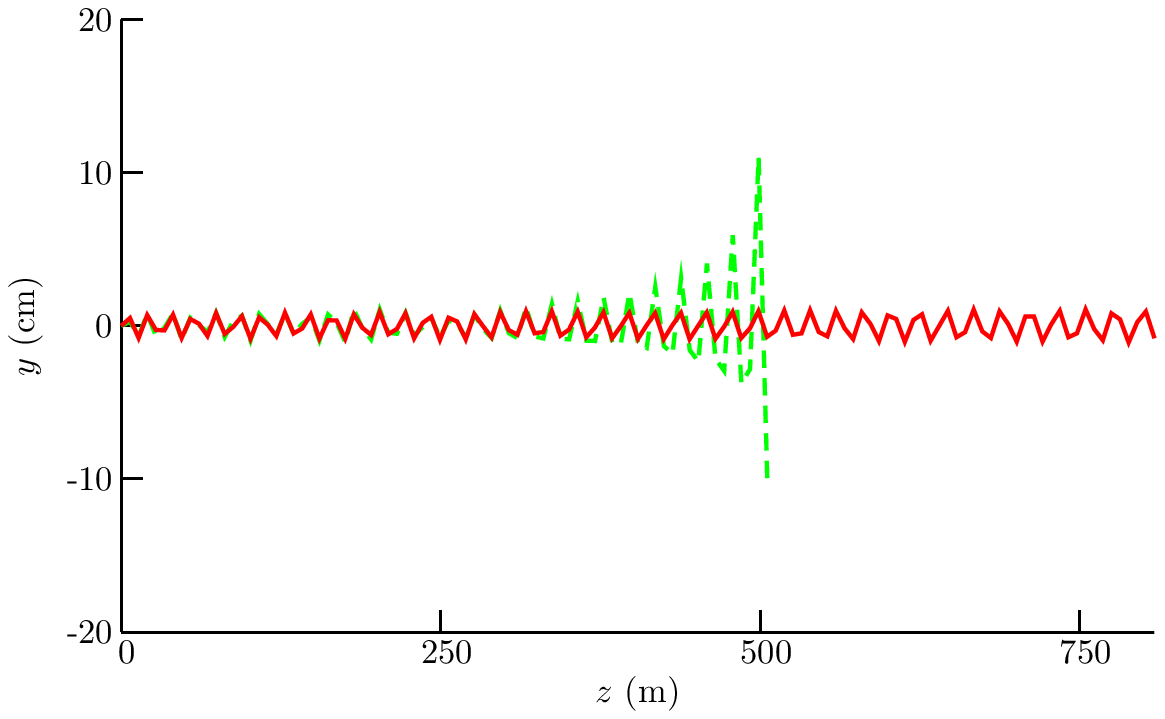}%
\includegraphics[width=0.5\textwidth]{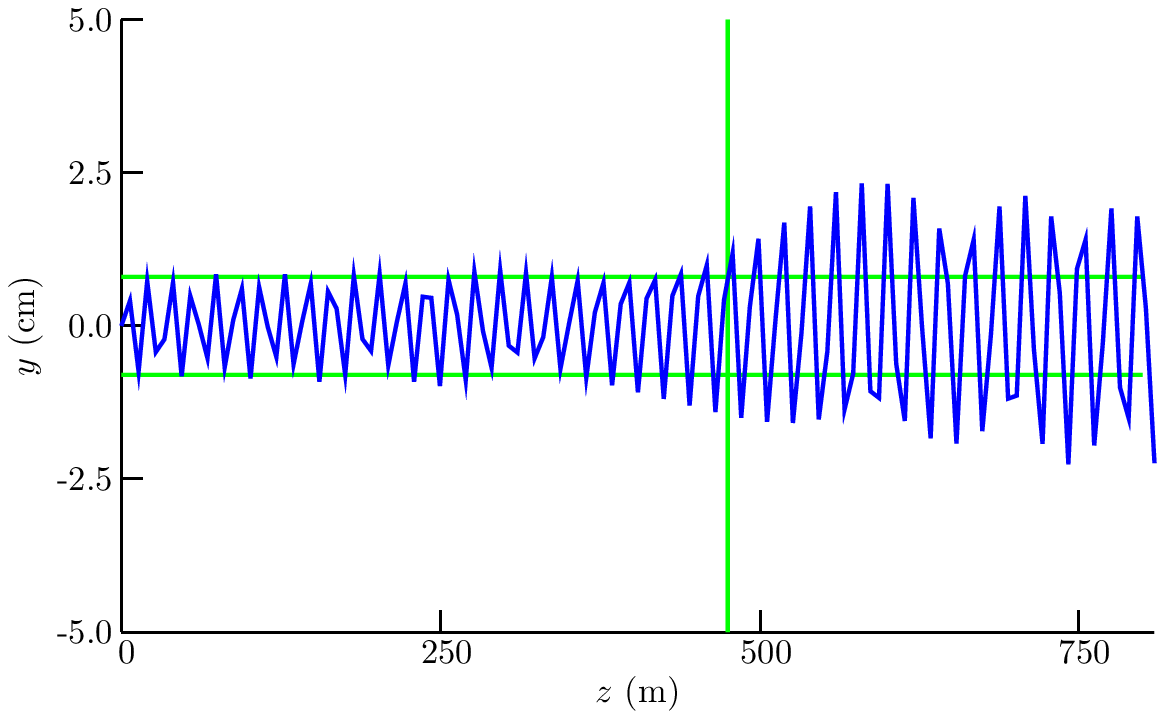}
\caption{(Color) Tracking of a particle at the edge of the acceptance with uniform acceleration
in the 5--10~GeV reference lattice. 
 On the left, the dashed line is without any
body sextupole, and the solid line is with the corrected body sextupole. On
the right, a smaller integrated sextupole correction is used (40\% instead
of 68\%), and significant emittance growth is observed.}
\label{fig:acc:trkacc}
\end{figure}
Injecting particles at the outer edge of the acceptance and tracking through
several cells, indicated a large third-order resonance at around
5.1~GeV/c as shown in Fig.~\ref{fig:acc:res3}. This resonance is
presumably being driven by the sextupole fields at the magnet ends. The
strength of the resonance can be reduced if the integrated sextupole in
the magnet is made zero. With some experimentation, it was found that if the
integrated body sextupole was set to 68\% of the integrated end
sextupoles, (see Fig.~\ref{fig:acc:sex}), the resonance was
eliminated (also shown in Fig.~\ref{fig:acc:res3}). When acceleration is
included, one sees particle loss when accelerating through the resonance if
there is no body sextupole correcting the end sextupoles, while there
appears to be almost none with the body correction included
(see Fig.~\ref{fig:acc:trkacc}). If the body correction is only partially
included, there is significant emittance growth, as seen in Fig.~\ref
{fig:acc:trkacc}. With these sextupole corrections, we can uniformly
accelerate over the entire 5--10~GeV energy range of the lower energy
reference FFAG without losing a high-amplitude particle or having its
amplitude grow by a large amount.

\begin{figure}[tbp]
\includegraphics[width=0.5\textwidth]{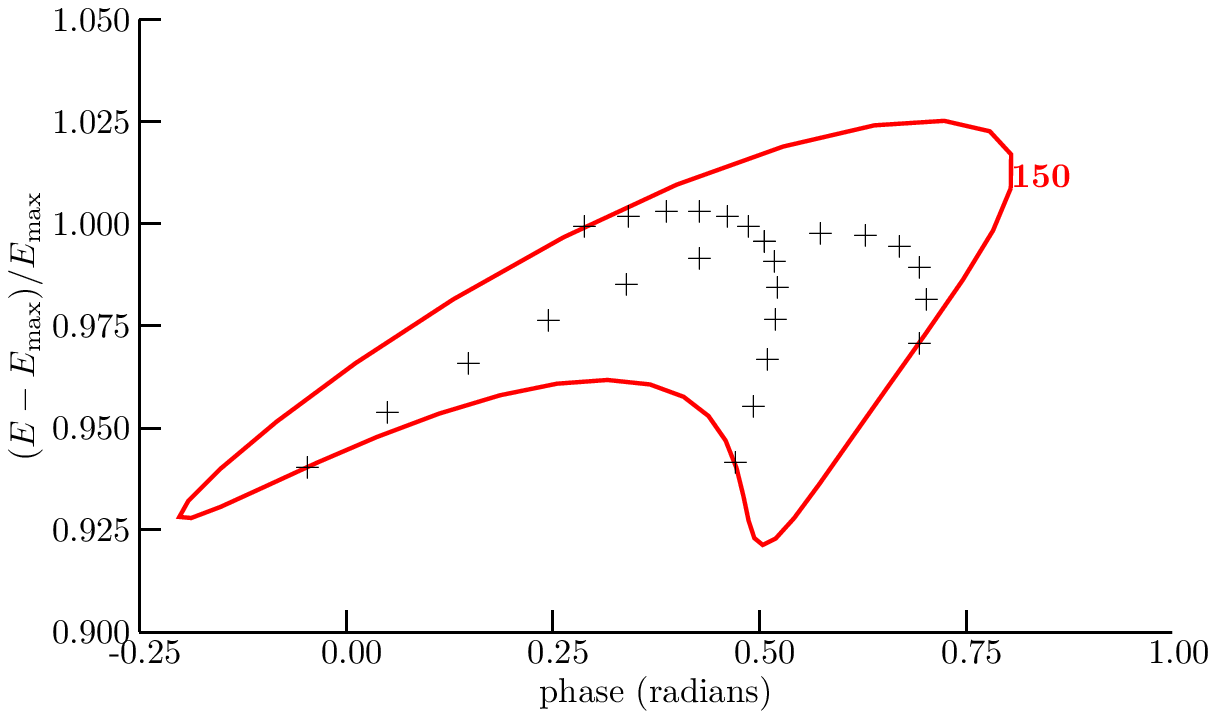}%
\includegraphics[width=0.5\textwidth]{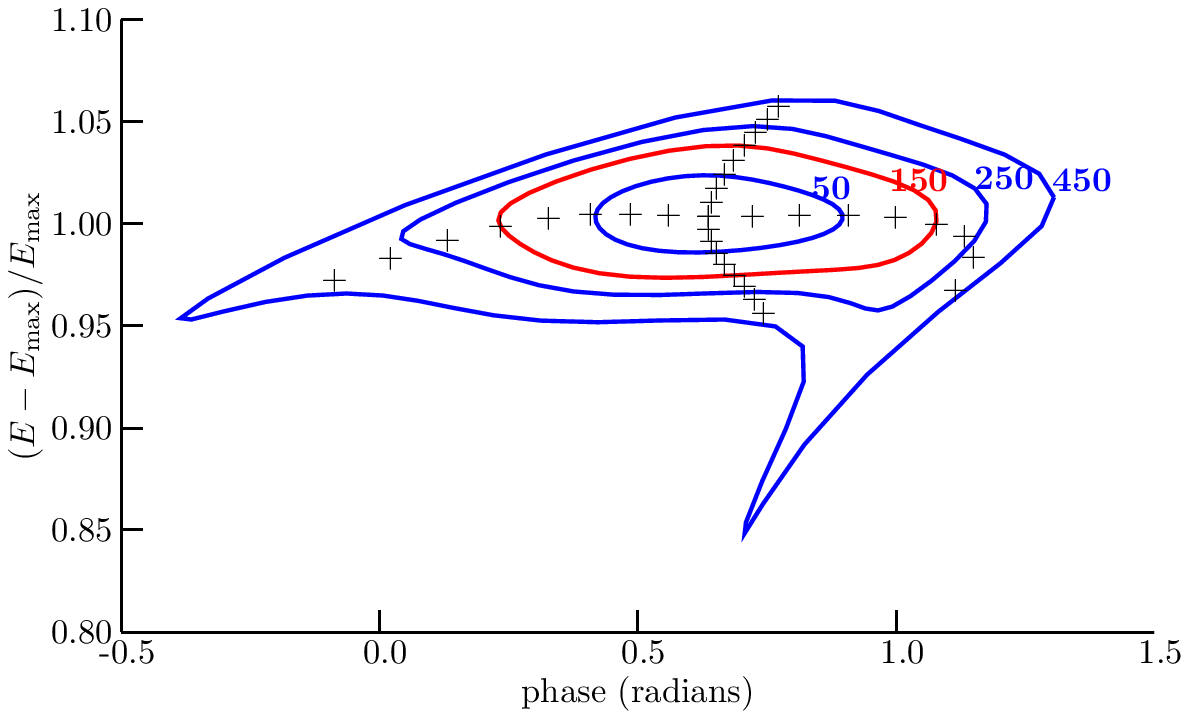}
\caption{(Color) Longitudinal tracking starting from an upright ellipse for the
5--10~GeV FFAG. On the left, with only 201.25~MHz rf. On the right, with
third-harmonic rf having voltage equal to 2/9 of the fundamental rf voltage.
Curves are labeled with their corresponding acceptance. Crosses for both
cases started out as horizontal and vertical lines in phase space.}
\label{fig:acc:longtrk}
\end{figure}

When tracking with rf is considered, the longitudinal
dynamics behavior is complex \cite{KoscielniakPAC03}. If one
begins with an upright ellipse, there is considerable emittance growth if
only the 201.25~MHz rf is used (see Fig.~\ref{fig:acc:longtrk}). Adding a 
third-harmonic rf considerably reduces the
emittance growth, as shown in Fig.~\ref{fig:acc:longtrk}. The amount of
third-harmonic rf required is substantial and that, combined with space 
considerations, makes this alternative unattractive. Tilting the
initial ellipse in phase space will reduce the emittance growth, but
a means to produce that tilt must be developed.
\subsubsection{FFAG Cost Optimization}
\begin{figure}[tbp]
\centering \includegraphics[width=0.65\textwidth]{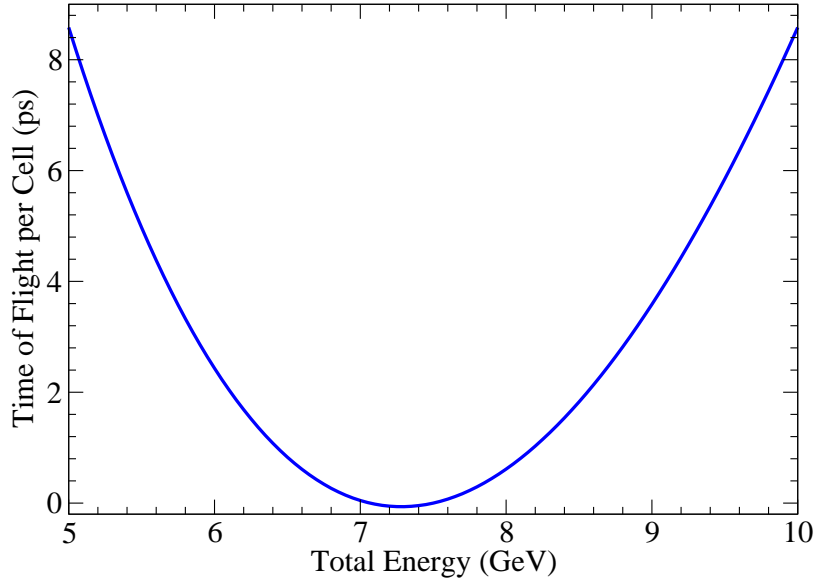}
\caption{(Color) Time-of-flight deviation per cell as a function of energy for the
5--10~GeV reference design.}
\label{fig:acc:tof}
\end{figure}
The designs for the FFAG lattices are chosen based on a cost optimization.
 The ``costs'' computed by our model
are not intended to be used in computing the cost of an actual machine, but
instead are used to find the cost of one design relative to another. For a given lattice type and a given energy range, the magnet lengths,
their dipole fields, and their quadrupole fields are allowed to vary to produce
the minimum cost for the lattice. Magnet apertures are determined by
finding a circle that encloses all of the beam ellipses within a magnet
(both at different energies and positions) for transverse amplitudes that
are equal to the transverse acceptance. The difference between the 
time-of-flight at the minimum energy and the minimum time-of-flight for all energies
(see Fig.~\ref{fig:acc:tof}) is constrained to be a certain value; similarly
for the time-of-flight at the maximum energy. That value is chosen to
make the quantity $V/\omega\Delta T\Delta E$ take on specific
values. These values depend on the energy of the lattice, as well as the
desired longitudinal acceptance. In the above equation, $V$ is the total rf
voltage in the ring as an energy gain, $\omega$ is the angular rf frequency, $\Delta T$ is the
time-of-flight difference described above, and $\Delta E$ is the energy
range over which the beam is accelerated.

We examined FFAG lattice designs that minimize the relative cost. 
 Some constraints were assumed in performing the optimization:
\begin{itemize}
\item  There is at least one 2-m-long drift in each cell to make room for a
superconducting rf cavity. This is substantially longer than the rf cavity,
but the extra length is needed to reduce the field from the magnets to
below 0.1~T at the cavity surface \cite{Ono99}.
\item  All shorter drifts in the cell are 0.5~m long. This is needed to
maintain sufficient space between magnets for cryostats and other
necessary equipment.
\item  The minimum and maximum energy of each accelerating stage are fixed.
\item  The type of lattice (doublet, triplet, FODO, etc.) is fixed.
\item  The time-of-flight on the energy-dependent closed orbit is the same
at the minimum and maximum energies (see Fig.~\ref{fig:acc:tof}). This
minimizes the deviation of the bunch from the rf crest and
therefore, presumably, maximizes the longitudinal acceptance of the lattice.
\item  The quantity $V/(\omega \Delta T\Delta E)$ is a fixed value that 
depends on the energy range being considered. This value characterizes the
longitudinal acceptance of the system.
\item  The rf gradient is not allowed to exceed a specified value. Here 
we take a conservative value of 10~MV/m. This corresponds to a value that 
has already been achieved~\cite{pac03:1309}.
\end{itemize}

\begin{table}[tbp]
\caption{Cost-optimum lattices with cavities in all but 8 cells.}
\label{tab:acc:shortopt}%
\begin{ruledtabular}
    \begin{tabular}{l|rrr|rrr|rrr}
      Minimum total energy (GeV)&\multicolumn{3}{c|}{2.5}&\multicolumn{3}{c|}{5}&\multicolumn{3}{c}{10}\\
      Maximum total energy (GeV)&\multicolumn{3}{c|}{5}&\multicolumn{3}{c|}{10}&\multicolumn{3}{c}{20}\\
      $V/(\omega\Delta T\Delta E)$&\multicolumn{3}{c|}{1/6}&\multicolumn{3}{c|}{1/8}&\multicolumn{3}{c}{1/12}\\
      Type                 &FD  &FDF &FODO&FD  &FDF &FODO&FD  &FDF &FODO\\
      No.\ of cells        &65  &60  &76  &79  &72  &91  &93  &85  &105 \\
      D length (cm)        &62  &96  &56  &82  &119 &77  &105 &143 &98  \\
      D radius (cm)        &13.6&16.5&16.0&10.2&12.7&11.7&7.8 &9.7 &8.7 \\
      D pole tip field (T) &3.7 &3.3 &1.9 &4.6 &4.2 &3.8 &5.8 &5.5 &5.0 \\
      F length (cm)        &99  &48  &93  &126 &64  &119 &162 &85  &151 \\
      F radius (cm)        &19.1&15.8&22.8&15.3&12.8&17.8&12.7&10.9&14.6\\
      F pole tip field (T) &2.2 &2.4 &1.7 &2.8 &3.1 &2.2 &3.5 &3.7 &2.8 \\
      No.\ of cavities     &57  &52  &68  &71  &64  &83  &85  &77  &97  \\
      RF voltage (MV)      &428 &390 &510 &533 &480 &623 &638 &578 &728 \\
      $\Delta E/V$         &5.8 &6.4 &4.9 &9.4 &10.4&8.0 &15.7&17.3&13.7\\
      Circumference (m)    &268 &295 &418 &362 &393 &543 &481 &521 &681 \\
      Decay (\%)           &6.8 &8.2 &8.8 &7.4 &8.9 &9.4 &8.5 &10.1&10.4\\
      Magnet cost (A.U.)\footnote{Arbitrary units} &36.4&41.6&49.6&32.8&37.4&40.0&34.1&39.2&38.4\\
      RF cost (A.U.)         &27.7&25.3&33.0&34.5&31.1&40.3&41.3&37.4&47.1\\
      Linear cost (A.U.)     &6.7 &7.4 &10.4&9.1 &9.8 &13.6&12.0&13.0&17.0\\
      Total cost (A.U.)      &70.8&74.3&93.1&76.3&78.3&93.8&87.4&89.6&102.5\\
      Cost per GeV (A.U.)&28.3&29.7&37.2&15.3&15.7&18.8&8.7 &9.0 &10.2
    \end{tabular}
  \end{ruledtabular}
\end{table}

To get an idea of what is achievable, we developed a  set of
cost-optimized lattices where a cavity is placed in every cell, except for 8
cells left open for injection and extraction hardware. The goal is to
minimize the time spent accelerating and therefore minimize the
decays. This drives the design toward a smaller ring but more rf. 
The results of this
optimization are shown in Table~\ref{tab:acc:shortopt}. The costs 
are  a significant improvement over the FS2 acceleration costs.

Table~\ref{tab:acc:shortopt} leads to several conclusions:
\begin{itemize}
\item  The doublet lattice is the most cost-effective design. The triplet
lattice requires less voltage, but has a higher magnet cost due to having more
magnets per cell.
\item  The cost per GeV of acceleration increases as the energy decreases.
The RLA from FS2 has a cost per GeV around 30 in the units of Table~\ref{tab:acc:shortopt}, so this in some sense
 sets a baseline for determining when an FFAG approach becomes cost
 effective. Thus, a 2.5--5~GeV FFAG is borderline in its cost effectiveness, while the
higher energy FFAGs are clearly cost effective.
\end{itemize}

\begin{figure}[tbp]
\includegraphics[width=0.65\textwidth]{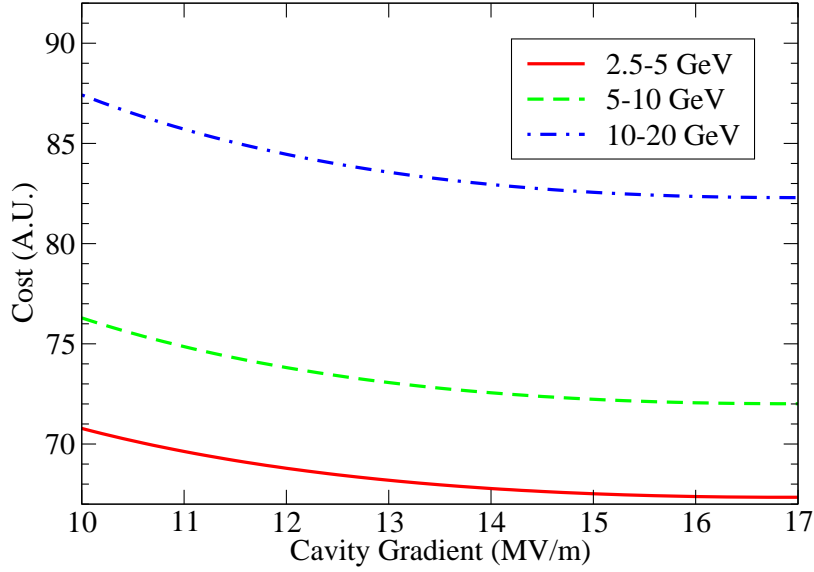}
\caption{(Color) Costs of the doublet lattices in Table~\ref{tab:acc:shortopt} with
the rf cost modified for higher cavity gradients.}
\label{fig:acc:shortvsv}
\end{figure}
 The effects of increasing the rf gradient on the costs of the doublet
 designs in Table~\ref{tab:acc:shortopt} are shown in Fig.~\ref{fig:acc:shortvsv}.
\begin{figure}[tbp]
\centering
\includegraphics[width=0.65\textwidth]{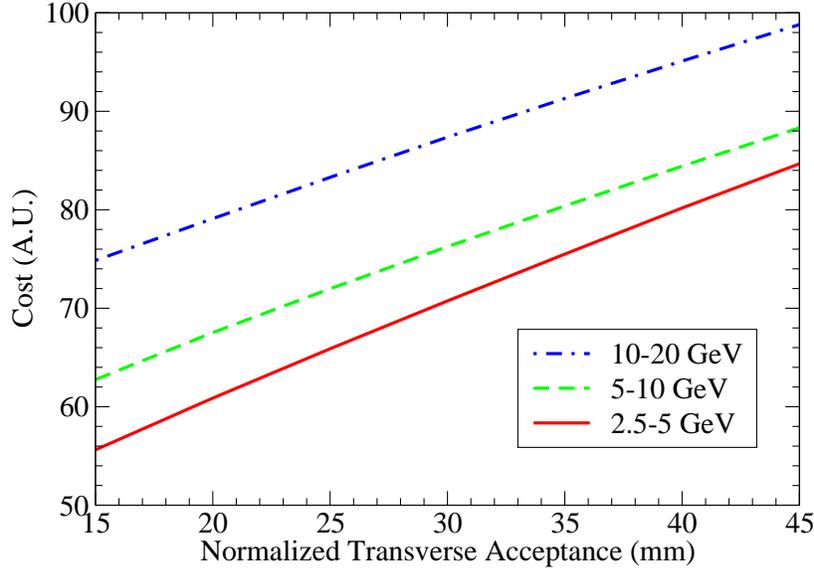}
\caption{(Color) Costs of the optimized doublet lattices as a function of the
transverse normalized acceptance.}
\label{fig:acc:accopt}
\end{figure}
The choice of the acceptance also has a strong effect on the optimized
cost, as shown in Fig.~\ref{fig:acc:accopt}.

\subsubsection{Design of Combined-Function Superconducting Magnet for FFAGs}
An initial design of a superconducting combined-function
(dipole--quadrupole) magnet has been developed~\cite{CaspiRef}.
A first-cut design of a superconducting combined function, dipole-quadrupole magnet, is outlined. 
The design is for one of the QD magnets requiring the highest field and gradient. The parameters of the QD cell are shown in Table~\ref{qdcell},
\begin{table}[htbp!]
\caption{Parameters of the QD cell}
\label{qdcell} 
\begin{ruledtabular}
\begin{tabular*}{10cm}{cc}
$E_{\text{min}}$ (GeV)& 10\\
$E_{\text{max}}$ (GeV) &20\\
$L_0$ (m) &2\\
$L_q$ (m) &0.5\\
Type & QD\\
$L$ (m)&1.762\\
$r$ (m)& 18.4\\
$X_0$ (mm)&1.148\\
$R$ (cm)&10.3756\\
$B_0$ (T)&2.7192\\
$B_1$ (T/m)&-15.495\\
\end{tabular*}
\end{ruledtabular}
\end{table}
where $L_0$ is the length of the long drift between the QF magnets, $L_q$ is the length of the short drift between QF and QD magnets; $L$ is the length of the reference orbit inside the magnet, $r$ is the radius of curvature of the reference orbit, $X_0$ is the displacement of the center of the magnet from the reference orbit, $R$ is the radius of the magnet bore, $B_0$ is the vertical magnetic field at the reference orbit, and $B_1$ is the derivative of the vertical magnetic field at the reference orbit.

The magnet design is based on a cosine-theta configuration with two double
layers for each function. The quadrupole coil is located within the dipole
coil  and both coils are assembled using key-and-bladder technology. All
coils are made with the same Nb--Ti cable capable of generating the
operating dipole field and gradient with about the same current of 1800~A
(a single power supply is thus possible with a bit of fine tuning). 
 The maximum central dipole field and gradient at short sample are 4.1~T and 26~T/m, as compared 
with the requirements of 2.7~T and 15.4~T/m, respectively. At this early
design stage, excess margin is left for safety and perhaps a field-rise in the magnet end region. The maximum azimuthal forces required for magnet 
pre-stress are of the order of 1~MN/m (assuming maximum safety). 
 The conductor strand size and cable parameters common to both dipole and quadrupole are listed in Table~\ref{nbti}.
\begin{table}[htbp!]
\caption{Nb--Ti conductor for dipole and quadrupole coils.\label{nbti}}
\begin{ruledtabular}
\begin{tabular}{lc}
Strand diameter (mm) &0.6477\\
Cable width, bare (mm) & 6.4\\
Cable thickness, insulated (mm) &1.35\\
Keystone angle  (deg.) &0.6814\\
Conductor type &Nb--Ti\\
Cu:SC ratio  & 1.8:1\\
Current density (at 5~T, 4.2~K) (A/{mm}$^2$)& 2850\\
Number of strands& 20\\
\end{tabular}
\end{ruledtabular}
\end{table} 

\begin{figure}[hbtp!]
\includegraphics*[width=4in]{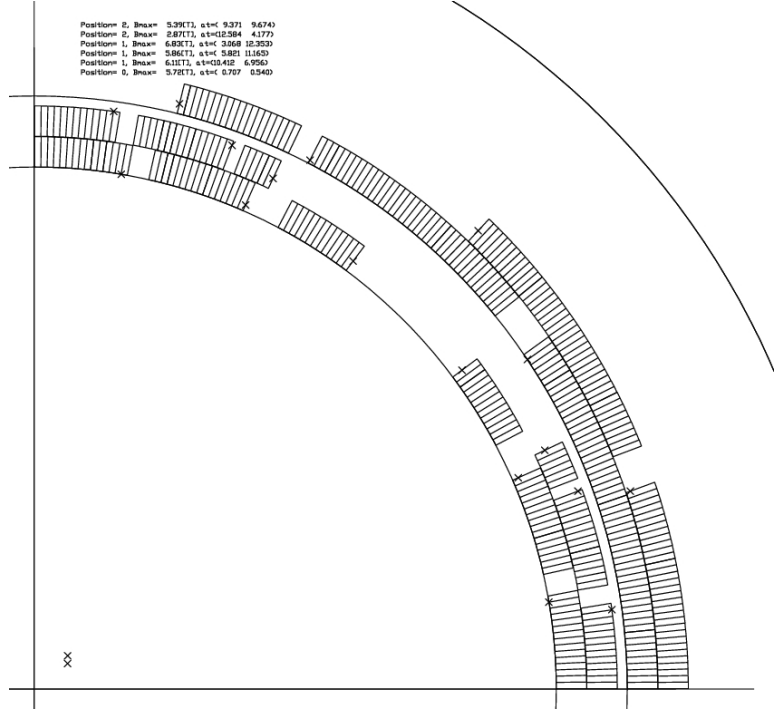}
\caption{First quadrant of the combined magnet cross section.}
\label{first-quad}
\end{figure}

\begin{figure}[hbtp!]
\includegraphics*[width=4in]{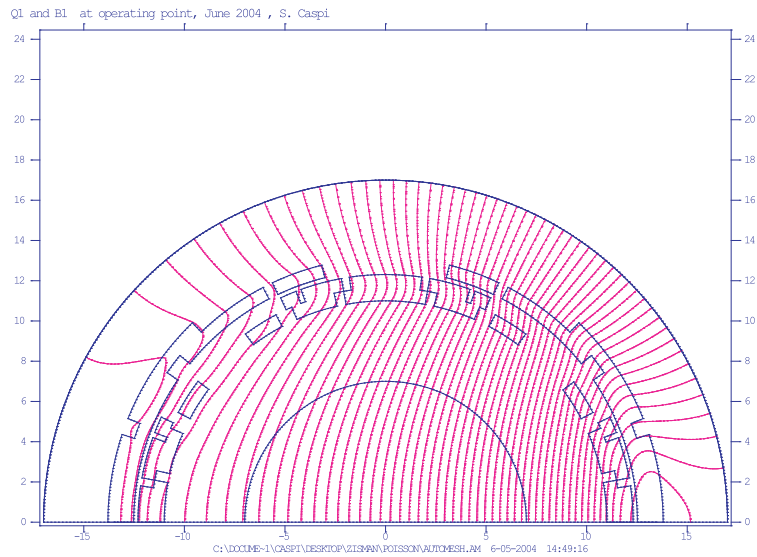}
\caption{(Color) Flux plot corresponding to a dipole field of 2.7~T and a gradient of 15~T/m.}
\label{first-dip}
\end{figure}

\begin{figure}[hbtp!]
\includegraphics[width=4in]{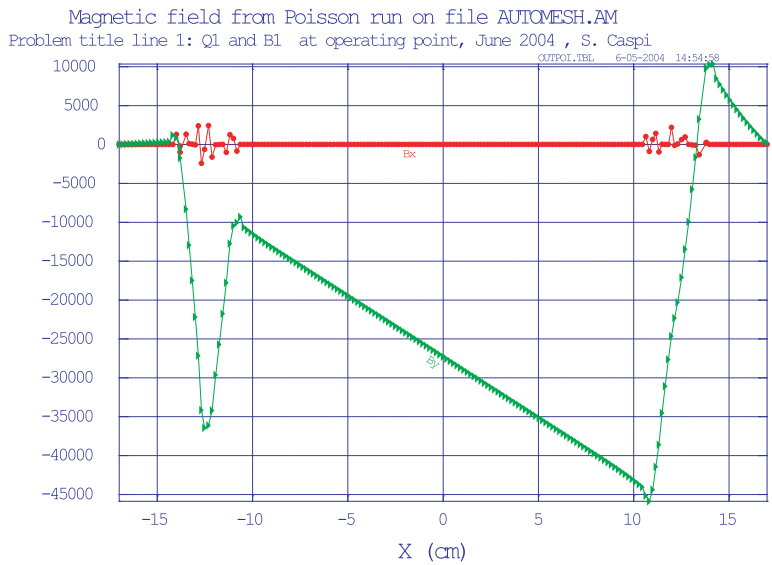}
\caption{(Color) B${}_y$ along the mid-plane showing a central 2.7~T field and a 15~T/m gradient.}
\label{first-field}
\end{figure}

\begin{figure}[hbtp!]
\includegraphics[width=4in]{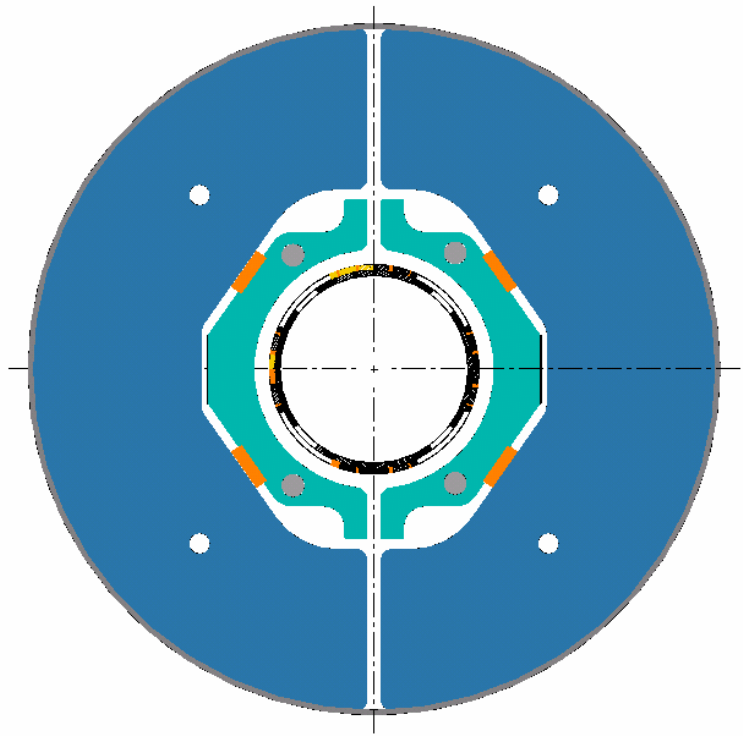}
\caption{(Color) End viw of magnet structure.}
\label{nbti-struc}
\end{figure}

\begin{figure}[hbtp!]
\includegraphics[width=4in]{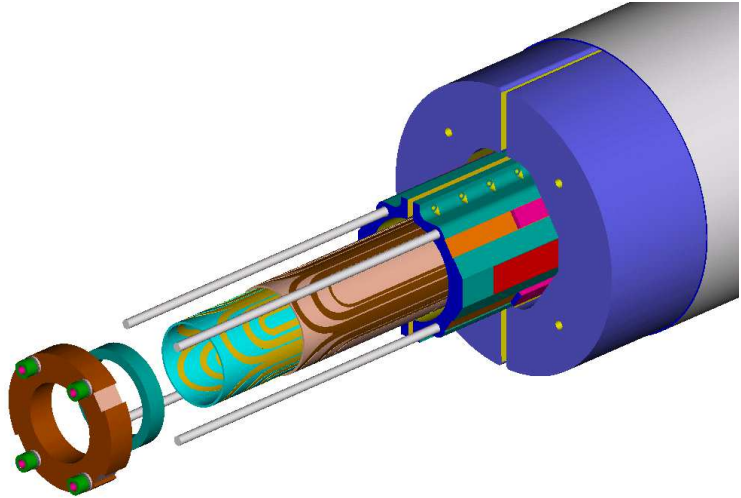}
\caption{(Color) Exploded view showing the two quadrupole layers (dipole coils not shown).}
\label{nbti-explo}
\end{figure}

\begin{figure}[hbtp!]
\includegraphics[width=4in]{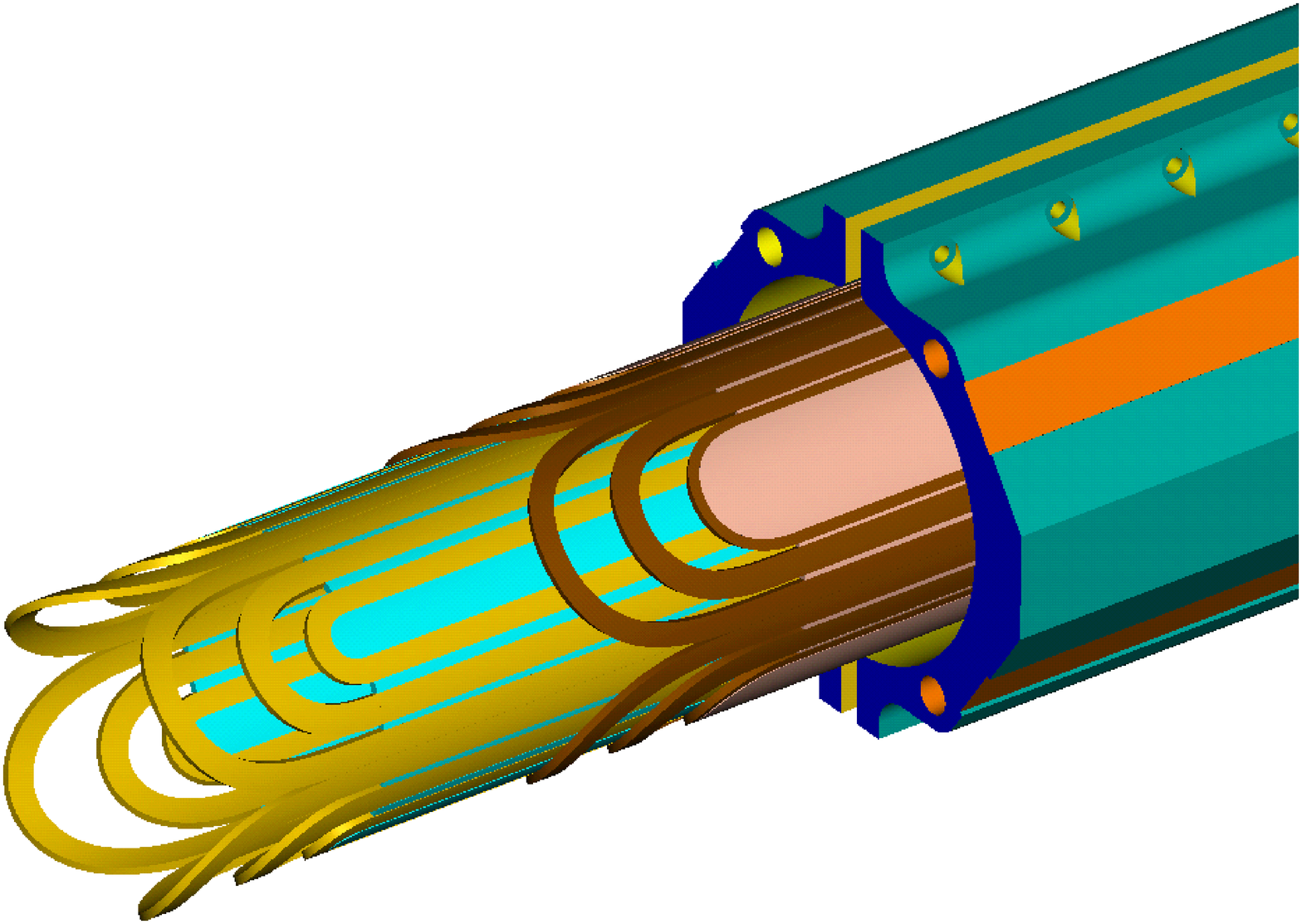}
\caption{(Color) Close-up of quadrupole coil return-end windings.}
\label{nbti-close}
\end{figure}

The initial cross sections of both dipole and quadrupole were designed to
give less than one part in one hundred units of  systematic 
multipole errors at a radius of 70~mm. It is straightforward to readjust
the design to cancel the end-field multipoles as proposed in
Section~\ref{sec5-sub2}. The combined cross section is shown in
Fig.~\ref{first-quad} for one quadrant. Figure~\ref{first-dip} shows 
the combined dipole-quadrupole magnetic flux.
The calculated mid-plane field profile of the magnet (plotted in
Fig.~\ref{first-field}) clearly shows the superposed dipole and quadrupole fields. A mechanical layout for the magnet has also been
developed, as shown in Figs.~\ref{nbti-struc}, \ref{nbti-explo}, and \ref{nbti-close}. 

\begin{table}[htbp!]
\caption{Coil current parameters.\label{tab:current}}
\begin{ruledtabular}
\begin{tabular}{ccc}
Current density (A/{mm}$^2$)&Central Field (T)&Gradient\\
\hline
730&2.5&15.4\\
800&2.7&17\\
1220 (maximum)&4.1&26\\
\end{tabular}
\end{ruledtabular}
\end{table}

\subsection{Considerations on a Beta Beam Facility in the U.S. \label{sec5-sub2-1}}
Motivated by the recent suggestion that a higher energy Beta Beam facility
might have considerable scientific merit~\cite{jj-beta}, we
consider here some of the possibilities of a U.S.-based scenario. There was
neither the time nor the effort available to carry out a study equivalent to
the baseline scenario prepared by the European Beta Beam Study Group, and we
make no pretense of having done so. Nonetheless, it was felt to be
interesting to look briefly at U.S. options that have the potential for
higher energy beams than likely to be available at CERN in the foreseeable
future. In particular, an ``intermediate'' beam energy of $\gamma =350,$
which corresponds well to the top energy available at the Tevatron, is
expected to have much better resolution than the CERN low-energy option in
terms of both \textsl{CP} violation and sensitivity to the mass hierarchy.

The two possible U.S. machines to consider are RHIC at BNL and the Tevatron
at Fermilab. Of these, the Tevatron looks more attractive due to its higher
energy reach. (RHIC has a top energy comparable to the SPS at CERN.) Both
BNL and Fermilab are interested in Superbeams, which complement the Beta
Beam in terms of physics reach, and both Laboratories are pursuing the
possibility of obtaining a high-intensity proton driver, a prerequisite for
 a Superbeam and helpful, but not critical, for a Beta Beam facility. It is
 worth noting, of course, that an intermediate-energy Beta Beam facility
 would require a very large decay ring to store a $\gamma=350$ beam. This would
 add substantially to the cost of implementing such a facility in the U.S. 
\subsubsection{Estimate of Decay Losses}
It is important to know how many ions survive the acceleration process
without decaying. This can be calculated as 
\begin{equation}
N=N_{0}e^{-\frac{1}{\tau }\int_{0}^{T}\frac{dt^{\prime }}{\gamma (t^{\prime })}}
\label{eqn:one}
\end{equation}
where $\tau $ is the decay time at rest. If the ramp rate is constant, the
integral (which represents the time passed in the rest frame) becomes 
\begin{equation}
\int_{t_{0}}^{t}\frac{dt^{\prime }}{\gamma (t^{\prime })}=\frac{T\ln (\gamma_{1}/\gamma_{0})}{\gamma_{1}-\gamma_{0}}.
\end{equation}
Here, $T$ is the ramp time, $\gamma_{0}$ the initial energy, and $\gamma_{1}$ final
energy. In most accelerators, the energy changes by an order of
magnitude ($\gamma_{1}\approx 10\gamma_{0}$), so  
\begin{equation}
\frac{T\ln (\gamma_{1}/\gamma_{0})}{\gamma_{1}-\gamma_{0}}\approx \frac{T}{0.4\gamma_{1}}.
\end{equation}
Substituting this back into Eq.~(\ref{eqn:one}) yields 
\begin{equation}
N\approx N_{0}e^{-\frac{T}{0.4\gamma_{1}\tau }}  \label{eqn:two}
\end{equation}
which is just the standard decay formula with the decay time calculated at
an ``effective'' or average energy equal to 40\% of the top energy. In
reality, the ramp rate is not constant. In particular, the start of the ramp
is usually slower, leading to a larger total decay than the above estimate
would predict. Nonetheless, Eq.~(\ref{eqn:two}) gives a useful first
approximation. If a more accurate result were needed, and the ramp function
is known, losses could be calculated directly from Eq.~(\ref{eqn:one}).

To calculate the Lorenz gamma of the ions, based on the corresponding gamma
of full energy protons in the same machine, we recall that the magnetic
rigidity is the same for both particles: 
\begin{equation}
B\rho =\frac{p}{q}\Big|_{\mathrm{proton}}=\frac{p}{q}\Big|_{\mathrm{ion}}
\end{equation}
which, since the momentum is 
\begin{equation}
p=m_{0}c\beta \gamma =m_{0}c\sqrt{\gamma ^{2}-1}
\end{equation}
gives 
\begin{equation}
\gamma _{\mathrm{ion}}=\sqrt{1+\left( \gamma _{\mathrm{proton}}^{2}-1\right)
\left( q/M\right) ^{2}}
\end{equation}
where $q/M$ is the charge-to-mass ratio of the ion.

Table~\ref{t:gammas} lists the U.S. machines relevant to Beta Beam
acceleration, and their maximum gamma $(\gamma_{\text{max}})$ for protons, $^{6}$He and $^{18}$Ne.
Table~\ref{t:losses} gives the approximate amount of beam loss due to beta
decay during acceleration. Due to its long ramp time, it is clear that RHIC
would not be very efficient at accelerating the ions in question. The
Tevatron, on the other hand, would be relatively efficient, as it was
originally designed as a fixed-target machine and hence has a reasonably
fast ramp. The Tevatron has the additional feature of being, for now, the
world's highest energy machine. Assuming that the LHC will be busy with
collider physics for the foreseeable future, the Tevatron would thus seem to
be the obvious candidate for generating an ``intermediate energy'' Beta Beam. There
remains one key question, however---we must assess how the Tevatron's
superconducting magnets would be affected by the decay products of the Beta
Beam. This, in turn, determines how many ions the machine could accelerate
to top energy in a single batch. If there is continued interest in exploring
the possibility of using the Tevatron for an intermediate energy Beta Beam
facility, this issue must be studied.
\begin{table}[tbph!]
\caption{Parameters of U.S. machines that could potentially be used to
accelerate ions for a Beta Beam.}
\label{t:gammas}
\begin{ruledtabular}
\begin{tabular}{ccccc}
Machine &Proton kinetic energy (GeV)&$\gamma (p)$&$\gamma (^{6}\mathrm{%
He}^{2+})$&$\gamma (^{18}\mathrm{Ne}^{10+})$ \\\hline
FNAL Booster & 8 & 9.5 & 3.3 & 5.4 \\ 
Main Injector & 150 & 161 & 64 & 89 \\ 
Tevatron & 980 & 1045 & 349 & 581 \\ \hline\hline
BNL Booster & 2 & 3.1 & 1.4 & 1.9 \\ 
AGS & 30 & 34 & 11 & 19 \\ 
RHIC & 250 & 268 & 89 & 149 \\
\end{tabular} 
\end{ruledtabular}
\end{table}

\begin{table}[tbp!]
\caption{Expected losses during acceleration, calculated using Eq.~\ref
{eqn:two}. For RHIC, the improved values in parentheses would require a
modest modification to the power supplies.}
\label{t:losses}
\begin{ruledtabular}
\begin{tabular}{cccc}
Machine&Ramp time (s)&$^{6}\mathrm{He}^{2+}$ loss (\%) & $^{18}\mathrm{Ne}^{10+}$ loss (\ \%) \\ \hline
FNAL Booster & 0.03 & 2 & 1 \\ 
Main Injector & 0.7 & 2 & 1 \\ 
Tevatron & 17 & 10 & 3 \\ \hline\hline
BNL Booster & 0.1 & 14 & 5 \\ 
AGS & 0.5 & 9 & 3 \\ 
RHIC & 100 (40) & 91 (62) & 50 (24) \\
\end{tabular}
\end{ruledtabular}
\end{table}

\subsubsection{Estimate of Power Deposition}
The total deposited power from decay products per unit machine length can be
written 
\begin{equation}
P=\frac{-\dot{N}E_{\mathrm{kin}}}{L}=\frac{N}{\gamma \tau }\frac{%
E_{0}(\gamma -1)}{L} 
\end{equation}
where, $\dot{N}$ is the number of decays per unit of time, $\tau$ is the decay time at rest, and $E_{0}$ is the rest
energy. For high energies (${\gamma\to \infty}$), the deposited power
$P\approx\frac{NE_{0}}{L \tau}$  is independent of gamma,
depending only on the number of ions per machine length. To obtain the
time-averaged power, one must multiply by the duty factor $f$ (fraction of
time with beam in the machine) 
\begin{equation}
\left\langle P\right\rangle \approx f\frac{NE_{0}}{L \tau}.
\end{equation}
For the Tevatron, this equation can be used to estimate the number of ions
that would generate 1~W/m from decay losses, which is about $1\times 10^{13}$
for both types of ions. For the lower energy machines, supplying the
Tevatron with this intensity would yield much lower power deposition, even
though their circumferences are smaller. This is because their duty factor
(ramp time divided by Tevatron cycle time) is very small. Anticipated levels
are about 0.05~W/m in the Main Injector, and 0.03~W/m in the Booster, based
on the simplified formula above.

\section{Neutrino Factory and Beta Beam R\&D \label{sec6}}

\label{r_and_d}
As should be clear from the design descriptions in Section~\ref{sec5}, both the
muon-based Neutrino Factory and the Beta Beam facility are demanding projects.
Both types of machine make use of novel components and techniques that are, in
some cases, at or beyond the state of the art. For this reason, it is critical
that R\&D efforts to study these matters be carried out. In this Section we
describe the main areas of R\&D effort under way in support of the two
projects. We give an overview of the R\&D program goals and list the specific
questions we expect ultimately to answer. We also summarize briefly the R\&D
accomplishments to date and give an indication of R\&D plans for the future.

Since neither of these projects is expected to begin construction in the near
future, it might be asked why it is necessary to pursue a vigorous R\&D
program now. One answer is that this R\&D is what allows us to
determine---with some confidence---both the expected performance and expected
cost of such machines. This information must be available in a timely way to
permit the scientific community to make informed choices on which project(s)
they wish to request at some future time. Experience has shown that large,
complex accelerator projects take many years of preparatory R\&D in advance of
construction. It is only by supporting this R\&D effort now that we can be
ready to provide a Neutrino Factory or Beta Beam facility when the proper time comes.
\subsection{Neutrino Factory R\&D}
Successful construction of a muon storage ring to provide a copious source of
neutrinos requires many novel approaches to be developed and demonstrated; a
high-luminosity Muon Collider, which might someday follow, would require an
even greater extension of the present state of accelerator design. Thus,
reaching the desired facility performance requires an extensive R\&D program.
Each of the major systems has significant issues that must be addressed by
R\&D activities. Component specifications need to be verified. For example,
the cooling channel assumes a normal conducting rf (NCRF) cavity gradient of
15 MV/m at 201.25 MHz, and the acceleration section demands similar
performance from superconducting rf (SCRF) cavities at this frequency. In both
cases, the requirements are beyond the performance reached to date for
cavities in this frequency range. The ability of the target to withstand a
proton beam power of up to 4 MW must be confirmed. Finally, an ionization
cooling experiment should be undertaken to validate the implementation and
performance of the cooling channel, and to confirm that our simulations of the
cooling process are accurate.
\subsubsection{R\&D Program Overview}
A Neutrino Factory comprises the following major systems: Proton Driver;
Target, (Pion) Capture, and (Pion-to-Muon) Decay Section; Bunching and Phase
Rotation Section; Cooling Section; Acceleration Section; and Storage Ring.
The R\&D program we envision is designed to answer first the key questions
needed to embark upon a Zeroth-order Design Report (ZDR). The ZDR will examine
the complete systems of a Neutrino Factory, making sure that nothing is
forgotten, and will show how the parts merge into a coherent whole. While it
will not present a fully engineered design with a detailed cost estimate,
enough detail will be presented to ensure that the critical items are
technically feasible and that the proposed facility could be successfully
constructed and operated at its design specifications.
By the end of the full R\&D program, it is expected that a formal Conceptual
Design Report for a Neutrino Factory could begin. The CDR would document a
complete and fully engineered design for the facility, including a detailed
bottom-up cost estimate for all components. This document would form the
basis for a full technical, cost, and schedule review of the construction
proposal, subsequent to which construction could commence (assuming strong
community support and government approval).
The R\&D issues for each of the major systems must be addressed by a mix of
theoretical, simulation, modeling, and experimental studies, as appropriate. A
list of the key physics and technology issues for each major Neutrino Factory
system is given below. These issues are being actively pursued as part of the
ongoing worldwide Neutrino Factory R\&D program, with participation from
Europe, Japan, and the U.S.

\textbf{Proton Driver}
\begin{itemize}
\item  Production of intense, short proton bunches, e.g., with space-charge
compensation and/or high-gradient, low frequency rf systems
\end{itemize}
\textbf{Target, Capture, and Decay Section}
\begin{itemize}
\item  Optimization of target material (low-\textit{Z} or high-\textit{Z}) and
form (solid, moving band, liquid-metal jet)
\item  Design and performance of a high-field solenoid ($\approx$20 T) in a
very high radiation environment
\end{itemize}
\textbf{Bunching and Phase Rotation Section}
\begin{itemize}
\item  Design of efficient and cost-effective bunching system
\item  Examination of alternative approaches, e.g., based upon combined rf
phase rotation and bunching systems or fixed-field, alternating gradient
(FFAG) rings
\end{itemize}
\textbf{Cooling Section}
\begin{itemize}
\item  Development and testing of high-gradient normal conducting rf (NCRF)
cavities at a frequency near 200 MHz
\item  Development and testing of efficient high-power rf sources at a
frequency near 200 MHz
\item  Development and testing of LH$_{2}$, LiH, and other absorbers for muon cooling
\item  Development and testing of candidate diagnostics to measure emittance
and optimize cooling channel performance
\end{itemize}
\textbf{Acceleration Section}
\begin{itemize}
\item  Optimization of acceleration techniques to increase the energy of a
muon beam (with a large momentum spread) from a few GeV to a few tens of GeV
(e.g., recirculating linacs, rapid cycling synchrotrons~\cite{rcsync}, FFAG rings)
\item  Development of high-gradient superconducting rf (SCRF) cavities at
frequencies near 200~MHz, along with efficient power sources (about 10~MW
peak) to drive them
\item  Design and testing of components (rf cavities, magnets, diagnostics)
that will operate in the muon-decay radiation environment
\end{itemize}
\textbf{Storage Ring}
\begin{itemize}
\item  Design of large-aperture, well-shielded superconducting magnets that
will operate in the muon-decay radiation environment
\end{itemize}
\subsubsection{Recent R\&D Accomplishments}
\paragraph{Targetry}
The BNL Targetry experiment, E951, has carried out initial beam tests~\cite{TgtRef} of both a solid carbon target and a mercury target at a proton
beam intensity of about $4\times 10^{12}$ ppp.
In the case of the solid carbon target, it was found that a carbon-carbon
composite having nearly zero coefficient of thermal expansion is largely
immune to beam-induced pressure waves. A carbon target in a helium atmosphere
is expected to have negligible sublimation loss. A program to verify this is
under way at ORNL~\cite{ORNLsublimation}. If radiation damage is the limiting
effect for a carbon target, the predicted lifetime would be about 12 weeks
when bombarded with a 1 MW proton beam.

\begin{figure}[hptb!]

\includegraphics[width=5.5in]{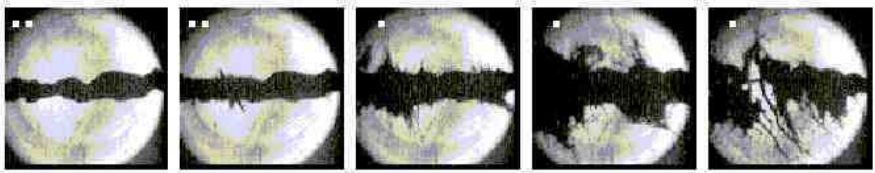}%

\caption{Disruption of Hg jet hit with AGS beam bunch containing $2\times10^{12}$ protons. Frames from left to right correspond to time steps of 0, 0.75, 2, 7, and 18~ms, respectively.}%

\label{fig:HgJet}%

\end{figure}

For a mercury jet target, tests with about 2 $\times$\ 10$^{12}$ ppp showed
that the jet is not dispersed until long after the beam pulse has passed
through the target (see Fig. \ref{fig:HgJet}). Measurements of the velocity of
droplets emanating from the jet as it is hit with the proton beam pulse from
the AGS ($\thickapprox$10 m/s for 25 J/g energy deposition) compare favorably
with simulation estimates. High-speed photographs indicate that the beam
disruption at the present intensity does not propagate back upstream toward
the jet nozzle. If this remains true at the higher intensity of 1.6 $\times
$\ 10$^{13}$ ppp, it will ease mechanical design issues for the nozzle.%

\begin{figure}[ptbh!]

\includegraphics[scale=0.3]{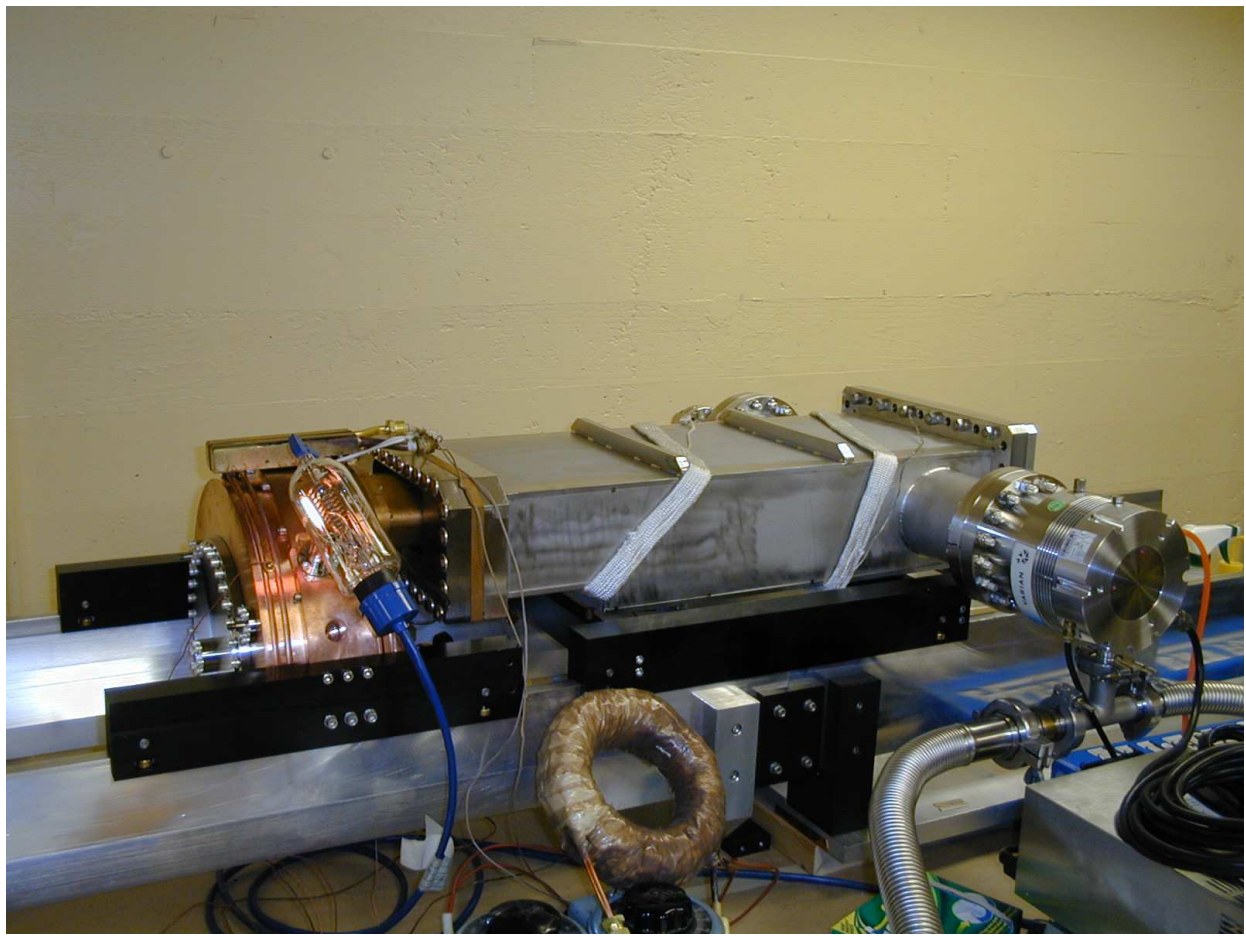}%

\caption{(Color) 805 MHz pillbox rf cavity used for testing. The cavity has removable 
windows to permit tests of different window materials, and a thin exit port to
permit dark current studies.}%
\label{fig:pillbox-805}%
\end{figure}

\paragraph{MUCOOL}
A primary effort has been to carry out high-power tests of 805-MHz rf cavities
in the Lab G test area at Fermilab. A 5-T test solenoid for the facility,
capable of operating either in solenoid mode (its two independent coils
powered in the same polarity) or gradient mode (with the two coils opposed),
was used to study the effects of magnetic field on cavity performance.
Most recently, a single-cell 805-MHz pillbox cavity (Fig.
\ref{fig:pillbox-805}) having Be foils to close the beam iris was tested.
This cavity permitted an assessment of the behavior of the foils under rf
heating and was used to study dark current effects \cite{DarkCurrentnote}. The
cavity reached 40 MV/m (exceeding its design specification) in the absence of
a magnetic field, but was limited by breakdown to less than 15 MV/m at high
magnetic field ($\thickapprox2$ T). Understanding the effects of the magnetic
field on cavity performance is crucial, as this is the environment required
for cavities in a muon cooling channel.%

\begin{figure}[ptbh!]
\includegraphics[width=4in]{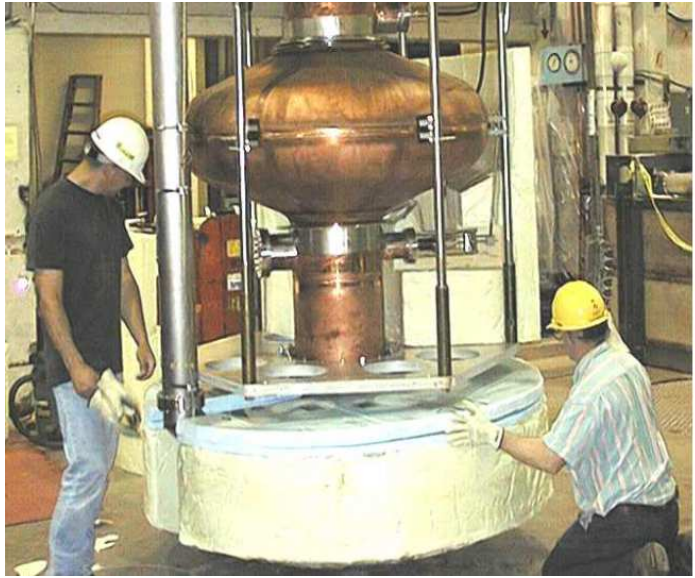}%

\caption{(Color) 201~MHz SCRF cavity being prepared for testing at Cornell.}%
\label{SCRFpict}%
\end{figure}

Development of a prototype LH$_{2}$ absorber, the material chosen for FS2 and
also for MICE~\cite{MICEref} (the Muon Ionization Cooling Experiment, see
Section~\ref{MICEtext}) is well along. Several large diameter, thin (125--350
$\mu$m) aluminum windows have been successfully fabricated by machining from
solid disks. These have been pressure tested with water and found to break at
a pressure consistent with finite-element design calculations
\cite{AbsorberRef}. Another absorber material that must be studied is LiH, the
material on which the cooling channel used in this report is based. In the new
scheme, the LiH serves both as an absorber and an rf window. This
configuration could be tested in the 805-MHz pillbox cavity described above.
A new area, the MUCOOL\ Test Area (MTA), is nearly completed at FNAL and will
be used for initial testing of the liquid-hydrogen absorbers. It will also
have access to both 805-MHz and 201-MHz high-power rf amplifiers for
continuing rf tests of the 805-MHz pillbox cavity and, soon, for testing a 
prototype 201-MHz cavity. The MTA is located at the end of the Fermilab proton
linac, and is designed to eventually permit beam tests of components and
detectors with 400 MeV protons.
\paragraph{Beam Simulations and Theory}
Subsequent to work on FS2, present effort has focused on further optimization
of Neutrino Factory performance and costs. The more cost effective front-end
design reported in this paper is a result of this work.
\paragraph{SCRF Development}
This work is aimed at development of a high-gradient 201-MHz SCRF cavity for
muon acceleration. (The choice of SCRF for a cooling channel is excluded
because of the surrounding high magnetic field; the acceleration system does
not suffer this limitation.) A test area of suitable dimensions was
constructed at Cornell (Fig.~\ref{SCRFpict}) and used to test a prototype
cavity fabricated for the Cornell group by CERN colleagues. The cavity reached
11~MV/m in initial tests, but exhibited a significant ``$Q$ slope'' as the
gradient increased~\cite{pac03:1309}. To better understand the origins of this phenomenon,
effort will shift to studies on a smaller 500~MHz cavity. Different coating
and cleaning techniques will be explored to learn how to mitigate the observed
$Q$ slope.%

\subsubsection{R\&D plans}
\begin{figure}[ptbh!]
\includegraphics[width=2in,angle=90]{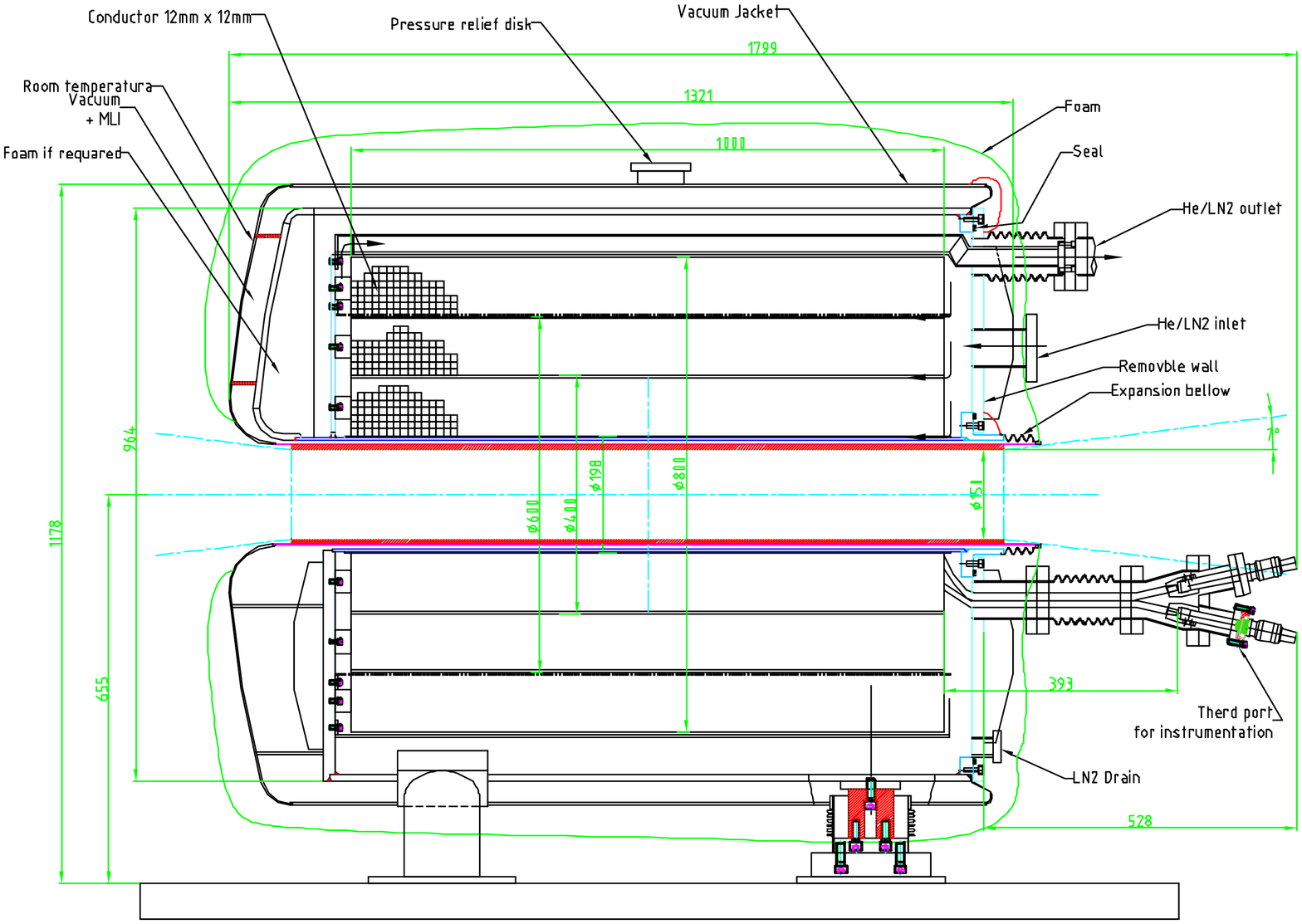}%
\includegraphics[width=2in]{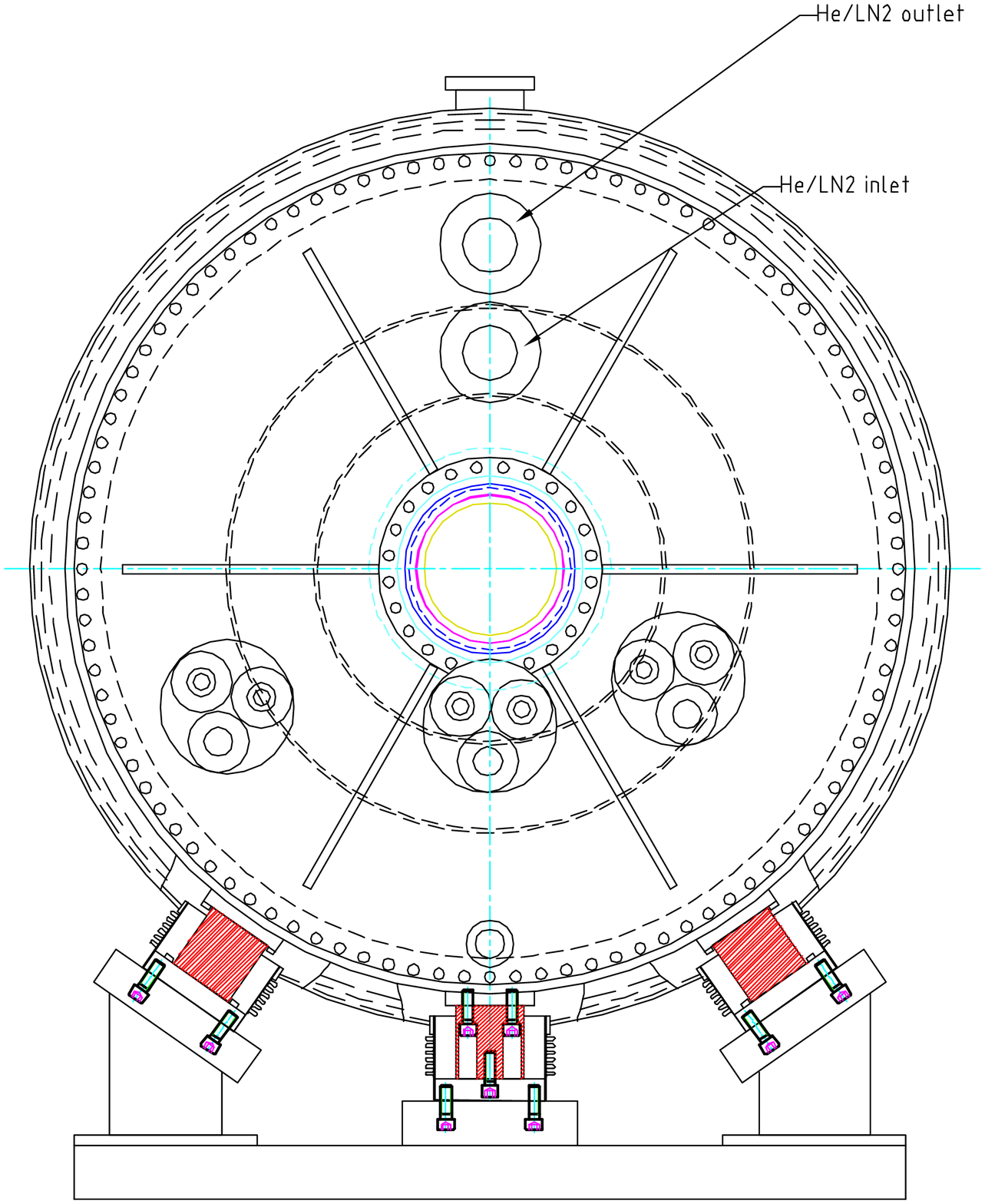}%

\caption{(Color) Design of targetry test magnet. The magnet has three nested coils
 that permit operation at 5, 10, and 15~T. The coils are normal conducting but
 cooled to liquid-nitrogen temperature to ease the requirements on the power
 supply.}%
\label{fig:TgtMag}%
\end{figure}

\paragraph{Targetry}
For the targetry experiment, design of a pulsed solenoid and its power supply
are under way. A cost-effective design capable of providing up to a 15~T field
has been developed (see Fig.~\ref{fig:TgtMag}).%
Tests of a higher velocity mercury jet (about 20~m/s velocity, compared with
about 2.5 m/s in the jet system initially tested), will be carried out. To
complement the experimental program, target simulation efforts are ongoing.
These aim at a sufficiently detailed understanding of the processes involved
to reproduce the observed experimental results both with and without a
magnetic field. Fully 3D magneto-hydrodynamics codes are being utilized for
this effort.
\paragraph{MUCOOL}
Further testing work for 805~MHz components will continue in the MTA. Work
will focus on understanding and mitigating dark current and breakdown effects
at high gradient. Many aspects of cavity design, such as cleaning and coating
techniques, will be investigated. In addition, tests of alternative designs
for window or grid electromagnetic terminations for the rf cavity will be
initially explored to identify the best candidates for the full-sized 201~MHz
prototype cavity. Fabrication of the 201 MHz cavity by a group from LBNL,
Jlab, and the University of Mississippi is nearly completed. This cavity will
also be tested in the MTA.
Thermal tests of a prototype absorber in the MTA are just getting under way.
Fabrication of other cooling channel components required for the initial phase
of testing will be carried out, including a large-bore superconducting
solenoid, and diagnostics that could be used for the experiment. With these
components, it will be possible eventually to assemble and bench test a full
prototype cell of a realistic cooling channel. Provision will be made to test
curved Be windows and grids in the 805~MHz cavity, followed by tests on the
201~MHz prototype.
As already noted, the site of the MTA was selected with the goal of permitting
beam tests of the cooling channel components with a high intensity beam of 400~MeV protons. While not the same as using an intense muon beam, such a test
would permit a much better understanding of how the cooling channel would
perform operationally, especially the high-gradient rf cavity and the LH$_{2}$
or LiH absorber.
\paragraph{Beam Simulations and Theory}
A major simulation effort will continue to focus on iterating the front-end
channel design to optimize it for cost and performance. Further effort will be
given to beam dynamics studies in the FFAG rings and storage ring, including
realistic errors. Work on optimizing the optics design will be done.
Assessment of field-error effects on the beam transport will be made to define
acceptance criteria for the magnets. This will require use of sophisticated
tracking codes, such as COSY~\cite{COSYref}, that permit rigorous treatment of
field errors and fringe-field effects.
In many ways, the storage ring is one of the most straightforward portions of
a Neutrino Factory complex. However, beam dynamics is an issue here as the
muon beam must circulate for many hundreds of turns. Use of a tracking code
such as COSY is required to assess fringe field and large aperture effects. As
with the FFAG rings, the relatively large emittance and large energy spread
enhance the sensitivity to magnetic field and magnet placement errors.
Suitable magnet designs are needed, with the main technical issue being the
relatively high radiation environment. Another lattice issue that must be
studied is polarization measurement. In the initial implementation of a
Neutrino Factory it is expected that polarization will not be considered, but
its residual value may nonetheless be important in analyzing the experiment.
Simulation efforts in support of MICE will continue. We also plan to
participate in a so-called ``World Design Study'' of an optimized Neutrino
Factory. This study, an international version of the two previous U.S.
Feasibility Studies, will likely be hosted in the UK by Rutherford Appleton
Laboratory (RAL), the site for the MICE experiment (see Section~\ref{MICEtext}%
). It will be organized jointly by representatives from Europe, Japan, and the U.S.
\paragraph{SCRF Development}
A prototype 500~MHz SCRF cavity will be used to study the $Q$ slope
phenomenon, with the goal of developing coating and cleaning techniques that
reduce or eliminate it. Detuning issues at 201~MHz associated with the very
large cavity dimensions and the pulsed rf system will be evaluated. Tests of
the 201 MHz SCRF cavity will include operation in the vicinity of a shielded
solenoid magnet, to demonstrate our ability to adequately reduce nearby
magnetic fields in a realistic lattice configuration.
If funds permit, design of a prototype high-power rf source will be explored,
in collaboration with industry. This source---presently envisioned to be a
multibeam klystron---must be developed for operation at two different duty
factors, because the cooling channel requires a duty factor of about 0.002
whereas the acceleration chain requires 0.045.
Magnet designs suitable for the FFAG rings and the muon storage ring will be
examined further. Both conventional and superconducting designs will be
compared where both are possible. With SC magnets, radiation heating becomes
an issue and must be assessed and dealt with.
\subsubsection{Cooling Demonstration Experiment}
\label{MICEtext}Clearly, one of the most important R\&D tasks that is needed
to validate the design of a Neutrino Factory is to measure the cooling effects
of the hardware we propose. Participation in the International Muon Ionization
Cooling Experiment (MICE) will accomplish this, and is therefore expected
eventually to grow into a primary activity. Unquestionably, the experience
gained from this experiment will be invaluable for the design of an actual
cooling channel.

\begin{figure}[ptbh!] 
\includegraphics*[bb=0 350 600 697]{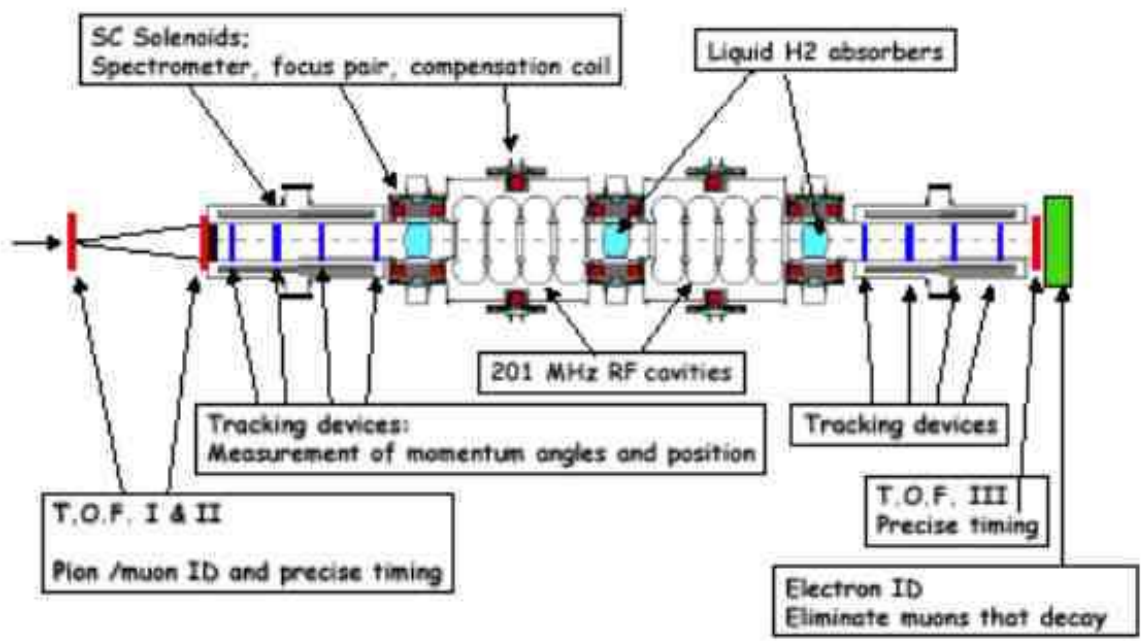}%

\caption{(Color) Schematic of the MICE layout.}%

\label{fig:MICElayout}%
\end{figure}

At the NUFACT'01 Workshop in Japan, a volunteer organization was created to
organize a cooling demonstration experiment that might begin as soon as 2004.
Membership in this group includes representatives from Europe, Japan, and the
U.S. The experimental collaboration now numbers some 140 members from the
three geographical regions. The MICE Collaboration has received scientific
approval for the experiment from RAL management, and is now in the process of
seeking funding.
The experiment will involve measuring, on a particle-by-particle basis, the
emittance reduction produced by a single cell of the FS2 cooling channel. A
schematic of the layout is shown in Fig.~\ref{fig:MICElayout}. The cooling
channel cell is preceded and followed by nearly identical detector modules
that accomplish particle identification and emittance measurement. Provision
for testing a series of absorber materials, including both LH$_{2}$ and solid
absorbers, has been made. A preliminary safety review of the liquid-hydrogen
system has been successfully passed, and permission to begin detailed
engineering has been granted.%
\subsection{Beta Beam R\&D}
Constructing a Beta Beam facility requires a number of new techniques to be
developed. In the CERN-based scenario, where the facility incorporates a number
of existing machines, a significant technical challenge is to ensure that the
proposed parameters for the Beta Beam case are compatible with the
capabilities of the present accelerators. If not, required modifications must
be identified and demonstrated. An additional constraint in the CERN
scenario---or a corresponding U.S. scenario based on existing machines---is
that the modifications to accommodate Beta Beams must maintain compatibility
with existing programs.
There are several significant challenges in providing Beta Beams of the
required intensity, and R\&D is required to validate the concepts proposed to
deal with them. These issues include:
\begin{itemize}
\item \textbf{Target:} To provide the required ion intensities, the target must be
able to handle the driver beam intensity for a reasonable lifetime. While the
production of $^{6}$He looks fairly straightforward, the production of
$^{18}$Ne is less so. The baseline production method for $^{18}$Ne relies on direct
bombardment of the target material with the proton beam. Determining what
intensity is acceptable to maintain a reasonable target lifetime must be done.
The alternative production technique suggested for $^{18}$Ne, namely the
$^{16}$O($^{3}$He,n) reaction, clearly requires a different incoming beam,
which might require an additional driver accelerator. In the scenario where
the outputs from three targets are combined, techniques for splitting the
driver beam into multiple paths and combining the target outputs into a single
ion source need to be worked out and tested. Since the ion source must
necessarily be remote from the highly active target area, good transport
efficiency for the nuclides of interest must be demonstrated.
\item \textbf{Ion Source:} With an ISOL-type production system, the ions are
produced continuously. To prepare the beam for acceleration, it must
then be bunched considerably, to below $20\,\mu$s. It is proposed to get the
required intensity and bunch structure by using a new ion source concept, with
state-of-the-art specifications. The source is an ECR source operating at high
frequency (60~GHz) with a high magnetic field (2--3~T) and high plasma density
$(n_{e}\thicksim 10^{14} \text{cm}^{-3}).$ Such a source has never been built,
though a development effort is now under way at Grenoble~\cite{sortais}. It
will be necessary to study the influence of the carrier gas on the
ionization efficiency. Also, the ability to produce fully stripped beams of
$^{18}$Ne must be demonstrated. Typically ECR sources produce high charge
states, but not so high as Ne$^{10+}.$ Even with an enhanced ion source of the type proposed, the expected~\cite{BouchezRef} antineutrino intensity is $2.1 \times 10^{18}$ per
``Snowmass year'' (1 $\times$\ 10$^{7}$ s), and that for neutrinos (assuming
that the outputs from three production targets are combined) is $3.5 \times 10^{17}.$
\item \textbf{Decay Losses:} The losses in the low-energy portion of the
accelerator chain are high because the beam is intense and its relativistic
$\gamma$ is low. In the CERN\ scenario, the PS is somewhat vulnerable as it
has already accumulated a high radiation dose over its operational lifetime
and its cycle time is not rapid. For $^{6}$He, the PS losses are estimated~\cite{LindroosRef} to be about 1.2~W/m while those in the storage ring are 28~W/m. In the latter case, specially designed superconducting magnets with no
coils in the midplane are required to avoid quenches. (A variant of this
approach is used for the muon storage ring of a Neutrino Factory, for the same
reason.) Building prototype magnets and measuring their field quality and
quench resistance should be undertaken to validate the proposed approach.
\item \textbf{Storage Ring Issues:} Since the storage ring must be frequently
topped off, injection requires the use of bunch-merging techniques. A concept
has been worked out for this, and an initial test was encouraging. Since many
of the problems with rf manipulation techniques are intensity dependent, it
will probably still be necessary to validate the proposed scheme under fully
realistic conditions. The recent proposal to run both $^{6}$He and $^{18}$Ne
beams simultaneously in the decay ring is another non-trivial complication. At
the same rigidity, the heavier beam has a relativistic $\gamma$ that is
$\frac{5}{3}$ that of the lighter beam, so the orbits must be adjusted to
provide the same revolution period. This may well have some beam dynamics
implications for the off-center beam, given that the proposed magnets may not
have ideal fields far off axis. It will certainly complicate beam
manipulations for the two beams, especially the injection and bunch merging.
Another issue that needs to be evaluated in detail is the influence of the beam
parameters (orbits, emittance, beta functions) on the neutrino spectrum at the detector.
\end{itemize}

\section{Summary \label{sec7}}
Two new types of facility have been proposed that could have a tremendous 
impact on future neutrino experiments---the Neutrino Factory and the 
Beta Beam facility. In contrast to conventional muon-neutrino beams, 
Neutrino Factory and Beta Beam facilities would provide a 
source of electron-neutrinos $(\nu_e)$ and -antineutrinos $(\bar{\nu}_e)$, 
with very low systematic uncertainties on the associated 
beam fluxes and spectra. The experimental signature for 
$\nu_e \to \nu_\mu$ transitions is extremely clean, with 
very low background rates. Hence, Neutrino Factories and Beta Beams 
would enable very sensitive oscillation measurements to be made. This 
is particularly true at a Neutrino Factory which not only provides 
very intense beams at high energy, but also provides muon-neutrinos $(\nu_{\mu})$ 
and -antineutrinos $(\bar{\nu}_{\mu})$ in 
addition to electron-neutrinos $(\nu_e)$ and -antineutrinos $(\bar{\nu}_e)$. This would facilitate 
a large variety of complementary oscillation measurements in a 
single detector, and dramatically improve our ability to test the 
three-flavor mixing framework, measure \textsl{CP} violation in the 
lepton sector (and perhaps determine the neutrino mass hierarchy), 
and, if necessary, probe extremely small values of the mixing angle $\theta_{13}$. 
 
At this time, we do not know the value of $\theta_{13}$. If $\sin^{2}%
2\theta_{13}<0.01$, much of the basic neutrino oscillation physics program 
will be beyond the reach of conventional neutrino beams. In this case 
Neutrino Factories
and Beta Beams offer the only known way to pursue the desired physics 
program.

The sensitivity that could be achieved at a Beta Beam facility presently looks
very promising, but is still being explored. In particular, the optimum Beta
Beam energy is under discussion. Low energy Beta Beam measurements would
complement Superbeam measurements, but would achieve a $\theta_{13}$ 
sensitivity that does
not appear to be competitive with that of a Neutrino Factory. Higher energy
Beta Beams may approach the sensitivity possible with a Neutrino Factory,
although systematics issues need further study. Thus, while a Beta Beam
facility may have a significant role to play in the future global neutrino
program,  more work must be done on its design, development, cost estimate,
and physics sensitivity to validate its potential. 
We note that, due to very limited resources, there has been no
significant activity in the U.S. on Beta Beams. Progress on Beta Beam development being made in Europe should be followed, 
especially if the higher energy solution continues to look favorable.

An impressive Neutrino Factory R\&D effort has been ongoing in the U.S. and
elsewhere over the last few years, and significant progress has been made
towards optimizing the design, developing and testing the required accelerator
components, and significantly reducing the cost, even during the current
Study. (Although a full engineering study is required, we have preliminary
indications that the unloaded cost of a Neutrino Factory facility based on an
existing Superbeam proton driver and target station can be reduced
substantially compared with previous estimates.) Neutrino Factory R\&D has
reached a critical stage in which support is required for two key
international experiments (MICE and Targetry) and a third-generation
international design study. If this support is forthcoming, a Neutrino Factory
could be added to the Neutrino Physics roadmap in about a decade. 

Given the present uncertainty about the size of $\theta_{13},$ \textit{it is critical to support an ongoing and increased U.S. investment in Neutrino Factory accelerator R\&D to maintain this technical option}. A Neutrino
Factory cannot be built without continued and increased
support for its development. We note that the 2001 HEPAP Report 
advocated an annual U.S. investment of \$8M on Neutrino Factory R\&D. The 
present support is much less than this. Since R\&D on the design of 
frontier accelerator facilities takes many years, 
support must be provided \textit{now} to have an
impact in about a decade.

\clearpage

\section{Recommendations \label{sec8}}

Accelerator R\&D is an essential part of the ongoing global neutrino program.
Limited beam intensity is already constraining the neutrino physics program,
and will continue to do so in the future. More intense and new types of
neutrino beams would have a big impact on the future neutrino program. A
Neutrino Factory would require a Superbeam-type MW-scale  proton source. We
thus encourage the rapid development of a Superbeam-type proton source. 

The Neutrino Factory and Beta Beam Working Group's specific recommendations are:
\begin{itemize}
\item \textbf{\textit{We recommend that the ongoing Neutrino Factory R\&D in the U.S.
be given continued encouragement and financial support.}} We note that the
HEPAP Report of 2001 recommended an annual support level of \$8M for Neutrino
Factory R\&D, and this level was considered minimal to keep the R\&D effort
viable.

In addition, and consistent with the above recommendation,
\begin{enumerate}
\item \textbf{\textit{We recommend that the U.S. funding agencies find a way to
support the international Muon Ionization Cooling Experiment (MICE), in
collaboration with European and Japanese partners.}} We note that MICE now has
scientific approval at the Rutherford Appleton Laboratory in the UK, and will
require significant U.S. participation. This has been identified as an
important experiment for the global Neutrino Factory R\&D program. A timely
indication of U.S. support for MICE is needed to move the experiment forward. 
\item \textbf{\textit{We recommend that support be found to ensure that the
international Targetry R\&D experiment proceeds as planned.}} We note that this
R\&D activity is crucial for the short-, medium-, and long-term neutrino
programs, and for other physics requiring high-intensity beams. 
\item \textbf{\textit{We recommend that a World Design Study, aimed at solidly
establishing the cost of a cost-effective Neutrino Factory, be supported at
the same level as Studies I and II.}} We note that the studies done here
suggest that the cost of a Neutrino Factory would be significantly less than
estimated for Studies I and II. This makes a Neutrino Factory a very
attractive ingredient in the global neutrino roadmap.
\end{enumerate} 
\item \textbf{\textit{We recommend that progress on Beta Beam development be
monitored, and that our U.S. colleagues cooperate fully with their EU
counterparts in assessing how U.S. facilities might play a role in such a
program.}} We note that there is no significant U.S. R\&D effort on Beta Beams
due to our limited R\&D resources. Insofar as an intermediate energy solution
is desirable, however, the Beta Beam idea is potentially of interest to the
U.S. physics community.
\end{itemize}

\begin{acknowledgments}
This research was supported by the U.S. Department of Energy under
Contracts {No. DE-AC02-98CH10886}, {No. DE-AC02-76CH03000}, and {No. DE-AC03-76SF00098}.
\end{acknowledgments}
\section{Appendix A \label{appendixa}}
\subsection{Cost Reduction}
Here we present the cost scaling we have done with respect to FS2 cost numbers~\cite{fs2}. Since there was neither time nor
engineering effort available to perform a bottom-up cost estimate for the new
systems we have developed during the present Study, we have based our costs on
the FS2 numbers and scaled them appropriately to derive the estimated savings
from our new technical approaches. For that reason, we quote the results as a
percentage of the original FS2 estimates, to avoid giving the impression that
this is anything more than a ``physicist's estimate'' at this point in time.

The method we employed was as follows:
\begin{itemize}
\item  Starting from the FS2 Work Breakdown Structure system costs, we derive
useful element costs per unit length, per integral rf voltage, or per unit acceleration.
\item  We then applied these scaling rules to the new parameters derived from
this Study (see Section~\ref{sec5}) to obtain a first approximation to the revised cost. Our
results are reported as costs relative to FS2. We have ignored minor
corrections, such as escalating the costs to FY2004 dollars, as these are small
compared with the precision of our estimate.
\item  Because it is expected that a future Neutrino Factory would be built as
an ``upgrade'' or follow-on to a Superbeam facility, we think it likely that
the Proton Driver---and quite possibly the Target facility as well---will
already exist at the time the Neutrino Factory construction commences. For
this reason, our costs are given with and without including the Proton Driver
or Target Station.
\end{itemize}

Based on the costing approach we used, we expect that the unloaded hardware
cost of the updated Neutrino Factory design will be reduced by about one-third
compared with the original FS2 cost estimate of \$1.8B (see Table~\ref{tab:FS2costs}).
\begin{table}[bhtp!]
\caption{Original (unloaded) costs from
  FS2.\label{tab:FS2costs}}
\begin{ruledtabular}
\begin{tabular*}{5in}[c]{lccc}
& \textbf{All} & \textbf{No Driver} & \textbf{No Driver, No Target}\\
& (\$M) & (\$M) & (\$M)\\\hline
TOTAL\footnote{No ``other'', no EDIA, no contingency.} & \textbf{1832} & \textbf{1641} & \textbf{1538}\\
\end{tabular*}%
\end{ruledtabular}
\end{table}%
\subsubsection{Proton Driver}
The cost basis for FS2 was an upgrade of the AGS to 1~MW beam power. The cost
used here is taken without change from FS2. As noted earlier, we anticipate
that this component would already be in place to support a prior Superbeam
experiment. In that case, it would not be part of the cost to construct a
Neutrino Factory. Since the Proton Driver tends to be the most site-specific
component of a Neutrino Factory, we expect the remaining costs to be largely
site independent.
\subsubsection{Target and Capture}
The updated Target and Capture system is almost the same as that in FS2, but
differs in the details. In particular, the region over which the field tapers
down is shorter by 5.5~m because it tapers only to 1.75~T rather than the
1.25~T used in FS2. The cost will thus be somewhat less. We estimated this
savings by subtracting the cost of 5.5~m of a 1.25~T transport channel, whose
cost per meter was taken from the drift region in FS2. This is a conservative
estimate, because the section eliminated had fields varying from 1.75~T to
1.25~T, whereas the savings are estimated assuming lower field transport at 1.25~T.
\subsubsection{Drift Region}
The first 18~m of drift is more expensive than later beam transport sections
because of the required radiation shielding. We therefore treat this first
18~m of drift separately from the subsequent transport. To evaluate the cost,
we took the FS2 costs for the first 18~m, and then made a correction due to
the higher solenoid field in the new channel compared with FS2 (1.75~T vs.
1.25~T). Specifically, the correction involved increasing the magnet, power
supply, and cryogenic costs using the second scaling formula from Green
\textit{et al.}~\cite{MAGref},
\begin{equation}
\text{Cost (in \$M)}\propto (BR^2L)^{0.577}.
\label{green2}
\end{equation} 

The subsequent 82~m drift requires less shielding and will thus be less
expensive. In FS2, there was no equivalent simple drift from which to scale
this cost. Therefore, we estimated the costs based on the magnets, power
supplies, and cryogenics included in the induction linac region of FS2. As
these costs were for 1.25~T magnets, we corrected for the higher 1.75~T field
using Eq.~\ref{green2}.
This estimate is quite conservative, because the transport magnets in the
induction linacs of FS2, which were introduced inside the induction linac
cores, had to meet more difficult requirements, and had more complicated cryostats.
\subsubsection{Buncher and Phase Rotation}
As discussed in Section~\ref{sec5}, the Buncher and Phase Rotation section adopted here
is quite different from the induction-linac-based system used in FS2. The
focusing now consists of an essentially continuous solenoid at 1.75~T, as in
the drift, but with a radius (65~cm) sufficient for it to be located outside
of the rf cavities. To estimate the cost of this solenoid, we again use the
FS2 induction linac transport magnets, scaled to the appropriate parameters
via Eq.~\ref{green2} (now correcting for both the higher field and the larger radius). This
estimate is again conservative, because it is scaled from the more difficult
transport solenoids inside the induction linacs of FS2.

As described in Section~\ref{sec5}, in place of induction linacs, the present study
uses a sequence of rf cavities at frequencies in the range of 200--300~MHz.
The cost of these cavities, and their required rf power supplies, are scaled
from the FS2 costs of cooling channel rf cavities. These costs are scaled for
the different average accelerating gradients as follows: cavity cost per GeV
$\propto\frac{1}{V}$, power supply cost per GeV proportional to $V$.
\subsubsection{ Cooling Section}
The rf system for the cooling channel used in the present study is essentially
identical to that in FS2, so the costs per GeV are taken to be the same. The
focusing lattice, however, is quite different---a simple alternating solenoid
array (FOFO) instead of the more complicated, and tapered, super-FOFO lattices
in FS2. We estimate the new magnet system cost by scaling from Lattice-1 of
the FS2 cooling channel, using the first scaling formula in
Ref.~\cite{MAGref}, which depends on the total stored energy $U$ (cost $\propto U^{0.662}$). The stored energy per unit length in the present study and the
earlier FS2 lattice are in the ratio of 189:382, so the new cost per meter is
taken to scale as $(\frac{189}{382})^{0.662}.$ The cryogenic system cost is also
scaled with the magnet costs, but based on the cryogenic costs of the FS2
phase rotation section. (We do not use cryogenic costs from the FS2 cooling
channel, as these are heavily biased by the cooling requirements of the
LH$_{2}$ absorbers.)

This is a quite conservative estimate, because the new lattice not only has a
smaller stored energy, it is simpler. In particular, the channel adopted in
this study uses only a single type of solenoid and, when powered, there are no
inter-coil forces. In contrast, the FS2 lattice employed two types of solenoid
magnets and had very large inter-coil forces between the ``focus'' coil pair.

No cost was included for the LiH absorbers in the present estimate, as we do
not yet have a good basis for one. Our expectation is that these rf windows
will cost no more than the Be windows they replace, but this is presently unverified.
\subsubsection{Match to Pre-Accelerator}
A section is required to match the (momentum-dependent) beta function in the
cooling channel to that in the pre-accelerator linac. In the FS2 case, beta
vs. momentum in the cooling lattice was highly non-linear, with low betas at
the upper and lower momentum limits and a maximum beta in the center, whereas
the beta functions in the pre-accelerator were approximately linear in
momentum. As failure to match the two would have resulted in significant
emittance growth and particle loss, we included a matching section using 18~m
of a modified 1.65~m cell (Lattice-2) cooling lattice. The optics of the first
two-thirds of the matching section was adjusted to adiabatically change the
beta vs. momentum shape, and raise the central beta from 20~cm to about 60~cm.
The final one-third of the matching section increased the beta function to
about 3~m to match the pre-accelerator optics.

In the present case, the match will be simpler and less expensive because
\textit{a)} the beta functions both before and after the match have similar
linear momentum dependence, \textit{b)} the match requires a smaller change in
beta function than was needed in FS2, and \textit{c)} the lattice on which it
will be based has a considerably lower stored energy per unit length
(189/1039). As a new matching section has not yet been designed, we correct
the cost only due to \textit{c)} by scaling the cost by this factor using the
first formula in Ref.~\cite{MAGref}. This too is a conservative approach, as
it is expected that the length of the new matching section will be much less
than the original one.
\subsubsection{Pre-Acceleration}
For the rf system and cryogenics, the Pre-Acceleration cost is scaled from
that in FS2 by the energy gain from the rf cavities. For magnets and vacuum,
we scaled with length.
\subsubsection{RLA}
The present study makes use of a dogbone RLA to accelerate from 1.5--5~GeV.
The RLA cost is scaled from the 2.5--20~GeV RLA in FS2. The number of passes
is 3.5, compared with 4 for FS2. Although we favor a dogbone geometry for ease
of the switchyard design, as opposed to the FS2 racetrack layout, costs per
unit length, or per unit energy gain, are expected to be very similar. We took
these to be the same. Similarly, the arcs are assumed to have the same average
bending field as the final FS2 arc, so the cost per unit length is taken to be
the same. The lengths of the arcs were chosen to provide an absolute bend
angle of $420^{\circ}$ at each end of the linac. Magnet costs for the special and transport magnets
were scaled with the final RLA energies.
\subsubsection{FFAG}
FFAG costs for all technical and conventional systems are taken from a cost
algorithm based on similar scaling arguments to those used above for the other
beam line sections. The algorithm we used, when applied to the FS2 RLA as a
``reality check'', gave a higher cost than determined in FS2. It thus appears
to be conservative in its cost estimation.

Injection and extraction kickers are assumed to be driven by typical
induction-linac pulsed power sources, and will contain similar amounts of
magnetic materials. Our estimated costs were based on a length of the FS2
induction linac having the same pulsed energy as required for the kickers.

Transfer line lengths are taken from Ref.~\cite{InjExtRef} and include lines
for both $\mu^{+}$ and $\mu^{-}$. The cost per meter of these transport lines
is based on RLA arcs (magnets, power supplies, and vacuum) from FS2.
\subsubsection{Storage Ring}
Storage ring costs are taken, without modification, from FS2. However, in that
case, there was a site-dependent constraint that no part of the downward
tilted ring should fall below the nearby water table. This constraint forced
the design to assume construction of the ring in an artificial hill, and also
to require unusually high (hence not cost optimized) bending fields to keep
the ring small. The cost at another site, without this constraint, would
likely be less.
\subsubsection{Overall Relative Costs}
The result of applying the scaling rules outlined in this Appendix is
summarized in Table~\ref{tab:FS2Acosts}. As can be seen, the present
design exercise, completed as part of the APS Neutrino Physics Study, has
maintained the original performance of the Neutrino Factory designed in FS2
for either muon sign, yielding either neutrinos or antineutrinos. Unlike
FS2, however, the present design will supply both $\mu^{+}$ and $\mu^{-}$
simultaneously (interleaved in the bunch train), thus effectively doubling
the performance compared with FS2. This has been accomplished while
reducing the cost of the facility by about 1/3.

While the present scaling estimate is not a
replacement for a detailed engineering cost estimate, we are confident that
the majority of the cost reductions identified here will survive a more
rigorous treatment.

We note that the design progress made in this Study is a direct result of the
funding made available to the \textit{Neutrino Factory and Muon Collider
Collaboration} for Neutrino Factory accelerator R\&D. Optimizing and refining
the design of state-of-the-art facilities such as this, as well as verifying
that component specifications can be met and that component costs are
realistic, is critical to allowing the high-energy physics community to make
sound technical choices in the future.%
\begin{table}[htbp!]
\caption{Scaled (unloaded) costs from the present
study, quoted as percentages of costs determined for FS2.\label{tab:FS2Acosts}}
\begin{ruledtabular}
\begin{tabular*}{5in}[c]{lccc}
& \textbf{All} & \textbf{No Driver} & \textbf{No Driver, No Target}\\
\hline
TOTAL\footnote{Percentages of the original FS2 costs summarized in Table~\ref{tab:FS2costs}.} (\%) & \textbf{67} & \textbf{63} & \textbf{60}\\
\end{tabular*}%
\end{ruledtabular}
\end{table}%

\subsubsection{Possible Further Savings}

\begin{itemize}
\item Earlier studies indicated small performance loss if the capture
  solenoid is reduced from 20 to 17--18~T. If we find that this is still the case, the field specified would be reduced and some savings made. 
\item Reducing the cooling channel length to 50~m would lower the cost of the channel 
  while reducing the performance by only about 15\%. With better
  optimization, some or all of this loss may be recoverable.
\item The expected shorter match from the Cooling Section to the
  Pre-Acceleration Section should reduce the cost somewhat.
\item Increasing the number of turns in the RLA, and lowering its injection
  energy, should reduce the costs of the early acceleration portion of the
  Neutrino Factory.
\item A lower field, larger storage ring should result in some savings.
\end{itemize}

The above list suggests that, while the collaboration's efforts have been effective in
reducing the costs of the major items (see Table~\ref{tab:FS2Acosts}), options still exist to reduce
the costs of the lesser items as well. Thus, we are hopeful that some further cost reductions are achievable.

\bibliography{NF-BB-WG}

\begin{thebibliography}{92}
\expandafter\ifx\csname natexlab\endcsname\relax\def\natexlab#1{#1}\fi
\expandafter\ifx\csname bibnamefont\endcsname\relax
  \def\bibnamefont#1{#1}\fi
\expandafter\ifx\csname bibfnamefont\endcsname\relax
  \def\bibfnamefont#1{#1}\fi
\expandafter\ifx\csname citenamefont\endcsname\relax
  \def\citenamefont#1{#1}\fi
\expandafter\ifx\csname url\endcsname\relax
  \def\url#1{\texttt{#1}}\fi
\expandafter\ifx\csname urlprefix\endcsname\relax\def\urlprefix{URL }\fi
\providecommand{\bibinfo}[2]{#2}
\providecommand{\eprint}[2][]{\url{#2}}

\bibitem[{aps(2004)}]{aps-study}
\emph{\bibinfo{title}{{APS Multi-Divisional Study of the Physics of
  Neutrinos}}}, \bibinfo{howpublished}{\url{http://www.aps.org/neutrino/},
  sponsored by the American Physical Society Divisions of: Nuclear~Physics,
  Particles and Fields, Astrophysics, Physics of Beams} (\bibinfo{year}{2004}).

\bibitem[{MC()}]{MC}
\bibinfo{note}{{The Neutrino Factory and Muon Collider Collaboration WEB page,
  \url{http://www.cap.bnl.gov/mumu/}}}.

\bibitem[{\citenamefont{{N.~Autin, ed.}}(2000)}]{nufact99}
\bibinfo{author}{\bibnamefont{{N.~Autin, ed.}}}, \bibinfo{journal}{{Nucl.
  Instrum. \& Meth.}} \textbf{\bibinfo{volume}{{A451}}} (\bibinfo{year}{2000}),
  \bibinfo{note}{{Proceedings of the ICFA/ECFA Workshop NUFACT'99: Neutrino
  Factories based on Muon Storage Rings, Lyon France}}.

\bibitem[{\citenamefont{{S.~Chattopadhyay, ed.}}(2001)}]{nufact00}
\bibinfo{author}{\bibnamefont{{S.~Chattopadhyay, ed.}}},
  \bibinfo{journal}{{Nucl. Instrum. \& Meth.}}
  \textbf{\bibinfo{volume}{{A472}}} (\bibinfo{year}{2001}),
  \bibinfo{note}{{Proceedings of the International Workshop NuFact'00: Muon
  Storage Ring for a Neutrino Factory, Monterey CA USA}}.

\bibitem[{\citenamefont{{S.~Machida and K.~Yoshimura, eds.}}(2003)}]{nufact01}
\bibinfo{author}{\bibnamefont{{S.~Machida and K.~Yoshimura, eds.}}},
  \bibinfo{journal}{{Nucl. Instrum. \& Meth.}}
  \textbf{\bibinfo{volume}{{A503}}} (\bibinfo{year}{2003}),
  \bibinfo{note}{{Proceedings of the 3rd International Workshop on Neutrino
  Factories based on Muon Storage Rings: NuFact'01, Tsukuba Japan}}.

\bibitem[{\citenamefont{{K.~Long and R.~Edgecock, eds.}}(2003)}]{nufact02}
\bibinfo{author}{\bibnamefont{{K.~Long and R.~Edgecock, eds.}}},
  \bibinfo{journal}{J. Phys. G: Nucl. Part. Phys.}
  \textbf{\bibinfo{volume}{29}} (\bibinfo{year}{2003}),
  \bibinfo{note}{{NuFact02--The 4th International Workshop on Neutrino
  Factories, London UK}}.

\bibitem[{nuf()}]{nufact03}
\bibinfo{howpublished}{to be published by AIP}, \bibinfo{note}{{NuFact03--The
  5th International Workshop on Neutrino Factories, Columbia University, New
  York, USA}}.

\bibitem[{\citenamefont{{N.~Holtkamp and D.~Finley, eds.}}(2000)}]{fs1}
\bibinfo{author}{\bibnamefont{{N.~Holtkamp and D.~Finley, eds.}}},
  \bibinfo{type}{Tech. Rep.} \bibinfo{number}{Fermilab-Pub-00/108-E},
  \bibinfo{institution}{{Fermilab}} (\bibinfo{year}{2000}),
  \bibinfo{note}{{\url{http://www.fnal.gov/projects/muon_collider/nu-factory/n%
u-factory.html}}}.

\bibitem[{\citenamefont{{S.~Ozaki, R.~Palmer, M.~Zisman, and J.~Gallardo,
  eds.}}(2001)}]{fs2}
\bibinfo{author}{\bibnamefont{{S.~Ozaki, R.~Palmer, M.~Zisman, and J.~Gallardo,
  eds.}}}, \bibinfo{type}{Tech. Rep.}, \bibinfo{institution}{{BNL-52623}}
  (\bibinfo{year}{2001}),
  \bibinfo{note}{{\url{http://www.cap.bnl.gov/mumu/studyii/FS2-report.html}}}.

\bibitem[{\citenamefont{Geer}(1998)}]{geer98}
\bibinfo{author}{\bibfnamefont{S.}~\bibnamefont{Geer}}, \bibinfo{journal}{Phys.
  Rev.} \textbf{\bibinfo{volume}{{D57}}}, \bibinfo{pages}{6989}
  (\bibinfo{year}{1998}), \bibinfo{note}{{\textit{ibid.} \textbf{59}, 039903E
  (1999)}}.

\bibitem[{\citenamefont{{M.~M.~Alsharo{\'{}}a~\textit{et
  al.}}}(2003)}]{status_report}
\bibinfo{author}{\bibnamefont{{M.~M.~Alsharo{\'{}}a~\textit{et al.}}}},
  \bibinfo{journal}{Phys. Rev. ST Accel. Beams} \textbf{\bibinfo{volume}{6}},
  \bibinfo{pages}{081001} (\bibinfo{year}{2003}).

\bibitem[{\citenamefont{{A.~Blondel \textit{et al.}}}()}]{blondel}
\bibinfo{author}{\bibnamefont{{A.~Blondel \textit{et al.}}}},
  \bibinfo{howpublished}{{CERN 2004-002 ECFA/CERN}}.

\bibitem[{\citenamefont{Zucchelli}(2002)}]{zucchelli}
\bibinfo{author}{\bibfnamefont{P.}~\bibnamefont{Zucchelli}},
  \bibinfo{journal}{Phys.Lett.} \textbf{\bibinfo{volume}{{B532}}},
  \bibinfo{pages}{166} (\bibinfo{year}{2002}).

\bibitem[{HEP(2000)}]{HEPAP}
\emph{\bibinfo{title}{{HEPAP White Paper: Planning for U.S. High-Energy
  Physics}}},
  \bibinfo{howpublished}{{\url{http://www.science.doe.gov/hep/hepap_reports.sh%
tm}}} (\bibinfo{year}{2000}), \bibinfo{note}{{DOE/SC-0027}}.

\bibitem[{UK()}]{UK}
\bibinfo{note}{{\url{http://hepunx.rl.ac.uk/neutrino-factory/}}}.

\bibitem[{CER()}]{CERN}
\bibinfo{note}{{\url{http://muonstoragerings.web.cern.ch/muonstoragerings/}}}.

\bibitem[{JAP()}]{JAPAN}
\bibinfo{note}{{\url{http://www-prism.kek.jp/nufactj/index.html}}}.

\bibitem[{\citenamefont{{B.~Autin \textit{et al.}}}(2003)}]{autin}
\bibinfo{author}{\bibnamefont{{B.~Autin \textit{et al.}}}},
  \bibinfo{journal}{{J. Phys.}} \textbf{\bibinfo{volume}{G29}},
  \bibinfo{pages}{1785} (\bibinfo{year}{2003}), \bibinfo{note}{{see also,
  \url{http://beta-beam.web.cern.ch/beta-beam/}}}.

\bibitem[{fna(2000)}]{fnal1}
\emph{\bibinfo{title}{{The Proton Driver Design Study}}},
  \bibinfo{howpublished}{{FERMILAB-TM-2136}} (\bibinfo{year}{2000}).

\bibitem[{fna(2002)}]{fnal2}
\emph{\bibinfo{title}{{Proton Driver Study II}}},
  \bibinfo{howpublished}{{FERMILAB-TM-2169 (Parts I and II)}}
  (\bibinfo{year}{2002}).

\bibitem[{\citenamefont{Mokhov}(2001)}]{mars1}
\bibinfo{author}{\bibfnamefont{N.}~\bibnamefont{Mokhov}}, in
  \emph{\bibinfo{booktitle}{{Proceedings of the 2001 Particle Accelerator
  Conference}}} (\bibinfo{year}{2001}), p. \bibinfo{pages}{745},
  \bibinfo{note}{see also \url{http://www-ap.fnal.gov/MARS/}}.

\bibitem[{\citenamefont{{J.R.~Miller \textit{et al.}}}(1994)}]{ITERmag}
\bibinfo{author}{\bibnamefont{{J.R.~Miller \textit{et al.}}}},
  \bibinfo{journal}{{IEEE Trans. Magn.}}
  \textbf{\bibinfo{volume}{{\textbf{30}}}}, \bibinfo{pages}{1563}
  (\bibinfo{year}{1994}).

\bibitem[{\citenamefont{{N.~Mokhov}}()}]{mars2}
\bibinfo{author}{\bibnamefont{{N.~Mokhov}}},
  \bibinfo{howpublished}{{\url{http://www-ap.fnal.gov/MARS/}}},
  \bibinfo{note}{{nucl-th/9812038}}.

\bibitem[{Bet()}]{BetaBeamWGpage}
\bibinfo{note}{{Home page of CERN Beta Beam:
  \url{http://beta-beam.web.cern.ch/beta-beam/}}}.

\bibitem[{\citenamefont{{H.L.~Ravn \textit{et al.}}}(2003)}]{EURISOLref}
\bibinfo{author}{\bibnamefont{{H.L.~Ravn \textit{et al.}}}},
  \emph{\bibinfo{title}{{Feasibility Study for A European Isotope-Separation
  On-Line Radioactive Ion Beam Facility: Appendix C}}},
  \bibinfo{howpublished}{{\url{http://www.ganil.fr/eurisol/Final_Report/APPEND%
IX-C.pdf}}} (\bibinfo{year}{2003}).

\bibitem[{\citenamefont{{H.L.~Ravn}}(2003)}]{helge}
\bibinfo{author}{\bibnamefont{{H.L.~Ravn}}}, \bibinfo{journal}{{Nucl. Instrum.
  \& Meth.}} \textbf{\bibinfo{volume}{B204}}, \bibinfo{pages}{197}
  (\bibinfo{year}{2003}).

\bibitem[{\citenamefont{{U.~K\"{o}ster \textit{et al.}}}(2003)}]{koster}
\bibinfo{author}{\bibnamefont{{U.~K\"{o}ster \textit{et al.}}}},
  \bibinfo{journal}{{Nucl. Instrum. \& Meth.}} \textbf{\bibinfo{volume}{B204}},
  \bibinfo{pages}{301} (\bibinfo{year}{2003}).

\bibitem[{\citenamefont{{H.L.~Ravn \textit{et al.}}}(1997)}]{ravn2}
\bibinfo{author}{\bibnamefont{{H.L.~Ravn \textit{et al.}}}},
  \bibinfo{journal}{{Nucl. Instrum. \& Meth.}} \textbf{\bibinfo{volume}{B126}},
  \bibinfo{pages}{176} (\bibinfo{year}{1997}).

\bibitem[{\citenamefont{{E.~M\'{e}tral, \textit{et al.}}}()}]{SPLref}
\bibinfo{author}{\bibnamefont{{E.~M\'{e}tral, \textit{et al.}}}},
  \bibinfo{howpublished}{{CERN-APB-2004-021-ABP}},
  \bibinfo{note}{{\url{http://doc.cern.ch/archive/electronic/cern/preprints/ab%
/ab-2004-021.pdf.}}}

\bibitem[{sor()}]{sortais}
\bibinfo{note}{{\url{http://moriond.in2p3.fr/radio/Moriond-Sortais_1.ppt}}}.

\bibitem[{Bun()}]{BunchMergeRef}
\bibinfo{note}{{\url{https://web3.cern.ch/beta-beam/References/ab-note-2003-08%
0.pdf.}}}

\bibitem[{\citenamefont{{M.~Mezzetto}}(2003)}]{mezzetto}
\bibinfo{author}{\bibnamefont{{M.~Mezzetto}}}, \bibinfo{journal}{{J. Phys.}}
  \textbf{\bibinfo{volume}{{G29}}}, \bibinfo{pages}{{1771}}
  (\bibinfo{year}{2003}).

\bibitem[{\citenamefont{Gaisser}(1990)}]{gaisser}
\bibinfo{author}{\bibfnamefont{T.}~\bibnamefont{Gaisser}},
  \emph{\bibinfo{title}{Cosmic Rays and Particle Physics}}
  (\bibinfo{publisher}{Cambridge University Press}, \bibinfo{year}{1990}).

\bibitem[{\citenamefont{{D.~MacFarlane \textit{et al.}
  (CCFR)}}(1984)}]{CCFRsigma}
\bibinfo{author}{\bibnamefont{{D.~MacFarlane \textit{et al.} (CCFR)}}},
  \bibinfo{journal}{Z. Phys.} \textbf{\bibinfo{volume}{{C26}}},
  \bibinfo{pages}{1} (\bibinfo{year}{1984}), \bibinfo{note}{{J.P.~Berge
  \textit{et al.} (CDHSW), Z. Phys. \textbf{C35}, 443 (1987), J.V.~Allaby
  \textit{et al.} (CHARM), Z. Phys. \textbf{C38}, 403 (1988), P.~Auchincloss
  \textit{et al.} (E701), Z. Phys. \textbf{C48}, 411 (1990), world average from
  J.~Conrad, M.~Shaevitz and T.~Bolton, Rev. Mod. Phys. \textbf{70}, 1341
  (1998)}}.

\bibitem[{num()}]{numi}
\emph{\bibinfo{title}{{MINOS Technical Design Report}}},
  \bibinfo{howpublished}{{\url{http://www.hep.anl.gov/ndk/hypertext/minos_tdr.%
html} }}, \bibinfo{note}{nuMI-L-337 TDR}.

\bibitem[{\citenamefont{{V.~Barger, S.~Geer and K.~Whisnant}}(2000)}]{bgw99}
\bibinfo{author}{\bibnamefont{{V.~Barger, S.~Geer and K.~Whisnant}}},
  \bibinfo{journal}{Phys. Rev.} \textbf{\bibinfo{volume}{{D61}}},
  \bibinfo{pages}{053004} (\bibinfo{year}{2000}),
  \bibinfo{note}{{\eprint{hep-ph/9906487}}}.

\bibitem[{\citenamefont{{C.~Crisan and S.~Geer}}(2000)}]{cg00}
\bibinfo{author}{\bibnamefont{{C.~Crisan and S.~Geer}}}, \bibinfo{type}{Tech.
  Rep.}, \bibinfo{institution}{FERMILAB-TM-2101} (\bibinfo{year}{2000}).

\bibitem[{\citenamefont{{J.~Burguet-Castell, D.~Casper, J.J.~Gomez-Cadenas,
  P.~Hernandez, F.~Sanchez}}(2004)}]{jj-beta}
\bibinfo{author}{\bibnamefont{{J.~Burguet-Castell, D.~Casper,
  J.J.~Gomez-Cadenas, P.~Hernandez, F.~Sanchez}}}, \bibinfo{journal}{Nucl.
  Phys.} \textbf{\bibinfo{volume}{{B695}}}, \bibinfo{pages}{217}
  (\bibinfo{year}{2004}), \bibinfo{note}{{\eprint{hep-ph 0312068}. All papers
  from the \textit{eprint} archive can be obtained from
  \url{http://arXiv.org/}}}.

\bibitem[{\citenamefont{{C.~Albright, \textit{et al.}}}(2000)}]{fn-692}
\bibinfo{author}{\bibnamefont{{C.~Albright, \textit{et al.}}}},
  \bibinfo{type}{Tech. Rep.}, \bibinfo{institution}{Fermilab}
  (\bibinfo{year}{2000}), \bibinfo{note}{{report to the Fermilab Director;
  Fermilab-FN-692, May 10; \eprint{hep-ex/0008064}}}.

\bibitem[{\citenamefont{{A.~De Rujula, M.~B.~Gavela and
  P.~Hernandez}}(1999)}]{DeRujula:1998hd}
\bibinfo{author}{\bibnamefont{{A.~De Rujula, M.~B.~Gavela and P.~Hernandez}}},
  \bibinfo{journal}{{Nucl. Phys.}} \textbf{\bibinfo{volume}{{B547}}},
  \bibinfo{pages}{{21}} (\bibinfo{year}{1999}),
  \bibinfo{note}{{\eprint{hep-ph/9811390}}}.

\bibitem[{\citenamefont{{A.~Cervera, A.~Donini, M.~B.~Gavela, J.~J.~Gomez
  Cadenas, P.~Hernandez, O.~Mena and S.~Rigolin}}(2000)}]{Cervera:2000kp}
\bibinfo{author}{\bibnamefont{{A.~Cervera, A.~Donini, M.~B.~Gavela, J.~J.~Gomez
  Cadenas, P.~Hernandez, O.~Mena and S.~Rigolin}}}, \bibinfo{journal}{{Nucl.
  Phys.}} \textbf{\bibinfo{volume}{{B579}}}, \bibinfo{pages}{17}
  (\bibinfo{year}{2000}), \bibinfo{note}{{\eprint{hep-ph/0002108}. Erratum:
  ibid.\ {\bf B593}, 731 (2001)}}.

\bibitem[{\citenamefont{{M.~Apollonio \textit{et al. }(CERN working group on
  oscillation physics at the Neutrino Factory)}}(2002)}]{Apollonio:2002en}
\bibinfo{author}{\bibnamefont{{M.~Apollonio \textit{et al. }(CERN working group
  on oscillation physics at the Neutrino Factory)}}},
  \emph{\bibinfo{title}{{Oscillation physics with a neutrino factory}}}
  (\bibinfo{year}{2002}), \bibinfo{note}{{\eprint{hep-ph/0210192}}}.

\bibitem[{\citenamefont{Geer}(2002)}]{comments}
\bibinfo{author}{\bibfnamefont{S.}~\bibnamefont{Geer}},
  \bibinfo{journal}{Comments Nucl. Part. Phys.}
  \textbf{\bibinfo{volume}{{A2}}}, \bibinfo{pages}{284} (\bibinfo{year}{2002}),
  \bibinfo{note}{{\eprint{hep-ph/0008155}}}.

\bibitem[{\citenamefont{{V.~Barger, S.~Geer, R.~Raja, and
  K.~Whisnant}}(2000{\natexlab{a}})}]{entry-level}
\bibinfo{author}{\bibnamefont{{V.~Barger, S.~Geer, R.~Raja, and K.~Whisnant}}},
  \bibinfo{journal}{Phys. Rev.} \textbf{\bibinfo{volume}{{D63}}},
  \bibinfo{pages}{{033002}} (\bibinfo{year}{2000}{\natexlab{a}}).

\bibitem[{\citenamefont{{V.~Barger, S.~Geer, R.~Raja, and
  K.~Whisnant}}(2000{\natexlab{b}})}]{bgrw99}
\bibinfo{author}{\bibnamefont{{V.~Barger, S.~Geer, R.~Raja, and K.~Whisnant}}},
  \bibinfo{journal}{Phys.Rev.} \textbf{\bibinfo{volume}{{D63}}},
  \bibinfo{pages}{113011} (\bibinfo{year}{2000}{\natexlab{b}}).

\bibitem[{\citenamefont{{H.~Minakata and H.~Nunokawa}}(2001)}]{Minakata}
\bibinfo{author}{\bibnamefont{{H.~Minakata and H.~Nunokawa}}},
  \bibinfo{journal}{JHEP} \textbf{\bibinfo{volume}{10}}, \bibinfo{pages}{001}
  (\bibinfo{year}{2001}), \eprint{{\eprint{hep-ph/0108085}}}.

\bibitem[{\citenamefont{{G.~L.~Fogli and E.~Lisi}}(1996)}]{Fogli:1996pv}
\bibinfo{author}{\bibnamefont{{G.~L.~Fogli and E.~Lisi}}},
  \bibinfo{journal}{{Phys. Rev.}} \textbf{\bibinfo{volume}{{D54}}},
  \bibinfo{pages}{3667} (\bibinfo{year}{1996}),
  \bibinfo{note}{{\eprint{hep-ph/9604415}}}.

\bibitem[{\citenamefont{{W.~Winter}}({2004})}]{Winter}
\bibinfo{author}{\bibnamefont{{W.~Winter}}}, \bibinfo{journal}{{Phys. Rev.}}
  \textbf{\bibinfo{volume}{{D70}}}, \bibinfo{pages}{033006}
  (\bibinfo{year}{{2004}}), \eprint{{\eprint{hep-ph/0310307}}}.

\bibitem[{\citenamefont{{J.~Burguet-Castell, M.~B.~Gavela, J.~J.~Gomez-Cadenas,
  P.~Hernandez and O.~Mena}}(2001)}]{Burguet-Castell:2001ez}
\bibinfo{author}{\bibnamefont{{J.~Burguet-Castell, M.~B.~Gavela,
  J.~J.~Gomez-Cadenas, P.~Hernandez and O.~Mena}}}, \bibinfo{journal}{Nucl.
  Phys.} \textbf{\bibinfo{volume}{{B608}}}, \bibinfo{pages}{301}
  (\bibinfo{year}{2001}), \bibinfo{note}{{\eprint{hep-ph/0103258}}}.

\bibitem[{\citenamefont{{V.~Barger, D.~Marfatia and
  K.~Whisnant}}(2002)}]{Barger:2001yr}
\bibinfo{author}{\bibnamefont{{V.~Barger, D.~Marfatia and K.~Whisnant}}},
  \bibinfo{journal}{{Phys. Rev.}} \textbf{\bibinfo{volume}{{D65}}},
  \bibinfo{pages}{073023} (\bibinfo{year}{2002}),
  \bibinfo{note}{{\eprint{hep-ph/0112119}}}.

\bibitem[{\citenamefont{{J.~Burguet-Castell, M.~B.~Gavela, J.~J.~Gomez-Cadenas,
  P.~Hernandez and O.~Mena}}(2002)}]{Burguet-Castell:2002qx}
\bibinfo{author}{\bibnamefont{{J.~Burguet-Castell, M.~B.~Gavela,
  J.~J.~Gomez-Cadenas, P.~Hernandez and O.~Mena}}}, \bibinfo{journal}{Nucl.
  Phys.} \textbf{\bibinfo{volume}{{B646}}}, \bibinfo{pages}{301}
  (\bibinfo{year}{2002}), \bibinfo{note}{{\eprint{hep-ph/0207080}}}.

\bibitem[{\citenamefont{{A.~Donini, D.~Meloni and
  F.~Migliozzi}}(2003)}]{donini}
\bibinfo{author}{\bibnamefont{{A.~Donini, D.~Meloni and F.~Migliozzi}}},
  \bibinfo{journal}{J. Phys.} \textbf{\bibinfo{volume}{{G29}}},
  \bibinfo{pages}{1865} (\bibinfo{year}{2003}),
  \bibinfo{note}{{\eprint{hep-ph/0209240}}}.

\bibitem[{\citenamefont{{D.~Autiero \textit{et al.}}}(2004)}]{synergy1}
\bibinfo{author}{\bibnamefont{{D.~Autiero \textit{et al.}}}},
  \bibinfo{journal}{{Eur. Phys. J.}} \textbf{\bibinfo{volume}{{C33}}},
  \bibinfo{pages}{243} (\bibinfo{year}{2004}),
  \bibinfo{note}{{\eprint{hep-ph/0305185}}}.

\bibitem[{\citenamefont{{P.~Huber and W.~Winter}}(2003)}]{huber}
\bibinfo{author}{\bibnamefont{{P.~Huber and W.~Winter}}},
  \bibinfo{journal}{{Phys. Rev.}} \textbf{\bibinfo{volume}{D68}},
  \bibinfo{pages}{037301} (\bibinfo{year}{2003}),
  \eprint{{\eprint{hep-ph/0301257}}}.

\bibitem[{\citenamefont{{P.~Huber, M.~Lindner, and W.~Winter}}(2002)}]{lindner}
\bibinfo{author}{\bibnamefont{{P.~Huber, M.~Lindner, and W.~Winter}}},
  \bibinfo{journal}{Nucl. Phys.} \textbf{\bibinfo{volume}{{B645}}},
  \bibinfo{pages}{3} (\bibinfo{year}{2002}),
  \bibinfo{note}{{\eprint{hep-ph/0204352}}}.

\bibitem[{\citenamefont{{A.~Donini, D.~Meloni and
  P.~Migliozzi}}(2002)}]{donini2}
\bibinfo{author}{\bibnamefont{{A.~Donini, D.~Meloni and P.~Migliozzi}}},
  \bibinfo{journal}{{Nucl. Phys.}} \textbf{\bibinfo{volume}{B646}},
  \bibinfo{pages}{321} (\bibinfo{year}{2002}),
  \eprint{{\eprint{hep-ph/0206034}}}.

\bibitem[{\citenamefont{{A.~Bueno, M.~Campanelli, A.~Rubbia}}(2000)}]{camp00}
\bibinfo{author}{\bibnamefont{{A.~Bueno, M.~Campanelli, A.~Rubbia}}},
  \bibinfo{journal}{{Nucl. Phys.}} \textbf{\bibinfo{volume}{{B589}}},
  \bibinfo{pages}{577} (\bibinfo{year}{2000}),
  \bibinfo{note}{{\eprint{hep-ph/0005007}}}.

\bibitem[{\citenamefont{{P.~Huber, M.~Lindner and W.~Winter}}()}]{globes}
\bibinfo{author}{\bibnamefont{{P.~Huber, M.~Lindner and W.~Winter}}},
  \bibinfo{note}{{\tt http://www.ph.tum.de/$^\sim$globes}}.

\bibitem[{\citenamefont{{M.~Maltoni, T.~Schwetz and
  J.W.F.~Valle}}(2003)}]{Maltoni:2002aw}
\bibinfo{author}{\bibnamefont{{M.~Maltoni, T.~Schwetz and J.W.F.~Valle}}},
  \bibinfo{journal}{Phys. Rev.} \textbf{\bibinfo{volume}{{D67}}},
  \bibinfo{pages}{093003} (\bibinfo{year}{2003}).

\bibitem[{\citenamefont{{T.~Ohlsson and W.~Winter}}(2003)}]{Ohlsson}
\bibinfo{author}{\bibnamefont{{T.~Ohlsson and W.~Winter}}},
  \bibinfo{journal}{{Phys. Rev.}} \textbf{\bibinfo{volume}{{D68}}},
  \bibinfo{pages}{{073007}} (\bibinfo{year}{2003}),
  \eprint{{\eprint{hep-ph/0307178}}}.

\bibitem[{\citenamefont{{J.A.~Aguilar-Saavedra, G.C.~Branco, and
  F.R.~Joaquim}}(2004)}]{radiative}
\bibinfo{author}{\bibnamefont{{J.A.~Aguilar-Saavedra, G.C.~Branco, and
  F.R.~Joaquim}}}, \bibinfo{journal}{Phys. Rev.}
  \textbf{\bibinfo{volume}{D\textbf{69}}}, \bibinfo{pages}{{073004}}
  (\bibinfo{year}{2004}), \bibinfo{note}{{\eprint{hep-ph/0310305}}}.

\bibitem[{lin()}]{lindner-beta}
\bibinfo{note}{These calculations have been performed by the authors of
  Ref.~\cite{lindner}}.

\bibitem[{\citenamefont{{A.~Donini, E.~Fernandez-Martinez, P.~Migliozzi,
  S.~Rigolin and L.~Scotto Lavina}}(2004)}]{Donini:2004hu}
\bibinfo{author}{\bibnamefont{{A.~Donini, E.~Fernandez-Martinez, P.~Migliozzi,
  S.~Rigolin and L.~Scotto Lavina}}}, \emph{\bibinfo{title}{{Study of the
  eightfold degeneracy with a standard beta-beam and a super-beam facility}}}
  (\bibinfo{year}{2004}), \bibinfo{note}{{\eprint{hep-ph/0406132}}}.

\bibitem[{\citenamefont{{F.~Terranova, A.~Marotta, P.~Migliozzi and
  M.~Spinetti}}(2004)}]{Terranova:2004hu}
\bibinfo{author}{\bibnamefont{{F.~Terranova, A.~Marotta, P.~Migliozzi and
  M.~Spinetti}}}, \emph{\bibinfo{title}{{High energy beta beams without massive
  detectors}}} (\bibinfo{year}{2004}),
  \bibinfo{note}{{\eprint{hep-ph/0405081}}}.

\bibitem[{\citenamefont{Neuffer}(2003{\natexlab{a}})}]{adiab1}
\bibinfo{author}{\bibfnamefont{D.}~\bibnamefont{Neuffer}},
  \emph{\bibinfo{title}{Exploration of the high-frequency buncher concept}},
  \bibinfo{howpublished}{{MUC-NOTE-269}} (\bibinfo{year}{2003}{\natexlab{a}}),
  \bibinfo{note}{all MUC-NOTE papers are available from
  \url{http://www-mucool.fnal.gov/notes/noteSelMin.html}}.

\bibitem[{\citenamefont{Neuffer}(2003{\natexlab{b}})}]{adiab2}
\bibinfo{author}{\bibfnamefont{D.}~\bibnamefont{Neuffer}},
  \emph{\bibinfo{title}{Beam dynamics problems of the muon collaboration:
  $\nu$-factory and $\mu^+ - \mu^-$ colliders}},
  \bibinfo{howpublished}{MUC-NOTE-266} (\bibinfo{year}{2003}{\natexlab{b}}).

\bibitem[{\citenamefont{Neuffer}(2000)}]{adiab3}
\bibinfo{author}{\bibfnamefont{D.}~\bibnamefont{Neuffer}},
  \emph{\bibinfo{title}{High-frequency buncher and phase rotation for the muon
  source}}, \bibinfo{howpublished}{MUC-NOTE-181} (\bibinfo{year}{2000}).

\bibitem[{\citenamefont{{D.~Neuffer and A.~Van Ginneken}}(2001)}]{adiab4}
\bibinfo{author}{\bibnamefont{{D.~Neuffer and A.~Van Ginneken}}},
  \bibinfo{howpublished}{{Proceedings of the 2001 Particle Accelerator
  Conference}} (\bibinfo{year}{2001}),
  \bibinfo{note}{{\url{http://accelconf.web.cern.ch/Accel/Conf/p01/PAPERS/TPPH%
162.pdf}}}.

\bibitem[{\citenamefont{{A.~Van Gineeken}}(2001)}]{adiab5}
\bibinfo{author}{\bibnamefont{{A.~Van Gineeken}}}, \bibinfo{type}{Tech. Rep.},
  \bibinfo{institution}{Fermilab} (\bibinfo{year}{2001}),
  \bibinfo{note}{{MUC-NOTE-220}}.

\bibitem[{\citenamefont{Fernow}(1999)}]{icool}
\bibinfo{author}{\bibfnamefont{R.}~\bibnamefont{Fernow}}, in
  \emph{\bibinfo{booktitle}{{Proceedings of the 1999 Particle Accelerator
  Conference}}}, edited by \bibinfo{editor}{\bibnamefont{{A.~Luccio and
  W.~MacKay}}} (\bibinfo{year}{1999}), p. \bibinfo{pages}{3020},
  \bibinfo{note}{{latest version available at
  \url{http://pubweb.bnl.gov/people/fernow/icool/readme.html}.}}

\bibitem[{\citenamefont{McDonald}(2001)}]{target}
\bibinfo{author}{\bibfnamefont{K.}~\bibnamefont{McDonald}}, in
  \emph{\bibinfo{booktitle}{Proceedings of the 2001 Particle Accelerator
  Conference}} (\bibinfo{year}{2001}), p. \bibinfo{pages}{1583},
  \bibinfo{note}{{also, H.G. Kirk \textit{et al.}, ibid. p. 1535 and Chapter 3
  in ~\cite{fs2}. All Particle Accelerator Conference papers can be obtained
  from \url{http://accelconf.web.cern.ch/accelconf/}}}.

\bibitem[{\citenamefont{Chao and Tigner}(1999)}]{multipac}
\bibinfo{editor}{\bibfnamefont{A.}~\bibnamefont{Chao}} \bibnamefont{and}
  \bibinfo{editor}{\bibfnamefont{M.}~\bibnamefont{Tigner}}, eds.,
  \emph{\bibinfo{title}{Handbook of Accelerator Physics and Engineering}}
  (\bibinfo{publisher}{World Scientific}, \bibinfo{year}{1999}).

\bibitem[{\citenamefont{{D.~Neuffer}}(2004)}]{NeufferRef}
\bibinfo{author}{\bibnamefont{{D.~Neuffer}}} (\bibinfo{year}{2004}),
  \bibinfo{note}{{presentation at APS Study Workshop, ANL;
  \url{http://www.cap.bnl.gov/mumu/study2a/notes/neuffer.pdf}}}.

\bibitem[{\citenamefont{{R.~Johnson \textit{et al.}}}(2003)}]{MuonsIncRef}
\bibinfo{author}{\bibnamefont{{R.~Johnson \textit{et al.}}}}, in
  \emph{\bibinfo{booktitle}{Proceedings of Particle Accelerator Conference}}
  (\bibinfo{year}{2003}), p. \bibinfo{pages}{1792}, \bibinfo{note}{{ Muon Inc.,
  MUC-NOTE-247 (2002); R.~Johnson \textit{et al.}, AIP Conf. Proc.
  \textbf{671}, 328 (2003)}}.

\bibitem[{\citenamefont{{C.~Johnstone \textit{et al.}}}(2004)}]{JohnstoneRef}
\bibinfo{author}{\bibnamefont{{C.~Johnstone \textit{et al.}}}},
  \bibinfo{journal}{{Nucl. Instrum \& Meth.}} \textbf{\bibinfo{volume}{A519}},
  \bibinfo{pages}{472} (\bibinfo{year}{2004}).

\bibitem[{\citenamefont{{R.L.~Geng \textit{et al.}}}(2003)}]{pac03:1309}
\bibinfo{author}{\bibnamefont{{R.L.~Geng \textit{et al.}}}}, in
  \emph{\bibinfo{booktitle}{Proceedings of the 2003 Particle Accelerator
  Conference}}, edited by \bibinfo{editor}{\bibnamefont{{J.~Chew, P.~Lucas, and
  S.~Webber}}} (\bibinfo{year}{2003}), p. \bibinfo{pages}{1309}.

\bibitem[{\citenamefont{{M.~Ono \textit{et al.}}}(1999)}]{Ono99}
\bibinfo{author}{\bibnamefont{{M.~Ono \textit{et al.}}}},
  \emph{\bibinfo{title}{{Magnetic field effects on superconducting cavity}}},
  \bibinfo{howpublished}{{9$^{\text{th}}$ Workshop on RF Superconductivity}}
  (\bibinfo{year}{1999}), \bibinfo{note}{{Los Alamos, NM, 2000), Los Alamos
  National Laboratory report, LA-13782-C.}}

\bibitem[{\citenamefont{{J.S.~Berg, C.~Johnstone, and
  D.~Summers}}(2001)}]{pac01:3323}
\bibinfo{author}{\bibnamefont{{J.S.~Berg, C.~Johnstone, and D.~Summers}}}, in
  \emph{\bibinfo{booktitle}{Proceedings of the 2001 Particle Accelerator
  Conference}}, edited by \bibinfo{editor}{\bibnamefont{{P.~Lucas and
  S.~Webber}}} (\bibinfo{year}{2001}), p. \bibinfo{pages}{3323},
  \bibinfo{note}{{D.J.~Summers, Snowmass 2001, \eprint{hep-ex/0208010}}}.

\bibitem[{ope()}]{opera3d}
\emph{\bibinfo{title}{{Vector Fields Inc., computer program OPERA-3d}}}.

\bibitem[{\citenamefont{{S.~Koscielniak and
  C.~Johnstone}}(2003)}]{KoscielniakPAC03}
\bibinfo{author}{\bibnamefont{{S.~Koscielniak and C.~Johnstone}}}, in
  \emph{\bibinfo{booktitle}{Proceedings of the 2003 Particle Accelerator
  Conference}}, edited by \bibinfo{editor}{\bibnamefont{{J.~Chew, P.~Lucas, and
  S.~Webber}}} (\bibinfo{year}{2003}), p. \bibinfo{pages}{1831}.

\bibitem[{\citenamefont{{S.~Caspi and R.~Hafalia}}(2004)}]{CaspiRef}
\bibinfo{author}{\bibnamefont{{S.~Caspi and R.~Hafalia}}},
  \emph{\bibinfo{title}{{A Combined Function Superconducting Magnet for
  Fixed-Field Muon Acceleration in an Alternating Gradient Ring: First-Cut}}},
  \bibinfo{howpublished}{{LBNL Report SC-MAG-839}} (\bibinfo{year}{2004}).

\bibitem[{\citenamefont{{D.J.~Summers, J.S~Berg, A.A.~Garren,
  R.B.~Palmer}}(2003)}]{rcsync}
\bibinfo{author}{\bibnamefont{{D.J.~Summers, J.S~Berg, A.A.~Garren,
  R.B.~Palmer}}}, \bibinfo{journal}{{J. Phys.}}
  \textbf{\bibinfo{volume}{{G29}}}, \bibinfo{pages}{1727}
  (\bibinfo{year}{2003}).

\bibitem[{\citenamefont{{A.~Hassenein \textit{et al.}}}(2003)}]{TgtRef}
\bibinfo{author}{\bibnamefont{{A.~Hassenein \textit{et al.}}}},
  \bibinfo{journal}{{Nucl. Instrum. \& Meth.}}
  \textbf{\bibinfo{volume}{{A503}}}, \bibinfo{pages}{70}
  (\bibinfo{year}{2003}).

\bibitem[{\citenamefont{{T.~Gabriel and J.~Haines}}(2004)}]{ORNLsublimation}
\bibinfo{author}{\bibnamefont{{T.~Gabriel and J.~Haines}}}
  (\bibinfo{year}{2004}), \bibinfo{note}{{private communication}}.

\bibitem[{\citenamefont{{J.~Norem, \textit{et al.}}}(2003)}]{DarkCurrentnote}
\bibinfo{author}{\bibnamefont{{J.~Norem, \textit{et al.}}}},
  \bibinfo{journal}{{Phys. Rev. ST Accel. Beams}} \textbf{\bibinfo{volume}{6}},
  \bibinfo{pages}{072001} (\bibinfo{year}{2003}), \bibinfo{note}{{also
  MUC-NOTE-226 (2001)}}.

\bibitem[{\citenamefont{Edgecock}(2003)}]{MICEref}
\bibinfo{author}{\bibfnamefont{R.}~\bibnamefont{Edgecock}},
  \bibinfo{journal}{J. Phys.} \textbf{\bibinfo{volume}{G29}},
  \bibinfo{pages}{1601} (\bibinfo{year}{2003}), \bibinfo{note}{{see also the
  MICE Proposal, \url{ http://mice.iit.edu/mnp/MICE0021.pdf.}}}

\bibitem[{\citenamefont{{M.A.~Cummings \textit{et al.}}}(2003)}]{AbsorberRef}
\bibinfo{author}{\bibnamefont{{M.A.~Cummings \textit{et al.}}}},
  \bibinfo{journal}{J. Phys.} \textbf{\bibinfo{volume}{{G29}}},
  \bibinfo{pages}{1689} (\bibinfo{year}{2003}).

\bibitem[{\citenamefont{{K.~Makino and M.~Berz}}(1999)}]{COSYref}
\bibinfo{author}{\bibnamefont{{K.~Makino and M.~Berz}}},
  \bibinfo{journal}{Nucl. Instrum. \& Meth.} \textbf{\bibinfo{volume}{{A427}}},
  \bibinfo{pages}{338} (\bibinfo{year}{1999}).

\bibitem[{\citenamefont{{J.~Bouchez, M.~Lindroos, and
  M.~Mezzetto}}()}]{BouchezRef}
\bibinfo{author}{\bibnamefont{{J.~Bouchez, M.~Lindroos, and M.~Mezzetto}}},
  \emph{\bibinfo{title}{{Beta-Beams: present design and expected
  performance}}}, \bibinfo{howpublished}{{Proc. of NuFact03, in press;
  \eprint{hep-ex/0310059}}}.

\bibitem[{\citenamefont{{M.~Lindroos}}()}]{LindroosRef}
\bibinfo{author}{\bibnamefont{{M.~Lindroos}}}, \emph{\bibinfo{title}{{The
  Acceleration and Storage of Radioactive Ions for a Beta-Beam Facility}}},
  \bibinfo{howpublished}{{CERN-AB-2003; see
  \url{https://web3.cern.ch/beta-beam/References/RNB-beta-beam.pdf.}}}

\bibitem[{\citenamefont{{M. A.~Green, R.~Byrns,
  S.J.St.~Lorant}}(1992)}]{MAGref}
\bibinfo{author}{\bibnamefont{{M. A.~Green, R.~Byrns, S.J.St.~Lorant}}},
  \bibinfo{journal}{{Advances in Cryo. Eng.}} \textbf{\bibinfo{volume}{37}},
  \bibinfo{pages}{637} (\bibinfo{year}{1992}), \bibinfo{note}{{LBNL-30824; see
  also Phys. Rev. \textbf{D66}, Review of Particle Physics, 010001-217
  (2002)}}.

\bibitem[{\citenamefont{{J. S.~Berg, R.~Fernow, and R.
  B.~Palmer}}(2004)}]{InjExtRef}
\bibinfo{author}{\bibnamefont{{J. S.~Berg, R.~Fernow, and R. B.~Palmer}}},
  \bibinfo{howpublished}{{FFAG Workshop, Vancouver, Canada}}
  (\bibinfo{year}{2004}), \bibinfo{note}{{see
  \url{http://www.triumf.ca/ffag2004/.}}}

\end{thebibliography}
\end{document}